\documentclass[a4paper,11pt,oneside,openright]{thesis}
\usepackage{setspace}
\usepackage{fancyhdr}
\usepackage{afterpage}
\usepackage{wrapfig}
\usepackage{geometry}
\usepackage{rotating}

\usepackage{notoccite}
\usepackage{microtype}
\usepackage[T1]{fontenc}
\usepackage{hyphenat}
\usepackage{graphicx}
\usepackage{amssymb}
\usepackage{color}
\usepackage{ulem}
\usepackage{enumerate}
\usepackage{multirow}
\usepackage{float}
\usepackage{pdfpages}
\usepackage[numbers,sort&compress]{natbib}
\usepackage[colorlinks = true,
            linkcolor = blue,
            urlcolor  = blue,
            citecolor = blue,
	    pagebackref=true,
            anchorcolor = blue]{hyperref}
\usepackage{backref}
\renewcommand*{\backrefalt}[4]{%
  \ifcase #1 %
  \or
    \quad $\leftarrow$ \textit{page #2.}% Single citation
  \else
    \quad $\leftarrow$ \textit{pages #2.}% Multiple citations
  \fi
}

\usepackage{lscape}
\usepackage{pslatex}
\usepackage{amsmath}
\usepackage{subfigure}
\usepackage{subfloat}
\usepackage{float}
\usepackage{multirow}
\usepackage{tabularx}
\usepackage{booktabs}
\usepackage{epsfig}
\usepackage{graphicx}
\usepackage{epstopdf}
\usepackage{cleveref}
\crefrangeformat{eqnarray}{eqs. #3(#1)#4--#5(#2)#6}
\usepackage{hypernat}
\usepackage{adjustbox}
\usepackage{makecell}
\usepackage{tikz}
\usepackage{ucs}
\let\oldAA\AA
\renewcommand{\AA}{\text{\normalfont\oldAA}}
\usepackage{emptypage}
\usepackage{xparse}
\usepackage{ifthen}
\newcommand\blfootnote[1]{%
  \begingroup
  \renewcommand\thefootnote{}\footnote{#1}%
  \addtocounter{footnote}{-1}%
  \endgroup
}
\renewcommand*{\thefootnote}{\fnsymbol{footnote}}
\usepackage[hang,flushmargin,multiple, para]{footmisc}
\usepackage[titletoc]{appendix}

\usepackage[only-used = true, list-style = tabular, sort=true]{acro}
\usepackage[para]{threeparttable}
\DeclareAcronym{MD}{ 
    short = MD, 
    long  = Molecular Dynamics,
    tag = abbrev
}
\DeclareAcronym{NAG}{ 
    short = NAG, 
    long  = Nucleation And Growth,
    tag = abbrev
}
\DeclareAcronym{HD}{ 
    short = HD, 
    long  = Hydro-Dynamics,
    tag = abbrev
}
\DeclareAcronym{Al}{ 
    short = Al, 
    long  = Aluminum,
    tag = abbrev
}
\DeclareAcronym{Cu}{ 
    short = Cu, 
    long  = Copper,
    tag = abbrev
}
\DeclareAcronym{Ni}{ 
    short = Ni, 
    long  = Nickel,
    tag = abbrev
}
\DeclareAcronym{EAM}{ 
    short = EAM,
    long  = Embedded Atom Method,
    tag = abbrev
}
\DeclareAcronym{Nb}{ 
    short = Nb,
    long  = Niobium ,
    tag = abbrev
}
\DeclareAcronym{Mo}{ 
    short = Mo,
    long  = Molybdenum ,
    tag = abbrev
}
\DeclareAcronym{PSO}{ 
    short = PSO,
    long  = Particle Swarm Optimization ,
    tag = abbrev
}
\DeclareAcronym{FSV}{ 
    short = FSV,
    long  = Free Surface Velocity ,
    tag = abbrev
}
\DeclareAcronym{NPT}{ 
    short = NPT,
    long  = Number of atoms-Pressure-Temperature,
    tag = abbrev
}
\DeclareAcronym{STGB}{ 
    short = STGB,
    long  = Symmetric Tilt Grain Boundary ,
    tag = abbrev
}
\DeclareAcronym{STwGB}{ 
    short = STwGB,
    long  = Symmetric Twist Grain Boundary ,
    tag = abbrev
}
\DeclareAcronym{PT}{ 
    short = PT,
    long  = Phase Transition ,
    tag = abbrev
}
\DeclareAcronym{GB}{ 
    short = GB,
    long  = Grain Boundary ,
    tag = abbrev
}
\DeclareAcronym{AVGFE}{ 
    short = AVGFE,
    long  = Average Void Growth in Fluid Element ,
    tag = abbrev
}
\DeclareAcronym{RH}{ 
    short = RH,
    long  = Rankine-Hugoniot ,
    tag = abbrev
}
\DeclareAcronym{SC}{ 
    short = SC,
    long  = Single Crystal ,
    tag = abbrev
}
\DeclareAcronym{BiC}{ 
    short = BiC,
    long  = Bi-Crystal ,
    tag = abbrev
}
\DeclareAcronym{FP}{ 
    short = FP,
    long  = First Principles ,
    tag = abbrev
}
\DeclareAcronym{FCC}{ 
    short = FCC,
    long  = Face Centered Cubic ,
    tag = abbrev
}
\DeclareAcronym{LAMMPS}{ 
    short = LAMMPS,
    long  = Large-scale Atomic Molecular Massively Parallel Simulator ,
    tag = abbrev
}
\DeclareAcronym{EOS}{ 
    short = EOS,
    long  = Equation of State ,
    tag = abbrev
}
\DeclareAcronym{1D-HD}{ 
    short = 1D-HD,
    long  = One-Dimensional Hydro-Dynamics ,
    tag = abbrev
}
\DeclareAcronym{GBE}{ 
    short = GBE,
    long  = Grain Boundary Energy ,
    tag = abbrev
}
\DeclareAcronym{FD}{ 
    short = FD,
    long  = Finite Difference ,
    tag = abbrev
}
\DeclareAcronym{SG}{ 
    short = SG,
    long  = Steinberg-Guinan ,
    tag = abbrev
}
\DeclareAcronym{PBC}{ 
    short = PBC,
    long  = Periodic Boundary Condition ,
    tag = abbrev
}
\DeclareAcronym{NVT}{ 
    short = NVT,
    long  = Number of atoms-Volume-Temperature ,
    tag = abbrev
}
\DeclareAcronym{NVE}{ 
    short = NVE,
    long  = Number of atoms-Volume-Energy ,
    tag = abbrev
}
\DeclareAcronym{RMSD}{ 
    short = RMSD,
    long  = Root Mean Square Deviation ,
    tag = abbrev
}
\DeclareAcronym{WSSR}{ 
    short = WSSR,
    long  = Weighted Sum of Squared Residuals ,
    tag = abbrev
}
\DeclareAcronym{HEL}{ 
    short = HEL,
    long  = Hugoniot Elastic Limit ,
    tag = abbrev
}
\DeclareAcronym{CNA}{ 
    short = CNA,
    long  = Common Neighbour Analysis ,
    tag = abbrev
}
\DeclareAcronym{DXA}{ 
    short = DXA,
    long  = Dislocation eXtraction Algorithm ,
    tag = abbrev
}
\DeclareAcronym{OVITO}{ 
    short = OVITO,
    long  = Open Visualization Tool ,
    tag = abbrev
}
\DeclareAcronym{ASTGB}{ 
    short = ASTGB ,
    long  = ASymmetric Tilt Grain Boundary ,
    tag = abbrev
}
\DeclareAcronym{ASTwGB}{ 
    short = ASTwGB ,
    long  = ASymmetric Twist Grain Boundary ,
    tag = abbrev
}
\DeclareAcronym{BCC}{ 
    short = BCC ,
    long  = Body Centered Cubic ,
    tag = abbrev
}
\DeclareAcronym{HCP}{ 
    short = HCP ,
    long  = Hexagonal Close Packed ,
    tag = abbrev
}
\begin{document}
% Title, dedication, table of contents, etc ..
\frontmatter
\pagenumbering{roman}
\thispagestyle{empty}
\afterpage{
	\newgeometry{bottom=20mm}

\begin{center}
%\textbf{\Large MULTI-SCALE SIMULATIONS OF RADIATION DAMAGE
%  	IN IRON SINGLE CRYSTALS DUE TO IRRADIATION}\\[1.8cm]
%\textbf{\Large EFFECT OF PLANE-WAVE SHOCK PROPAGATION AND ASSOCIATED
%  	SPALL FRACTURE IN SINGLE AND BICRYSTAL SYSTEM OF ALUMINUM: 
%  	A MULTISCALE APPROACH} \\[1.5cm]
\textbf{\Large Multi-scale modeling of high strain rate deformation and spall fracture in poly-crystalline metals} \\ [1.5 cm]

by\\
\textbf{\large S MADHAVAN} M. Sc. \\
Bhabha Atomic Research Centre, Atchutapuram Mandal (Primary affiliation)\\
Department of Nuclear Physics, College of Science and Technology, Andhra University (Secondary affiliation)\\
Visakhapatnam, INDIA\\[1.5cm]
\vspace{1cm}
{\includegraphics[width=0.32
\textwidth]{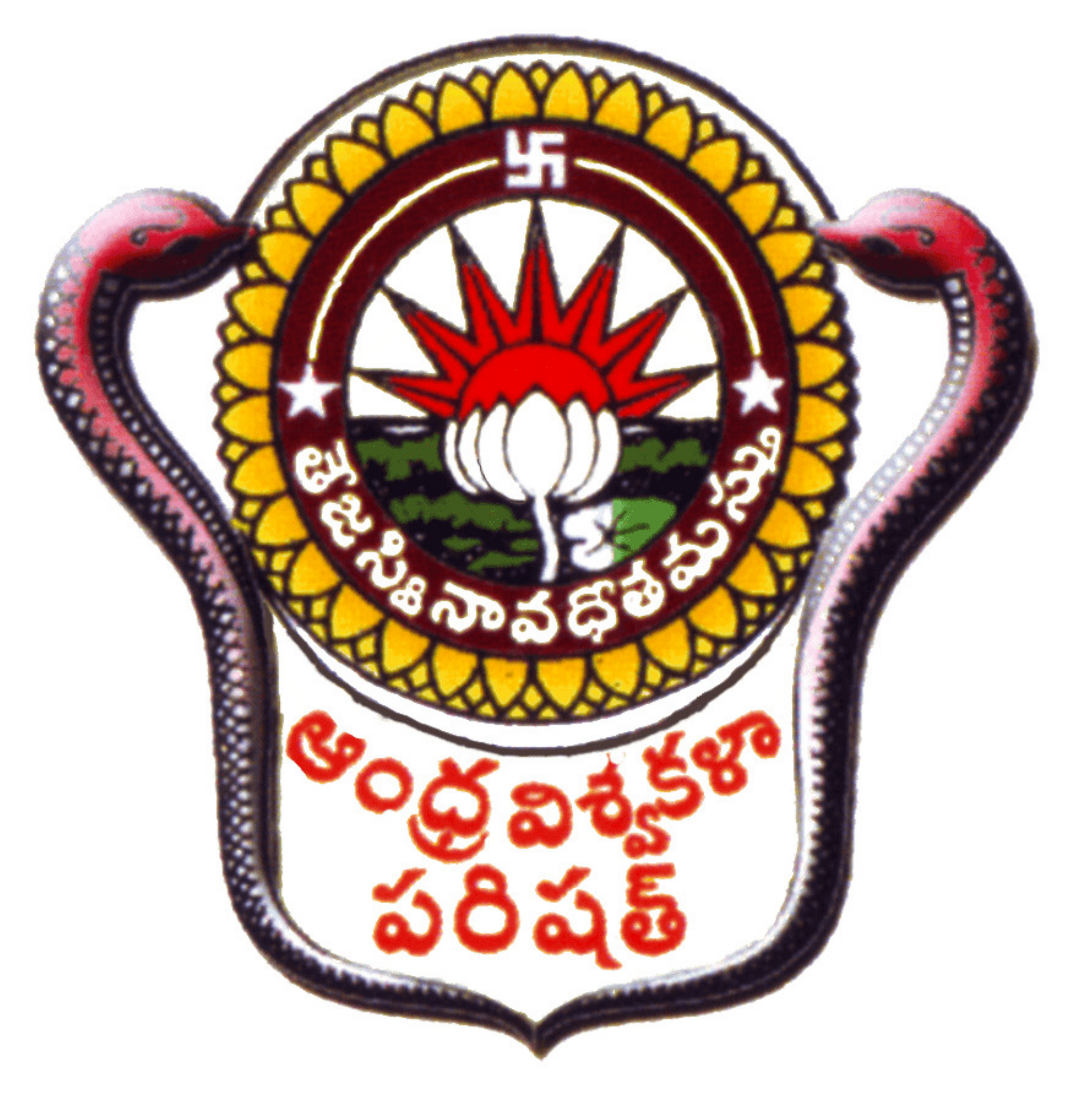}}\\[1.5cm]

\vspace{1.0cm}
THESIS SUBMITTED TO THE ANDHRA UNIVERSITY, VISAKHAPATNAM\\
IN PARTIAL FULFILMENT OF THE REQUIREMENTS \\FOR THE DEGREE OF\\
\textbf{\large DOCTOR OF PHILOSOPHY}\\
%IN \large{MATERIALS SCIENCE}\\
IN \large{NUCLEAR PHYSICS}\\
\textbf{April 2023}
\end{center} 

\begin{tikzpicture}[remember picture,overlay]
\draw[gray!60,line width=0.4pt]
    ([xshift=25mm,yshift=18mm]current page.south west)
    --
    ([xshift=75mm,yshift=18mm]current page.south west);

\node[
    anchor=south west,
    text=gray!70,
    font=\footnotesize,
    align=left
] at ([xshift=25mm,yshift=8mm]current page.south west)
	{A note on \copyright\ is available in page number~\pageref{copyrightNotePage}.};
\end{tikzpicture}

\clearchapter
\restoregeometry
}

  \thispagestyle{empty}

\vspace*{2.0in}

\noindent
%\hspace{1cm}{\Large {\bf {\it I dedicate this work to \textbf{my parents} and \\
\hspace{0.5cm}{\Large {\bf {\it I dedicate this work to my\\ \\
%\noindent		
%\hspace*{1cm}to my:\\
\hspace*{2cm}Mother: Smt. Vedhavalli, \\
\hspace*{2cm}Father: Vedha Re. Veera. Srinivasagopalachariyar,\\ \\
%\hspace*{2cm}Maternal \& Paternal Grandparents\\
%\vspace{0.5cm}
%\hspace*{4cm}my Siblings,\\
%\hspace*{4cm}Wife, Son, Daughter,\\
%\hspace*{4cm}nephews \& niece\\
%\hspace*{4cm}cousins\\
%\hspace*{4cm}Relatives \& Friends\\
%%\hspace*{4cm}and Close friends\\
%\vspace{2cm}
%\noindent
%\hspace{0.5cm}for the care, love, support and encouragement...
}}}

\clearchapter
 
\singlespacing
%\thispagestyle{empty}
%%%%%%%%%%%%%%
%% For print version: Self sign to be done.
%  \include{declaration}
%% For Soft Version : Self signed - scanned pdf file included, as in the following line
\includepdf[fitpaper,angle=0]{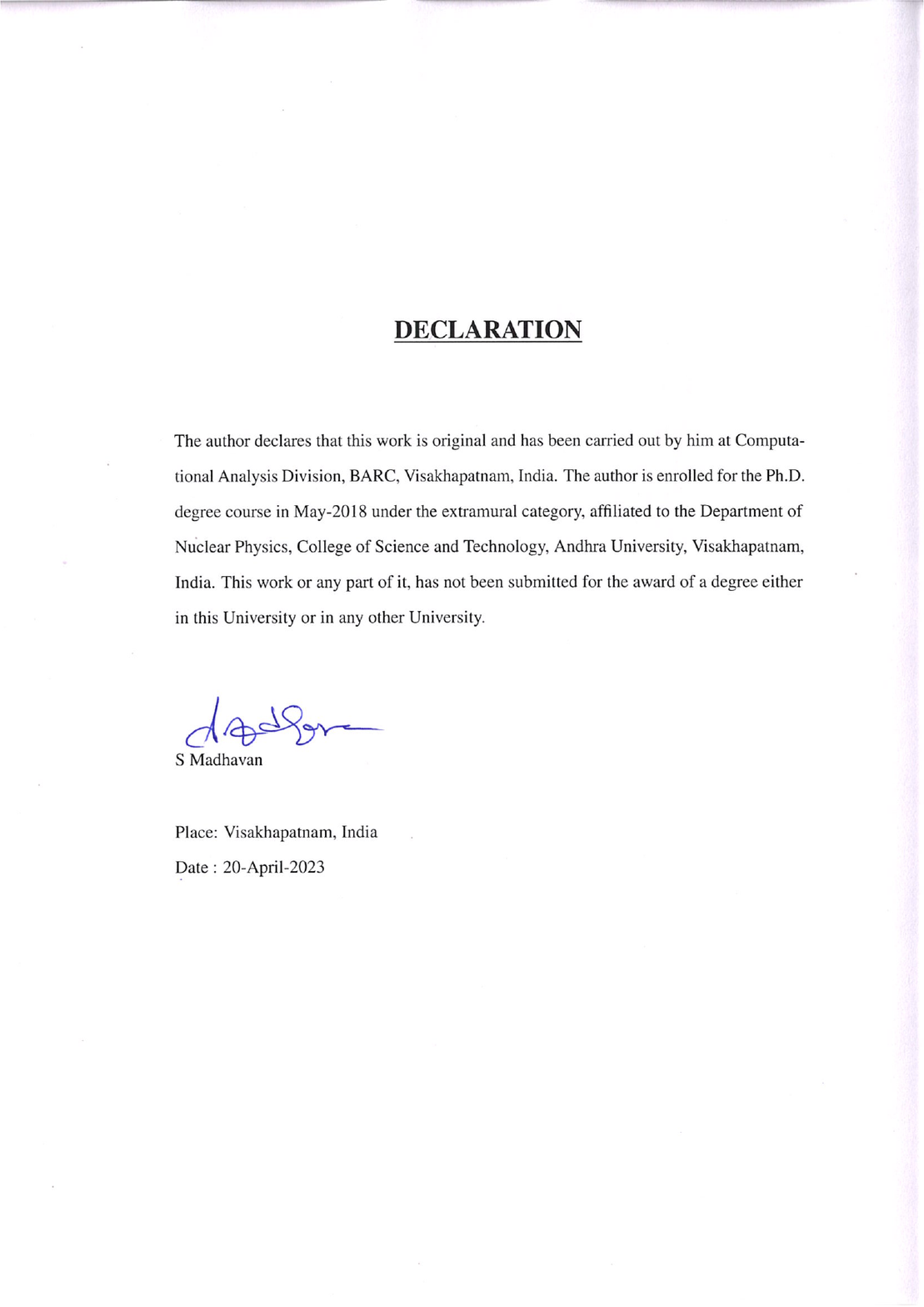}
%%%%%%%%%%%%%%
%\thispagestyle{empty}
%  \include{certificate}
\cleardoublepage
\includepdf[fitpaper,angle=0]{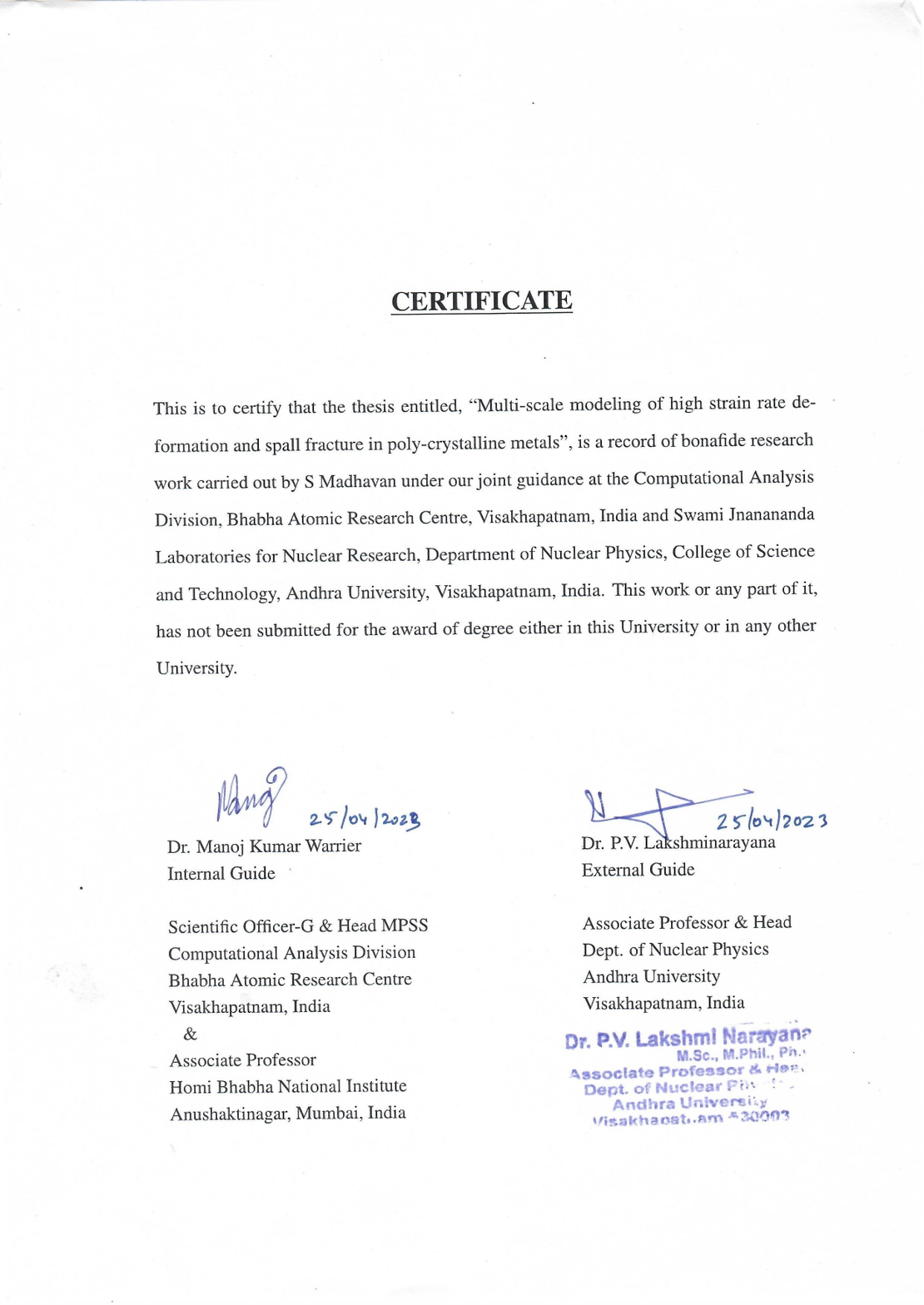}
\cleardoublepage
\includepdf[fitpaper,angle=1]{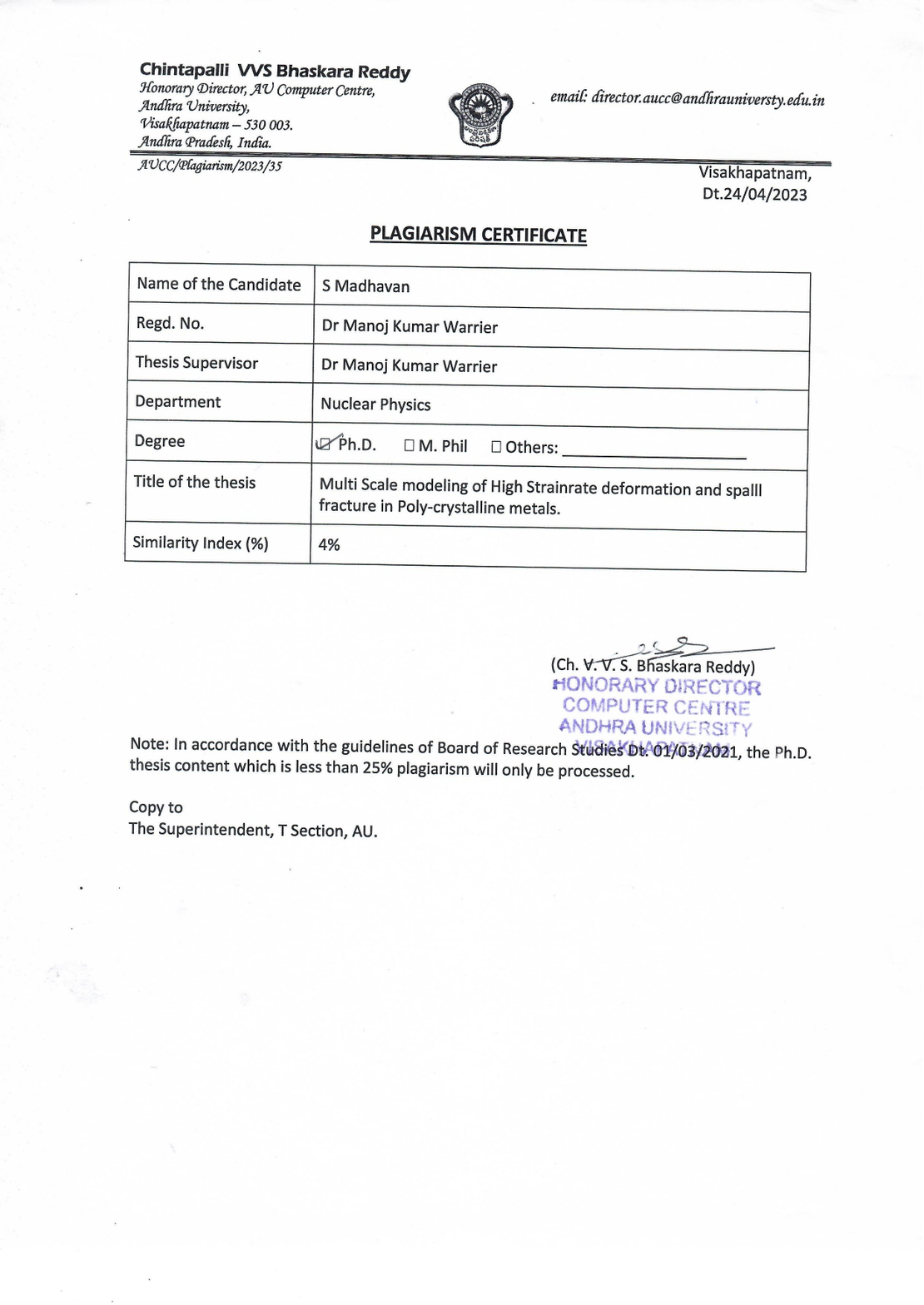}
\cleardoublepage
\cleardoublepage

%\setcounter{tocdepth}{3}
%\addcontentsline{toc}{chapter}{Contents}
%\pagenumbering{roman}\setcounter{page}{1}
%\tableofcontents

\pagenumbering{roman}
%\setcounter{page}{1}	% with twoside pages
%\addtocontents{toc}{\protect\afterpage\protect\cleardoublepage}
%\addtocontents{toc}{}
\addtocontents{toc}{\protect\thispagestyle{empty}}
\tableofcontents
\cleardoublepage
%%%%%%%%%%%%%
%%\setcounter{page}{6}	% with twoside page
\addtocontents{lof}{\protect\addcontentsline{toc}{chapter}{List of Figures}}
%% either above or the following
%\clearpage
%\phantomsection
%\addcontentsline{toc}{chapter}{\listfigurename}
%\listoffigures
%%%%%%%%%%%%%

%\addcontentsline{toc}{chapter}{List of Figures}
%\pagenumbering{roman}\setcounter{page}{7}
\listoffigures
%\cleardoublepage
%\clearchapter

%%\pagenumbering{roman}
%\setcounter{page}{9}	% with twoside page
\addtocontents{lot}{\protect\addcontentsline{toc}{chapter}{List of Tables}}
%%\pagenumbering{roman}
%\setcounter{page}{11}
%\addcontentsline{toc}{chapter}{List of Tables}
%\pagenumbering{roman}\setcounter{page}{11}
\listoftables
\cleardoublepage
\clearchapter

\markboth{}{}
\begin{spacing}{1}
\footnotesize
\twocolumn
%\printacronyms[include-classes=abbrev,name=List of Abbreviations/Acronyms, template=lot,display=all,sort=true]
%\addcontentsline{toc}{chapter}{List of Abbreviations/Acronyms}
%\chapter*{List of Abbreviations}
\phantomsection
\protect\addcontentsline{toc}{chapter}{List of Abbreviations}
%{\protect\addcontentsline{toc}{chapter}{List of Abbreviations}}
%\printacronyms[include-classes=abbrev,name={List of Abbreviations \& Acronyms}]
\printacronyms[name={List of Abbreviations \& Acronyms}]
%\printacronyms %% this worked for listing but not added to the TOC
%\listofacronyms
%\printacronyms{}
%\addtocontents{lot}{\protect\addcontentsline{toc}{chapter}{List of Abbreviations2}}
%\printglossary[type=\acronymtype,title={\centering\normalsize\normalfont\MakeUppercase{\acronymname}},toctitle=\acronymname
\onecolumn
\end{spacing}
\normalsize
%\printacronyms[include-classes=abbrev,name=Abbreviations]
%\printacronyms[include-classes=nomen,name=Nomenclature]
\cleardoublepage
\clearchapter

%\fancyhead[LE,RO]{\bfseries\thepage}	% \thepage -- page number on Left-side for even and Right-side for odd pages.
%\fancyhead[LO]{{\it \bfseries\rightmark}}	% \rightmark - section heading
%\fancyhead[RE]{ {\it \bfseries\leftmark}}	% \leftmark  - chapter heading
%\addtocontents{lot}{\protect\addcontentsline{toc}{chapter}{Abstract}}
%\include{pubs}
  %{\it \chapter{Detailed list of Publications \label{pubsFul}}}
{\it \chapter{List of Publications \label{pubsFul}}}
\normalsize
\vspace{-2.0 cm}
\blfootnote{All are Indexed in Scopus \& UGC-CARE apart from others as indicated by the symbols:\\}
\begin{spacing}{1}
%\section*{List of publications in refereed journals}
\section*{In refereed journals after peer-review:}
\begin{enumerate}[1]
\item \textbf{S. Madhavan}, H. Hemani, P.V. Lakshmi Narayana, V.R. Ikkurthi and M. Warrier.
``Effect of Symmetric Tilt and Twist Grain Boundaries on the Void Nucleation, Growth and
Spall in polycrystalline Al : Multiscale modelling''.
{\it Computational Materials Science}\footnote{SCIE\label{f1}} :
\textbf{211}:111543 (2022); Elsevier Publication. \url{https://doi.org/10.1016/j.commatsci.2022.111543}

\item \textbf{S. Madhavan}, V. Mishra, P.V. Lakshmi Narayana and M. Warrier.
``Dynamic response of single crystal Al, Cu \& Ni upon impact : MD and ab-initio calculations''.
{\it Journal of Dynamic Behavior of Materials}\footnote{ESCI\label{f5}} :
\textbf{9(1)}; pp 24-35; (2023), Springer publication; \url{https://doi.org/10.1007/s40870-022-00356-5}

\item \textbf{S. Madhavan}, P.V. Laxmi Narayana and M Warrier.
``Spall Fracture in Aluminum Bicrystals : Molecular Dynamics Study''.
{\it Materials Today Proc.}\footnote{CPCI\label{f4}} :
\textbf{87}; pp 164-169; (2023); Elsevier Publication.
\url{https://doi.org/10.1016/j.matpr.2023.03.281}

\item \textbf{S. Madhavan}, V. Mishra, P.V. Laxmi Narayana and M Warrier.
``On the Relationship between Shock and Particle Velocities in Single and Bicrystal
Systems of Aluminum: A Molecular Dynamics Study''.
{\it Materials Today Proc.}\textsuperscript{\ref{f4}} :
\textbf{87}; pp 204-209; (2023); Elsevier Publication.
\url{https://doi.org/10.1016/j.matpr.2023.04.543}

\section*{\noindent Manuscript published post-PhD award:}
\item \textbf{S. Madhavan}, V. R. Ikkurthi and M. Warrier.
``Strain Rate Effects on Shock-Induced Fracture in Symmetric Grain Boundaries of Aluminum: Insights from Molecular Dynamics'',
%%%%%%%%%%%%
{\it Physica Scripta}\textsuperscript{\ref{f1}}: 
~\textbf{101(32)}:325907 (2026); IOP Publication. \url{https://doi.org/10.1088/1402-4896/ae92df} \\
%%%%
%2026, Physica  Scripta, IOP publication: in press \url{https://doi.org/10.1088/1402-4896/ae92df}\\
%%%%%%%%%%%%
\begin{spacing}{0.9}
\item[] \textit{\small \textbf{Editorial Note:} Originally listed as ``To be submitted'' in the official Ph.D. thesis (Jan 2024), this manuscript was peer-reviewed, accepted, and published in Physica Scripta in 2026. The final work stands as an expanded, distinct body of analysis and simulation developed post-Ph.D. award. This article introduces a rigorous information-theoretic statistical model selection (AIC/BIC metrics) for strain-rate scaling, cross-validates shock-induced polymorphic phase transformations using parallel CNA and PTM filters, better links the dislocation loops to quantify defect kinetics, and incorporates a comprehensive 24 nm target-size sensitivity validation to eliminate system-size artifacts. Consequent to these substantial scientific extensions, the title proposed in the original thesis has been appropriately modified.}
\end{spacing}
\end{enumerate}

\pagebreak
%\section*{Work Presented at International/National Conferences}
\section*{Work Presented at Conference/Symposium: \label{copyrightNotePage}}
\begin{enumerate}[1]

\item \textbf{S. Madhavan}, V.R. Ikkurthi, P.V. Laxmi Narayana and M Warrier.
``Multiscale modelling to estimate spall parameters in metallic single crystals''.
{\it Proc. of ICONS-2018: International Conference on Structural Integrity},
IIT-Madras Chennai, India, Dec. 14-17, 2018. arXiv DOI: \url{https://doi.org/10.48550/arXiv.2112.00368}

\item \textbf{S. Madhavan}, P.V. Laxmi Narayana and M Warrier.
`` Spall Fracture in Aluminum Bicrystals : Molecular Dynamics Study''.
{\it IMPLAST-2022: The 13th International Symposium on Plasticity and Impact Mechanics},
IIT-Madras Chennai, India, Aug. 21-26, 2022

\item \textbf{S. Madhavan}, V. Mishra, P.V. Laxmi Narayana and M Warrier.
``On the Relationship between Shock and Particle Velocities in Single and Bicrystal
Systems of Aluminum: A Molecular Dynamics Study''. 
{\it IMPLAST-2022: The 13th International Symposium on Plasticity and Impact Mechanics},
IIT-Madras Chennai, India, Aug. 21-26, 2022

\item \textbf{S. Madhavan}, V. Mishra, P.V. Laxmi Narayana and M Warrier.
``Modelling of Shock-Hugoniot for single crystal Ni using Molecular Dynamics:
Comparison with ab-initio and Experimental results''.  
{\it Proc. of DAE SSPS-2022: 65th DAE Solid State Physics Symposium},
BARC-Mumbai, India, Dec. 15-19, 2021. pp 551-552.
\url{https://www.daessps.in/ssps2021/DAE\%20SSPS\%20Proceedings\%202021.pdf}
\end{enumerate}
\end{spacing}

\vspace{2cm}
\begin{spacing}{0.9}
\noindent\textit{\textbf{Note on \textcopyright :} The copyright of this thesis is vested in the author.
Portions of this thesis have previously appeared in peer-reviewed journal articles. Those publications are cited throughout the thesis. The published versions remain subject to the copyright policies of their respective publishers, where applicable.}
\end{spacing}

\vspace{2cm}
\begin{spacing}{0.9}
	\noindent\textit{\textbf{Additional Editorial Note (arXiv Edition):} This arXiv edition is a presentation-enhanced version of the official Ph.D. thesis deposited in the Shodhganga repository.\footnote{Official repository web links:\\\hspace{1cm}(i) \url{http://hdl.handle.net/10603/550672};\\(ii) \url{https://shodhganga.inflibnet.ac.in/handle/10603/550672}} The core scientific content, results, conclusions, and chapter organization remain entirely unchanged. This edition incorporates improvements solely to the document's presentation. These include a single, searchable PDF format with internal hyperlinks, optimized file and page (single line spaced) compression, and enhanced navigation features (such as bidirectional text-to-reference switching) for reader convenience. To ensure maximum readability, no watermarks have been applied to the main content.}
\end{spacing}

\clearchapter

%\fancypagestyle{plain}{}
\phantomsection
\protect\addcontentsline{toc}{chapter}{Acknowledgement}
\chapter*{Acknowledgments\label{ch:thanks}}
\begingroup
\normalsize
\begin{spacing}{1}

First and foremost, I thank the Almighty for giving me the strength, knowledge, 
ability and opportunity to undertake this research work. This work is dedicated
to my parents who were my first teachers. They made me learn a lot in life.
They never tried preaching but made me learn from their good deeds.\\

%This work would not have been possible without the support of my thesis supervisors
%(1) Internal: Dr. Manoj Kumar Warrier\footnote[1]{SO-G \& Head MPSS CAD BARC-Vizag,
%}\footnote[2]{Associate Prof., HBNI-Mumbai\\}
%and (2) External: Dr. P.V. Lakshmi Narayana\footnote[3]{Associate Prof. \& HOD, Dept. of Nuc. Phy., College
%of Science \& Tech., Andhra University-Vizag\\}.\\

It gives me immense pleasure to acknowledge my deep sense of gratitude and indebtedness to
%Dr. Manoj Kumar Warrier\footnote[1]{SO-G \& Head MPSS CAD BARC-Vizag,}\footnote[2]{Associate Prof., HBNI-Mumbai\\},
Dr. Manoj Kumar Warrier\footnote[5]{Scientific Officer G, Head MPS Section CAD BARC-Vizag \& Associate Professor, HBNI-Mumbai},
internal supervisor, for his guidance. He taught me molecular dynamics and how to use
LAMMPS, an open-source package used for this work. He has helped me in understanding the subject and deeply
enriched my knowledge through discussions and valuable comments. He has always been a source of
inspiration to me and I greatly benefited from the discussions with him academically and otherwise.
Though he is one of my senior colleagues, our discussions were mostly very casual due to his
humble attitude, which deserves great appreciation.\\

I thank Dr. P.V. Lakshmi Narayana\footnote[3]{\noindent Associate Prof. \& HOD, Dept. of Nuclear
Physics, College of Sci. \& Tech., Andhra University (AU)\\}, external supervisor for his
guidance. His encouraging tendency is highly appreciated. His help started even before my
enrollment/registration, continued during the entire course of my research period and went up to the
time of thesis submission.  He helped me in my academic and as well as administrative work. Interaction
with him made me realise his amicable friendliness and my appreciation for the same.\\

I thank my colleagues who co-authored the published papers related to this research work.
Their contributions added value to the research papers that emerged from this work.
I thank the doctoral committee members and the faculty members of the Nuclear Physics Department,
Andhra University (AU), for their valuable comments. I thank Dr. Srinivas Bukkuru for the
administrative help during my enrollment time. I thank the co-scholars, Mr. K. Surendra,
Mr. Sathyanarayana and Mr. Sivakumar for providing me with academic and administrative information
from time to time. I thank the non-teaching staff members of the Department of Nuclear Physics (AU)
for their administrative help.\\

%%I thank Shri. Unmesh D Malshe\footnote[4]{Head, CAD BARC-Vizag}\footnote[5]{Head, ED\&DD BARC-Mumbai}, Outstanding Scientist, for his
%I thank Shri. Unmesh D Malshe\footnote[4]{Head, CAD BARC-Vizag and Head, ED \& DD BARC-Mumbai},
%Outstanding Scientist, for his encouragement throughout this research work. I thank my colleague
I thank Shri. Unmesh D Malshe, Outstanding Scientist of BARC, Head CAD (Vizag), Head ED \& DD (Mumbai),
for his encouragement throughout this research work. I thank my colleague Shri. N. Sakthivel and his
team members for the provision of the High Power Computing System facility. I thank my colleagues of
CAD, BARC-Vizag for the valuable discussion related to this work.\\

I thank Dr. Shashank Chaturvedi, Distinguished Scientist \& Director IPR, Gandhinagar (Formerly: Head, CAD BARC-Vizag).
I leant shockwave physics from him. His deep knowledge and strong shock-speed actions
inspired me a lot. I thank my physics teachers Prof. Philominathan (at my graduation and M.Sc)
and Prof. Ulaganathan (School final years) who inspired me to develop an interest in the field of
physical science.\\

My pleasure to thank the contributors and the developers of LAMMPS, Linux-OS, GNU applications
and TeX/LaTeX for making computing and documenting a pleasure experience. I thank the online service
provider of `https://app.grammarly.com' for spelling and grammar checks. It helped to enhance the English
language of this thesis document.\\

Family support is a great compliment to pass through the tough phase.
I thank my wife Smt. Hemamalini Madhavan, my son Shri. Padmanabhan Madhavan and my daughter
Saubhagyavati Nithyasri Madhavan for their encouragement as well as their help in reading the proof of the thesis
and the research papers for publication. I thank my brothers, sisters and their spouses for their blessings,
appreciation and encouragement. My thanks to my nephews \& nieces for their cheerful encouragement. I thank my
cousins and my brother-in-law for their appreciation upon my publications online.
It is my pleasure to acknowledge the blessings of my parents-in-law. The encouragement received from one and all 
is acknowledged.\\

Last but not least, as a mark of respect for the language, Tamil \& its rich literature, couple of poems
%``{\it Naalaayira Divya-Prabandham}''
from the ``{\it Tamil literature}'' are quoted here.
\vspace{-2mm}
%\begin{centering}
%{\includegraphics[width=0.6
%        \textwidth]{Figures/sampleThanksGiving}}\\[1.5cm]
%\end{centering}
\begin{figure}[ht!]
\centering
%\subfigure {\includegraphics[width=0.7\textwidth,natwidth=610,natheight=642]{Prabandham/nalayiram.jpg}}
%\subfigure {\includegraphics[width=0.62\textwidth,natwidth=610,natheight=642]{Prabandham/unniththu.jpg}}
%\subfigure {\includegraphics[width=0.62\textwidth,natwidth=610,natheight=642]{Prabandham/meiyarkke.jpg}}
%{\includegraphics[width=0.75\textwidth,natwidth=610,natheight=642]{Prabandham/tamilLit3.jpeg}}
%{\includegraphics[viewport=0mm 0mm 72mm 72mm, scale=0.5]{Prabandham/tamilLit3.jpeg}}
%{\includegraphics[bb=650 600 3in 4in, scale=0.38]{Prabandham/tamilLit3.jpeg}} This works on 6th Oct 2023
{\includegraphics[bb=650 600 3in 4in, scale=0.45]{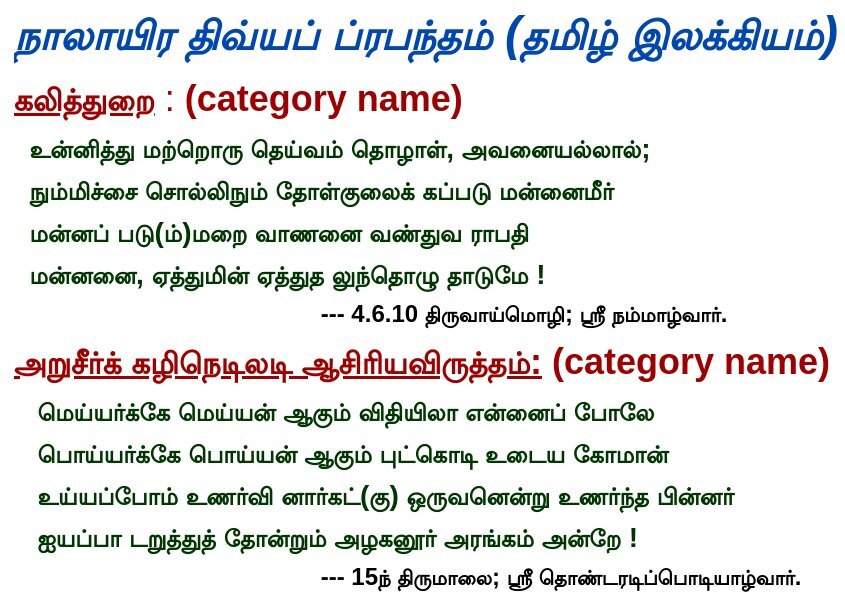}}
%{\includegraphics[width=11 cm]{Prabandham/tamilLit.jpg}}
\end{figure}

\end{spacing}
\endgroup
%\clearchapter
 
%{\protect\addcontentsline{toc}{chapter}{List of Abbreviations}}
\cleardoublepage
\clearchapter
\chapter{Abstract\label{ch:abstract}}
Shock-induced dynamic response of materials has been studied over decades by
high-velocity impact experiments, theoretical and computational methods.
In these studies, mostly macroscopic and microscopic physics were examined.
\ac{MD} methods are
useful to understand the underlying physics in the dynamic response of materials
at atomic scales. The emergence of MD methods along with 
the growing power of computing, facilitates the modeling and simulation of shock propagation
in materials at atomistic spatial and temporal scales. Associated dynamic responses
of materials due to high strain rates are being studied using MD simulations.
%Such calculations certainly help to gain in-depth knowledge of various physical
%phenomena involved in the dynamic failure of materials under shock loading.

Generally, a material fails under compressive as well as tensile stresses. 
Tensile pressure can be created by a flyer-target impact system in which the
one-dimensional strain shock state is generated. Upon impact, the compressive
shock wave is generated and it propagates into the flyer and the target
in opposite directions. These waves reach the respective free surface of the
flyer-target system and they reflect as the tensile/rarefaction wave.
These oppositely travelling tensile waves meet and create 
high tensile pressure. When the magnitude of the tensile pressure exceeds a critical value namely
'spall strength', the materials fail. Such a process is called 'spallation'.
The spall strength is a material dependant parameter. Over the years
experiments at the macroscopic levels have generated spallation data.
Details of these phenomena can be obtained from MD simulations.
Shock and spall data are available for conventional materials from experiments.
Suitable and the most accurate diagnostic facilities are critical in evaluating
the dynamic material properties. Setting up the experiments and deploying accurate
diagnostic methods are very sophisticated. Modeling and simulations are
relatively easier to set up for repeated studies. For the materials of interest
and new alloys, \ac{MD} methods can be used to understand the underlying
physics as well as to generate materials shock and spall data. From these aspects, this
research work is initiated and the following problems are studied:
(i) MD simulations to study the shock-dynamic response in single crystals.
(ii) Multiscale method that involves MD-based calculation at the atomic scales
and \ac{HD} based calculations at the macroscopic scales is developed.
This method is applied to single crystals for obtaining the
\ac{NAG} fracture-model parameters. Subsequently, the NAG parameter so obtained,
are used to obtain spall parameters of single crystals. (iii) Shock and high strain
rate phenomena are initiated and the material response is studied across
%individual bi-crystals of \ac{Al}. (iv) Multiscale method is applied to the
%individual bi-crystals of Al to calculate the spall parameters of an Al polycrystal.
individual bi-crystals of aluminum. (iv) Multiscale method is applied to the
individual bi-crystals of aluminum to calculate the spall parameters of an Al polycrystal.
The objective and the outcome of this research work are briefly given below.

\vspace{4mm}
{\bf (A) The Objective of problem 1}:
To initiate shock and study its propagation in single crystals of \ac{Al},
\ac{Cu} and \ac{Ni} using MD. To find the isothermal $P$-$V$ and the shock-Hugoniot
relation between $U_s$ and $U_p$ as well as between $P$ and $V$  at high strain rates
 $(\dot{\epsilon}>10^6\ s^{-1})$ and high pressures ($\simeq$ 100 GPa).
To add shock Hugoniot data at the higher compression region for the nickel.

\vspace{2mm}
{\bf Outcome from this work}:
For three single crystals (Cu, Al \& Ni), the parameters relating
$U_s$ to $U_p$ are obtained, when they are subjected to impact shock loading.
$P$-$V$ relationship under isothermal and dynamic shock conditions is
calculated from MD simulations. They are compared with published experimental results.
The relative errors between the MD simulation results and experiments
are mostly \textless 6\%. Only in the case of the shock Hugoniot parameter $S$
for Ni, the relative error in the MD simulation is 16\%. Stiffening the pair interaction
component of the Foiles' \ac{EAM} potential did not change the results.
%The EAM potentials by Foiles for Ni used in this
%study yields the closest match with experiments, compared to four other inter-atomic
%potentials independently used by Choi, Jarmakani and Liu hai. The
EAM potentials used in this study for Cu and Al can be readily used for MD simulations
at high strain rates, $10^7$-$10^8\ s^{-1}$. % that corresponds to high pressures.
For Ni, a trend of deviation is observed at higher impact velocity where
very few experimental data are available for comparison. Shock Hugoniot
values obtained using Foiles' EAM potential used in this study seem to
show the least deviation from experimental results,
in comparison to six other EAM potentials of Ni.

\vspace{4mm}
{\bf (B) The Objective of problem 2}:
To set up multiscale modeling and validate the spall strength calculated
for Cu, \ac{Nb} and \ac{Mo}. To improve the accuracy of the results for Nb and Mo in comparison
to the earlier works. To obtain the spall strength of single-crystal
aluminum using multiscale modelling

\vspace{2mm}
{\bf Outcome from this work}:
MD simulations of tri-axial tensile deformation are performed and time-varying void volume
fraction and pressure data are obtained. This method simulated the void nucleation,
void growth and coalescence at a constant strain rate ($\dot{\epsilon}$ = 5$\times{10}^9\ s^{-1}$).
The \ac{PSO} method developed for this work was found to be efficient and fast in convergence. The NAG fracture
parameters obtained for the single crystals are in the right range.
For copper, niobium and molybdenum, the spall strength and spall thickness are estimated using
the multiscale model. The \ac{FSV} profiles obtained from this work match the experiments better
than other MD results published in the literature. For aluminum, NAG parameters are obtained
by PSO fitting and the temporal FSV profile is obtained from hydrodynamic
calculation. Spall strength estimated from FSV for `Al', is found to be matching well with
experiments. Calculated spall strength was found to deviate around 8\% in comparison to the
experiments. The multiscale method is found to meet the objective of spall strength calculation
for the single crystals (Cu, Nb, Mo \& Al). The period of oscillation of the FSV profile was found
to deviate from the experiments. Further investigations are required to resolve this deviation.

\vspace{4mm}
{\bf (C) The Objective of problem 3}:
To set up several bi-crystal arrangements in MD. To calculate shock and spall
parameters from individual grain boundaries, using the temporal FSV
profile. To calculate corresponding peak tensile stress. To compare peak
tensile stress and FSV method in predicting the spall strength

\vspace{2mm}
{\bf Outcome from this work}:
Ten Al bicrystals with different tilt and twist angles have been created and equilibrated in the MD simulations
with constant \ac{NPT}.
The target and the flyer sizes were chosen so that the tensile spall occurs at/near the grain boundary.
Flyer velocities in the range of 1 to 3.5 km/s are used in this work. From the temporal FSV profile obtained
from the MD simulations, the following are the observation:
No spall occurs at $U_p$ = 0.5 km/s for all \ac{STGB} and \ac{STwGB} cases.
For $U_p$ = 0.75 km/s, no spall is observed for higher misorientation angle STGBs
($53.2^o$ and $77.3^o$). For all the cases STwGB at this $U_p$, spall is observed.
Overall, the values of the spall strength obtained for the individual STGB and STwGB
cases are higher than that of the polycrystalline Al.
The calculated strain rates from the MD simulations are in the range $10^{10}$-$10^{11}$ $s^{-1}$.
The pattern seen in the spall strength calculated from MD simulations at these strain
rates is consistent with experimental results of polycrystalline Al at lower
strain rates.
Coefficients of two different functional forms that relate spall strength with strain rate are estimated
by fitting the ${\sigma_{sp}}(\dot\epsilon)$ data of polycrystal Al from experiments
and MD simulation of Al bi-crystal.
\ac{PT} occurs near \ac{GB} upon shock propagation and it influences the free
surface velocity profile. At higher particle velocities ($U_p$) due to phase transition,
the spall strength gets reduced.
If dislocations occur due to shock-induced grain boundary plasticity, pull-back
is prolonged and the time to spall is delayed.
This leads to higher spall strength value (eg. STwGB, $14.2^o$ for $U_p$ at 1 km/s).
Spall strength obtained from FSV was found to be lower than the peak tensile strength.
Values of spall strength and the peak tensile strength are not the same for Al bicrystal
systems.

\vspace{4mm}
{\bf (D) The Objective of problem 4}:
To extend the multiscale model from single crystal to bi-crystal.
To obtain NAG fracture parameters for various symmetric grain boundaries of aluminum.
To devise a methodology that accounts for the individual effects of various grain boundaries on
the NAG fracture parameters. The primary objective here
is to obtain the spall strength of polycrystalline aluminum from MD simulations of
various bi-crystal systems.

\vspace{2mm}
{\bf Outcome from this work}:
A multi-scale model to obtain spall parameters at high strain rates for
polycrystalline materials using MD simulations at the atomistic scales and
hydrodynamic simulations at the macro-scales has been developed.
In a bi-crystal or a polycrystal, the grain boundaries are the weakest
location and voids can nucleate there when subjected to tensile deformation.
The grain boundaries in the MD simulations are mimicked using eleven symmetric
tilt (STGB) and twelve symmetric twist (STwGB) grain boundaries. The nucleation,
growth and coalescence of voids due to isotropic expansion were monitored as a
function of pressure in the MD simulation. In all the cases the voids nucleated
at the grain boundaries as expected since the bonds are weakest at the grain
boundaries. Using particle swarm optimization to fit a nucleation and growth model of voids
with the MD results the NAG parameters were obtained for each of the 23
bi-crystals. An \ac{AVGFE} model is proposed to
simulate the effect of the various grain boundaries in a simulation cell of the
hydrodynamic simulations. Using these NAG parameters, one-dimensional
hydrodynamic simulations are performed for flyer impact simulations at three
different flyer velocities, 518, 1588 and 2275 m/s reported in
experiments for pure Al. The free surface velocity curve vs time calculated from
the multiscale model is found to match the experiments. The spall strength
calculated from the multiscale simulations for three different impact
velocities shows a good match with experiments.  The relative deviations are
well within 2.15\%, 2.9\% and 6.0\% respectively for the three impact velocities.
The spall thicknesses were found to deviate within 2.4\% to 4\%, in comparison to
the experiments.

\clearchapter

%\fancyhead[LO]{\underline {\it \bfseries\rightmark}}	%  LO top_left of the page: \rightmark - section heading

%\include{notations}

% Various chapter tex files are included here
\mainmatter

% Intro_madhavan : Chapter 1. Introduction, Research already done, missing issues, motivation, objective, formulation of the problem, way to execute.
\pagenumbering{arabic}
  \chapter{Introduction\label{chapter_Intro}}
%Shock-induced dynamic response of materials
%% above line replace by the next line: Modified on 6th Nov 2023 to avoid same statements w.r.to abstract 1st line
The dynamic response of materials to high velocity impact
has been studied for
years~\cite{grahamBook, meyersBook}. Experiments on high-velocity impact, shock
and fracture studies in a material are necessary to ascertain its quality and strength.
Over the years, shock data of specific metals are generated from experiments
and are available in the literature~\cite{JC_fracture, fortov1991,
spall_moshe1998, spall_moshe2000, Kanel2001JAPSingleCrysFoilAl}.
Experimental, theoretical and computational studies on the dynamic shock propagation
and associated fracture studies are one among the leading research topics for over five 
decades~\cite{seamanCompmodelDuctBrit_JAP,
curran_nag1, johnsonPulseDurationSpallFrac_JAP, ikkurthi1_nag1, RINGDALENVATNE2013306,
IJF_AlBucklingSM}. Such calculation started even before the emergence of
modern computational hardware and software. These were mostly micro-macroscopic in nature.
Material fails under compressive as well as tensile stresses.
Spallation is a process in a material in which failure occurs due to tensile stress.
Most of the experimental works on material failure under tensile stress are primarily
for conventional materials such as copper, aluminum, niobium, molybdenum, nickel
and some alloys of steel~\cite{Anton_spalldata}.

Generation of material spall data from experiments involves a lot of effort as this
needs to be repeated for every new material of interest. Suitable and
the most accurate diagnostic facilities are critical in evaluating the
dynamic material properties.
Setting up the experiments and deploying accurate diagnostic methods are
very sophisticated. The underlying physical phenomena at the atomic scales are
difficult to measure experimentally. For eg. the atoms when disturbed by an external
source, come back to their original positions when the disturbance ceases and is said to be
in the elastic regime. But, under high-strain rate impact, the lattice will undergo
an irreversible state, that includes permanent defects, melting, phase change,
resolidification and even solid structural phase change. These changes substantially
influence the thermodynamic states of the materials. The availability of experimental data
concerning the nucleation of dislocation, crystal defects, nucleation of voids and their
growth under dynamic shock propagation conditions is limited. Two approaches are
mentioned by Lorenzana et al~\cite{shockExptChallanges} for getting detailed insight
involved in the associated physical phenomenon. One of the approaches emphasizes
microstructural analysis. But direct probing of the samples under dynamic loading is
restricted for practical reasons.  In the second approach, the bulk response can be
recorded using fast diagnostics methods such as velocimetry, VISAR~\cite{VISAR1974};
laser Doppler velocimeters, ORVIS~\cite{ORVIS1983}; and internal stress gauges~\cite{duff1957},
in a real-time scales. These approaches are proved to be valuable in understanding the
underlying physical phenomena to some extent. However such continuum-level methodologies
have the limitation that these cannot provide lattice-level processes under dynamic
conditions~\cite{shockExptChallanges}.

The emergence of Molecular Dynamic (MD) methods along with growing power of computing,
facilitates modeling and simulation of shock propagation in materials
at atomistic spatial and temporal scales. Associated dynamic response
of materials due to high strain rates are being studied using MD
simulations~\cite{srinivasan_MdSpallSingleNiXl, srinivasan2007,
rawat1, JIANG2022114474, Influence_Fensin, ZHU2022110923}.
Such calculations certainly help to gain in-depth knowledge
of various physical phenomena involved in the dynamic failure of materials
under shock loading.

Shock loading in metals is a high strain rate ($\dot{\epsilon}$)
phenomena~\cite{meyersBook, hemp_wilkins}. The plate impact system produces
$\dot{\epsilon}$ in the range $10^5$-$10^6\ s^{-1}$~\cite{fortov1991}. Higher strain
rates ($\sim10^7\ s^{-1}$) are achieved by nanosecond laser pulses~\cite{spall_moshe1998},
while ultra-high strain rates ($\sim10^8\ s^{-1}$) can be obtained by femto-sec laser
pulses~\cite{spall_moshe2000}. High-power laser-pulsed experiments have reported
extremely high strain rates over $10^{10}\ s^{-1}$~\cite{Murphy_2010}.

In this research work, the flyer-target impact system is simulated to understand the
physical phenomena involved in dynamic impact-shock propagation in materials.
Molecular dynamics (MD) methods are used at atomic scales of length and time.
Associated phenomena in macroscopic scales are also examined using Hydro-Dynamic
(HD) methods in sequence.
%Results from ab-initio/first principle (FP) methods and
%experiments available in the literature are used for validating shock
%parameters obtained from MD and HD.

%\section{Flyer-target system to generate dynamic shock}
\section{Dynamic impact shock by a Flyer-target system~\label{sec1.1}}
The Dynamic response of metals can be examined by the plate-target impact system.
Flyer-plate impact system is relatively easier to setup for shock and impact studies.
Such systems are widely used in experiments~\cite{meyersBook, hemp_wilkins} and
modeling~\cite{ikkurthi1_nag1, rawat2_hydro, ramana2_hydro, icons2018_madhavan,
madhavan_spall_multiscale, UsUp_Madhavan2022}.
With a suitable arrangement, flyer plate impact system can generate
uniaxial strain by launching one-dimensional plane-wave shocks.

%\label{sec-1.1}
\begin{figure*}[h]
\centering
\includegraphics[width=4.3in]{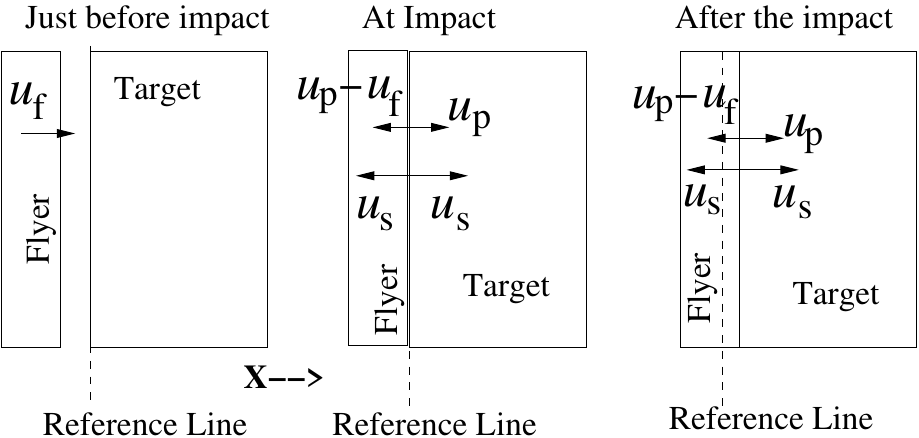}
\caption[Schematic: Flyer-Target Impact system]{Schematic of flyer-target impact
system to generate a planar shock wave. The shock wave is formed in the YZ plane and
propagates along the X-direction. Here, the YZ plane is perpendicular to the plane
of the paper. This figure is made from the knowledge gained from
Meyers~\cite{meyersBook}.  (Fig:Ref.~\cite{UsUp_Madhavan2022}).}
\label{Chap1_2dview}
\end{figure*}
Figure~\ref{Chap1_2dview} shows a schematic view of the flyer-target impact system.
It shows the scene just before the impact, at the time of impact and the moment
just after the impact. The flyer with a velocity $U_f$ upon impacting a target
generates shock waves. These waves travel through the flyer
as well as the target materials but in opposite directions. At the interface of
the flyer and the target, the thermodynamic pressure and associated particle velocity are
continuous. Hence $P_{flyer}=P_{target}$. The atoms at the flyer-target interface
are set to motion with a velocity whose magnitude is $U_{p}$, but they start
moving in opposite directions from the plane of impact. The Shock
impedance of a material is the product of its
normal density ($\rho$)
%\ac{rho}
and
the
shock velocity ($U_s$).
%\ac{Us}.
If the flyer and target are made of the same material,
the shock impedance matches at the interface. This makes the kinetic energy to be
distributed equally to the neighbouring atoms on either side of the impact surface.
In such a case the atoms on both sides of the interface are set to motion with a
velocity whose magnitude will be equal to $U_f/2$~\cite{meyersBook}. When these two
shock waves reach the respective ends of the material they return as a rarefaction wave,
which is also called a tensile or release wave. When these two tensile plane waves meet,
the tensile stress of the materials accumulates around that region.
There exists a critical value for the stress beyond which the material fails. This
failure process is called `spallation'. The value of the stress at which `spallation' occurs
is called spall strength and this is a material dependant parameter.
Dynamic response of the material at a high strain rate and associated spall effects draw
the attention in materials science because it helps to identify the suitability and
sustainability of materials in engineering design. Modelling of mechanical behavior at
atomic length and time scales is of great interest in the context of material
and metallurgical research. As mentioned above, modeling and simulations at the atomic scales
provide lattice-level understanding of the physical processes, which is not practically
possible from experimental studies.

Shock propagation in materials follows mass, momentum and energy conservation.
Mathematically a shock-state is expressed by three relations known as %Rankine-Huginiot (RH)
\ac{RH} equations.
These are derived from basic conservation principles~\cite{meyersBook}.
The parameters that appear in RH shock equations can be
calculated using MD simulations of a plate impact system with different impact
velocities. The relation between shock velocity ($U_s$) and particle velocity ($U_p$)
is mostly linear in nature for metals when phase transition does not occur. These parameters
are material dependant.  These can be obtained by fitting experimental or calculated
shock data. More details on this are given later in chapter~\ref{chapter_SCImpactShock}.

% Introduce literature survey to make a Brief note on the present scenario in this field \\
\section
[Shock and Spall in materials]
{Shock and Spall in materials: Motivation from previous works and Objective of present work
~\label{sec1.2}}
Even from 1970, impact shock experiments were being conducted with sophisticated
diagnostic methods. Also, there were hand calculations based on known conservation
principles and available theoretical knowledge. During the early years
of the present century the availability of high-power computing facilities made it possible 
to study various underlying physical phenomena of shock propagation even at atomic 
length and times scales.

\subsection{Literature survey in MD shock propagation and motivation for this research study
\label{sec1.2.1}}
Various groups have performed MD high strain rate shock
studies~\cite{brad, bringa, srinivasan2007} in single crystals.
Jarmakani {\it et al.}~\cite{Ni_md_Jarmakani},
Liu Hai {\it et al.}~\cite{Ni_md_Liu_Hai} and
Choi {\it et al.}~\cite{Ni_md_Choi} have independently shown the
variation of $U_s-U_p$ in Nickel (Ni), with different Embedded Atom Method (EAM) potentials.
When these simulation results are compared with experiments,
different levels of deviation are found. The availability of experimental
and MD simulations shock data at higher compression for nickel is limited.
Even among the available results, substantial levels of deviation are found when MD
results~\cite{Ni_md_Jarmakani, Ni_md_Liu_Hai, Ni_md_Choi} are compared with
experiments. Thus shock studies in single crystal Ni at higher compression
(over 150 GPa) require a systematic study.

Results from high strain rate MD simulations of tensile deformation are used
to fit void Nucleation And Growth (NAG) fracture model~\cite{SUGANDHI2015113}
by minimization of pressure-void volume relation. The NAG parameters obtained
from such methods are used in hydrodynamic simulations to predict spall strength and
spall thickness of single crystals at macroscales~\cite{rawat2_hydro, ramana2_hydro}.
For niobium calculated spall strength was found to be deviated by 44\% when
the potential is not stiffened and the stiffened potential yielded 11\% deviation
when compared with experiments~\cite{ramana2_hydro}.

In nucleation dislocation, Grain Boundary (GB) structure plays an important
role because GB and interface atoms assist plastic deformation by providing
varying degrees of mobility to the atoms~\cite{ADLAKHA2014345}. As shock propagation across
individual bicrystal grain boundary induces grain boundary plasticity,
it is an important aspect and is being studied using MD
simulations~\cite{xueyang_GBonShockInduPhTransinFeBiCryst, LONG2020109411}.
Crack growth and associated fracture involving Grain Boundaries (GB) are also
studied using MD simulations by various groups~\cite{LIN201294,Luo_bicrys_asymm_sigma3_110TiltGB,
Han_spall_shockCuBiC,Fensin_GBonplasticdefor,ADLAKHA2014345,ZHOU2014116,Barrles_MSEA}.

Among various published spall simulations using MD methods, peak tensile stress (${\sigma}_{T}$)
is referred to as the spall strength for single crystal Aluminum (Al)~\cite{JIANG2022114474}, bicrystal
Copper (Cu) in tilt~\cite{Influence_Fensin} and twist~\cite{LONG2020109411} systems. Only a few researchers
have calculated the spall strength ($\sigma_{sp}$) from Free Surface
Velocity (FSV)~\cite{srinivasan2007, ZHU2022110923, LIN201294}.
For nano Cu/Ni bimetallic multi-layer case~\cite{ZHU2022110923}, the
spall-strength obtained from FSV was almost equal to the values of
peak tensile strength under different loading conditions.
But for single crystal Ni~\cite{srinivasan2007} and Cu bicrystal with $\Sigma$3
asymmetric tilt GB~\cite{LIN201294} at various impact loading, the spall strength
obtained from FSV are higher than peak tensile stress~\cite{LIN201294}.
There is a need to confirm if the peak tensile stress can be referred to as the spall strength
or not. 

%High strain rate phenomena introduce dislocations. When the dislocations
%propagate, they create plastic deformation and make void-nucleation.
%create plastic deformation and void to nucleat.
High strain rate phenomena introduce dislocations.
%When the dislocations
%propagate, they create plastic deformation and make void nucleation.
Tensile deformation breaks bonds leading to void formation.
When these voids coalesce
and grow, material failure is initiated
% which is the fundamental mechanism in
leading to
dynamic fracture. It is important to understand these phenomena in fracture modelling
of strained materials~\cite{wwpang2014}. Grain boundaries play an important
role in dislocation propagation, void nucleation and growth. In this context,
it is noteworthy to mention that the MD simulation domains were generated based on the voronoi method,
to study the tensile deformation of nano-crystallites of Al by
Swygenhoven {\it et al.}~\cite{HVSwygenhoven2005}. Spearot {\it et al.}
simulated uni-axial tension in a set of Symmetric Tilt Grain Boundaries
(STGB), in which they found that full dislocation loops are nucleated
near the grain boundary~\cite{DESpearot2005}. MD impact simulations are
reported by Dremov {\it et al.}~\cite{VDremov_2006} in nanocrystals and
polycrystals. It has been shown that the direction of shock loading
determines the spall planes in monocrystals, while the void nucleation
occurs along the grain boundaries for polycrystalline materials. Long
{\it et al}~\cite{LONG2020109411} have carried out flyer impact on target
simulations using MD in bi-crystal Cu. The size of the flyer and target
was chosen such that the material fails by `spallation' at the vicinity
of the grain boundary. Calculated spall strength is also reported there-in.

However, all these shock and spall studies were for specific grain boundaries
only. It is not straightforward to predict the spall parameters for 
a polycrystalline material from MD simulations of shock propagation
across the individual grain boundaries. Thus it is necessary to devise
a method that accounts for individual grain boundary effects under
shock loading to calculate spall parameters for polycrystalline materials.
Motivated by such a necessity, this research work is carried out.
%This is the main motive of this research work.

\subsection
{Objective of this research work
\label{sec1.2.2}}
%Primarily, the objective of this research work is proposed to be executed
%by bringing them under two categories, namely (i) single crystal and (ii) bi-crystal
%systems. In each of these two categories, two types of computational studies are
%planned. In one type, only MD simulations are performed to initiate shock-propagation,
%and to calculate the hugoniot and the spall parameters. In the second type, multiscale
%approach is followed to obtain NAG fracture parameters and spall strength.
%Thus the task is split up in to four problems and executed independently.
%Each of the problems is defined briefly in the following list.
%%%%%%%%%%%%%%%%
%Primarily, this research work is categorized into two, namely (i) single crystal
%and (ii) bi-crystal systems. For each category, two types of computer simulations are
%planned for execution. In one of the two types, only MD simulations will be performed
%to initiate shock-propagation and to calculate the hugoniot and the spall parameters.
%In the second type, a multiscale approach will be followed to obtain NAG fracture
%parameters and spall strength. Thus the task is split-up into four problems. Each of
%these four problems is to be executed independently. The four problems are defined
%briefly in the following list.
%%%%%%%%%%%%%%%%
Primarily, this research work is categorized into dynamic shock propagation in the
(i) \ac{SC} and (ii) \ac{BiC} systems. Experimental spall parameters
are available separately for single crystals and poly-crystals. The experimental
results are used for comparing the simulations results. For each category, two types
of computer simulation are planned for execution. In one of the two types, only MD
simulation of plate impact will be performed. This will launch shock-wave in the
material. From the propagation of shock, the shock-Hugoniot parameters can be
calculated. From the dynamics of free surface motion, the spall parameters can be
calculated. In the second type, a multiscale approach will be followed to obtain NAG
fracture-parameters and spall-parameters.

When a single crystal and bi-crystals are independently subjected to impact-shock propagation, shock-Hugoniot,
the peak-tensile stress and the temporal FSV profile can be obtained. By introducing
defects in a perfect single crystal, multiscale spall simulation can be done. Upon
successfully performing these steps, shock and spall parameters of single crystals
are calculated and compared with available experiments.  Recent MD impact-shock
simulations of individual bi-crystals~\cite{LONG2020109411} have shown that the peak
tensile stress matches with the spall strength of polycrystal Cu. This approaoch was
adapted in this research work for `Al' bi-crystals. But, the MD simulation results
indicate that the peak tensile stress is much higher than the spall strength obtained
from FSV profiles. Hence multiscale model is also used for `Al' bi-crystals 
to determine the spall parameters of a polycrystal.  Thus, the task is split up into four problems.
Each of these four problems has to be executed independently. The four problems are
defined briefly in the following list.

\begin{itemize}
\item To initiate shock and study its propagation in single crystals of Aluminum (Al),
Copper (Cu) and Nickel (Ni) using MD. To find the isothermal $P$-$V$ and the shock-Hugoniot
relation between $U_s$ and $U_p$ as well as between $P$ and $V$  at high strain rates
 $(\dot{\epsilon}>10^6\ s^{-1})$ and high pressures ($\simeq$ 100 GPa).
To add shock Hugoniot data at the higher compression region for nickel.

\item To setup multiscale modelling and validate the spall strength calculated
for Cu, Nb and Mo. To improve the accuracy of the results for Nb and Mo in comparison
to the earlier works~\cite{ramana2_hydro}. To obtain the spall strength of single-crystal
aluminum using multiscale modelling.

\item To setup several bi-crystal arrangements in MD. To calculate shock and spall
parameters from individual grain boundaries, using the temporal Free Surface Velocity (FSV)
profile. To calculate corresponding peak tensile stress. To compare peak
tensile stress and FSV method in predicting the spall strength.

\item To extend the multiscale model from single crystal to bi-crystal.
To obtain NAG fracture parameters for various symmetric grain boundaries of aluminum.
To devise a methodology that accounts for the individual effects of various grain boundaries on
the NAG fracture parameters. The primary objective here
is to obtain the spall strength of polycrystalline aluminum from MD simulations of
various bi-crystal systems.
\end{itemize}

%{Literature survey on shock and spall studies, Motivation and Objective of this research work}
%2. Motivation to come up \\
%3. Objective to take up this work.

\section{Thesis outline: Chapter-wise}
%plan of the Thesis presentation in sequence to connect - chapter-wise brief note }
The computational methods developed in-house for executing this research work are
Particle Swarm Optimization (PSO) and Hydro-Dynamic (HD). Details of these
methods and their validation are presented in chapter~\ref{chapter_compMethods}.
%Results from the available ab-initio/first principle (FP) method~\cite{vmishraPhDThesis}
Results from the available ab-initio/\ac{FP} method~\cite{UsUp_Madhavan2022}
are used for verifying the shock parameters obtained from MD methods.

Shock propagation in single crystal \ac{FCC} materials: Aluminum (Al), Copper (Cu) and 
Nickel (Ni) are carried out using the MD method in this research work.
%Large-scale Atomic/Molecular Massively Parallel Simulator~\cite{lammps} is an open source
\ac{LAMMPS}~\cite{lammps} is an open source
MD package LAMMPS is used in all the MD calculations for this
research work. Shock parameters obtained from MD simulations are compared with the 
results from ab-initio simulations  and published experimental values available from the literature. Details
of this study along with the explanation for the deviation in the case of single crystal
nickel are presented in chapter~\ref{chapter_SCImpactShock}. A brief description of the \ac{EOS}
is also given in chapter~\ref{chapter_SCImpactShock}.

Single crystal Copper (Cu), Niobium (Nb), Molybdenum (Mo) and Aluminum (Al) are subjected to triaxial
tensile deformation using molecular dynamics (MD) simulations. MD simulations are carried out in
atomic length ($\AA$) and time (pico-sec or ps) scales. Material-dependant coefficients that appear in
the nucleation and growth (NAG) fracture model are obtained by PSO methods.
Subsequently, \ac{1D-HD} simulations are carried out at the macro-scales using
the values of NAG coefficients obtained at the atomic scales, to calculate the spall strength of a single crystal
material. This is a multi-scale approach. Details of multi-scale modeling to find the spall 
strength for single crystals metals are discussed in
chapter~\ref{chapter_SCSpall-Multiscale}.

Bi-crystal configuration of aluminum for various symmetric tilt and twist grain boundaries
are created and subjected to shock loading by flyer-target impact mechanism.
The individual effect of the grain boundary upon shock propagation and associated
dislocation motion, void nucleation and growth are studied at atomic scales.
From free surface velocity measurement, the spall strength of individual
grain boundaries is calculated. Peak tensile stress experienced by the
individual cases is also calculated and compared with respective spall strength
values. Details of the MD simulations and the major conclusion drawn from
this work are discussed in chapter~\ref{chapter_spallMDonly}.

The effect of individual grain boundary of an aluminum bi-crystal configuration
under tri-axial tensile deformation, on the nucleation and growth of voids,
is simulated using the MD method. Using the multi-scale approach a detailed
study is carried out for various individual bicrystal configurations of aluminum.
The NAG fracture model coefficients for various bicrystal configurations of
aluminum are calculated by the optimization fitting (PSO) method. The combined
effect of the various grain boundaries on the NAG model parameters is obtained
by proposing a novel method namely, ``Average Void Growth in Fluid Element''
(AVGFE). Using this novel method in one-dimensional hydrodynamic simulation,
the spall strength and the spall thickness of poly-crystalline aluminum are obtained.
Details of this work are presented in chapter~\ref{chapter_BiCSpall-Multiscale}.

%Chapter~\ref{chapterSummary}, contains the summary of observation and
%the main outcome from this research work. Possible future works in continuation,
%are also discussed in chapter~\ref{chapterSummary}.
%%%%%%%%%%%%%%%%
Important information gained from this computational study is summarized along with the
respective observations in chapter~\ref{chapterSummary}. The overall outcome of this
research work is briefly presented. In continuation of the present research, the possible
future works are also outlined in chapter~\ref{chapterSummary}.

%observation and results from these
%computational studies are presented. The main outcome from this work
%as well as the possible future works
%are also presented there-in.
%are also listed in chapter~\ref{chapterSummary}.
%% possible future work
%% (1) Nicket MD simulations with improvising potentials that suits dynamic shock conditions
%% (2) MS at low strani rates.. 1e6, 1e7, 1e8 etc. for SC
%% (3) same as 2 but for bi-crystals of Al
%% (4) AVGFE for other polycrystals for eg. Cu, Ni, Mo, Nb, Iron etc.,
%% (5) AVGFE for other composits
%% (6) Direct FSV from MD for other bi-crystals of Ni, Mo, Nb, Iron etc,

\clearchapter

% CompMethods : Chapter 2. MD - PSO and 1D Hydro, methods-development-validation
  \chapter{Computational Methods for Multiscale Modeling\label{chapter_compMethods}}
To carry out this research work on impact shock propagation and spall parameter calculations,
three mathematical methods are used viz. (1) Molecular Dynamics (MD), (2) Particle
Swarm Optimization (PSO) and (3) One-dimensional Hydro-Dynamics (HD) methods. Details of these
computational methods are discussed in this chapter. These three methods are used in
sequence to obtain the spall parameters in a multi-scale approach~\cite{rawat2_hydro,
ramana2_hydro, rawat_thesis}. The computational work is divided into three parts
and the sequential flow of the calculations is shown in figure~\ref{subfig:StepsInMSModel}
as a flowchart. The main objective of this work is to study the effect of shock
propagation involving grain boundaries in specific and to predict associated materials failure.
Thus in MD simulations grain boundary is set up and validated by calculating
\ac{GBE}. This step is also mentioned in the flowchart (fig.~\ref{subfig:StepsInMSModel}).
Whereas in the case of multi-scale modeling with a single crystal, the grain boundary energy calculation
is not involved. For this case, the first three steps in part 1 (refer fig.~\ref{subfig:StepsInMSModel}),
are (a) setup atoms positions, (b) remove half plane to introduce imperfection, (c) equilibrate in
NPT ensemble. Rest of the steps as shown in fig.~\ref{subfig:StepsInMSModel} are followed
for single crystal MD simulation. Figure~\ref{subfig:StepsInMSModel} is given at the end of
this chapter because it is felt that the description of the mulstiscale method in detail
may be useful for the reader to follow the flow-chart.

\section{Molecular Dynamics : LAMMPS\label{lammpsMDsec2.1}}
MD methods are available as open-source packages, for instance, LAMMPS~\cite{lammps}.
%LAMMPS is the acronym for ``Large-scale Atomic/Molecular Massively Parallel Simulator''.
It is well documented and validated with many benchmark problems~\cite{lammpsBenchmarkpg}.
%The user manual and tutorials help to set up and simulate impact-shock systems.
This package is well-established and widely used by the research community.
The details of various modules available in LAMMPS can be obtained from LAMMPS~\cite{lammps}.
However in this thesis report, the methods particular to high strain rate deformation,
tensile stress propagation, impact, shockwave generation \& propagation, spall \&
peak tensile strength calculations are presented with appropriate validation.
In brief, LAMMPS solves atoms' motion under intrinsic and applied external
forces. %The equation solved in LAMMPS is governed by Newton's inertial law of force (F).
MD method is used to evolve a system of N particles under a given
dynamical condition. The classical equations of motion are integrated to generate
the trajectory of a system of N particles. The N-body force equation is given by,
%The force ($F_i$) on $i^{th}$ atom among `N' number of atoms with an individual mass `$m_i$' with respective position
%vector $\vec{r_i}$ is given by eqn.\ref{eqnofMotionLammps}, where $\ddot{\vec{r}}$ represents
%the acceleration vector, second order time derivative of $\vec{r}$.
\begin{equation}
%\begin{subequations}
%\begin{align}
F_i = m_i\ddot{\vec{r_i}} = \sum_{j=1,j\neq{i}}^N \left(-\nabla{\phi}_{ij}\right) \label{eqnofMotionLammps}  % \ \ \ i\ne j \label{eqnofMotionLammps} \\
%\text{For EAM}, \phi & = f_i\left(\rho(\vec{r})\right)+ \sum_{j=1}^N V\left(\vec{r}_{ij}\right) \label{eampotEqn}
%\end{align}
%\end{subequations}
\end{equation}
where for the $i^{th}$ particle, $\vec{r_i}$ is its position vector, $m_i$ is its mass and $\phi$
is the total potential energy of the system.
%Here $\nabla{\phi}$ is the force exerted between two atoms $i$ and $j$.
%$\phi$ is the potential energy.
Thus the force ($F_i$) is governed by the potential energy function $\phi$.
Embedded Atom Method (EAM) is used to calculate the atomic energetics~\cite{CuNi_eam}.
According to the quantum mechanical density functional theory, the ground state energy and
properties of a system are determined by the electron density. As per this approach,
the energy of a system is computed as the energy obtained by embedding an atom into local
electron density in presence of other atoms in the system.
In addition to this, there is an electrostatic interaction potential. Therefore, the total
energy of a system is given by,
\begin{equation}
%\phi = f_i\left(\rho(\vec{r})\right)+ \sum_{j=1}^N V\left(\vec{r}_{ij}\right) \label{eampotEqn}
%	\phi = \sum_i{f_i} \left(\sum_{i=1,j{\ne}i}^N{\rho}_j(r_{ij})\right) + \frac{1}{2}\sum_{i{\ne}j}\phi_{ij}(r_{ij}) \label{eampotEqn}
	E = \sum_i{f_i}({\rho}_i) + \frac{1}{2}{\sum_i}\sum_{j,j{\ne}i}\phi_{ij}(r_{ij}) \label{eampotEqn}
%	\phi = \sum_i{f_i}({\rho}_i)  \label{eampotEqn}
%	\left(\sum_{i=1,j{\ne}i}^N{\rho}_j(r_{ij})\right) + \frac{1}{2}\sum_{i{\ne}j}\phi_{ij}(r_{ij}) \label{eampotEqn}
\end{equation}
where $\phi_{ij}$ the pair energy between atoms $i$ and $j$ separated by a distance
$r_{ij}$, and $f_i$ is the embedding energy associated with embedding an atom $i$
into a local site with an electron density ${\rho}_i$. If $\rho_j(r_{ij})$ the electron
density at the site of atom $i$ arising from atom $j$ at a distance $r_{ij}$ away, then
the electron density can be calculated using,
\begin{equation}
\rho_i=\sum_{j,j{\ne}i}\rho_j(r_{ij}) 
\end{equation}
%with $\rho_j(r_{ij})$ the electron density at the site of atom $i$ arising from atom $j$ at a distance $r_{ij}$ away.

For a pure element $a$, the EAM potential is composed of
three functions: the pair energy $\phi_{ij}$, the electron density ($\rho_i$), and
the embedding energy $f_i(\rho_i)$.
%where `$f_i$' is the embedding energy function for an atom `$i$'.
%$r_{ij}$ is the distance between atoms $i$ and $j$.
%$\phi_{ij}$ is the pair potential between atom $i$ and its neighboring atom $j$. The electron density $\rho_j$
%is the sum of the electron densities of all the atoms $j$ with $j{\ne}i$.
In MD, this energy is used to determine the force on each atom.
The interaction force between atoms $i$ and $j$ along $\alpha$ direction is given by,
\begin{equation}
	F^{\alpha}_{ij}=\frac{\partial \phi}{\partial r_{ij}} \frac{r^{\alpha}_{ij}}{r_{ij}} \label{termForce}
\end{equation}
The virial definition is used calculate the stress tensor in MD given by,
\begin{equation}
	\sigma_{\alpha\beta}=-\frac{1}{V}\left[\sum_i \frac{p^{\alpha}_i p^{\beta}_i}{m_i}+\sum_i\sum_{j>i}r^{\alpha}_{ij}F^{\beta}_{ij}\right] \label{virialStressEqn}
\end{equation}
where $p^{\alpha}_i$ is the momentum of the $i^{th}$ atom in the $\alpha$ direction, $m_i$
is the mass of the $i^{th}$ atom, $r_{ij}^{\alpha}$ is the distance between the $i^{th}$
and $j^{th}$ atoms along the direction of $\alpha$, $F_{ij}^\beta$ is the force on the
$j^{th}$ atom from $i^{th}$ atom in the $\beta$ direction and $V$ is the volume
of the system.
The first term on the right hand side of eqn.~\ref{virialStressEqn} is the kinetic contribution to
the stress and the second term gives the virial potential stress. The stress for an
atomistic system is defined as the volume average of per-atom tensor

%This function defines the energy required to place an atom
%in a volume of atoms defined by a specific radius with respect to the $i^{it}$ atom in consideration.
%This function depends on the local electron density $\rho(\vec{r}_i)$.
%The second term in eqn.~\ref{eampotEqn} is a pair potential term to account for
%the pair interaction of atom `$i$' with the other atoms in the neighbourhood
%`$j$'. The pair interaction depends only on the distance between the two
%pairs of atoms under consideration.
Even for a small number of atoms or for any complicated  potential function, $\phi_{i}$,
MD solves eqn.~\ref{eqnofMotionLammps} numerically because this equation has no analytical
solution. The numerical method follows `velocity-Verlet' formulation.
LAMMPS MD simulations often use interatomic
potentials that are available from interatomic potentials
repository~\cite{potential_website}. Details of various potentials can be
obtained from the references there-in~\cite{potential_website}.

\subsection{Validation of the EAM potentials used for Al, Cu, Mo \& Ni\label{EAMValidatedChap2.1.1}}
\begingroup
\begin{spacing}{1}

\begin{table}[ht]
\footnotesize
\centering
\begin{threeparttable}[b]
\caption
[Validation of EAM potentials of Al, Cu, Mo \& Ni]
%{Validation of the EAM potentials used in this work for Al, Cu, Mo \& Ni:
%Calculated elastic constants ($B_0$, $G_0$, $Y_0$ \& $\nu$)~\cite{sspKittel},
%cohesive energy ($E_{coh}$)~\cite{sspKittel} and vacancy formation energy
%($E_{vf}$)~ref.~\cite{vacancyFormEnrAl, MCGERVEY197353, PhysRevB.80.224104, WOLFF19974759}
%with experimental/theoretical values reported in the literature.
%}
{
%Validation of the EAM potentials of Al, Cu, Mo \& Ni:
%Calculated (A) elastic constants:
%(1) $B_0$~\cite{sspKittel},
%(2) $G_0$~\cite{mechEnggHBDanBManghitu},
%(3) $Y_0$~\cite{mechEnggHBDanBManghitu},
%(4) $\nu$~\cite{mechEnggHBDanBManghitu} and
%(B) Energies :
%(5) Cohesive Energy ($E_{coh}$) for Al, Cu, Mo \& Ni~\cite{sspKittel} and
%(6) Vacancy Formation Energy ($E_{vf}$) for:
%Al~\cite{vacancyFormEnrAl},
%Cu~\cite{MCGERVEY197353},
%Mo~\cite{PhysRevB.80.224104},
%Ni~\cite{WOLFF19974759},
%are compared with experimental/theoretical values available in the 
%literature~\cite{sspKittel, mechEnggHBDanBManghitu, vacancyFormEnrAl, MCGERVEY197353,
%PhysRevB.80.224104, WOLFF19974759}.
Validation of the EAM potentials of Al~\cite{Michin_Al99.eam.alloy}, Cu~\cite{CuNi_eam},
Mo~\cite{Zhou_MoEAM} \& Ni~\cite{CuNi_eam}: Calculated elastic constants,
$B_0$, $G_0$,  $Y_0$, $\nu$, Cohesive-Energy ($E_{coh}$) and  Vacancy Formation Energy ($E_{vf}$)
are compared with experiments (references are indicated with special symbol).
%as mentioned in square
%brackets: $B_0$~\cite{sspKittel}, $G_0$~\cite{mechEnggHBDanBManghitu},
%$Y_0$~\cite{mechEnggHBDanBManghitu}, $\nu$~\cite{mechEnggHBDanBManghitu},
%Cohesive Energy ($E_{coh}$)~\cite{sspKittel} and
%Vacancy Formation Energy ($E_{vf}$) of Al~\cite{vacancyFormEnrAl},
%Cu~\cite{MCGERVEY197353}, Mo~\cite{PhysRevB.80.224104}, Ni~\cite{WOLFF19974759}.
}
\vspace{-0.3cm}
\begin{tabular}{cccccccccc}
\toprule
\multirow{2}{*}{\parbox{0.2cm}{\centering{Sl. No.}}} &
\multirow{2}{*}{\parbox{1.4cm}{\centering{Physical constants}}} &
      \multicolumn{2}{c} {Al} &
      \multicolumn{2}{c} {Cu} &
      \multicolumn{2}{c} {Mo} &
      \multicolumn{2}{c} {Ni}\\
\cmidrule(lr){3-4} \cmidrule(lr){5-6} \cmidrule(lr){7-8}\cmidrule(lr){9-10}
& & {Exp.}&{MD}&{Exp.}&{MD}&{Exp.}&{MD}&{Exp.}&{MD}\\
\midrule
1&$B_0$ (GPa)&\tnote{\dag}\ \ 72.2&82.73&\tnote{\dag}\ \ 137&134.53&\tnote{\dag}\ \ 272.5&300&\tnote{\dag}\ \ 186&175\\  % From SSP kittel pg 52
%\hdashline
2&$G_0$ (GPa)&\tnote{*}\ \ 26.2&27.00&\tnote{*}\ \ 44.7& 37.45&\tnote{*}\ \ 117&114& \tnote{*}\ \ 76& 70.92\\  % From Hand book, BAN.B. Marghitu pg 130
%\hdashline
3&$Y_0$ (GPa)&\tnote{*}\ \ 71&73.06&\tnote{*}\ \ 119&102.82&\tnote{*}\ \ 331&305&\tnote{*}\ \ 200&187\\  % From Hand book, BAN.B. Marghitu pg 130
%\hdashline
4&$\nu$&\tnote{*}\ \ 0.334&0.35&\tnote{*}\ \ 0.326&0.37&\tnote{*}\ \ 0.31&0.33&\tnote{*}\ \ 0.31&0.32\\  % From wikipedia
%% Makes the following row in a single line 
%5&$E_{coh}$ (eV/atom)&\tnote{\dag}\ \ 3.39 &3.36 &\tnote{\dag}\ \ 3.49 &3.529 &\tnote{\dag}\ \ 6.82 &6.4 &\tnote{\dag}\ \ 4.44 &4.45\\  % From SSP Kittel pg 50
%\hdashline
%% Makes the following row in a Double lines
	\multirow{2}{*}{\centering{5}} &
	$E_{coh}$ &
	\multirow{2}{*}{\centering{\tnote{\dag}\ \ 3.39}} &
	\multirow{2}{*}{\centering{3.36}} &
	\multirow{2}{*}{\centering{\tnote{\dag}\ \ 3.49}} &
	\multirow{2}{*}{\centering{3.529}} &
	\multirow{2}{*}{\centering{\tnote{\dag}\ \ 6.82}} &
	\multirow{2}{*}{\centering{6.4}} &
	\multirow{2}{*}{\centering{\tnote{\dag}\ \ 4.44}} &
	\multirow{2}{*}{\centering{4.45}} \\
  & (eV/atom) & & & & & & & & \\
  %%%%%%% above section for double row for E_coh
%\hdashline
6&$E_{vf}$ (eV) &\tnote{\S}\ \ 0.69$\pm$0.03&0.625 &\tnote{\P}\ \ 0.98$\pm$0.07&1.28 &\tnote{\ddag}\ \ 3$\pm$0.2&3.25 &\tnote{\pounds}\ \ 1.73$\pm$0.07&1.63\\

\bottomrule
\label{elasticTable}
\end{tabular}
\vspace{-0.3cm}
 \begin{tablenotes}
    \item[\dag]Ref.~\cite{sspKittel} \item[*]Ref.~\cite{mechEnggHBDanBManghitu} \item[\S]Ref.~\cite{vacancyFormEnrAl} \item[\P]Ref.~\cite{MCGERVEY197353}
    \item[\ddag]Ref.~\cite{PhysRevB.80.224104}
    \item[\pounds]Ref.~\cite{WOLFF19974759}
 \end{tablenotes}
 \end{threeparttable}
%brackets: $B_0$~\cite{sspKittel}, $G_0$~\cite{mechEnggHBDanBManghitu},
%$Y_0$~\cite{mechEnggHBDanBManghitu}, $\nu$~\cite{mechEnggHBDanBManghitu},
%Cohesive Energy ($E_{coh}$)~\cite{sspKittel} and
%Vacancy Formation Energy ($E_{vf}$) of Al~\cite{vacancyFormEnrAl},
%Cu~\cite{MCGERVEY197353}, Mo~\cite{PhysRevB.80.224104}, Ni~\cite{WOLFF19974759}.
\end{table}

\end{spacing}
\endgroup

The EAM potentials used in this work for Al~\cite{Al_eam, Michin_Al99.eam.alloy}, Cu~\cite{CuNi_eam}, Mo~\cite{Zhou_MoEAM} \& Ni~\cite{CuNi_eam}
are validated by calculating the elastic constants viz.
bulk modulus ($B_0$), shear modulus ($G_0$), Young's modulus ($Y_0$),
Poisson's ratio ($\nu$), cohesive energy ($E_{coh}$) and
vacancy formation energy ($E_{vf}$).
Calculated values are listed in table~\ref{elasticTable}
along with experiment/theoretical values. For comparision, the elastic constants
($B_0$, $G_0$, $Y_0$, \& $\nu$) data and the cohesive energy for various metals
available in the text book~\cite{sspKittel} and the handbook~\cite{mechEnggHBDanBManghitu} are used.
The vacancy formation energy values are compared with
ref.~\cite{vacancyFormEnrAl} for Al,
ref.~\cite{MCGERVEY197353} for Cu,
ref.~\cite{PhysRevB.80.224104} for Mo and
ref.~\cite{WOLFF19974759} for Ni.
%% added on 30th Nov 2023
%The literature references for the EAM potentials are available in chapter~\ref{chapter_SCSpall-Multiscale}
%for Al, Cu \& Mo and in chapter~\ref{chapter_SCImpactShock} for Ni.

\subsection{Choice of time step in updating the position \& velocity}
The time steps used in MD simulations are mostly in femtoseconds or a fraction
of it~\cite{rawat_thesis}. Taylor's expansion of position ($\vec{r}$), velocity ($\vec{v}$) and
acceleration ($\vec{a}$) of the atoms from a time step `$t$' to `$t+\delta{t}$'
are given by,
\begin{eqnarray}
\vec{r}(t + \delta{t}) = \vec{r}(t) + \vec{v}(t)\delta{t} + \frac{1}{2} \vec{a}(t)(\delta{t})^2 + ....\\
\vec{v}(t + \delta{t}) = \vec{v}(t) + \vec{a}(t)\delta{t} + \frac{1}{2} \vec{j}(t)(\delta{t})^2 + ....\\
\vec{a}(t + \delta{t}) = \vec{a}(t) + \vec{j}(t)\delta{t} + \frac{1}{2} \vec{s}(t)(\delta{t})^2 + ....
\end{eqnarray}
where $\vec{j}$ \& $\vec{s}$ are higher-order time derivatives of position known as
`jerk' and `snap' respectively.
%Here, $\vec{s}$ is not to be confused with displacement vector.
%However these higher order terms are used mostly in cosmological physics~\cite{jerkSnapRef}
%and can be neglected with respect to the present context.
%Moreover when the acceleration of the particle does not change substantially in a given
%time step, higher-order terms can be neglected.
%%%%%%
Higher order terms are neglected with respect to the present context, since
the acceleration of the particle does not change substantially in a given
time step. Also the Verlet algorithm allows the numerical integration of momentum equations
with an error that is of the O($\delta{t}^3$) while the energy is conserved.
%%%%%%
The acceleration depends upon the gradients of the potential energy of the system
with respect to the position of the individual atom.
Therefore the timestep must be small enough such that these gradients
do not change much during a single timestep $\delta{t}$.
Thumb rule suggests a time step in which a particle  moves only one-twentieth (1/20$)^{th}$
of the inter-atomic distance between its nearest neighbour~\cite{KaiNotesPkg2timestepMD}.
One can also check the suitability of the timestep by ensuring energy conservation
in an isolated system. In such a system no energy exchange is expected to take place.
The overall energy change should not be significant for the chosen timestep.

\subsection{Ensembles used in MD simulations :}
The collection of all possible systems with different microscopic states but have
an identical macroscopic state is called an ensemble. Ensembles are classified
based on their characteristics. Following are the three ensembles used in
this research work.
\begin{itemize}
\item Microcanonical ensemble (NVE) : The state of the system in which
the number of atoms (N) and volume (V) and energy (E) do not change
and remain constant throughout the process.
\item Canonical Ensemble (NVT): The state of the system corresponds
to a fixed number of atoms (N), volume (V) and temperature (T).
Room temperature 300 K is an example of this system in which
though the system can exchange the thermal energy with a thermostat to
maintain the temperature, no exchange of particles occurs.
\item Isobaric-Isothermal Ensemble (NPT): This ensemble is characterized by
a fixed number of atoms (N), a fixed pressure (P) and a fixed temperature (T).
This indicates that the system volume changes to keep the pressure
constant whilst the thermostat maintains the temperature.
\end{itemize}

\subsection{MD details specific to this work in brief}
Shock propagation in single crystals is simulated using the MD method.
Details of the computational domain size, boundary conditions, setting up the
flyer-impact system, launching impact shock, calculation of shock Hugoniot
parameters (Rankine-Hugoniot) and validation of the MD calculations by comparing
them with the results of experiments as well as First Principles (FP) methods are presented in chapter
chapter~\ref{chapter_SCImpactShock}.

The Multi-scale approach is used for spall parameter calculations for single crystal
as well as bicrystal systems. Details of the computational methods involved in single
crystal MD simulations are presented in chapter~\ref{chapter_SCSpall-Multiscale}.
While they are presented in chapter~\ref{chapter_spallMDonly}
and~\ref{chapter_BiCSpall-Multiscale} for bicrystal systems. Computational
domain details and the method of Grain Boundary Energy (GBE) calculation
are also provided in chapters~\ref{chapter_spallMDonly}
and~\ref{chapter_BiCSpall-Multiscale} with validation.

\section
[Particle Swarm Optimization (PSO)]
{Particle Swarm Optimization: Minimization method for solving multi-parameter function
\label{PSOsec.2.2}}
An evolutionary algorithm such as Particle Swarm Optimization (PSO) is effective
in multi-parameter optimization~\cite{Yshi_PSO, SUGANDHI2015113}. Each of the unknown
parameters is treated as the element of a multi-dimensional set.  A guess
range of values is made initially that bounds the solution for each of the
`n' number of unknown parameters.

\subsection{PSO Methodology}
%For `n' unknown problem, the PSO solver defines
%`N' number of particles in `n' dimensional space.
For fitting a function with `N' unknown parameters, a particle is defined with
components in the `N' dimensional parameter space.
Each particle is assigned 
an initial value within the initialy specified range. The objective function
is evaluated for every particle. At every iteration, the particle that gives
the best solution is tagged as ``best-particle''. This is then compared with
the ``global-best-particle'' that is being tracked from the very first iteration.
In each iteration ``global-best-particle'' is reassigned if any particle
in that iteration is found to be better than the current ``best-particle''.
The particle positions are updated
in `n' dimensional parameter space as explained below.

Let the $i^{th}$ particle during the current iteration be 
represented as $X_i=(x_{i1},x_{i2},...,x_{in})$ for a `n' parameter 
problem. If its previous best position is represented as 
$P_i=(p_{i1},p_{i2},...,p_{in})$ and the global best position among all 
particles is represented by $P_g=(p_{g1},p_{g2},...,p_{gn})$, the new 
incremental value of each independent variable, required to move closer to the 
solution is given by the equations~\ref{eqn_vidChap2} and ~\ref{eqn_xidChap2}.
\begin{subequations}
\begin{align}
v_{id} & = v_{id} + c_{1}r_{1}(p_{id} - x_{id}) + c_{2}r_{2}(p_{gd} - x_{id})
           \label{eqn_vidChap2} \\
x_{id} & = v_{id} + x_{id} \label{eqn_xidChap2}
\end{align}
\end{subequations}
where, $c_1$ and $c_2$ are constants and $r_1$ and $r_2$ are uniform
random numbers between 0 to 1. The second term in RHS of the equation~\ref{eqn_vidChap2},
is called the `cognition' part. This term drives the particle to
its best position in each iteration. Globally, the best position is tracked and updated
by the third term in eqn.~\ref{eqn_vidChap2} as per the algorithm~\cite{Yshi_PSO}.

\subsection{PSO Validation}
\begin{figure}[ht!]
%        \centering
            \subfigure[Rosenbrock's Fn. (2 variables)]
            {
\includegraphics[width=0.45\textwidth]{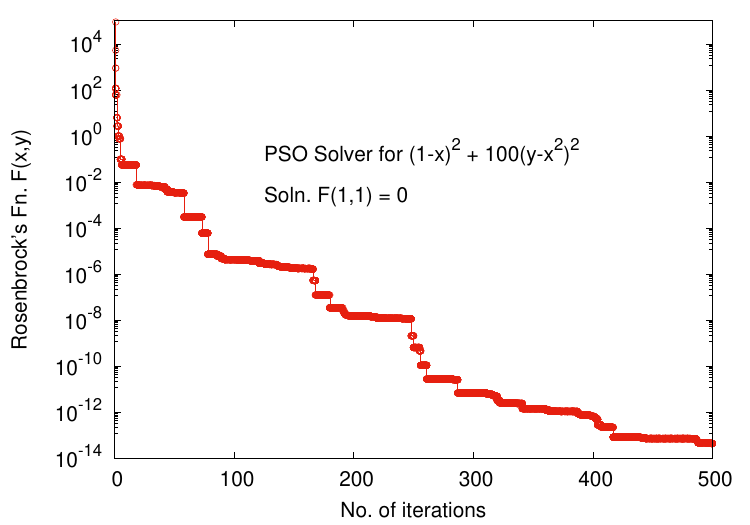}
		\label{rosenbrokfnfig}
            } 
            \subfigure[Rastrigin's fn. (5 variables)]
            {
\includegraphics[width=0.45\textwidth]{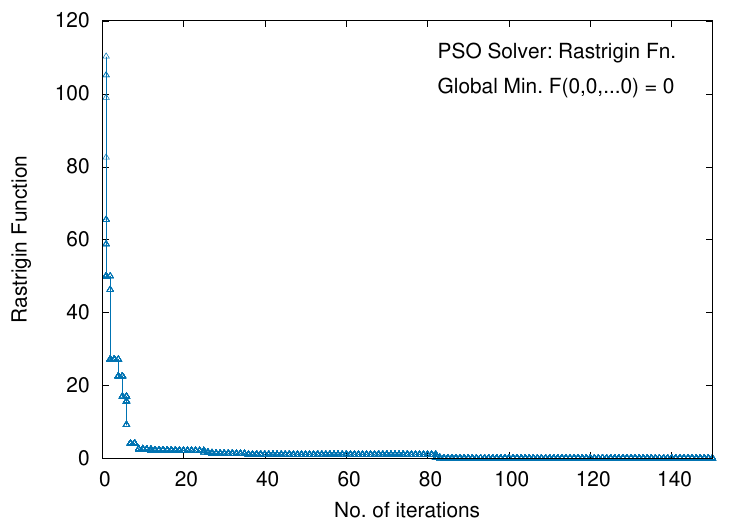}
		\label{rastriginfnfig}
            }
            \subfigure[Spherical fn. (10 variables)]
            {
\includegraphics[width=0.45\textwidth]{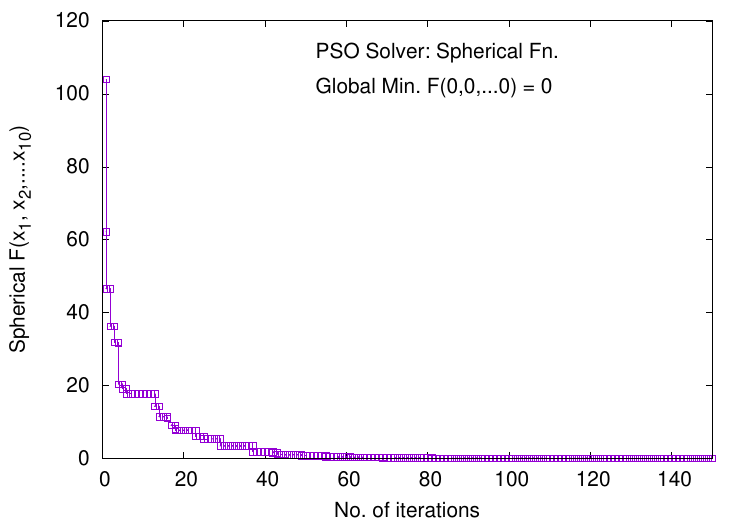}
		\label{sphericalfnfig}
            } 
\hfill
            \subfigure[Ackley's fn. (10 variables)]
            {
\includegraphics[width=0.45\textwidth]{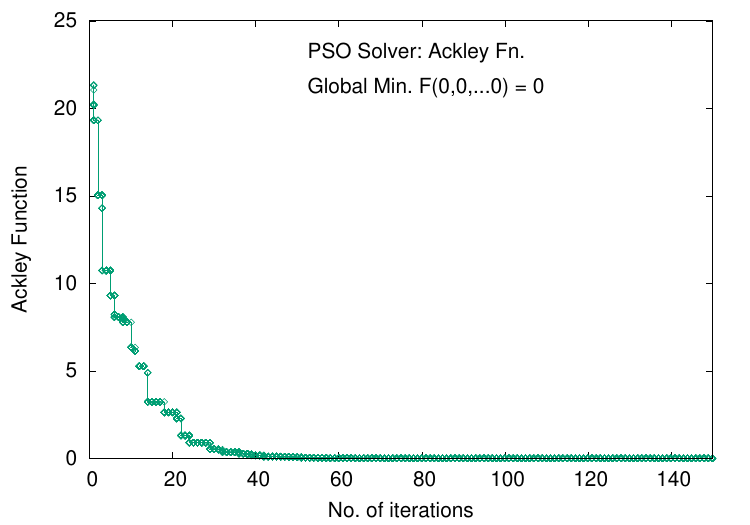}
		\label{ackleyfnfig}
            } 
\caption
[Validation of PSO method (developed in-house)]
{Validation of PSO method (developed in-house) with standard multi-dimension
mathematical functions as indicated in figures ($a$) to ($d$).}
\label{splfunctionsfig}
\end{figure}

A PSO code is developed in Fortran language and it is validated with standard functions
viz. (a) Rosenbrock's function, (b) Rastrigin's function, (c) Spherical functions,
and  (d) Ackley's function~\cite{book_optimisation}. Figures~\ref{rosenbrokfnfig}-
%~\ref{rastriginfnfig} %~\ref{sphericalfnfig}
~\ref{ackleyfnfig} show the function values obtained at each of the iterations
for Rosenbrock, Rastrigin, Spherical and  Ackley functions respectively.
Corresponding functional forms are given in equations~\ref{rosenbrokfneqn}-~\ref{ackleyfneqn}.

\subsubsection{Rosenbrock's function}
\begin{equation}
f(x,y) = (a-x)^2 + b(y-x^2)^2
\label{rosenbrokfneqn}
\end{equation}
\subsubsection{Rastrigin's function}
\begin{equation}
f(x_1,x_2,...x_N) = N\times A + \sum_{i=1}^N \left[x_i^2 - A\cos(2 \pi x_i)\right]
\label{rastriginfneqn}
\end{equation}
\subsubsection{Spherical Function}
\begin{equation}
f(x_1,x_2,...x_N) =  \sum _{i=1}^N x_i^2
\label{sphericalfneqn}
\end{equation}
\subsubsection{Ackley's function}
\begin{equation}
f(x_1,x_2,...x_N) =
-20 {}\exp\left[-0.2\sqrt{\frac{1}{N}\sum _{i=1}^N x_i^2}\,\right] 
 {} -\exp\left[\frac{1}{N}\sum _{i=1}^N \left(\cos 2\pi x_i \right)\right] + e + 20
\label{ackleyfneqn}
\end{equation}

The parameter $N$ that appears in eqns.~\ref{rastriginfneqn},~\ref{sphericalfneqn}
and~\ref{ackleyfneqn} is the number of independent variables used to define the
function values for Rastrigin's, Spherical and Ackley's respectively. `$x_N$' represents
`$N$' number of independent variables. The value used for $A$ in Rastrigin function
(eqn.~\ref{rastriginfneqn}) is 10. For Rosenbrock's function (eqn.~\ref{rosenbrokfneqn})
$a=1$ and $b=100$ are used, while $x$ and $y$ are the two independent variables that
defines the function, $F(x,y)$. The global optimum point is $f(0,0) = 0$ for
Rastrigin, spherical and Ackley functions. For Rosenbrock's function, the global
minimum is $f(x,y)=0$ at $(1,1)$.

Number of independent parameters used in the validation of PSO are different
for different functions because of the nature of their convergence-behavior.
For (i) Rosenbrock, (ii) Rastrigin, (iii) Spherical and (iv) Ackley functions
the no. of independent parameters used are 2, 5, 10 and 10 respectively.
For Ackley's function, the number of particles used are 50, while 25 particles
are used
for the other three functions. The accurate solution is achieved for these
functions iteratively as shown in figures~\ref{rosenbrokfnfig} to~\ref{ackleyfnfig}.
Rastrigin, Spherical and Ackley's functions get converged with in 60-80 iterations
while Rosenbrock's function requires 100 iterations to get the function
value less than $10^{-5}$. For all these functions the converged results
from the PSO method, showed high accuracy when compared with the theoretical results.
%Validation of the PSO method developed in-house is carried out this way
%by comparing the results with theoretical solutions.

\section
[One-dimensional hydrodynamics (1D-HD)]
{One-dimensional hydrodynamics (1D-HD) for elastic-plastic flow and shock wave propagation
\label{1Dhydroec.2.3}}
%% Copied from file 'bk-up-pcharge_rep3.tex' in the directory
% '/home/madhavpc2008/FromP C_HP2008/Linux/madhavan/Madhavan1-bkup-110308/Linux/Bkup/Pvm1/Linux/Pcharge-Post-trial/Latex/31Oct/27thNov02'
\subsection{Physical phenomena included in model}
%Hydrodynamic simulations used in this research work is an in-house developed 
%one-dimensional (1-D), time-dependent, lagrangian hydrodynamics code, 
A one-dimensional, time-dependent, Lagrangian HD simulation code was developed based on
the formalism of the KO~\cite{hemp_wilkins, wilkinsMethodsInCompPhys}.
The full form of `KO' is not known, and is widely mentioned as `KO' only.
According to Wilkins~\cite{wilkinsMethodsInCompPhys}, for the calculation of elastic-plastic flow
the important requirement is a formulation of the equation of state,
which must describe elastic, elastic-plastic, and hydrodynamic flow.

It is formulated using \ac{FD} method in Lagrangian frame of
reference. In such a formulation, the hydrodynamic flow of the object
is fixed to a well defined computational grid or cell. This makes the initially
assigned mass of each computational grid to remain constant throughout
the temporal evolution of the system. Thus the instantaneous volume of each grid
is calculated for change in density upon the hydrodynamic flow.
Corresponding local pressure and temperature are obtained using `Equation Of State'
(EOS) in terms of density and temperature. For materials with mechanical strength,
appropriate strength models viz. Johnson-Cook, Steinberg-Guinan, Zerilli-Armstrong
are available. The code supports plane and symmetric geometries
(cylindrical and spherical) and includes the following major features:
It can handle shock waves in solids and fluids.
It can use either tabulated or analytical EOS data~\cite{wilkinsMethodsInCompPhys}.
This also includes the effect of melting.
The von Mises yield criterion is used for solids.
The \ac{SG} high-strain-rate model has been incorporated
for calculating the variation of yield strength $Y$ and shear
modulus $G$ with temperature T \& pressure P, and also the variation
of $Y$ with equivalent plastic strain~\cite{SG_strength,STEINBERG1987603}.
Nucleation and Growth (NAG) fracture model has been used in this work
for aluminum~\cite{curran_nag1}.
The model has been extensively validated with
available experimental results~\cite{hemp_wilkins}.
%%%%%%%%%%%
%\begin{enumeration}
%\item	It can handle shock waves in solids and fluids.
%
%\item	It can use either tabulated or analytical equation-of-state
%	(EOS) data. This also includes the effect of melting.
%
%\item	The von Mises yield criterion is used for solids.
%
%\item	The Steinberg-Guinan (SG) high-strain-rate model has been incorporated
%	for calculating the variation of yield strength $Y$ and shear
%	modulus $G$ with temperature T \& pressure P, and also the variation
%	of $Y$ with equivalent plastic strain~\cite{SG_strength,STEINBERG1987603}.
%
%\item	Two different fracture mechanics models have been included. 
%	These include the Void Growth model~\cite{johnsonPulseDurationSpallFrac_JAP}
%	and the Nucleation and Growth model~\cite{NAGModel}.
%\end{enumeration}
%
\subsection{Steinberg-Guinan: Dynamic strength model}
	This model assumes the following variation of Y and G~\cite{SG_strength,STEINBERG1987603}:
\begin{eqnarray}
	Y\ & =& \ Y_{pl} 
	\times \left[ 1 + \left(\frac{{Y_p}^{'}}{Y_0}\right) \frac{P}{{\rho^{'}}^{1/3}}
	+ \left(\frac{{Y_T}^{'}}{Y_0}\right) (T - 300) \right] \label{SGY}\\
	G\ & =& G_0 \times \left[ 1 + \left(\frac{{G_p}^{'}}{G_0}\right)
	\frac{P}{{\rho^{'}}^{1/3}} + \left(\frac{{G_T}^{'}}{G_0}\right) (T - 300)\right]\label{SGG}
\end{eqnarray}
where $Y_{pl}\ = \ Y_{0} \times [1 + \beta ( \epsilon + {\epsilon}_i)]^n$
represents the strain hardening term, and we impose the condition:
$Y_{pl}\ \le\ Y_{max}$. Here, $\rho^{'}$ is the compression equal to the ratio of
instantaneous density ($\rho$) to initial density ($\rho_0$). $\beta$  and $n$ are work-hardening
parameters. $\epsilon_i$ is the initial equivalent plastic strain,
normally equal to zero. $P$ is the pressure, having the same dimensions as 
$G$ and $Y$, and $T$ is the temperature in Kelvin. The subscript `$0$' refers
to the reference state (T = 300 K, P = 0, $\epsilon$ = 0). $G_T^{'}$ and 
$G_P^{'}$ are the rates of change of shear modulus with temperature and 
pressure, respectively. $Y_T^{'}$ and $Y_p^{'}$ are defined in the same way.
An explanation of how these coefficients are obtained is available in reference~\cite{SG_strength}.
The coefficients for several materials, including aluminum, are available there-in.

\subsection{Ductile fracture (NAG) model for Aluminum\label{NAG_intro}}
This model is ductile in nature and is called `DFRACT'. In this work, `DFRACT' fracture
model is incorporated into the hydrodynamic calculation of spall strength for aluminum.
DFRACT is a ductile fracture model and is one of the Nucleation and Growth (NAG)
micro-physical models developed at Stanford Research Institute. It deals with
the processes of nucleation and growth of many micro-voids. It also describes
the effect of the growing damage on stress-strain relations. The void volume
($V_v$) computed in this model has contributions from nucleation ($V_{vn}$) and 
growth ($V_{vg}$) processes. If `$V_{v0}$' is the initial void volume, then
`$V_v$' is computed using formulae given in eqn.~\ref{V_vEqn0},
\begin{equation}
	V_v = V_{v0} + \Delta{V_{vn}} + \Delta{V_{vg}}
\label{V_vEqn0}
\end{equation}
%where $V_{v0}$ is the initial void volume.
\clearpage
\noindent{\bf Nucleation of Voids:}\\
New voids are created when the tensile pressure ($P_s$) in the solid
material exceeds the nucleation threshold ($P_{n0}$) level of the material. The
rate of creation of voids is given by
\begin{eqnarray}
\dot{N} & = & \dot{N_0}\ V_{sim} \ exp\left[\frac{P_s-P_{n0}}{P_1}\right],\hspace*{.5in} for\ P_s > P_{n0}
\label{voidNucEqn1}
\end{eqnarray}
where $\dot{N_0}$ is the nucleation rate of voids (m$^{-3}$s$^{-1}$) and $P_1$ the
pressure sensitivity for nucleation. Both are material-dependent parameters.
$V_{sim}$ is the volume obtained from MD simulations when the material is
subjected to tensile pressure.  Subscript `$sim$', is the short term for simulation.
The void volume nucleated in a single time step($\Delta{t}$) is given by
\begin{equation}
\Delta{V}_{vn} = 8\:\pi\:\dot{N}\:\Delta{t}\:R_n^3
\label{voidNucEqn2}
\end{equation}
where $R_n$ is a material parameter called the `nucleation size parameter'
and is taken to be the size of a unit cell. ${\Delta}t$ is the time-step
used in the hydrodynamic model.
\vspace{0.2in}\\
{\bf Growth of Voids:}\\
The existing voids grow if the tensile pressure exceeds the
material threshold for growth ($P_{g0}$).  The change in void volume due
to the growth of the voids during a single time step is given by
\begin{equation}
%%%\Delta\!V_{vg} = V_{v0}\:\exp\left(\frac{3}{4}\ \frac{(P_s-P_{g0})}{\eta} \ \Delta\!t\right) - V_{v0}
%\Delta{V_{vg}} = V_{v0}\:\exp\left(\frac{3}{4}\ \frac{(P_s-P_{g0})}{\eta} \ \Delta{t}\right) - V_{v0}
\Delta{V_{vg}} = V_{v0}\:\exp\left(\frac{3}{4}\ \frac{(P_s-P_{g0})}{\eta} \ \Delta{t}\right)
\label{voidGrowthEqn1}
\end{equation}
where $\eta$ is the material viscosity. The five unknown material parameters
used in NAG fracture model are $\dot{N}_0,\ P_1,\ P_{n0}, P_{g0}$ and $\eta$.
These are calculated by fitting the NAG fracture function by optimization methods
such as PSO.
\vspace{0.2in}\\
{\bf Material moduli with damage:}\\
Nucleation and growth of voids affect both the yield strength (Y) and the
effective shear modulus (G) of the material. In the present model, both Y \& G are
re-calculated using the equations~\ref{Ycorrection} and~\ref{Gcorrection}.
This accounts for the effect of damage parameters on the elastic moduli (Y \& G).
\begin{eqnarray}
  Y & = & Y\ \left(1-4\:V_v\right)\label{Ycorrection}\\
  G & = & G \left(1-V_v\:F \right)\label{Gcorrection} \\
  F & = & 15\ \left(\frac{1-\nu}{7-5\:\nu}\right) 
\end{eqnarray}
$Y$ and $G$ that appear in RHS of the above equations are the yield strength and shear
modulus respectively that are calculated from Steingberg-Guinan dynamic strength
model (eqn.~\ref{SGY} \&~\ref{SGG}) and $\nu$ is the Poisson's ratio of the material.
%In our simulation, the G and Y values calculated from the SG model are used
%in place of $G_0$ and $Y_0$ in Eqs.(11-13).
The average pressure in a computational cell is also affected by the void
volume hence it is also modified accordingly.
\vspace{0.2in}\\
{\bf Equation of State (EOS):}\\
Equation of state relates the thermodynamic parameters (i) Pressure ($P$),
(ii) Density ($\rho$), (iii) Energy ($E$) and (iv) Temperature ($T$).
A simple equation of state is a polynomial expression in which the pressure
depends only upon the density, $P(\rho)$. According to
Wilkins~\cite{wilkinsMethodsInCompPhys}, it can be expressed conveniently as:
\begin{equation}
P\left(\frac{\rho}{\rho_0}\right) = a\left(\frac{\rho}{\rho_0} -1\right)+b{\left(\frac{\rho}{\rho_0} -1\right)}^2+c{\left(\frac{\rho}{\rho_0} -1\right)}^3 
\label{pofetawilklinseos}
\end{equation}
Here, $\rho$ is instantaneous density and $\rho_0$ is  
normal density of the material. The constants `a, b, and c' are material dependent.
Winkins in ref.~\cite{wilkinsMethodsInCompPhys} shown that $P\left(\frac{\rho}{\rho_0}\right)$ + (2/3)$Y^0$ reproduces the
Hugoniot for shocks above the elastic limit.

\subsection
[Validation of hydrodynamic method]
{Validation of hydrodynamic method with realistic EOS, dynamic strength \& fracture model:
%%Change made in Nov'23 'Al' is new inplace of 'Aluminum' below
% Aluminum plate impact system} \% changed to Al as written below.
 Al plate impact system}
\label{secAlpp}
One dimensional hydrodynamic model is validated by simulating a metallic plate impact system.
The setup consists of an aluminum (Al) flyer plate striking an aluminum (Al) target material
that is kept initially at rest.
%Materials parameters give in equation~\ref{constAl} are used for Al.
For aluminum EOS, pressure (in megabar) is calculated using eqn.~\ref{pofetawilklinseos}
with the values $a=0.73$, $b=1.72$ and $c=0.4$ in megabar ($10^{11}$ $N/m^2$).
The yield strength $Y_0=2.976\times 10^8$ $N/m^2$, shear modulus
$G_0=248\times 10^8$ $N/m^2$. Normal density $\rho_0 = 2700$ $kg/m^3$.
The thickness of the flyer plate is 5 mm and that of the target aluminum
is 45 mm. The other two dimensions are assumed to be infinite.
This ensures one-dimensional plane wave shock is launched upon the flyer
impacting the target. Hugoniot jump conditions are valid only for
the principle, one-dimensional shock without any reflection. The flyer plate velocities used
in the validation are 800 m/s and 2000 m/s.

\begin{figure}[ht!]
\label{Fig4}
\includegraphics[width=0.9\textwidth]{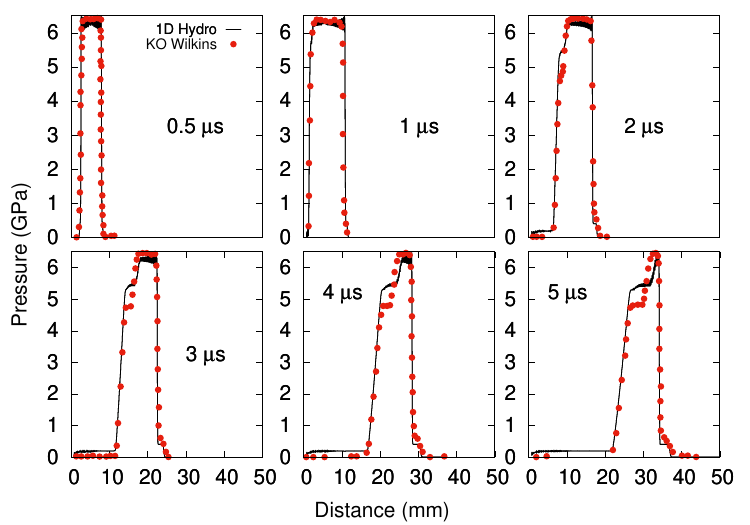}
\caption
[Pressure vs Shock propagation distance for flyer plate velocity 800 m/s]
{Pressure vs Shock propagation distance for flyer plate velocity 800 m/s.
Continuous lines are from the 1D Hydrodynamic code developed in-house;
Points are from KO code by Wilkins~\cite{wilkinsMethodsInCompPhys}
}
\label{800mpersecAlImpactKO}
\end{figure}

\begin{figure}[ht!]
\label{Fig6}
\includegraphics[width=0.9\textwidth]{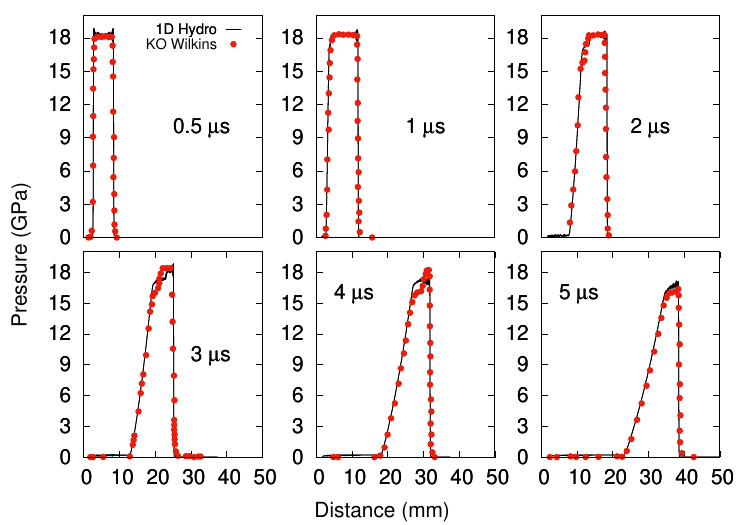}
\caption
[Pressure vs Shock propagation distance for flyer plate velocity 2000 m/s]
{Pressure vs Shock propagation distance for flyer plate velocity 2000 m/s.
Continuous lines are from the 1D Hydrodynamic code developed in-house;
Points are from KO code by Wilkins~\cite{wilkinsMethodsInCompPhys}
}
\label{2000mpersecAlImpactKO}
\end{figure}

%One dimensional hydrodynamic (1D-HD) model developed in-house is validated
%with aluminum plate-impact simulation.
The results are shown in
figures~\ref{800mpersecAlImpactKO} and~\ref{2000mpersecAlImpactKO}.
The spatial pressure profile at various times is shown in these figures.
The shock-wave structure at any instant of time can be obtained
from these simulations, as presented in figures~\ref{800mpersecAlImpactKO}
and~\ref{2000mpersecAlImpactKO}, for the impact velocities 800 m/s \& 2000 m/s
respectively. Elastic pre-cursor appears in the shock structure for lower
impact velocity (800 m/s) as seen in Figure~\ref{800mpersecAlImpactKO}.
% shows the presence of elastic pre-cursor at lower impact-velocity (800 m/s),
While for the impact velocity of 1200 m/s, the elastic pre-cursor is overtaken
by the shock wave due to higher shock propagation velocity. Hence it does not
appear in figure~\ref{2000mpersecAlImpactKO}. For comparison, results published
by Wilkin's using KO code~\cite{wilkinsMethodsInCompPhys} are also shown with
points, for two impact velocities viz. 800 m/s and 2000 m/s. This is a qualitative
measure of the validation.


\begin{figure}
     \centering
	\subfigure[$U_s$-$U_p$ Hugoniot for Al from 1D-HD]
            {
         \includegraphics[width=0.495\textwidth]{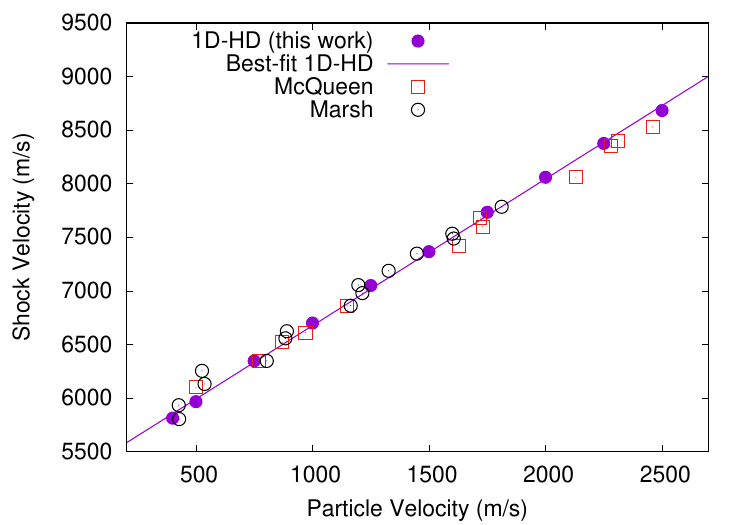}
                \label{1dHDValidUsUp}
            }
	\subfigure[$P$-$V$ Hugoniot for Al from 1D-HD]
            {
         \includegraphics[width=0.495\textwidth]{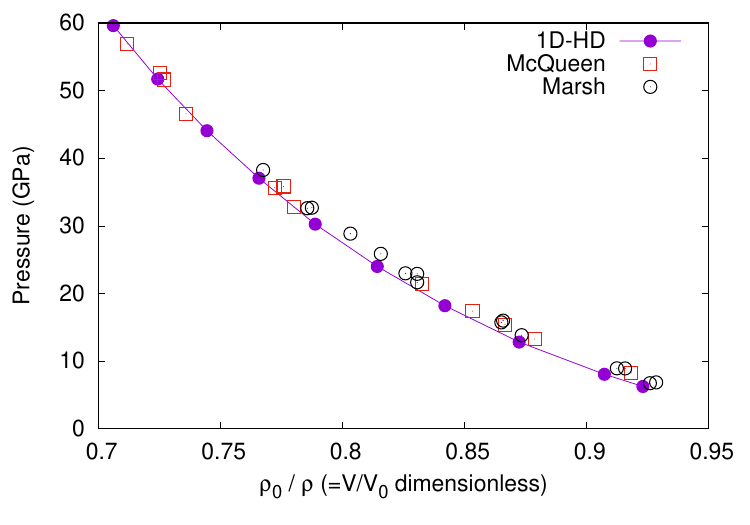}
                \label{1dHDValidVolPrs}
            }
	\subfigure[$P$-$U_p$ Hugoniot for Al from 1D-HD]
            {
         \includegraphics[width=0.495\textwidth]{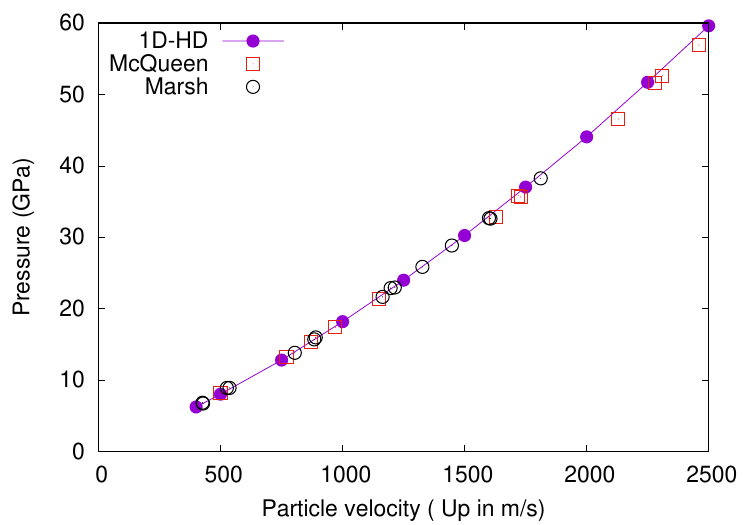}
               \label{usupNi}
                \label{1dHDValidUsPrs}
            }
 \caption[Validation of Hugoniot calculated from 1D-Hydrodynamic  for Al]
 {Validation of Hugoniot calculated from 1D-Hydrodynamic code for Al : Comparing the simulations
results with the experimental work of McQueen~\cite{mcqueen1970equation} and Marsh~\cite{lasl_shockdata}}
     \label{fig:1dHDvalidation}
%%%%%%%%%%%
\end{figure}

\begin{figure}[ht!]
\centering
%\includegraphics[scale=0.5, natwidth=640, natheight=480]{FlowChart/FlowChartMSdrawio.jpg}
%\includegraphics[scale=0.33, natwidth=100, natheight=200]{FlowChart/FlowChartMSdrawioDetailed.jpg}
%\includegraphics[scale=0.31, natwidth=100, natheight=200]{FlowChart/FlowChartMSdrawioDetailed.jpg}
%{\includegraphics[bb=50 50 400 100, scale=0.40]{FlowChart/FlowChartMSdrawioDetailed.jpg}}
%{\includegraphics[bb=10 1200 20cm 10cm, scale=0.31]{FlowChart/FlowChartMSdrawioDetailed.jpg}}
%{\includegraphics[bb=0 0 20cm 10cm, scale=0.31]{FlowChart/FlowChartMSdrawioDetailed.jpg}}
%\includegraphics[scale=0.31,width=\textwidth,height=0.7\textheight]{FlowChart/FlowChartMSdrawioDetailed.jpg}
\includegraphics[bb=0 0 1500 1710, scale=0.3]{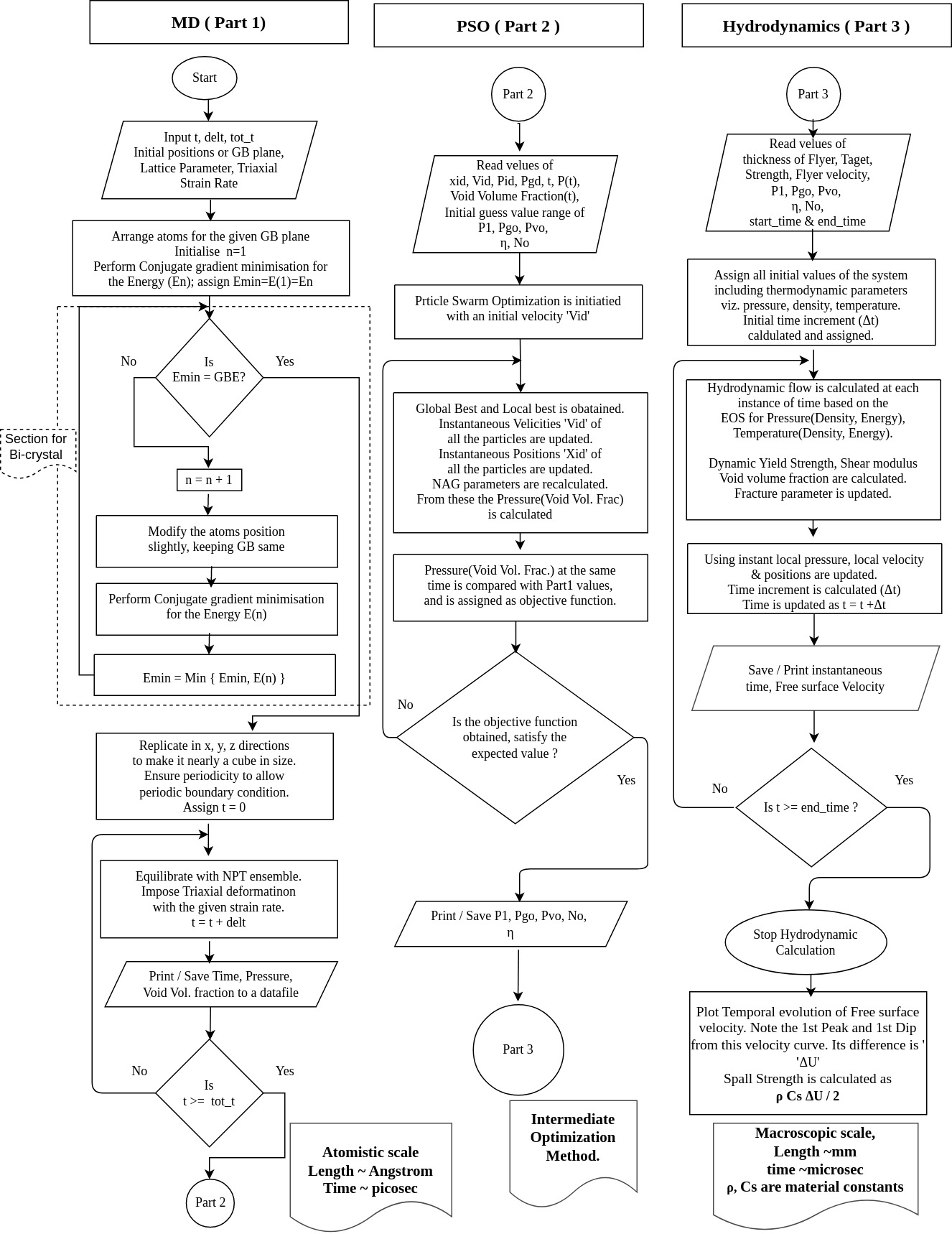}
\caption{Flowchart- Sequential steps in the Multiscale modelling: (1) MD $\longrightarrow$ (2) PSO $\longrightarrow$ (3) HD}
\label{subfig:StepsInMSModel}
\end{figure}

Quantitative validation of 1D-HD code is carried out by calculating the shock-Hugoniot
for aluminum in a plate-impact arrangement. Shock-Hugoniot is a well-established set of
relations based on mass, momentum and energy conservation principles. A brief note on
shock Hugoniot is presented in chapter~\ref{chapter_SCImpactShock}. However, it is
worth mentioning here that the three shock Hugoniot relations namely,
(i) Shock-Particle velocity ($U_s$-$U_p$ plane), (ii) Pressure-Volume ($P$-$V$ plane) and
(iii) Pressure-Particle velocity ($P$-$U_p$), are often used to validate shock propagation
in a material. Data obtained from 1D-HD calculation are presented pictorially in
figures~\ref{1dHDValidUsUp}, ~\ref{1dHDValidVolPrs} and ~\ref{1dHDValidUsPrs}.
These three figures represent the shock-Hugoniot in (i) $U_s$-$U_p$, (ii) $P$-$V$ 
and (iii) $P$-$U_p$ planes, respectively. Shock velocity, pressure, volume and
density for various particle velocities ($U_p$) that are calculated from 1D-HD
are quantitatively compared with experimental data~\cite{mcqueen1970equation, lasl_shockdata}.
The agreement observed from the comparison in figs.~\ref{1dHDValidUsUp}-~\ref{1dHDValidUsPrs},
qualifies the veracity of 1D-HD model. The flow of the multiscale method is summarised and is
shown pictorially in figure~\ref{subfig:StepsInMSModel}.

%\newpage

\clearchapter

% Shock : Chapter 3. Us-Up and PV Hugoniot calculations for single crystal, FCC: Al, Cu and Ni : JDBM-2022 publication
  \chapter
%[Shock Hugoniot for FCC single crystals]
{Shock-Hugoniot calculation for single crystal, FCC: Al, Cu and Ni\label{chapter_SCImpactShock}}
\nopagebreak
\makeatletter
{\renewcommand{\@makefntext}[1]{\noindent #1}%
\footnotetext{%
  \fontsize{10pt}{12pt}\selectfont%
  The content of this chapter is fully adapted from the author's published work~\cite{UsUp_Madhavan2022}. Reproduced with permission from Springer Nature. A note on \copyright\ is available in page number~\pageref{copyrightNotePage}.}}
\makeatother

\label{sec-1}
\section{Background of the problem~\label{sec1-Chap3}}
When a flyer hits a target with a high velocity, a shock-wave is generated. Shock propagation
in materials creates spatial discontinuity in the thermodynamic parameters such as
density ({$\rho$), temperature (T), pressure (P), etc.  In the shocked region
the value of these parameters are suddenly increased, while the themodynamic
status of the unshocked region remains at the normal state. A shockwave, when
created by a planar impact, is generally plane-wave in nature (plane-wave shocks)
and it introduces one-dimensional strain in the materials as it propagates.
The interface between the shocked and unshocked region is very thin. Hence, spatially
the material undergoes a sudden jump in the values of thermodynamic parameters.
The speed at which this interface moves is the shock velocity ($U_s$).
The thermodyamics parameters between the shocked and unshocked regions
are related by Rankine-Hugoniot~\cite{zel2002physics} equations (RH eqn.).
The pressure, energy in the shocked region are generally denoted by ${P_H}$,
${E_H}$~\cite{hemp_wilkins, meyersBook, alcuta_usup_Mitchell, alcu_high_prs_eos}.
The mathematical relations that connect the thermodynamic parameters between
the shocked ($\rho$, $P_H$, $E_H$) and unshocked ($\rho_0$, $P_0$, $E_0$) regions
are easily derived from the mass, momentum and energy conservation
principles~\cite{zel2002physics, hemp_wilkins, meyersBook}. The equations often
used in shock dynamics are (i) Rankine-Hugoniot equations (R-H),
(ii) the form of Equation of State (EOS) and (iii) the relationship
between shock and particle velocities. These are given by,
\begin{eqnarray}
\rho(U_s-U_p) & = & \rho_0(U_s-u_0) \label{rho_hugo} \\
P_H-P_0 & = & \rho_0 (U_s-u_0) (U_p-u_0) \label{press_hugo} \\
E_H-E_0 & = & \left[(P_H+P_0)(V_0-V)\right]/2 \label{energy_hugo}\\
P & = &  P(\rho, T) \label{Peos}\\
E & = & E(\rho, T) \label{Eeos} \\
U_s & = & C + S \times U_p \label{us_up}
\end{eqnarray}
where, $P$, $E$, $\rho$, $V$  etc are the pressure, specific internal energy, density 
and specific volume ($1/\rho$) of the system. The subscripts `0' and `$s$'
represent the initial and the shocked states respectively, while `H' represents
the Hugoniot state. $U_s$ represents the
shock velocity, while $u_0$ and $U_p$ are the particle velocities of unshocked
and shocked regions.  $S$ and $C$ can be obtained by fitting experimental 
%, theoritical or computed
shock data ($U_s$, $U_p$). The equations~\ref{rho_hugo}-~\ref{us_up} form a complete set
to represent shocked-state in a material under dynamic shock-compression.
%There are five shock parameters $P$, $E$, $\rho$ (or $1/V$), $U_s$ and $U_p$.
%With two EOS equations~\ref{Peos} \&~\ref{Eeos}, these shock parameters
%can be separated into 10 pairs. These in turn provide 20 equations
When $u_0$ and $P_0$ are very small compared to $U_p$ and $P_H$ respectively,
%as in the case of flyer impact shock dynamics,
they can be neglected.
Hence the shocked pressure ($P_H$) and the specific volume
($V$=$1/\rho$) relation is reduced to eqn.~\ref{PhofV}, using eqns.~\ref{rho_hugo},~\ref{press_hugo}
and~\ref{us_up}. From eqns.~\ref{press_hugo} and \ref{us_up}, the relation between
$P_H$ and $U_p$ is obtained (eqn.~\ref{PhofUp}).
\begin{eqnarray}
P_H & = & \frac{{C}^2(V_0-V)}{[V_0-S(V_0-V)]^2} \label{PhofV} \\
P_H & = & {\rho_0}(C U_p + S{U_p}^2) \label{PhofUp}
\end{eqnarray}
The relations represented by eqns.~\ref{us_up},~\ref{PhofV} and~\ref{PhofUp} are
often used to show the shocked-state in $U_s$-$U_p$, $P_H$-$V$ and $P_H$-$U_p$ planes, respectively.

\subsection
%[Motivation of studying shock propagation in single crytals]
{Motivation of studying shock propagation in single crytals~\label{sec1.1-Chap3}}
%{Brief history of the earlier research in dynamic shock propagation associated with single crytals~\label{sec1.1-Chap3}}
Steinhauser {\it et al.}, have covered various computational methods
methods viz. Molecular Dynamics (MD), Hydro-Dynamics (HD) and First Principles (FP)
to study shock wave phenomena in a review paper~\cite{Martin}.
Modeling of shock wave propagation at atomic scales is reported in the literature
using MD simulations by many
researchers~\cite{brad, bringa, Ni_md_Jarmakani, Ni_md_Liu_Hai, Ni_md_Choi, al_md_shock_melting, md_shockcompress_metals_confProc}.
MD simulations can be performed to  calculate the Hugoniot-shock parameters
(eqns.~\ref{rho_hugo}-\ref{us_up}) from a flyer-plate-taget impact system.
From MD impact-shock simulations with different impact velocities, corresponding
shock velocities are calculated.
Thus we get $U_s$-$U_p$ data.
%If the flyer and the target materials are same,
%due to their impedance-match ($\rho_0 U_s$), the particle velocity is calculated
%as half of the flyer or impact velocity~\cite{meyersBook}.
Using eqn.~\ref{us_up}, a best fit can be made to determine the values, `$C$' and `$S$'
of the target material.

Jarmakani {\it et al.}~\cite{Ni_md_Jarmakani}, Liu Hai {\it et al.}~\cite{Ni_md_Liu_Hai} and
Choi {\it et al.}~\cite{Ni_md_Choi} have calculated the variation of
$U_s-U_p$ in nickel using MD methods. The former two have used different EAM potentials while
Choi {\it et al.} have demostrated it for four other EAM potentials. Comparison of their
MD results with experiments, revealed different levels of deviation. Specifically for Ni,
MD and experimental data at high pressure shock-compression region are limited. Among the available
shock-data, MD results for Ni are found to deviate substantially from the experiments.
Thus shock data of single crystal Ni at higher compression requires a systematic study.
%Here, first principles and molecular dynamics methods are carried out to explore
%the static and dynamical response of FCC single crystals at these pressures.

First principles calculations based on the density functional 
theory are also useful in estimating various material properties in shocked and 
unshocked conditions~\cite{eos_alcu_etc, eos1stprins_Joshi, vinayak_eos_papers}.
%The cold contribution of EOS of crystalline solids can be calculated
%directly from first principles.
%Pressure induced structural transformations which can change the cold EOS
%significantly, can also be investigated by using first principles
%calculations~\cite{balvin, arora, jiang}. High temperature 
%EOS contributions of ions and electrons can be added separately using available 
%theoretical models in the literature to get the total EOS~\cite{more}.
MD simulations are carried out in this work, to obtain the coefficients
that appear in $U_s$-$U_p$ relations (eqn.\ref{us_up}) for single crystals of Al and Cu.
For Ni, as the shock-data results from earlier MD simulation~\cite{Ni_md_Liu_Hai,
Ni_md_Jarmakani, Ni_md_Choi} were found to deviate with
experiments, it is revisited using another potential~\cite{CuNi_eam}.
EAM potentials are used by the MD method, whereas FP method calculates EOS and uses them
in eqns.~\ref{rho_hugo}-\ref{us_up}. $P$-$V$ isotherms are also generated by these two
methods and they are compared with the results from diamond anvil cell (DAC) experiments.
%High strain rate shock simulations are carried out in metallic single crystals,
%which take them to extreme thermodynamic state of pressure (Mbar regime).
The computational details are given in section~\ref{sec-2}. The results are discussed
in section \ref{sec-3} and concluding remarks are presented in section~\ref{sec-4}.

\section{Computational aspects\label{sec-2}}
\subsection{Calculation of P-V isotherm}
P-V isotherm is the variation of pressure as a function of volume at a constant
temperature.  To obtain P-V isotherm, it is necessary to initially obtain the
equilibrium configuration at room temperature with zero pressure.
Then in the next step, the number of atoms
are kept the same and the lattice parameter is decreased slightly. This corresponds
to a reduced volume than the previous step. The pressure value at this reduced volume
is calculated. Subsequently in each step of the reduced volume, the pressure
is calculated to get P-V isotherm curve.

\subsubsection{MD simulations for the P-V Isotherm:}
\label{sec-2z}
MD simulations are carried out for an isothermal-isobaric (NPT) ensemble, using
$20\times20\times20$ unit cells containing 32000 atoms with 0 bar pressure at 300 K.
%Cubes with sides 80.9, 72.3 \& 70 $\AA$ for Cu, Al \& Ni are constructed accordingly.
%This step is executed separately for the individual Cu, Al and Ni single crystals.
All the boundaries are constrained by
%periodic boundary conditions (PBCs). With a time-step of 1 femtosec, the 
\ac{PBC}. With a time-step of 1 femtosec, the 
simulations are carried out for 10000 steps. The system typically attains 
equilibrium by 6 ps. Subsequently the system is subjected to
a canonical ensemble with constant \ac{NVT} under PBC. In this case also, a time step
of 1 femtosec is used. The simulation is carried out for 10 ps.
Time avaraged pressure of the system is recorded between
the duration 6 ps to 10 ps. In these simulations the volumes were
reduced to 66\%, 65\% and 72\% of their
initial volumes at zero pressure for Cu, Al \& 
Ni respectively, to cover the pressure range from zero to 135 GPa (1.35 Mbar) for Cu,
85 GPa (0.85 Mbar) for Al and 105 GPa (1.05 Mbar) for Ni.

\subsection{Computation of $U_s$-$U_p$ relation:}
Flyer-target system as shown schematically in figure~\ref{Chap1_2dview},
sec.~\ref{sec1.1} of chapter~\ref{chapter_Intro} is used in MD simulation
to determine the variation of $U_s$ as a function of $U_p$.

\subsubsection{MD Simulations of Shock Propagation due to Flyer Impact:}
The LAMMPS code \cite{lammps} is used to carry out the impact-shock simulations
in FCC single crystals (Al, Cu \& Ni). The EAM potential used for Cu and Ni,
is from the literature work of Foiles {\it et al.}~\cite{CuNi_eam}. For aluminum
the work from Jacobsen {\it et al.}~\cite{Al_eam} is used. The potentials are validated at
normal temperature and pressure by calculating the bulk modulus, shear modulus,
Young's modulus, Poisson's ratio , vacancy formation energy and cohesive energy
for Al, Cu and Ni as listed in table~\ref{elasticTable}
(Chap.~{\ref{chapter_compMethods}, sec.~\ref{EAMValidatedChap2.1.1}).
%shows the calculated values for the various EAM potentials used in this research work
%for Al, Cu, Mo and Nb.
This table also lists the experimental values for comparison.

%After confirming the consistancy to get the same values with in
Plate impact MD simulations were first carried with an impact velocity
2 km/s on three different computational-domain-sizes viz.
$99\times40\times40$, $99\times30\times30$ \&
$99\times20\times20$ unit cells each along X$\times$Y$\times$Z directions
to check the finite size effects.
Observed shock propagation, calculated shock velocity, pressure etc.,
do not vary significantly for these three sets.
After these trial simulations, computational domain with 99$\times$20$\times$20 unit cells,
along X$\times$Y$\times$Z directions is generated for the main simulation. This size
accommodates nealy 158400 atoms. Also, the number of unit cells
in the transverse directions are chosen to eliminate finite size effects
following Rawat {\it et al.}~\cite{rawat1}. The lengths of the simulation
box along the shock propagation direction are 358 $\AA$, 402 $\AA$
\& 346 $\AA$ for Cu, Al \& Ni respectively. The size along X-direction
is much longer than Y and Z dimensions. This enables the shock to propagate
along $<$100$>$ crystal direction. The simulation domain along the X-axis
is split into two parts by a ratio 3:7, with the smaller fraction
representing the flyer and the larger fraction representing the target.
The atoms in the flyer are assigned an initial velocity along X-direction
corresponding to the flyer velocity. Periodic boundary conditions are
imposed along the Y and Z axes, while the two boundaries on the X axis
are free to move. The period boundary conditions ensure that no reflection
appears from these boundaries.
%The pressure obtained from all the three
%sets of trials runs with different number of unit cells, as mentioned above, are similar. Also
No atom or energy or momentum is lost in the
simulations.
%Thus calculated pressure (eqn.10) by means of one dimensional
%bins along shock propagation direction (X) follows the governing laws and is
%not affected.
For these impact simulations, the number of atoms and
total energy are constant, corresponding to a micro-canonical ensemble.
The time-step for the simulations is chosen to be 0.1 femtosec to sample
the interatomic potential with a sufficient resolution ~\cite{rawat1}.
Flyer velocities in the range 1000-6000 m/s in steps of 500 m/s for Cu,
Al and Ni are used.

\subsubsection{Calculation of the shock velocity from the MD simulations:}
\label{sec-2b}
The target region is divided into bins of 3 unit cells each along the X 
direction (i.e direction of shock propagation). The time and space averaged
pressure in each bin is monitored over 100 time steps. After the impact,
the pressure wave travels in opposite directions from the flyer-target
interface, along the X-axis and
builds up to a shock wave in both, the flyer and the target. These shock
waves reflect from the free surfaces as rarefaction waves. The pressure in
each bin is obtained from the space and time average of the diagonal
components of the stress tensor given by,
\begin{equation}
\label{press_stress}
P=-\frac{1}{3}{\left[\sigma_{xx}+\sigma_{yy}+\sigma_{zz}\right]}
\end{equation}

The spatial pressure profile at different times is shown in figure 
\ref{shock_calculation_prs_profile} for Cu crystal with a flyer velocity of 3.0
km/s. The corresponding particle velocity is 1.5 km/s. The shock velocity is 
calculated from this figure using the distance moved by the shock front 
in a given time. The uncertainty in the calculated shock velocity,
($\Delta{U}$) is estimated from~\cite{velocity_error_estimation},
\begin{eqnarray} 
\Delta{U} & = & \left( \frac{\Delta{L}}{L} + \frac{\Delta{t}}{t} \right) U 
\label{error_velocity}
\end{eqnarray}
where `$L$' is the total distance the shock travelled in a time `$t$' and 
$\Delta{L}$ is half of the bin size that is used to calculate the spatially 
averaged pressure. $\Delta{t}$ is the uncertainty involved in the time 
calculation and is given by the number of MD time-steps used in the time 
averaging. In our MD simulations, $\Delta{L}/{L}$ is $\sim 3\%$ and 
$\Delta{t}/{t}$ is $\sim 1\%$. Thus for various flyer velocities $U_f$, the 
corresponding particle $U_p$ and shock $U_s$ velocities are calculated.
The particle velocity $U_p$ in the target is indeed verified to be half
the value of the flyer velocity $U_f$.

\begin{figure}[h]
\centering
\includegraphics[width=3.5in]{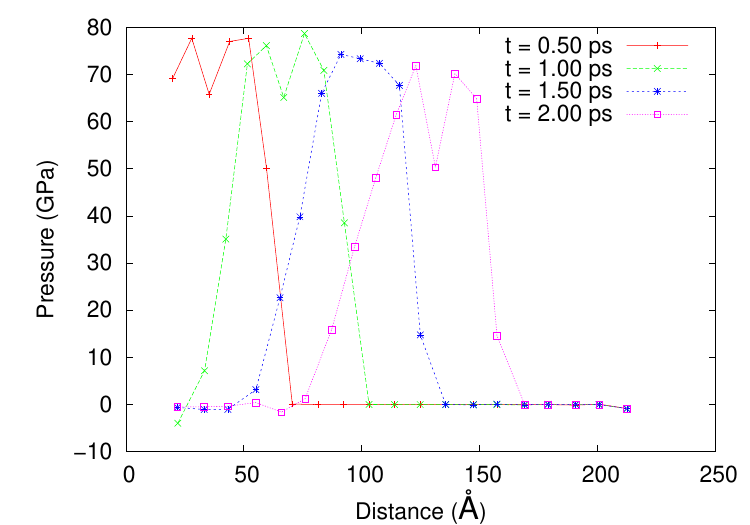}
%\caption{Spatial Pressure Profile along X-direction for Cu flyer with an impact 
%velocity of 3.0 km/sec. This shows Shock wave position at different times}
\caption[Spatial pressure profile for Cu single crystal under shock loading]
{Spatial Pressure Profile along X-direction for Cu flyer with an impact 
velocity of 3.0 km/sec. This shows Shock wave position at different times.
 (Fig:Ref.~\cite{UsUp_Madhavan2022})}
\label{shock_calculation_prs_profile}
\end{figure}

\section{Results and Discussion}
\label{sec-3}
The results from MD calculations to obtain the $U_s-U_p$ shock
Rankine-Hugoniot (RH) parameters are presented first. Then the results for the
variation in pressure as a function of volume at both, isothermal and shocked
states are presented. This is followed by a discussion on the variation of results
among various EAM potentials for Ni, when compared to the experimental results,
at extreme pressure states.

\subsection[Comparison of $U_s$-$U_p$: MD, FP \& Expt.]{Comparison of shock parameters in $U_s$-$U_p$, that are calculated 
from MD with First principle calculations and with Experimental results:}
\label{sec-3b}
Shock parameters are obtained for single crystal copper, aluminum and nickel by 
performing MD simulations as described in Section \ref{sec-2b}.
The simulations yield the shock 
velocity ($U_s$) for different flyer velocities ($U_f$). Flyer impact
velocities simulated are in the range 1-6 km/s which correspond to particle velocities ($U_p$) 
of 0.5-3 km/s. A comparison of our simulated (MD) results with FP simulation data and with experimental
data is shown for $U_s-U_p$ in the figures~\ref{usupCu},~\ref{usupAl} 
and~\ref{usupNi} for single crystal Cu, Al and Ni respectively. Good agreement 
is observed with experimental results for Cu and Al.
For Ni,
%the first principles calculations match the experiment 
%values~\cite{meyersBook, lasl_shockdata} very well, but
the MD simulation 
results match~\cite{meyersBook, lasl_shockdata} with experimental results, only for $U_p\leq$ 1600 m/s.
 %% Multiplot 1
\begin{figure}
     \centering
\subfigure[Cu with Expt. results from Mitchell~\cite{alcuta_usup_Mitchell}]
            {
         \includegraphics[width=0.495\textwidth]{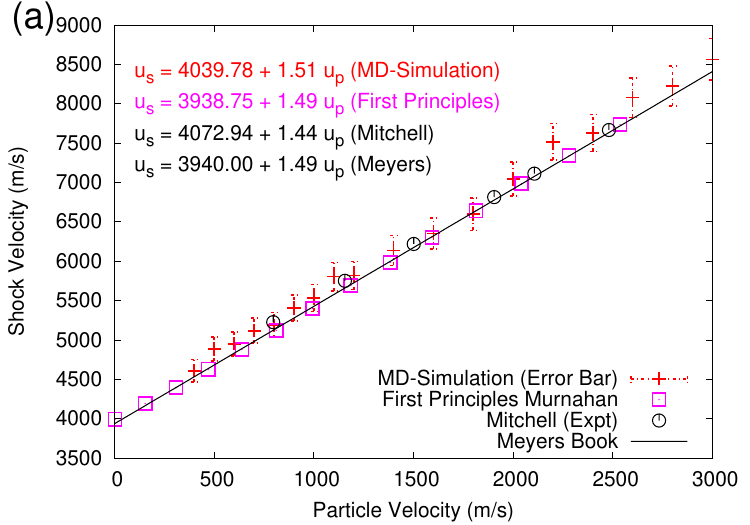}
	        \label{usupCu}
            }
\subfigure[Al with Expt. results from Mitchell~\cite{alcuta_usup_Mitchell}]
            {
         \includegraphics[width=0.495\textwidth]{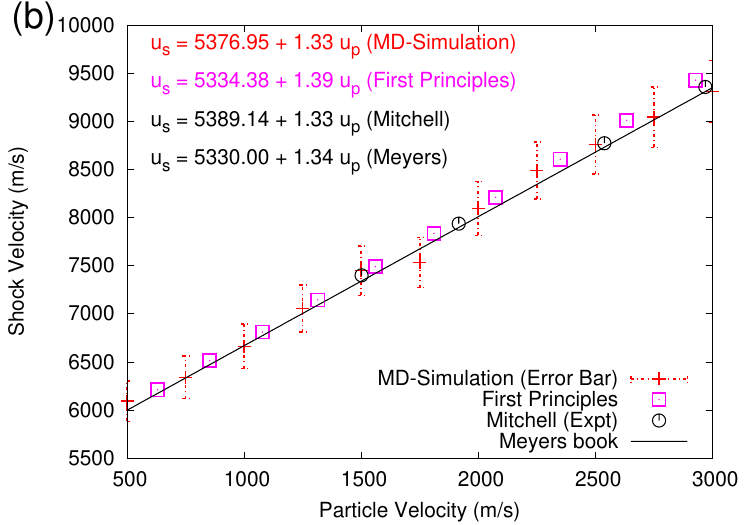}
	       \label{usupAl}
            }
\subfigure[Ni with Expt. results from Marsh~\cite{lasl_shockdata}]
            {
         \includegraphics[width=0.495\textwidth]{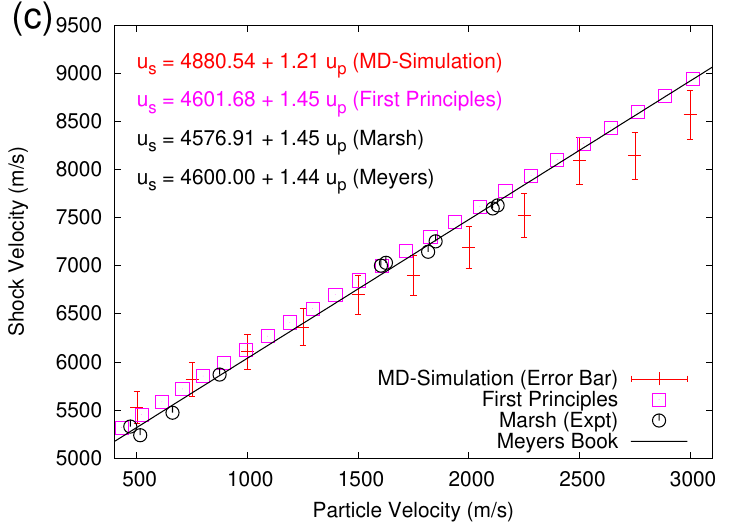}
	       \label{usupNi}
            }

%%     \begin{subfigure}[b]{0.60\textwidth}
%         \centering
%         \includegraphics[width=0.495\textwidth]{Figures/UsUp_Figs/fig4a_Comparison_Us_Up_Cu}
%%         \caption{Cu with Expt. results from Mitchell~\cite{alcuta_usup_Mitchell}}
%	        \label{usupCu}
%%     \end{subfigure}
%     
%%     \begin{subfigure}[b]{0.60\textwidth}
%         \centering
%         \includegraphics[width=0.495\textwidth]{Figures/UsUp_Figs/fig4b_Comparison_Us_Up_Al}
%%         \caption{Al with Expt. results from Mitchell~\cite{alcuta_usup_Mitchell}}
%	       \label{usupAl}
%%     \end{subfigure}
%
%%     \begin{subfigure}[b]{0.60\textwidth}
%         \centering
%         \includegraphics[width=0.495\textwidth]{Figures/UsUp_Figs/fig4c_Comparison_Us_Up_Ni}
%%         \caption{Ni with Expt. results from Marsh~\cite{lasl_shockdata}}
%	       \label{usupNi}
%%     \end{subfigure}
%%%%%%%%%%%%%
%     \caption{Shock velocity($U_s$) Vs Particle velocity($U_p$) data obtained 
     \caption[$U_s$ Vs $U_p$ from MD for Cu, Al \& Ni (single crystal)]
     {Shock velocity($U_s$) Vs Particle velocity($U_p$) data obtained 
     from MD are compared with the results of first principles~\cite{UsUp_Madhavan2022} and experiments
     for Cu (a), Al (b) \& Ni (c). MD results are shown with error bars to 
     represent the uncertainty involved in the shock velocity calculation. 
     $U_s-U_p$ relationship given in Meyers~\cite{meyersBook} is shown as a 
     continuous straight line.
 (Fig:Ref.~\cite{UsUp_Madhavan2022})}
     \label{fig:UsUp_hugo_cualni}
%%%%%%%%%%%
\end{figure}

A linear fit to the shock velocity as a function of the particle velocity,
yields the parameters $C$ and $S$, appearing in equation~\ref{us_up}. Fitted
$C$ and $S$ values from the MD simulation are listed in table-\ref{usup_table1}
along with first principles calculation and experimental results for comparison.

\begin{table}[!h]
\centering
%\caption{Shock Hugoniot parameters $C$ \& $S$ obtained from our first principle calculations (FP) \& MD
\caption
[Values of $C$, $S$ calculated from MD method for Cu, Al \& Ni]
{Shock Hugoniot parameters $C$ \& $S$ obtained from our MD
simulations for Cu, Al and Ni. These are compared with published results
(i) Meyers compilation of shock data ref.~\cite{meyersBook}(pg.133) and
(ii) Expt.(ref.~\cite{alcuta_usup_Mitchell} for Cu \& Al; ref.~\cite{lasl_shockdata} 
	for Ni) and (iii) First Principles (FP) ref.~\cite{UsUp_Madhavan2022}*}
\vspace{0.2cm}
    \label{usup_table1}
    \begin{tabularx}{0.92\textwidth}{@{    } *{9}{c} @{    } }
\toprule
\multirow{2}{*}{\parbox{1.2cm}{\centering{\bf Metal}}} & 
      \multicolumn{4}{c} {\bf $C$ (m/s)} &
      \multicolumn{4}{c} {\bf Slope, $S$  (dimensionless)} \\
\cmidrule(lr){2-5} \cmidrule(lr){6-9}
%& {\bf Ref~\cite{meyersBook}} &{\bf Expt.} & {\bf FP~\cite{UsUp_Madhavan2022}} &{\bf MD} & 
%  {\bf Ref~\cite{meyersBook}} &{\bf Expt.} & {\bf FP~\cite{UsUp_Madhavan2022}} & {\bf MD}\\
& {\bf Ref~\cite{meyersBook}} &{\bf Expt.} & {\bf FP*} &{\bf MD} & 
  {\bf Ref~\cite{meyersBook}} &{\bf Expt.} & {\bf FP*} & {\bf MD}\\
\midrule
    Cu    & 3940 & 4028 & 3939 & 4040 & 1.49 & 1.47 & 1.49 & 1.51 \\
    Al    & 5330 & 5426 & 5334 & 5777 & 1.34 & 1.32 & 1.39 & 1.33 \\
    Ni    & 4600 & 4577 & 4601 & 4881 & 1.44 & 1.45 & 1.45 & 1.21 \\
\bottomrule
    \end{tabularx}
\end{table}
It may be noted that in most of the cases in table~\ref{usup_table1}, the relative error 
between the MD simulation
and experiment is 
\textless 6\%. Only in the case of the slope ($S$) for Ni, the 
relative error in the MD simulation is 16\%.
%The relative error in the 
%first principles calculations in all cases are \textless 6\%.

\subsection[$P$-$V$ Isothermal \& Shocked States]{Results for the $P$-$V$ variation under isothermal and shocked 
conditions:}
\label{sec-3a}
In MD simulations, the $P$-$V$ variation is obtained by varying the lattice
parameter from its equilibrium value as described in sec.~\ref{sec-2z}.

The {\it P-V} data for the metals Cu, Al and Ni from MD are compared with first-principles
calculations~\cite{UsUp_Madhavan2022} and with diamond anvil cell (DAC) 
experiments~\cite{Cu_vol_prs_phyRewB70, Ni_vol_prs_phyRewB78,Cu_vol_prs_phyRewlet}
in figs.~\ref{fig:PVIsot_Cu}, \ref{fig:PVIsot_Al} and \ref{fig:PVIsot_Ni} respectively.
These simulations have been performed for the specific volumes compressed upto approximately 1.5
times the equilibrium specific volume.
%The simulations are compared with the
%experimental results of P-V isotherms obtained from diamond anvil cell (DAC)
%experiments~\cite{Cu_vol_prs_phyRewB70, Ni_vol_prs_phyRewB78,Cu_vol_prs_phyRewlet}.

Note from Fig.~\ref{fig:PVIsot_Al} that the experimental results of Akahama
{\it et al.} ~\cite{Cu_vol_prs_phyRewlet} and Dewaele {\it et al.} 
~\cite{Cu_vol_prs_phyRewB70}, slightly deviate with each other for higher 
pressures. The MD results for aluminum are in good agreement with FP
simulations~\cite{UsUp_Madhavan2022} and the experimental results of Akahama {\it et al.} 
~\cite{Cu_vol_prs_phyRewlet}. The results  for aluminum slightly deviate with 
the experimental results of Dewaele {\it et al.}~\cite{Cu_vol_prs_phyRewB70} at 
higher pressure. This is reasonable as the two experimental results themselves 
have similar amount of deviation from each other. Similarly, in the case of Ni 
also our results are in reasonable agreement with the experimental results of 
Dewaele {\it et al.}~\cite{Ni_vol_prs_phyRewB78}. In the case of Cu, our MD results 
are in good agreement with the experimental results.
% whereas our first principles
%calculations results show some deviation from the experimental results. The 
%deviation increases with pressure. This is because of the exchange correlation 
%function GGA, used in our first-principles calculations.
%Several types of exchange correlations potentials are available in VASP -
%such as local density approximation (LDA), generalized gradient approximation (GGA)
%and meta-GGA etc. We have used PBE parameterization of GGA in these calculations as
%it is generally considered  very accurate in predicting equilibrium volume and bulk
%modulus for most of the materials, as can be verified in~\cite{MISHRA2013509}
%and the references therein. However, in the case of Cu neither
%LDA nor GGA is very accurate, LDA under-estimates while GGA over-estimates the
%equilibrium volume of Cu at ambient conditions~\cite{PhysRevB.65.064302}.
%This over estimation can increase at higher pressures as the 
%incompressibility increases with pressure.

%%%%%%%%%%%% P-V Hugoniot %%%%%%%%%%%
% Section for PV isotherm \& hugoniot Multiplot 2 :
\begin{figure}
     \centering
\subfigure[PV Isotherm - Cu]
            {
         \includegraphics[width=0.47\textwidth]{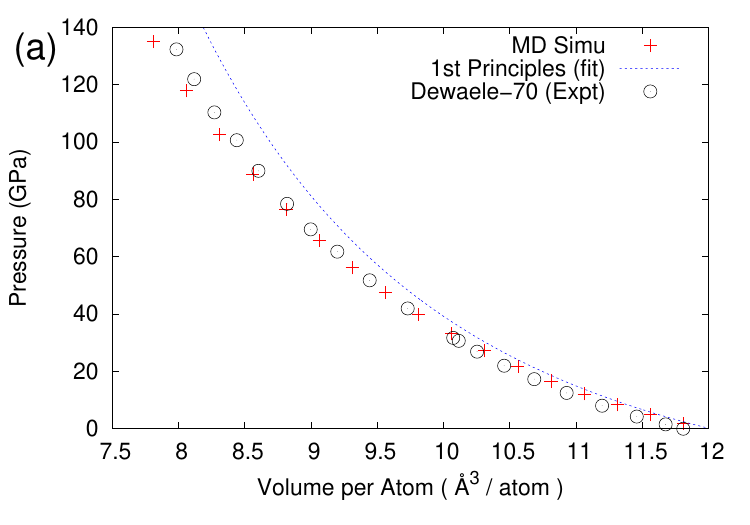}
         \label{fig:PVIsot_Cu}
            }
\subfigure[PV Hugoniot - Cu]
            {
         \includegraphics[width=0.47\textwidth]{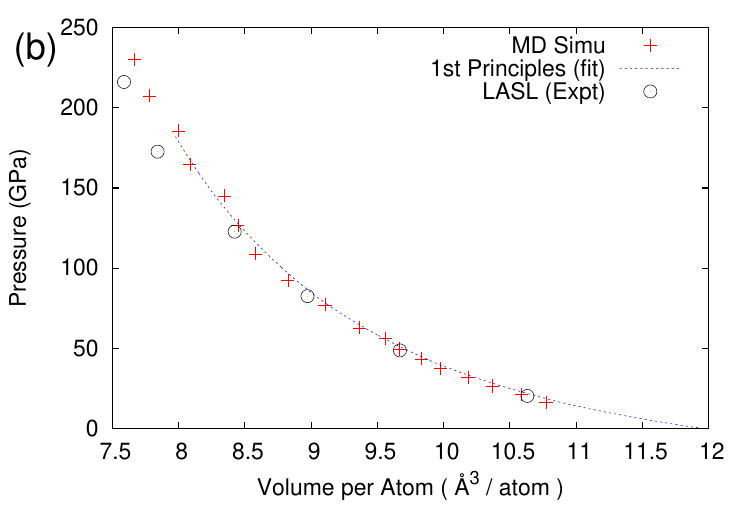}
         \label{fig:PVHugo_Cu}
            }
\subfigure[PV Isotherm - Al]
            {
         \includegraphics[width=0.47\textwidth]{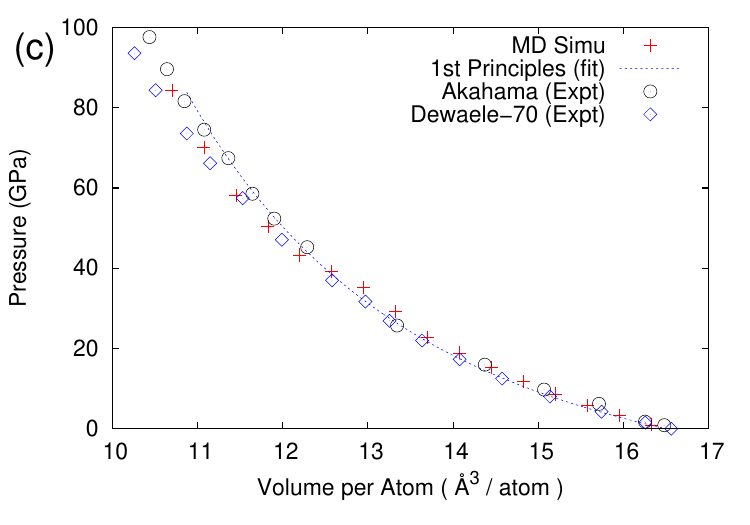}
         \label{fig:PVIsot_Al}
            }
\subfigure[PV Hugoniot - Al]
            {
         \includegraphics[width=0.47\textwidth]{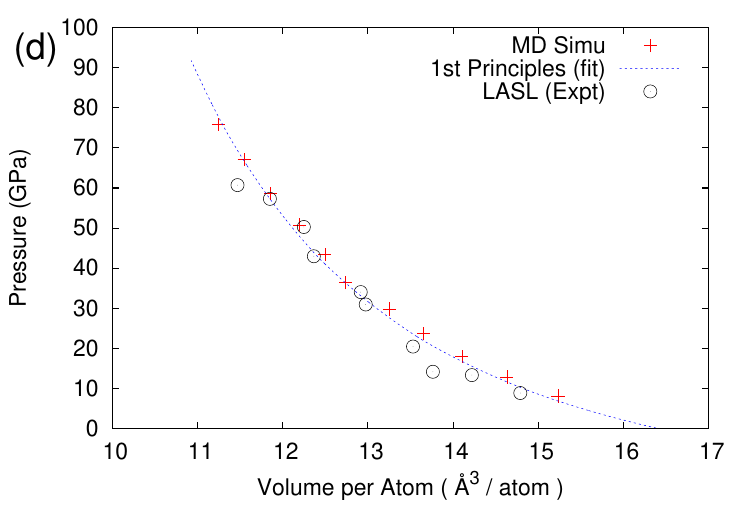}
         \label{fig:PVHugo_Al}
            }
\subfigure[PV Isotherm - Ni]
            {
         \includegraphics[width=0.47\textwidth]{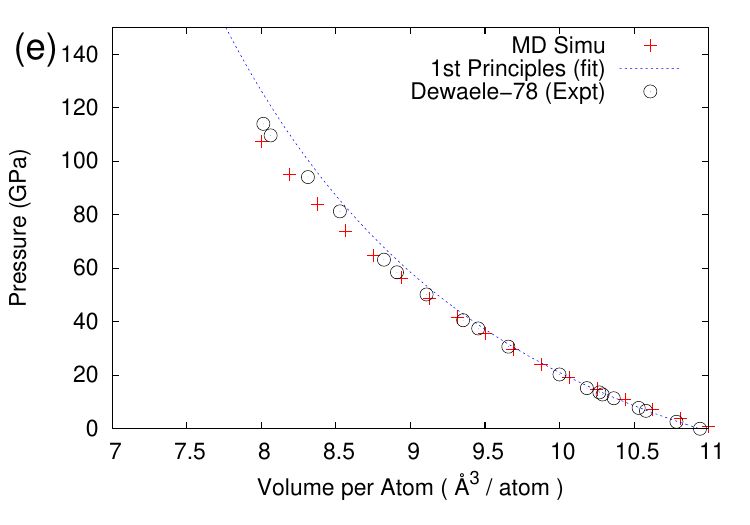}
         \label{fig:PVIsot_Ni}
            }
\subfigure[PV Hugoniot - Ni]
            {
         \includegraphics[width=0.48\textwidth]{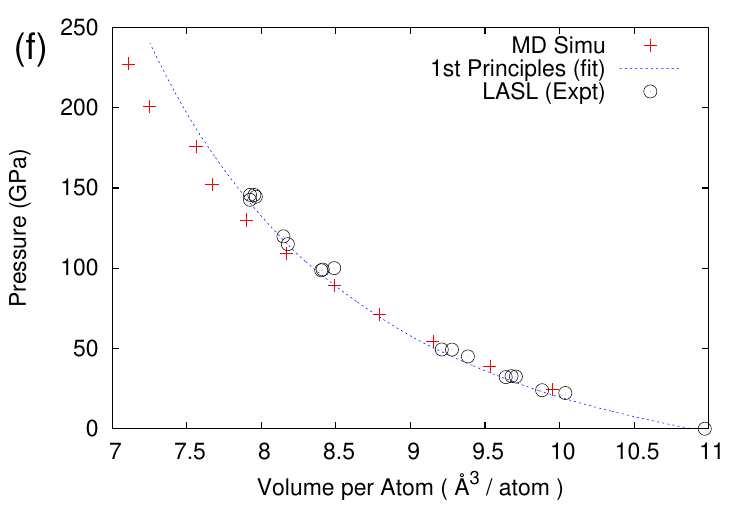}
         \label{fig:PVHugo_Ni}
            }

     \caption
[Comparison of P($\rho_0/\rho$): Isothermal and Shock-Hugoniot of Cu, Al \& Ni]
{Pressure vs Specific Volume obtained under Isothermal and 
     Dynamic Shock-Hugoniot States from MD calculation are compared with
     published results of first principles calculation and experiments;
     For isotherm states, the comparisons are made for Cu with Akahama Ref.\cite{Cu_vol_prs_phyRewlet},
     Cu \& Al with Dewaele(Vol.70) Ref.~\cite{Cu_vol_prs_phyRewB70}, Ni with Dewaele(Vol.78) 
     Ref.\cite{Ni_vol_prs_phyRewB78}.
     For shock-Hugoniot states, the comparisons are made for Cu, Al \& Ni with Marsh-LASL
     Shock Hugoniot data Ref.~\cite{lasl_shockdata}.
 (Fig:Ref.~\cite{UsUp_Madhavan2022})}
     \label{fig:PV_iso_hugo_cualni}
\end{figure}

\subsection{MD calculation of shock velocity and shock Hugoniot pressure}
From the results listed above, it is evident that the EAM potentials used for
Cu and Al are suitable for both, static and high strain rate MD simulations
even at a very high pressure regime, $230\ GPa\ (2.3\ Mbar)$ for Cu and
$100\ GPa\ (1\ Mbar)$ for Al. This is confirmed by the isotherm and shock
Hugoniot $PV$ relations represented in the figures \ref{fig:PVIsot_Cu},
\ref{fig:PVHugo_Cu},~\ref{fig:PVIsot_Al} \&~\ref{fig:PVHugo_Al} respectively.
%Hugoniot generated from the first principle calculations show very good match
%with experiments for single crystals of Cu \& Al.

Comparison of figures~\ref{fig:PVIsot_Ni} \&~\ref{fig:PVHugo_Ni} indicate that
in case of Ni the PV isotherm obtained from MD sumulations show
good match with FP simulations and with  the experiments. PV-shock Hugoniot
relation also shows good match for both MD \& FP simulations with experiments,
for the volume/atom range, 8 to 11 $\AA^3/atom$. At higher compression region
(volume/atom $<8\ \AA^3/atom$) MD results are found to deviate from the experiments,
%while FP calculations continue to match with experiments.
This deviation is consistent with the MD results for $U_s$-$U_p$ which is shown in
figure~\ref{usupNi}. The deviation corresponds to 16\% relative error in
the slope $S$, of the $U_s$-$U_p$ fitted straight line.

Root Mean Square (RMS) for $U_s$-$U_p$ data, is calculated to know
the dominant range of $U_p$ where the MD results show more deviation
in comparison to the experimental results.  
RMS = $\sqrt{\frac{1}{n}\sum\limits_{i=1}^n}{x_i}^2$, where `n' is the
number of $U_p$ values, and $x_i$ is the difference between the
shock velocity $U_s$ calculated from MD and experiments for each $i^{th}$ 
$U_p$ value. Calculated RMS value for the range of
$U_p$=500 to 3000 m/s is 236.16 m/s and for $U_p$=500 to 2500 m/s
RMS is 188.53 m/s. These two RMS values confirm that the
deviation is primarily from the higher $U_p$ region.

\subsection
[Deviation at higher pressure for Ni: Exploration:]
{Exploring the reasons for the deviation at higher pressure region for Ni:}
\label{otherEAM_description}
The mismatch seen at higher pressure for Ni in MD simulations, could be due
to the fact that EAM potential parameters were fitted using the data close
to equilibrium conditions~\cite{CuNi_eam}. During a shock simulation, the atoms in the compressed
region come close to each other. For example at an impact velocity of 6 km/s,
single crystal Ni shows a compression of $\sim1.5$ times its normal density as
observed from our MD simulations. The corresponding closest inter atomic
distance is $\sim$2.145 \AA. 

It is a standard practice in radiation damage simulations to `stiffen' an EAM potential to account for
individual energetic atoms coming close to each other. Stiffening is done by
modifying the pair interaction of the EAM potential at short distances of
interaction to match the universal repulsive potential fitted by
Zeigler {\it et al.}~\cite{ziegler1985stopping}. We have stiffened the EAM
potential for Ni and validated the stiffened potential by comparing the bulk
modulus, elastic constants, vacancy formation energy and lattice parameter of
%Ni at equilibrium with the standard values. No change is observed in the MD
%%% Modified by Madhavan on 22 Sep'22
Ni at equilibrium with the standard values. It has been observed that
the potential energy obtained from Foiles'~\cite{CuNi_eam} EAM potential
and from stiffened (Foiles' EAM) potential remain the same for inter atomic
distance $>0.8\AA$. This distance is much smaller than the closest inter atomic
distance observed in the MD simulation of Ni even with an impact velocity
of 6 km/s. Thus no change is observed in the MD
%%% Modified by Madhavan on 22 Sep'22
shock simulations results obtained using stiffened potential, in comparison
to non-stiffened potential.

Belashchenko~\cite{Belashchenko_EAM} showed that the inclusion of
thermal-electron energy contribution to the embedded part of EAM potential
does not make any difference in $P$-$V$ Hugoniot. Embedded part may be explored
further for other physical parameters, however that is out of scope of the
present work.

A mention about specific observation of the various EAM potentials of
Nickel, used for MD high strain rate shock simulations is noteworthy.
In figures~\ref{fig:Ni_UsUpHugo_DiffEAMS} \& \ref{fig:Ni_PVHugo_DiffEAMS},
$U_s-U_p$ and $P$-$V$ Hugoniot of Ni obtained from our MD simulations are
presented along with the MD simulations results of six other EAM potentials.
Calculated shock velocity data from published results for six other EAM
potentials are available graphically from published literatures. Those pictorial
graphs are digitized to generate data for comparison in this work.
Among the six results available in the literature one each are from the work
of Jarmakani {\it et al.}~\cite{Ni_md_Jarmakani} and
Liu Hai {\it et al.}~\cite{Ni_md_Liu_Hai}, whose results
are marked as `Jarmakani (2008) and Hai(2017)' in
figures~\ref{fig:Ni_UsUpHugo_DiffEAMS} \& \ref{fig:Ni_PVHugo_DiffEAMS}.
Other four MD results are from Choi {\it et al.}~\cite{Ni_md_Choi}.
These potentials are abbreviated as RW (1999), DWY (2013), ZJP (2004)
\& MFMP (1999) by Choi in his work to represent the initials of the developers.
Hence the same abbreviations are used in this work to represent them in the
figs.~\ref{fig:Ni_UsUpHugo_DiffEAMS} \& \ref{fig:Ni_PVHugo_DiffEAMS}.
Choi in his work~\cite{Ni_md_Choi} mentioned that ZJW is a generalized EAM
potential for 16 metals. The EAM parameters includes sublimation energy
and heat of solution. It also describes the dislocation line energy as a
function of number of Ni layers. DWY is a ternary potential model for nickel
alloy, especially putting an emphasis on the alloy properties such as the
planar defects of Ni$_3$Al. MFMP coefficients include phonon-dispersion curve,
the vacancy formation, migration energies, the stacking fault energies, and
the surface energies. RW is based on Voter-Chen EAM, fitted to reproduce an
experimental value of stacking fault energy of Ni.

In figures~\ref{fig:Ni_UsUpHugo_DiffEAMS} \& \ref{fig:Ni_PVHugo_DiffEAMS},
shock Hugoniot results available from the literature
~\cite{Ni_md_Jarmakani, Ni_md_Liu_Hai, Ni_md_Choi}
are shown along with the results of Foiles potential used in this work.
Results from experiments~\cite{lasl_shockdata} are also shown along with.
For higher compression Hugoniot pressure region (Mbar range), available
experiment data are very little for Ni.
Using the available $U_s$-$U_p$ experimental data in
the Hugoniot equations \ref{rho_hugo} and \ref{press_hugo}, $P(V/V_0)$ data is
generated by extrapolation for higher pressure region. $V/V_0$ is the reciprocal
of $\rho_0/\rho$. The high pressure region is encircled in
fig.~\ref{fig:Ni_PVHugo_DiffEAMS} to highlight the region of interest.
It is clear from this figure that among the results of various EAM potentials,
%%%%%%%%% !! best is replaced by closest
%Foiles potential~\cite{CuNi_eam}, used in this work shows the best match with
Foiles potential~\cite{CuNi_eam}, used in this work yields the closest match with
%%%%%%%%% !! best is replaced by closest
experimental result~\cite{lasl_shockdata}.
It may be noted that Liu Hai~{\it et al.} used the earlier version of EAM potential
developed by ZJW in 2001.
%%, while Jarmakani used similar version of MFMP (1999).
%%Hence the results of Jarmakani~\cite{Ni_md_Jarmakani} and Choi's MFMP(1999)~\cite{Ni_md_Choi}
%%are similar to each other in the figures~\ref{fig:Ni_UsUpHugo_DiffEAMS}
%%\& \ref{fig:Ni_PVHugo_DiffEAMS}.

\begin{figure}[ht!]
     \centering
\subfigure[$U_s-U_p$ of Ni with diff. EAM potentials]
            {
         \includegraphics[width=0.475\textwidth]{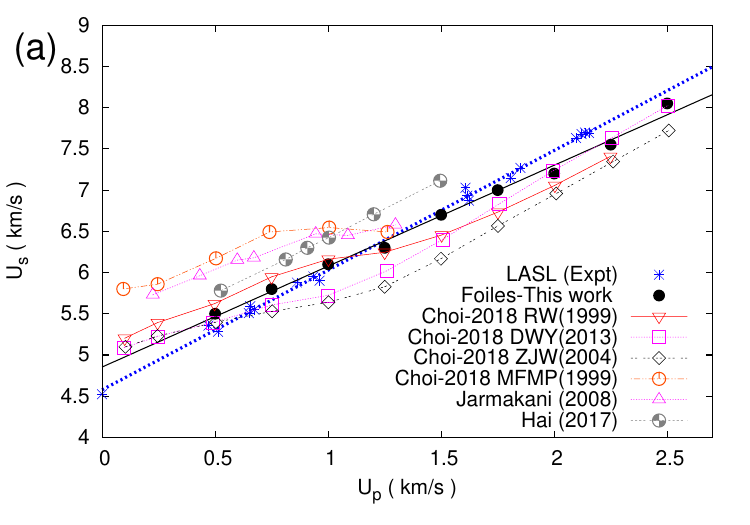}
         \label{fig:Ni_UsUpHugo_DiffEAMS}
            }
\subfigure[$P$-$V$ of Ni with diff. EAM potentials]
            {
         \includegraphics[width=0.475\textwidth]{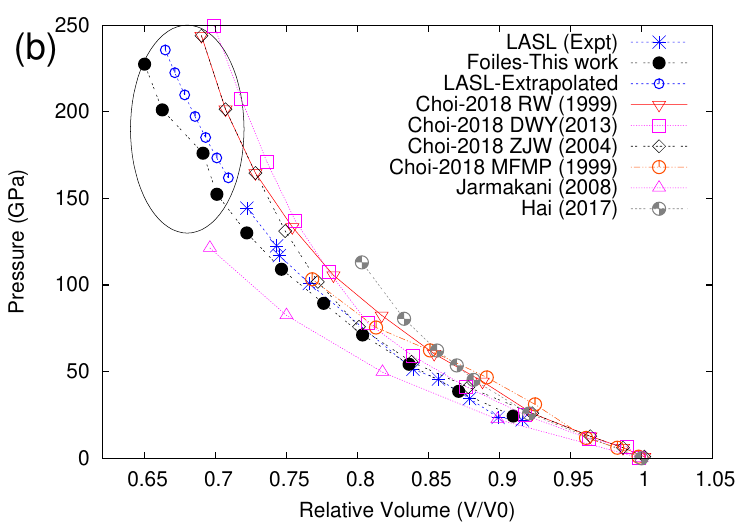}
         \label{fig:Ni_PVHugo_DiffEAMS}
            }

%%     \begin{subfigure}[b]{0.8\textwidth}
%         \centering
%         \includegraphics[width=0.8\textwidth]{Figures/UsUp_Figs/fig7a_Ni_Us_Up_manypotentials}
%%         \caption{$U_s-U_p$ Hugoniot of Ni for different EAM potentials}
%         \label{fig:Ni_UsUpHugo_DiffEAMS}
%%     \end{subfigure}
%     \hfill
%%     \begin{subfigure}[b]{0.8\textwidth}
%         \centering
%         \includegraphics[width=0.8\textwidth]{Figures/UsUp_Figs/fig7b_Ni_PV_manypotentials}
%%         \caption{P-V Hugoniot of Ni for different EAM potentials}
%         \label{fig:Ni_PVHugo_DiffEAMS}
%%     \end{subfigure}
     \caption
[$U_s-U_p$ \& P-V Hugoniot obtained for Ni using Foiles' EAM]
{$U_s-U_p$ \& P-V Hugoniot obtained for Ni from this work
     using Foiles' EAM potential~\cite{CuNi_eam} are compared with MD
     simulations reported independently by Choi, {\it et al.}~\cite{Ni_md_Choi}
     with four different potentials(RW, DWY, ZJW, MFMP);
     Jarmakani {\it et al.}~\cite{Ni_md_Jarmakani}; Liu Hai ~{\it et al.}\cite{Ni_md_Liu_Hai}.
     The experiment results reported by Marsh~\cite{lasl_shockdata} are also
     shown for comparison. Extrapolated region of the experimental data is encircled for clarity (Blue open circles).
 (Fig:Ref.~\cite{UsUp_Madhavan2022})}
\end{figure}

\subsection
[Ni: Comparison of results between MD \& Experiments]
{Comparison of MD results of Ni with Experimental results : other aspects}
In a recent experiment and theoretical work by Dolgoborodov {\it et al.}~\cite{dolgoborodov_nNi_porous},
shock Hugoniot of porous nano-sized nickel(n-Ni) is presented, where porosity ($m$) is defined as the
ratio of the density of monolithic sample to the density of the sample to be studied.
$m = \rho_0/\rho_s$, where $\rho_0$ is the normal density of the material
and $\rho_s$ is the density of the sample that is used in the experiment.
By this definition porosity($m$) increases with the decrease in the density($\rho_s$)
of the sample under study. If a material has $\rho_s=71.42\%(\rho_0$) and another
material has $\rho_s=50\%(\rho_0$) then corresponding values of $m$ are 1.4 and 2
respectively. From experiments~\cite{dolgoborodov_nNi_porous}, the shock Hugoniot
($U_S-U_P$) is obtained for n-Ni with a porosity of 50\%. This is compared with the
perfect SC of nickel (SC-N) obtained from our MD simulations in fig.~\ref{fig:nAndSC_Ni_Us_Up}.
It indicates that the slope as well as the value of the intercept ($C_0$) are very different.
For n-Ni with 50\% porosity the slope ($S$) is 1.77, for a polycrystal (experiment) the slope ($S$)
is 1.44, while for perfect single crystal Ni, the calculated slope is 1.21 (MD simulation).
Thus a trend is seen among: perfect crystal - polycrystalline - polycrystalline with varying
porosity in increasing order.

\begin{figure}[ht!]
     \centering
\subfigure[$U_s-U_p$ Hugoniot of perfect SC-Ni \& porous n-Ni]
            {
         \includegraphics[width=0.475\textwidth]{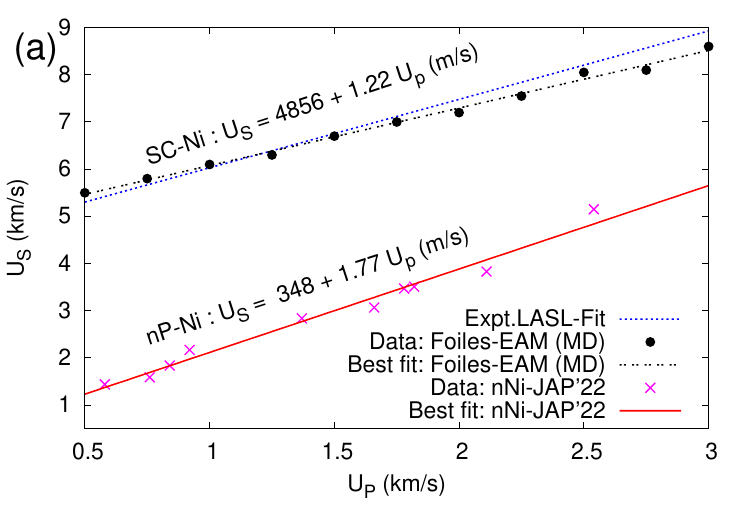}
         \label{fig:nAndSC_Ni_Us_Up}
            }
\subfigure[$P$-$\rho$ Hugoniot of perfect SC-Ni \& porous Ni]
            {
         \includegraphics[width=0.475\textwidth]{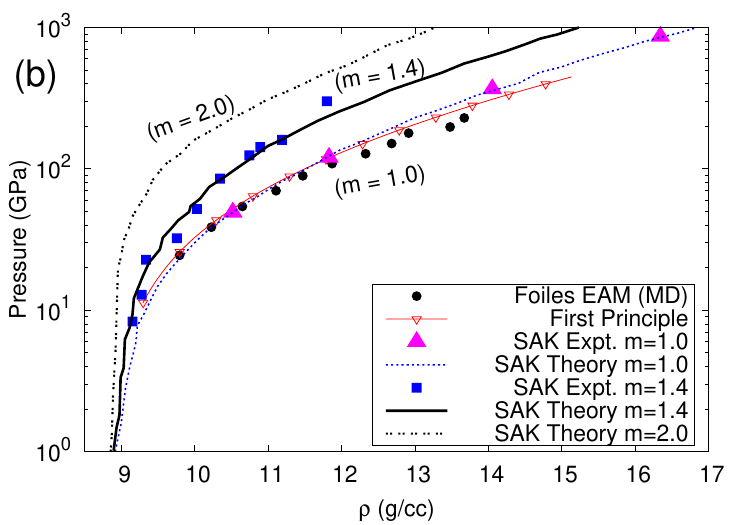}
         \label{fig:nAndSC_Ni_P_rho_log}
            }
\subfigure[$P$-$\rho$ Hugoniot of perfect SC-Ni (MD, FP) \& polycrystal Ni (Expt.)]
            {
         \includegraphics[width=0.55\textwidth]{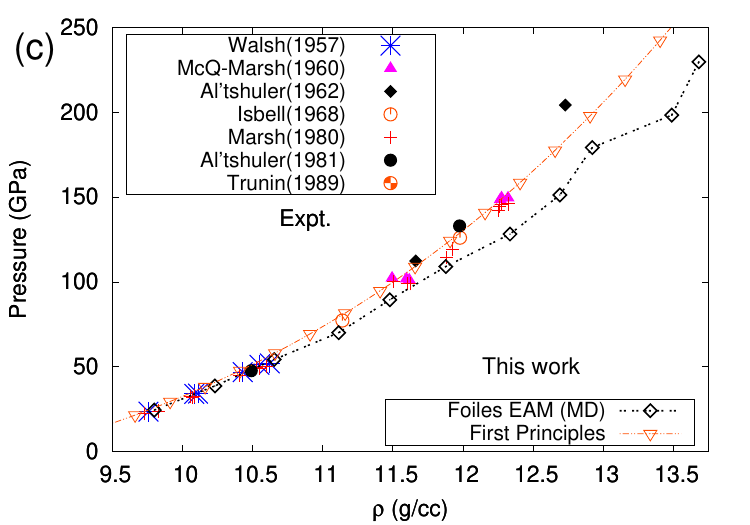}
         \label{fig:polyAndSC_Ni_P_rho}
            }
%%     \begin{subfigure}[b]{0.8\textwidth}
%         \centering
%         \includegraphics[width=0.8\textwidth]{Figures/UsUp_Figs/fig8a_UsUp_NI_withPorous_nNi}
%%         \caption{$U_s-U_p$ Hugoniot of perfect SC-Ni \& porous n-Ni(m-2)}
%         \label{fig:nAndSC_Ni_Us_Up}
%%     \end{subfigure}
%     \hfill
%%     \begin{subfigure}[b]{0.8\textwidth}
%         \centering
%         \includegraphics[width=0.8\textwidth]{Figures/UsUp_Figs/fig8b_PofRho_Ni_SAKComp}
%%         \caption{$P$-$\rho$ Hugoniot of perfect SC-Ni \& polycrystal Ni (m=1, 1.4 \& 2)}
%         \label{fig:nAndSC_Ni_P_rho_log}
%%     \end{subfigure}
%%     \begin{subfigure}[b]{0.8\textwidth}
%         \centering
%         \includegraphics[width=0.8\textwidth]{Figures/UsUp_Figs/fig8c_Expt_FoilesComp}
%%         \caption{$P$-$\rho$ Hugoniot of perfect SC-Ni \& polycrystal Ni from experiments}
%         \label{fig:polyAndSC_Ni_P_rho}
%%     \end{subfigure}
     \caption
[Comparison of Hugoniot P($\rho$) for Ni using Foiles' EAM potential with Expt. result]
{MD shock Hugoniot obtained for Ni from this work using Foiles' EAM potential~\cite{CuNi_eam}
     are compared (a) for $U_s-U_p$ with experimental results reported by~\cite{dolgoborodov_nNi_porous}
     for n-Ni (m=2), (b) for $P$-$\rho$ with experimental results reported
     by~\cite{SAK_high_porosityNi} for Ni with porosity m=1, 1.4 \& 2 (Y-axis in log scale) and
     (c) for $P$-$\rho$ with experimental
     results
~\cite{shock_repository_rusbank_Online}
sourced from LASL shock data,
Walsh~{\it et al.}, McQueen~{\it et al.}, Alt`shuler(1962, 1981)~{\it et al.},
Isbell~{\it et al.}, Trunin~{\it et al.}.
 (Fig:Ref.~\cite{UsUp_Madhavan2022})}
\end{figure}

$P$-$V$ Hugoniot of the MD results are compared with FP results and various other
experimental results of polycrystalline nickel~\cite{SAK_high_porosityNi}.
A comparison is shown in fig.~\ref{fig:nAndSC_Ni_P_rho_log} for three different
porosity (m=1, 1.4 and 2) along with theoretical calculations from SA Kinelovskii
{\it et al.}~\cite{SAK_high_porosityNi}. It is found that the MD
results are matching with the results of FP method, the experiments and theory~\cite{SAK_high_porosityNi}
for m=1. However some intrinsic porosity due to grains, grain boundary
and imperfection are likely in polycrystalline nickel, that are used in various experiments.
This may also be the reason for the mismatch in $P$-$\rho$.
This figure is shown with log-scale for the y-axis to accommodate different orders of
magnitude in the pressure values. Fig.~\ref{fig:polyAndSC_Ni_P_rho} shows the comparison
of the shock Hugoniot curves $P$-$\rho$ for SC-Ni obtained from the simulations (MD),
FP method and the experimental results. The experimental shock-wave data is available
as a repository~\cite{shock_repository_rusbank_Online}.
This repository contains experimental shock data of Ni sourced from LASL,
Walsh (1957)~{\it et al.}, McQueen (1960)~{\it et al.}, Alt`shuler
(1962, 1981)~{\it et al.}, Isbell (1968)~{\it et al.}, Trunin (1989)~{\it et al}.
%Details of these experiments
%can be obtained from the references listed along with the shock-wave
%data~\cite{shock_repository_rusbank_Online}.
%Overall, figure.~\ref{fig:polyAndSC_Ni_P_rho} shows very good match with
%experiment for the first principles calculation.
Comparison of experimental shock Hugoniot with the MD simulations
(figs.~\ref{fig:Ni_UsUpHugo_DiffEAMS} \& ~\ref{fig:Ni_PVHugo_DiffEAMS})
indicate that among the various EAM potentials that are described in
sec~\ref{otherEAM_description}, Foiles' EAM potential~\cite{CuNi_eam}
employed in this study appears
%to be the most efficient.
to show the least deviation from experimental results.

This work aims to explore MD shock simulations in single crystal FCC materials.
The extent of deviation seen in previously available published data of single crystal
Ni is found to be reduced by employing an existing EAM potential~\cite{CuNi_eam}.
%First principle calculations are also carried out at higher compression of
%shocked Ni, which show a good match with experimental data.

\section{Conclusions}
\label{sec-4}
Shock-Hugoniot and the isothermal compressibility for single FCC 
crystal of Cu, Al and Ni are simulated using MD methods.
The parameters relating $U_s$ to $U_p$ have been obtained.
The relationship obtained for $U_s$-$U_p$ \& $P$-$V$ under isothermal and dynamic
shock conditions, are compared with published experimental results.
%The relative error in the first principles calculations in all cases are \textless 6\%.
The relative errors between the MD simulation results and
experiments~\cite{meyersBook, lasl_shockdata, alcuta_usup_Mitchell}
are mostly \textless 6\%. Only in the case of the shock Hugoniot parameter $S$ 
for Ni, the relative error in the MD simulation is 16\%. Stiffening the pair interaction
component of the Foiles EAM potential did not change the results.
The EAM potentials by Foiles {\it et al.}~\cite{CuNi_eam} for Ni used in this
study yields the closest match with experiments, compared to four other interatomic
potentials used by Choi {\it et al.}\cite{Ni_md_Choi} as well as the MD
simulations results of Jarmakani {\it et al.}~\cite{Ni_md_Jarmakani}
and Liu Hai {\it et al.}~\cite{Ni_md_Liu_Hai}.
We conclude that the EAM potentials used in this study for Cu~\cite{CuNi_eam}
and Al~\cite{Al_eam}, can be readily used for MD simulations at high
strain-rates, $10^7$-$10^8\ s^{-1}$ that corresponds to high pressures,
up to 2.3 Mbar in Cu and 0.8 Mbar in Al.
For Ni, a trend of deviation is observed at higher impact velocity where
very few experimental data are available for comparison. Using available shock
velocity and $P$-$V$ experimental data for Ni and applying Hugoniot
relations~\cite{meyersBook}, $P$-$V$ curve is extrapolated. Shock Hugoniot
values obtained using Foiles EAM potentials in this study shows the closest
match with the experiments for the higher pressure range also
(fig.~\ref{fig:Ni_PVHugo_DiffEAMS}). These figures also indicate
that at high-compression shock regime where the pressure reaches Mbar in magnitude,
Foiles EAM potential used in this study seems to
show the least deviation from experimental results,
in comparison to six other EAM potentials of Ni.

%\bibliography{madhavan_usup}
\clearchapter

% MScaleSpall_SC : Chapter 4. Multiscale spall calculation for Single Crystal Cu,Mo,Nb and Al : ICONS 2018 / arXiv paper
  \chapter
%[Multiscale modeling in Single Crystal: To find NAG and Spall parameters]
%{Multiscale modeling to find NAG fracture parameters for Metallic Single Crystals and to estimate their spall strength\label{chapter_SCSpall-Multiscale}}
%{Multiscale modeling to find NAG and Spall parameters of Single Crystals\label{chapter_SCSpall-Multiscale}}
{Multiscale Modeling to obtain the Dynamic Spall Strength of Single Crystals\label{chapter_SCSpall-Multiscale}}
\section{Background of the Problem}
To model dynamic impact shock and associated fracture, hydrodynamic codes solve basic conservation laws
of mass, energy, and momentum in space and time~\cite{wilkinsMethodsInCompPhys, meyersBook, hemp_wilkins, intro_hydro_zukas}.
It uses appropriate models to account for the elastic-plastic deformation~\cite{hemp_wilkins, intro_hydro_zukas},
equation of state~\cite{eliezer, alcuta_usup_Mitchell, condmat4030071}, dynamic material
strength~\cite{SG_strength, STEINBERG1987603, JC_fracture} and fracture~\cite{seamanCompmodelDuctBrit_JAP,
johnsonPulseDurationSpallFrac_JAP, curran_nag1, ikkurthi1_nag1} as described in
chapter~\ref{chapter_compMethods}, sec.~\ref{1Dhydroec.2.3}. Among various dynamic damage models,
Nucleation And Growth (NAG) of voids is a micro-physical model~\cite{curran_nag1}. It computes the
amount of damage in the material by including the physical processes such as nucleation, growth and
coalescence of voids. The Nucleation threshold and the growth threshold are the two important factors
that play a main role in the NAG model to find out the degree of damage in materials under dynamic impact.
These parameters are fitted using experimentally observed void volume
distributions~\cite{seamanCompmodelDuctBrit_JAP, johnsonPulseDurationSpallFrac_JAP, curran_nag1}.
However, with cost-effective computing facilities, these parameters can be obtained
using molecular dynamics (MD) simulations of void nucleation and growth, by fitting
`temporally evolving P and void volume fraction' MD-data~\cite{rawat_thesis, ramana2_hydro}.
% optimization of temporally evolving pressure and void volume fraction~\cite{rawat_thesis, ramana2_hydro}.
Using these parameters in hydrodynamic simulations, dynamic spall parameters can
be calculated~\cite{ikkurthi1_nag1}. The space and time scales vary from atomic to macro levels, hence
this is called a multi-scale approach.
The Multi-scale procedure is executed with the following steps.
MD simulations are carried out for cubic metallic single crystals by subjecting them to
isotropic triaxial deformation at a constant strain rate ($\dot{\epsilon}$).
Time-varying pressure of the system and the void volume fracture are recorded.
These results are used in Particle Swarm Optimization (PSO) method, to obtain
NAG parameters for materials of specific interest. Using these parameters in a 1D hydrodynamic
code developed in-house and fracture parameters such as spall strength and thickness are obtained.
The results are validated with published experimental data~\cite{Anton_spalldata,
Kanel2001JAPSingleCrysFoilAl, chen_asay}. This chapter describes the multi-scale model to obtain
the NAG and spall parameters of single crystals. The results are compared with published experimental
results in single crystal Al.

\section{Computational aspects~\label{compAspectsChap4}}
\begin{figure}[!htb]
\centering
\includegraphics[width=0.45\linewidth]{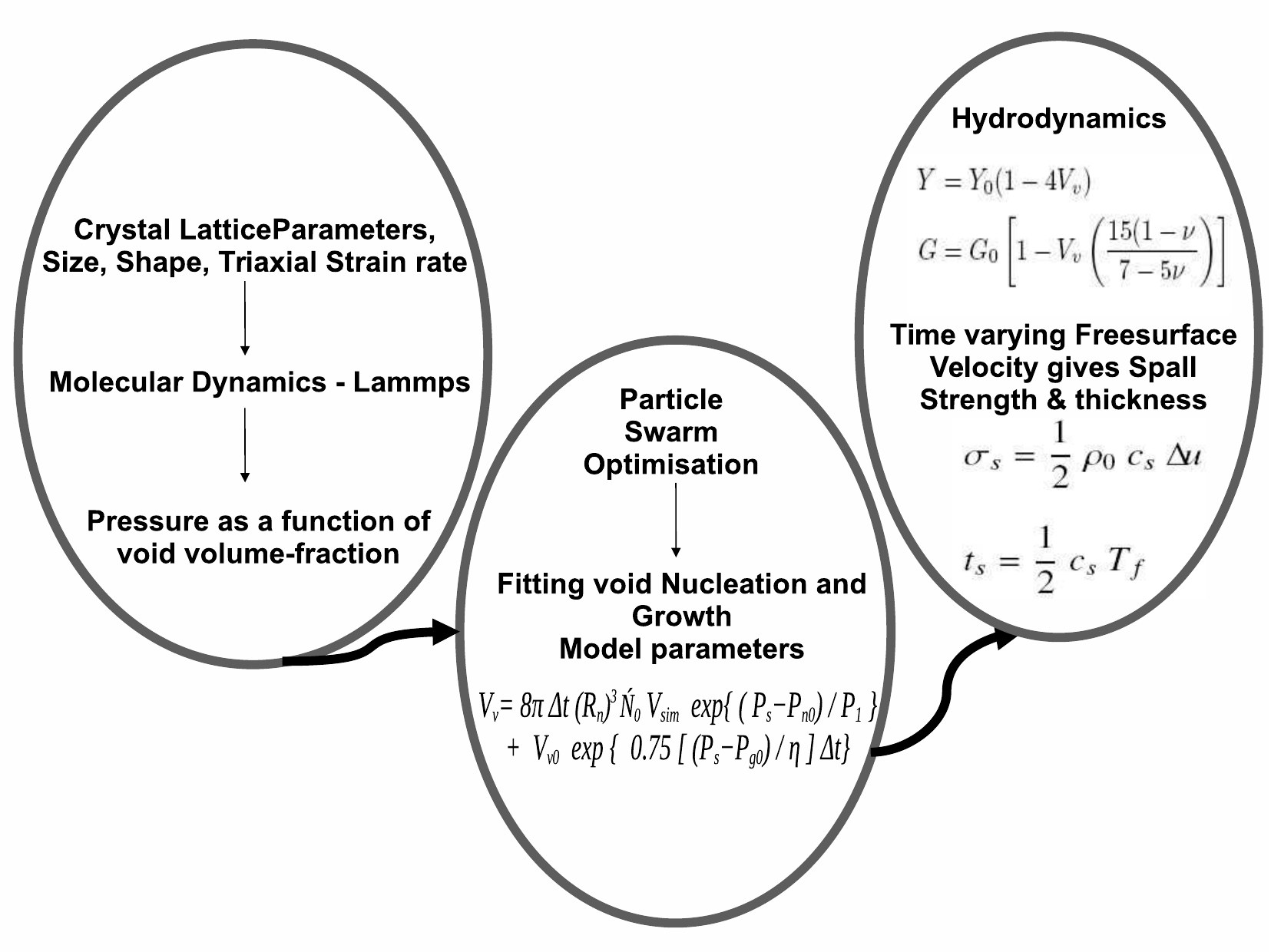}
\vspace{-0.2cm}
\caption
[Multi-scale for Single Crystal : Simplified work flow]
{Workflow of Multi-scale modelling consisting of MD simulations of
triaxial deformation of bi-crystals to obtain the void nucleation and growth as
a function of tensile pressure. This information is used by a particle swarm
optimization (PSO) code to fit the parameters of the macroscopic void
nucleation and growth (NAG) model. The NAG model is used in hydrodynamic
simulations of flyer plate impact on targets to obtain the free surface
velocities and therefore the spall strength. (Fig:Ref.~\cite{madhavan_spall_multiscale}).}
\label{figChap4:flowchart_multi-scale}
\end{figure}
Detailed flowchart of the multiscale method is already presented in
chapter~\ref{chapter_compMethods}:figure~\ref{subfig:StepsInMSModel}.
Compared to the figure~\ref{subfig:StepsInMSModel}, which includes the process
of setting up a grain boundary and validating its energy, for a single crytal since
there is no grain boundary, these steps are not required.
Figure~\ref{figChap4:flowchart_multi-scale} is the simplied flow of multiscale
model for a single crystal. NAG and spall parameters equations are also mentioned
explicitely in this flow-chart (figure~\ref{figChap4:flowchart_multi-scale}).

\subsection{Molecular Dynamics simulation: Tri-axial deformation~\label{chap4:sec4.2.1}}
The computational domain used in LAMMPS to simulate the isotropic, tri-axial
deformation consists of 100$\times$100$\times$100 unit cells of a single crystal.
Appropriate EAM potentials are used for the independant tri-axial high strain rate
deformation in respect of Mo~\cite{Zhou_MoEAM}, Nb~\cite{Fellinger_NbEAM}% !! Fellinger for Niobium
, Cu~\cite{CuNi_eam} and Al~\cite{Al_eam}. The potentials were validated by
carrying out MD simulations to obtain the bulk modulus ($B_0$),
shear modulus ($G_0$), Young's modulus ($Y_0$), Poisson's ratio ($\nu$),
cohesive energy ($E_{coh}$) and vacancy formation energy ($E_{vf}$).
The calculated values for the materials used in this work, are listed in
table~\ref{elasticTable} of chapter~\ref{chapter_compMethods}, along with
the values obtained from experiments. It may be noted that the single crystals
used in spall-experiments are very likely to have defects and
impurities, when they are grown under laboratory conditions. For simulating with
some crystal defect, an edge dislocation is created by removing a half-plane of the
crystal in MD computational domain. Then the atoms are equilibrated using an NPT ensemble.

A constant time-step of 1 femto sec (fs) is used with periodic boundary conditions
along X, Y and Z. The simulations are performed in two steps. The first step is to carry out
the NPT simulation at 0 bar pressure and 300 K temperature, to equilibrate the system 
in which half-plane of atoms are removed. Subsequently in the second step, NVE simulations are carried
out by imposing tri-axial expansion with a constant strain
rate ($\dot{\epsilon}=5\times{10}^{9\ }s^{-1}$). The strain rate
$5\times10^9$ s$^{-1}$ is applied by changing the box size subjected to the condition
$L(t) = L_0(1 + \dot{\epsilon} t)$. Here $L_0$ is the initial box size,
$\dot{\epsilon}$ is the strain rate and `t' is the time step. Once in every
100 time steps the box size is changed.
For each resizing of the box, 
%For a specific duration of increasing strain
the atoms have 0.1 picosec (ps) to relax before any further deformation.
%% These simulations were carried out during May 2018 in Tej cluster machines.
% I don't remember to have saved these data. Hence the following may be removed
% on account of not having the data in my hand. 26 Feb'23 (comment made)
The pressure time history obtained by varying the time interval between
these deformations from 1 to 100 time steps, do not vary.

%\subsection{Void Volume ($V_{sim}$) calculation from MD for fitting the NAG parameters\label{chap4:sec4.2.2}}
\subsection{Void Volume ($V_{sim}$) calculation from MD\label{chap4:sec4.2.2}}
The space and time averaged data of atoms position, the computational domain size, the volume,
the pressure, etc., that are obtained from MD simulation, are processed using voxel based method~\cite{void_vol_harsh}
to find the pressure versus void volume fraction. This algorithm can handle simulation size
with billion atoms. It is also fast and efficient.

This is relatively simple to implement. The three dimensional computational domain
is divided in to voxels. Voxel is the 3D equivalent  of a 2D pixel. One voxel size
is considered to be equal to 1 unit cell of the crystal. Every voxel is assumed
to be empty initially. Depending upon the position of the atoms
% at any desired frequency of the time step,
the voxel domain is assigned a status empty or filled.
The sum of the empty voxels gives the number of voids. The ratio of the number of
voids to total number of voxels is the void volume fraction. Voxels are treated as
connected when any two voxels share a face with each other. This is done recursively
and the voxel size gets accumulated over time. Thus the time varying number
of independant voids, the total void volume fraction and void size distributions are
calculated. The void volume fraction is `$V_{sim}$' that appears in NAG fracture model
equation~\ref{voidNucEqn1}. Time varying void volume fraction-pressure data
obtained from MD simulations is used in NAG model
equations~\ref{V_vEqn0}-~\ref{voidGrowthEqn1} (section~\ref{1Dhydroec.2.3}
of chapter~\ref{chapter_compMethods}). PSO method described in
section~\ref{PSOsec.2.2} of chapter~\ref{chapter_compMethods}
is used to obtain the five unknown material parameters by fitting
the NAG model equations to the MD results.

\subsection{One dimensional hydrodynamic simulation~\label{1dHD-Chap4.2.3}}
%To calculate spall fracture from hydrodynamic method to be written
The one dimensional hydrodynamic code described in section~\ref{1Dhydroec.2.3} of
chapter~\ref{chapter_compMethods} is used to calculate the spall parameters
as the last part of the multiscale model.
High velocity flyer-target impact produces shock-waves. Shockwave is characterized
by its non-linear propagation and spatial discontinuity. Thermodynamic parameters
like pressure, density etc are discontinous between the shocked and un-shocked region.
Solution is not expressible due to such discontinuity in the discretized computational
grids. Addition of artificial viscosity helps to smooth out the shock around the
discontinuity region. One dimensional hydrodynamic model used in this work has
this feature to handle shock-wave.

In hydrodynamics shock-impact simulations, two zones namely the impactor and target are made.
The computational grids of both the zones are assigned with an initial co-ordinate position,
density, pressure, energy and temperature. The impactor zone grids are initially given an unidirectional
velocity directed towards target at the start of the simulation.
%% Following sentence is modified since Monoj commented,
%% "Gives and impression that the two ends of the impactor and two ends of the target are given fbcs!!"
The impact surfaces evolve in time, as governed by the conservation of the mass, the momentum and
the energy, while the other ends of the impactor and the targets
%The two ends of the zones
are given freely moving boundary conditions.
The simulation starts with the moment of
impact where the two zones are kept in contact. For simulating high strain rate shock
propagation in aluminum using a flyer-target impact system, (i) a density ($\rho$) based polynomial EOS
(eqn.~\ref{pofetawilklinseos}), (ii) the NAG fracture model (eqn.~\ref{V_vEqn0}-\ref{voidGrowthEqn1}),
and (iii) A strength of materials defined by equations~\ref{SGY}-\ref{SGG} are included.
The stength of materials are also modified in accordance with the
equations~\ref{Ycorrection}-\ref{Gcorrection}, to account for the effect change
in void volume fraction.

When void volume fraction ($V_v$ in eqn.~\ref{V_vEqn0}) reaches a value equal
to 0.25, as per eqns.\ref{Ycorrection} and~\ref{Gcorrection} the yield strength
equal to `0' and shear modulus reduces to $\simeq{G_0}/2$. Beyond
this value, the material does not support further shear stress. Further
growth in void volume makes yield strength negative which is unphysical
otherwise.  This indicates the status of material failure. Thus when the
void volume fraction at any of the computational cell exceeds 0.25, the cell
is marked fractured. When fracture occurs at a location in the shock induced
target, the leading front part detaches from the other parts. This is reflected in the
oscillating nature of the Free Surface Velocity (FSV) profile as shown in
figures~\ref{FSVCuOnly},~\ref{FSVMoNb} and~\ref{FSVAl3Cases}.
%figures~\ref{FSVAl3Cases},~\ref{FSVMoNb} and~\ref{FSVCuOnly}.

%The one-dimensional hydro-dynamic code calculates the instantaneous free surface
The hydro-dynamic simulation calculates the instantaneous free surface
velocity of the target.  From the temporal free surface velocity,
the spall strength (${\sigma}_{sp}$) and the spall thickness ($d_{sp}$)
%can be calculated using the equations~\cite{curran_nag1, ikkurthi1_nag1},
can be calculated~\cite{curran_nag1, ikkurthi1_nag1} using the equations,
\begin{subequations}
\begin{align}
{\sigma}_{sp} & = \frac{1}{2}\ {\rho}_0 C_0 {\Delta}U \label{eqn_sp_strengthChap4} \\
d_{sp} & = \frac{1}{2}\ C_0 T_f \label{eqn_sp_thickChap4}
\end{align}
\end{subequations}
where ${\rho}_0$ is the normal density of the target material. $C_0$ is the
bulk sound speed in that material. ${\Delta}U$ is the pull-back velocity in the
free-surface velocity profile and $T_f$ is the period of oscillations of the
spall signal. The pull-back velocity is the velocity difference between the first
peak to the first dip seen in the free surface velocity profile.
%Near the time around which the dip occurs, the material fails by spall fracture.
Little before the time around which the dip occurs, the material fails by spall fracture.
%The results are compared with experimental results.

\section{Results and Discussion\label{resultsAndDiscus}}
\subsection{Results from MD simulations}
\begin{figure}[!htb]
        \centering
        \subfigure [Nucleated voids in Al]
        {
        \includegraphics[width=0.42\textwidth] {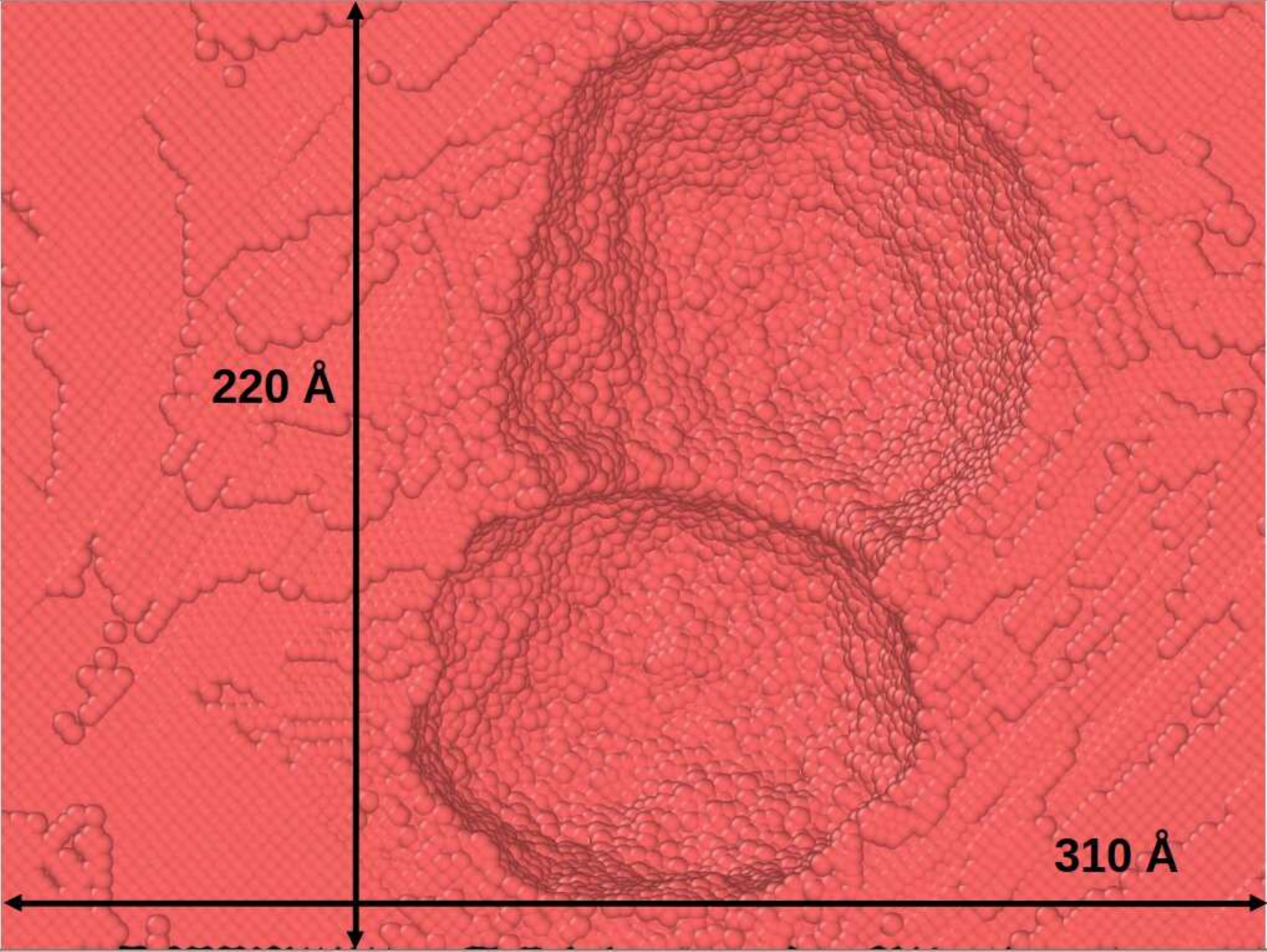}
                 \label{chap4Fig2asnapShotVoidNuc}
        }
        \subfigure [Pressure (t)]
        {
        \includegraphics[width=0.42\textwidth]{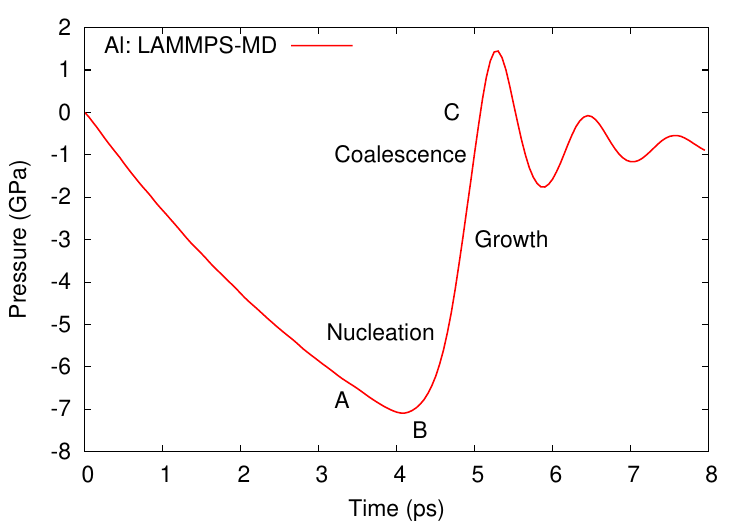}
        \label{chap4Fig2bPofT}
        }
\vspace{-0.2cm}
 \caption
        [Nucleated Voids \& Pressure profile of Al-SC under triaxial tensile force]
        {MD simulation of single crystal Al subjected to high strain rate triaxial
	deformation ($\dot{\epsilon}$ = 5$\times{10}^9\ s^{-1}$). Fig.(a) Image of Al-single crystal with nucleated Void,
	figure (b) Temporal pressure profile that is used to find the NAG parameters of Al.}
    \label{snapShotVoidsAndpofTinAl}
\end{figure}
Figure~\ref{chap4Fig2asnapShotVoidNuc} shows nucleated voids for aluminum single
crystal (Al-SC) from MD simulations. This snap-shot is taken at about 4.75 ps
from an Al-SC that is subjected to tri-axial high strain rate deformation.
The figure shows the nucleated voids are grown further and about to coalesce.

Figure~\ref{chap4Fig2bPofT} shows the temporal evolution of the space and
time averaged pressure of aluminum single crystal. This is the result
of triaxial deformation of the single crystal at room temperature.
The applied strain rate is $5\times 10^9\ s^{-1}$. Under tri-axial tension,
the magnitude of tensile pressure increases continuously up to 7 GPa after
which it turns around and decreases with further expansion of the box in
all the three directions.  When the stress relaxes due to nucleation and growth of voids, the
pressure turns around as seen in figure~\ref{chap4Fig2bPofT}. The pressure-time profile
shown for aluminum single crystal from this work can be divided into three regions
as described in the literatures for single crystal copper~\cite{rawat2_hydro, rawat_thesis}.
These three regions are marked in figure~\ref{chap4Fig2bPofT} and all the discussion
written in this context below are with reference to this figure.
\begin{enumerate}
\vspace{-0.2cm}
\item Region of Void nucleation (AB) :
Voids nucleate (or form) during this early duration of tensile pressure.
Voids nucleate, when the thresohold ($P_{n0}$) limit of void nucleation
is exceeded. When a void is formed about the time marked `A', local relaxation occurs
due to the relaxation of atomic bonds around the region of nucleation. There is
a time-elapse between the onset of local relaxation and the time
the system is equilibrated on account of the local pressure drop. Because it takes
some time for this change in pressure to travel the whole volume of the sample.
As the fig.~\ref{chap4Fig2bPofT} shows for `Al', this duration time is $\sim$1 ps.
This is typically the vibration period of an atom. Around the time `B', the
total pressure of the system includes the effect of such a local relaxation.
Also already nulceated voids grow by then along with the nucleation of more voids.
This decreases the stress in the system further and hence the tensile pressure
gets reduced even further.
\vspace{-0.2cm}
\item Growth region BC :
During this duration, the voids grow hence this region is referred as `Growth region'.
Void volume increases primarily due to the expansion of voids rather than nucleation
in BC region. Because the tensile pressure falls below the threshold for void nucleation.
This leads to the slow down of void nucleation and it almost stops eventually.
Also the threshold for void growth is lower than the tensile pressure, thus
void growth continues.
\vspace{-0.2cm}
\item Coalescence region- Region beyond C :
This region corresponds to coalescence of voids. During this duration, void volume increases
primarily because already nucleated voids grow or expand further. Expansion of two 
or more voids leads to the coalescence and void volume increases further.
During this time, the number of voids comes down for two reasons: (1) nucleation
of void stops and (2) voids begin to combine with each other. But, at the same time 
void volume continues to increase. In this region coalescence of the voids dominates
over growth of voids.
\vspace{-0.2cm}
\end{enumerate}
%\vspace{-0.5cm}

%\subsection{Results from PSO: NAG coefficients obtained by fitting.}
\subsection{Calculation of NAG model coefficients by fitting MD data}.
%\vspace{-1.2cm}
\begin{figure}[!htb]
        \centering
        \subfigure [`Mo' single crystal]
        {
        \includegraphics[width=0.47\textwidth] {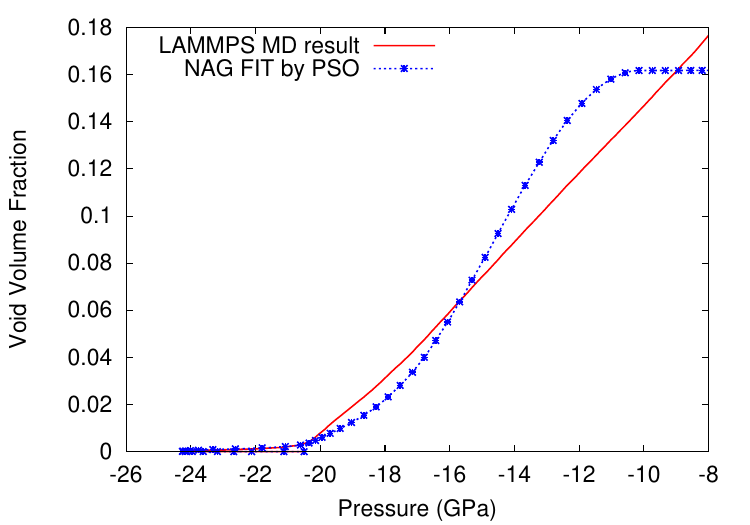}
                 \label{chap4Fig3aVvofPMo}
        }
        \subfigure [`Al' single crystal]
        {
        \includegraphics[width=0.47\textwidth]{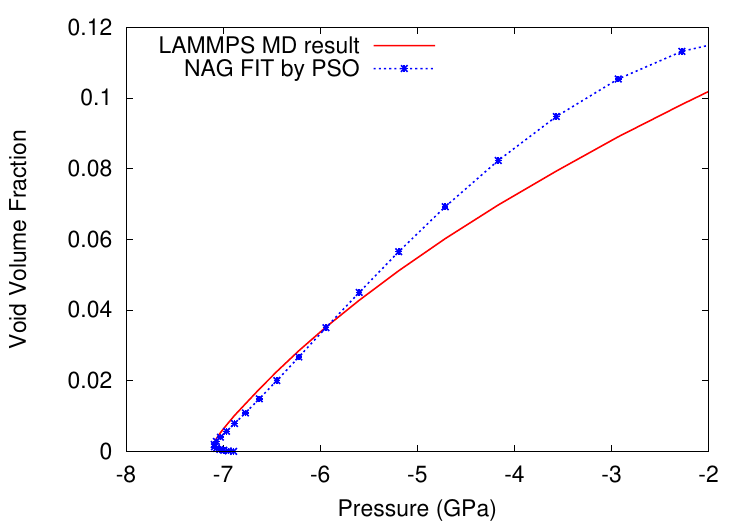}
        \label{chap4Fig3bVvofPAl}
        }
%\vspace{-0.2cm}
\vspace{-0.2cm}
 \caption
        [Void volume fraction Vs Pressure: Al-SC under tensile formation]
        {Void Volume fraction as a function of Pressure. Also shown the line fitted to to MD 
	data using PSO method.}
    \label{VoidVolFracVsPresMoAl}
\end{figure}

%Under applied tri-axial tensile deformation, aluminum single crystal found
%have increase in void volume as a function of tensile stress/pressure.
Under tri-axial tensile force the void volume fraction and the net
pressure of single crystal material vary. This phenomena is captured by
MD simulations of molydbenum and aluminum.
%Figures~\ref{chap4Fig3aVvofPMo} and~\ref{chap4Fig3bVvofPAl} show the
%time evolving void volume fraction as a function of pressure.
The NAG fracture equation relates the void volume fraction with
pressure (ref. equations~\ref{V_vEqn0} to~\ref{voidNucEqn2}). This model
has five coefficients that are material dependant.
These coefficients are determined by fitting the MD results of void nucleation
and growth with the pressure, using the PSO
algorithm described in chapter~\ref{chapter_compMethods}, section~\ref{PSOsec.2.2}.
The NAG parameter fitted curves are shown in figures~\ref{chap4Fig3aVvofPMo}
and figure~\ref{chap4Fig3bVvofPAl} for molybdenum and aluminum respectively.
For comparison, the values from MD simulation are also shown along with in each case.
The relative errors obtained in the fitting are 4\%  for molybdenum and 8\% aluminum.
The fitted curve for Mo (fig.~\ref{chap4Fig3aVvofPMo}) has a plateau because
$V_v$ does not grow further (NAG model equation~\ref{V_vEqn0}) due to coalescence of voids and thereby fracture.
Also, the fitting for Al (fig.~\ref{chap4Fig3bVvofPAl}) has increasing deviation for $P$ > -5 GPa
because it enters coalescence region as can be seen in figure~\ref{chap4Fig2bPofT}.
The material constants of the NAG fracture model that are calculated
from the present work for molybdenum and aluminum are presented in
table~\ref{tableSingleCrystalNAGvalues}. The table also contains NAG coefficients
calculated for niobium by V.R. Ikkurthi {\it et al.}~\cite{ramana2_hydro}  and
for copper by Rawat~\cite{rawat_thesis}, using MD simulations. It is worth to
mention that only the nucleation and growth part are used in the fitting,
ignoring the coalescence.
%\vspace{-0.2cm}
%However these errors are lesser than the earlier fits
%reported by V.R. Ikkurthi {\it et al.}~\cite{ramana2_hydro}.
%% !! Introduce spall parameters accuracy in Comp. methonds of MD/Hydro and
%% refer them here. as a sentence given below.
% The accuracy or the deviation involved in MD and hydro simulations are already
% mentioned in chapter~\ref{chapterkey}, section~\ref{section}.
%%%%%%%%%%%%%%%%%%%%%%%%%%%%%%%%%%%%
%\begin{landscape}
\begin{table}[!ht]
\centering
\vspace{-0.2cm}
\caption {NAG parameters calculated for single crystals of Cu, Nb, Mo \& Al}
\begin{tabular}{cccccccc}
\toprule
\multirow{2}{*}{\parbox{1.0cm}{\centering{\bf Sl.No.\  }}}
&
\multirow{2}{*}{\parbox{1.0cm}{\centering{\bf Sample\  }}}
 & ${\bf \ \dot{N}_0 (\times 10^{35})}$ & ${\bf P_{1}}$ & ${\bf P_{n0}}$ & ${\bf P_{g0}}$ & ${\bf \eta\ (\times 10^{-2})}$ &
\multirow{2}{*}{\parbox{1.0cm}{\centering{\bf $\nu$}}} \\
&  &
$(m^{-3}s^{-1})$ &
$(GPa)$ &
$(GPa)$ &
$(GPa)$ &
$(Pa-s)$ & \\
\midrule
1 & Nb & 3.6 & -1.43 & -16.9 & -6.36 & 1.9 & 0.39 \\
2 & Cu & 7.1x$10^ {-17}$ & -0.0167 & -16.04 & -2.12 & 0.34 & 0.33 \\
3 & Mo & 26.4234 & -17.5225 & -1.9131 & -10.0509 & 0.8965 & 0.29 \\
4 & Al & 13.0600 & -3.6880 & -10.1188 & -0.9323 & 0.7038 & 0.33 \\ 
\bottomrule
\end{tabular}
\label{tableSingleCrystalNAGvalues}
\end{table}

%\vspace{-1.1cm}
\subsection{Results from Hydrodynamic simulations}
%\vspace{-0.25cm}
Results from hydrodynamic simulations of flyer plate-target impact system for single
crystals of  copper (Cu), niobium (Nb), molydbenum (Mo), and aluminum (Al) are
presented here. All the simulations are performed such that the shock is loaded along
$<$100$>$ direction of the single crystals. Because in all the single crystal
experiments cited here for comparison, the shock loading direction is $<$100$>$ only.
The NAG fracture parameters listed in table~\ref{tableSingleCrystalNAGvalues}.
are used in the hydrodynamic simulations. The dimension of the flyer plate and the targets,
impact velocities are listed for these impact simulations in
table~\ref{tableSingleCrystalSpallVal}.
These specifications
%and the expt. IDs (B70, B71, B80, B82, B99 \& B101),
are adapted from the 
spall data experiments, compiled by T.H. Anton {\it et al.}~\cite{Anton_spalldata} for Cu,
Nb, and Mo while the experimental specifications of Kanel {\it et al.}~\cite{Kanel2001JAPSingleCrysFoilAl}
and Asay {\it et al.}~\cite{chen_asay} are used for Al. Table~\ref{tableSingleCrystalSpallVal}
also contains the accuracy involved in the experimental measurements such as impact velocity,
estimation of spall strength and spall thickness. Only for one experiment with
`Al`~\cite{Kanel2001JAPSingleCrysFoilAl}, the accuracy is not mentioned explicitly.
However it is mentioned that the temporal free surface velocity profiles are recorded
using VISAR~\cite{VISAR1974} and ORVIS~\cite{ORVIS1983} laser Doppler veloci-meters.
For VISAR and ORVIS, the time resolution are 3 ns and 1 ns
respectively~\cite{Kanel2001JAPSingleCrysFoilAl}.

In all these cases, aluminum flyer plate is used. The target material for expt. IDs B70 and B71,
is `Cu'~\cite{Anton_spalldata}. For B99 and B101 `Nb' is the target~\cite{Anton_spalldata},
while for IDs B80 and B82 `Mo' is the target~\cite{Anton_spalldata}. In case of Sl.No. 7 to 9
listed in table~\ref{tableSingleCrystalSpallVal}, `Al' target is
used~\cite{Kanel2001JAPSingleCrysFoilAl, chen_asay}. In case of `Mo' (Sl.No.5), 2 mm thick `Al'
base plate is also used in-between the impactor and the target~\cite{Anton_spalldata}.

\begin{landscape}
\clearpage % Forces a clean page break
\vspace*{\fill} % Pushes the table down from the top
\begin{table}[!ht]
%\footnotesize
\centering

% 1. Increase vertical padding between rows (default is 1.0)
\renewcommand{\arraystretch}{1.3}
% 2. Increase horizontal padding between columns (default is 6pt)
\setlength{\tabcolsep}{10pt}

%\parbox{22cm}{ % Increased parbox width to match a landscape layout width
\vspace{-0.2cm}
%\caption[...]
%{...}
%\label{tableSingleCrystalSpallVal}
%}

\parbox{22.5cm}{
\caption
[Spall parameters of SC (Cu, Nb, Mo \& Al), compared with experiments]
{The spall parameters obtained from the multiscale model for the single crystals of
Cu, Nb, Mo \& Al are compared with experimental results: For the comparison,
the experiments results of the spall data compiled by T.H. Anton {\it et al.}
are used for Cu, Nb \& Mo (Ref.~\cite{Anton_spalldata}); Experiments with ID (i)
`Baumang' (Ref.~\cite{Kanel2001JAPSingleCrysFoilAl}), (ii) `28:Asay' \& (iii) `8:Asay'
(Ref.~\cite{chen_asay}) are used for `Al'; In case of Sl.No.5, 2 mm thick `Al' base plate is also
used between the `Al' impactor and `Mo' target. The errors associated with the experimentally
measured parameters are also provided in this table with `$\pm$' marks at respective places,
where `NA' means `Not available'.
%The limitation of the accuracy
%involved in the experimental measurement are also given using the
%symbol $\pm$ at
%appropriate places.\\
\\ \\
}
\label{tableSingleCrystalSpallVal}
}
%\begin{tabular}{ccccccccc}
\begin{tabular}{lllllllll}
\toprule
\multirow{2}{*}{\parbox{1.0cm}{{\bf Sl. No.}}} &
\multirow{2}{*}{\parbox{1.0cm}{\centering{\bf Material/ Expt.ID.}}} &
      \multicolumn{2}{c} {\bf Thickness (mm)} &
\multirow{2}{*}{\parbox{1.5cm}{\centering{\bf Impact Velocity (m/s)}}} &
      \multicolumn{2}{c} {\bf Spall Strength (GPa) } &
      \multicolumn{2}{c} {\bf Spall thickness (mm) } \\
\cmidrule(lr){3-4} \cmidrule(lr){6-7} \cmidrule(lr){8-9}
 & 
 & {\bf Flyer }&{\bf Target}& &{\bf Simu.}&{\bf Expt.}&{\bf Simu.} & {\bf Expt.}\\
\midrule
%1 & Cu / B-70 & 0.40 & 4.30 & \ \ 660 $\pm$ 20 & \ \ 4.05 & \ \ 4.49 $\pm$ 0.1 & 0.34 & 0.33 \ \ $\pm$ 10\% \\
%2 & Cu / B-71 & 0.40 & 4.50 & \ \ 660 $\pm$ 20 & \ \ 3.91 & \ \ 3.95 $\pm$ 0.1 & 0.36 & 0.30 \ \ $\pm$ 10\% \\
%3 & Nb / B-99 & 0.05 & 0.49 & 4100 $\pm$ 150 & \ \ 8.98 & \ \ 8.66 $\pm$ 0.2 & 0.025 & 0.029 $\pm$ 10\% \\
%4 & Nb / B-101 & 0.05 & 0.44 & 4100 $\pm$ 150 & 10.76 & 11.15 $\pm$ 0.2 & 0.033 & 0.029 $\pm$ 10\% \\
%5 & Mo / B-80 & 0.40 & 1.38 & 1250 $\pm$ 50 & \ \ 4.57 & \ \ 5.41 $\pm$ 0.2 & 0.35 & 0.29 \ \ $\pm$ 10\% \\
%6 & Mo / B-82  & 0.05 & 0.275 & 4100 $\pm$ 150 & 10.69 & 11.49 $\pm$ 0.2 & 0.027 & 0.022 $\pm$ 10\% \\
%8 & Al / 28:Asay  & 1.99 & 5.685 & 2338 $\pm$ 12    & \ \ 1.59 & \ \ 1.74 $\pm$ 1\%  & 1.62 & 1.52 \ \ $\pm$ 1\% \\
%9 & Al / \ 8:Asay  & 3.197 & 5.881 & \ \ 543 $\pm$ 2      & \ \ 1.21 & \ \ 1.12 $\pm$ 1\%  & 1.82 & 2.60 \ \ $\pm$ 1\% \\
1 & Cu / B-70 & 0.40 & 4.30 & \ \ 660 $\pm$ 3.0\% & \ \ 4.05 & \ \ 4.49 $\pm$ 0.1 & 0.34 & 0.33 \ \ $\pm$ 10\% \\
2 & Cu / B-71 & 0.40 & 4.50 & \ \ 660 $\pm$ 3.0\% & \ \ 3.91 & \ \ 3.95 $\pm$ 0.1 & 0.36 & 0.30 \ \ $\pm$ 10\% \\
3 & Nb / B-99 & 0.05 & 0.49 & 4100 $\pm$ 3.7\% & \ \ 8.98 & \ \ 8.66 $\pm$ 0.2 & 0.025 & 0.029 $\pm$ 10\% \\
4 & Nb / B-101 & 0.05 & 0.44 & 4100 $\pm$ 3.7\% & 10.76 & 11.15 $\pm$ 0.2 & 0.033 & 0.029 $\pm$ 10\% \\
5 & Mo / B-80 & 0.40 & 1.38 & 1250 $\pm$ 4.0\% & \ \ 4.57 & \ \ 5.41 $\pm$ 0.2 & 0.35 & 0.29 \ \ $\pm$ 10\% \\
6 & Mo / B-82  & 0.05 & 0.275 & 4100 $\pm$ 3.7\% & 10.69 & 11.49 $\pm$ 0.2 & 0.027 & 0.022 $\pm$ 10\% \\
7 & Al / Baumang & 0.40 & 2.90 & \ \ 700 $\pm$ NA & \ \ 2.01 & \ \ 2.21 $\pm$ NA & 0.55 & 0.40 \ \ $\pm$ NA \\
8 & Al / 28:Asay  & 1.99 & 5.685 & 2338 $\pm$ 0.5\%& \ \ 1.59 & \ \ 1.74 $\pm$ 1\%  & 1.62 & 1.52 \ \ $\pm$ 1\% \\
9 & Al / \ 8:Asay  & 3.197 & 5.881 & \ \ 543 $\pm$ 0.25\%& \ \ 1.21 & \ \ 1.12 $\pm$ 1\%  & 1.82 & 2.60 \ \ $\pm$ 1\% \\
\bottomrule
\end{tabular}
\end{table}
\vspace*{\fill} % Pushes the table down from the top
\clearpage % Forces a clean page break
\end{landscape}
% %  for Nb101, the values for expt are 1700-1070 = 630 and Multiscale is 1750-1180=570, (630-570)/630 = 9.5 %
% %  for Nb101, the values thickness are expt : 1700-1070 = 630 and Multiscale 15 ns==>  mm nearly 9 % deviation it seems
%%%%%%%%%%%%%%%%%%%%%%%%%%%%%%%%%%%%

The propagation of shock wave along the flyer and target material for the case of `Al'
(Expt.ID `8:Assay' in table~\ref{tableSingleCrystalSpallVal})
from one dimensional hydrodynamic simulation is shown in figure~\ref{PresProfOf1dx}.
%This hydrodynamic simulation is done for the case of 
%% Indicate which expt is corresponding to this figure.
Figure~\ref{hydroAlAsay08EarlyT} shows the shock wave propagation at very early times.
Shock wave is created upon the flyer impacting a target and it travels in both the
directions from the point of impact which is at a location 2 mm in figure~~\ref{hydroAlAsay08EarlyT}.
The left moving wave reflects from the free surface of the flyer
and a rarefaction wave is formed. Figure~\ref{hydroAlAsay08EarlyT} also shows
the elastic pre-cursor that is formed and visible because at this low impact velocity
(543 m/s), the elastic pre-cursor travels faster than the shock wave. The shock front is also marked in 
fig.\ref{hydroAlAsay08EarlyT}. This spatial profile of pressure at various
times is used to locate the shock front and to calculate the shock velocity.

Figure~\ref{hydroAlAsay08LateT} shows the reflected shock wave from the free surface of
the target. This is a left going rarefaction wave. When this wave and the right coming
rarefaction wave from the free surface of the flyer meet, tensile stress starts growing.
This causes the spallation in materials.

\begin{figure}[!htb]
        \centering
        \subfigure [Early time: Shock wave]
        {
        \includegraphics[width=0.46\textwidth] {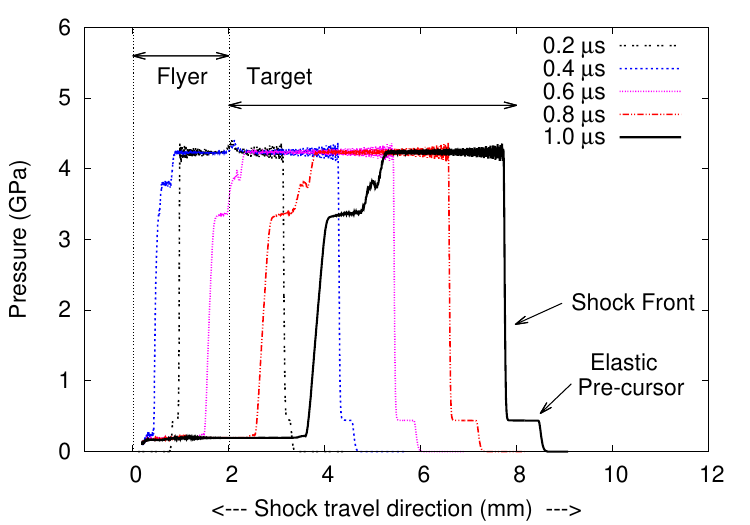}
        \label{hydroAlAsay08EarlyT}
        }
        \subfigure [Later time: Rarefaction wave]
        {
        \includegraphics[width=0.46\textwidth] {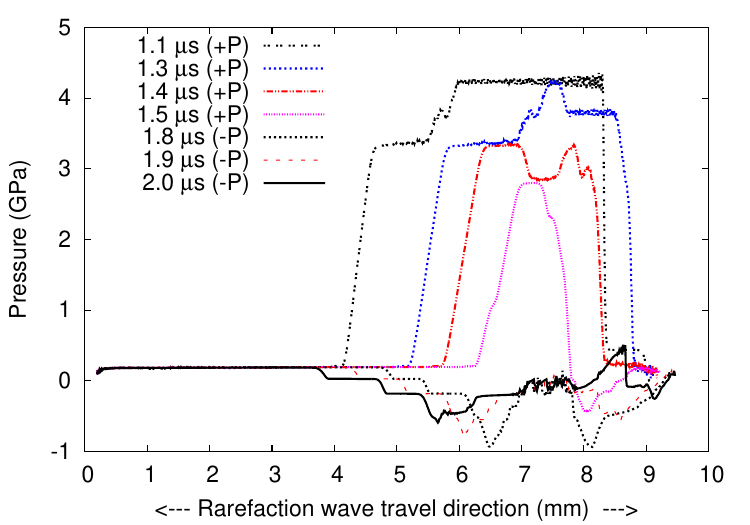}
        \label{hydroAlAsay08LateT}
        }
\vspace{-0.2cm}
 \caption
        [Spatial pressure profile of shockwave in `Al' from hydrodynamic simulations]
        {Spatial pressure profile of shockwave in `Al' (Expt.ID `8:Asay', Ref.table~\ref{tableSingleCrystalSpallVal})
	from hydrodynamic simulations at various
	 moments: (a) Shock wave propagating before reflection from the target free surface. As time
	progresses, rerefaction wave from the free surface of the flyer starts and travels
  	from left to right hand side. (b) Rarefaction wave from the free surface of the target is
	generated and it starts travelling from right to left hand side. `+P' represents the pressure under compression;
	`-P' represents tensile pressure.
	}
    \label{PresProfOf1dx}
\end{figure}

\begin{figure}[!htb]
        \centering
        \subfigure [Copper (Cu) B70]
        {
        \includegraphics[width=0.46\textwidth] {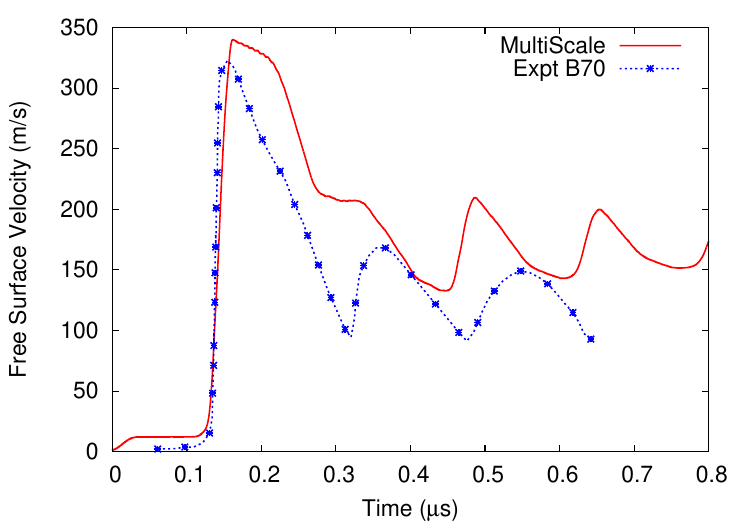}
        \label{chap4Fig4aPofTCuB70}
        }
        \subfigure [Copper (Cu) B71]
        {
        \includegraphics[width=0.46\textwidth] {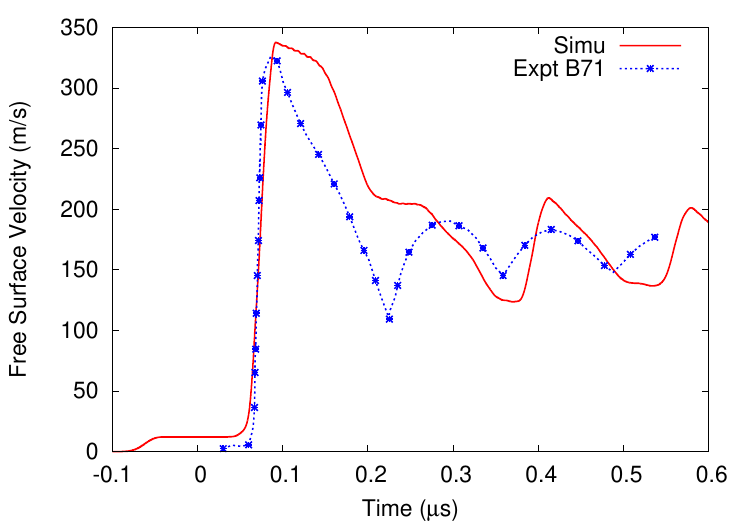}
        \label{chap4Fig4bPofTCuB71}
        }
\vspace{-0.2cm}
 \caption
        [FSV profile for Single Crystal of Cu : Compared with Expt.]
        {Comparison of the temporal evolution of Free Surace Velocity (FSV) from `Multiscale' simulations
	  for single crystal of Cu with expt. from ref.~\cite{Anton_spalldata} }
    \label{FSVCuOnly}
\end{figure}
\begin{figure}[!htb]
	\centering
        \subfigure [Niobium (Nb) B99]
        {
        \includegraphics[width=0.46\textwidth] {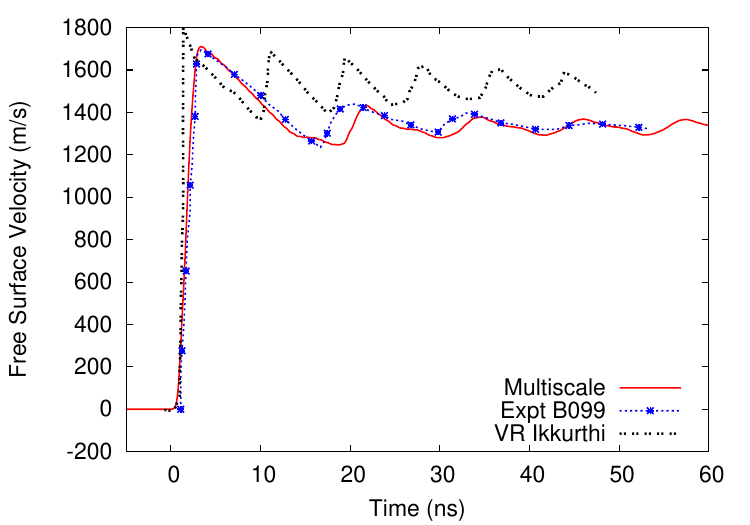}
        \label{chap4Fig4ePofTNbCuB099}
        }
        \subfigure [Niobium (Nb) B101]
        {
        \includegraphics[width=0.46\textwidth] {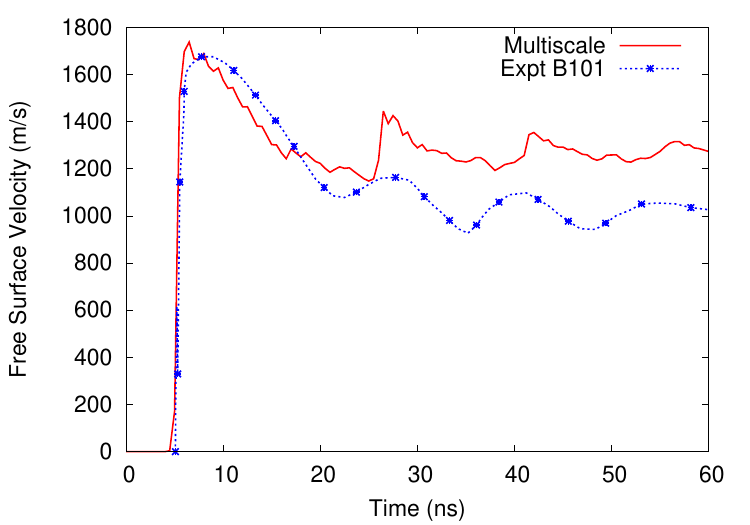}
        \label{chap4Fig4fPofTNbB101}
        }
        \subfigure [Molybdenum (Mo) B80]
        {
        \includegraphics[width=0.46\textwidth]{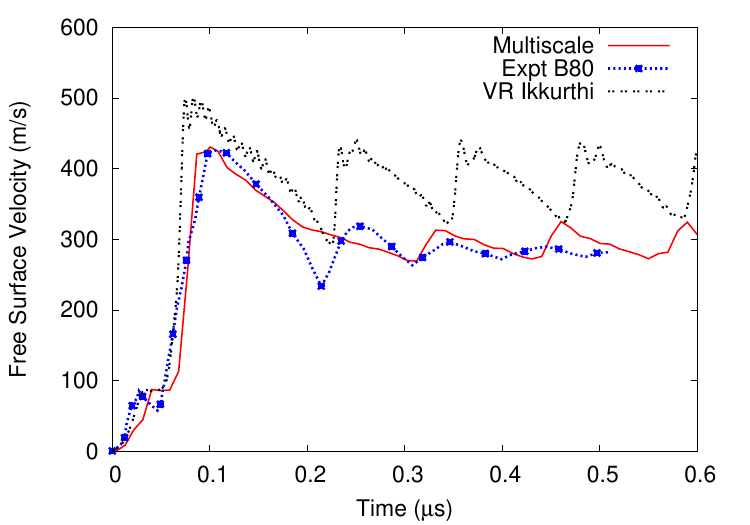}
        \label{chap4Fig4cPofTMoB80}
        }
        \subfigure [Molybdenum (Mo) B82]
        {
        \includegraphics[width=0.46\textwidth]{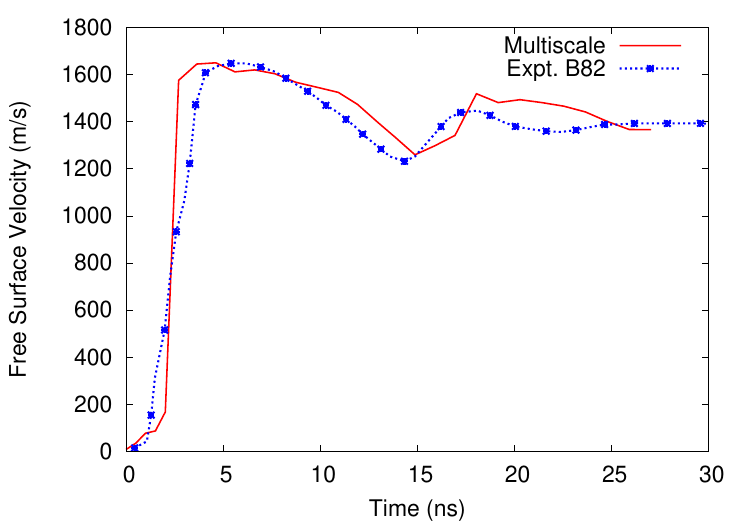}
        \label{chap4Fig4dPofTMoB82}
        }
\vspace{-0.2cm}
 \caption
        [FSV profile for sinble crystals of Nb \& Mo: Compared with Expt.]
        {Comparison of Temporal evolution of Free Surace Velocity (FSV) from `Multiscale' simulations
	  for single crystal of  Nb \& Mo with expt. from ref.~\cite{Anton_spalldata}.
	  For Nb (B99) and Mo (B80), multiscale results reported by VR Ikkurthi {\it et al.} from
	  ref.~\cite{ramana2_hydro} is also presented. Multiscale method is improvised in terms
	  of PSO and hydrodynamic methods to get a better match with experiements when compared
	  to the results, previously reported by VR Ikkurthi {\it et al.}.
	}
    \label{FSVMoNb}
\end{figure}

NAG parameters are obtained newly in this research work for Mo and Al, while for
Cu and Nb, they are adapted from the previous work of V.R Ikkurthi {\it et al.}~\cite{ramana2_hydro}.
However, hydrodynamic impact-shock simulations are performed for these materials afresh
and reported here. The temporal evolution of the free surface velocity calculated
(i) for two cases of Cu are shown in figures~\ref{chap4Fig4aPofTCuB70} and ~\ref{chap4Fig4bPofTCuB71},
(ii) for two cases of Nb are shown in figures~\ref{chap4Fig4ePofTNbCuB099} and \&~\ref{chap4Fig4fPofTNbB101},
(iii) for two cases of Mo are shown in figures~\ref{chap4Fig4cPofTMoB80}\&~\ref{chap4Fig4dPofTMoB82}.
%For Al, three different cases of experiments are simulated.
%One is for the experimental
%parameters of Kanel~\cite{Kanel2001JAPSingleCrysFoilAl}, while the other two are for the
%experimental parameters of Asay~\cite{chen_asay}. Time varying free surface velocities for
%these three cases of Al, are shown in figures~\ref{chap4Fig5aPofTAl1}-~\ref{chap4Fig5cPofTAl3}.

\begin{figure}[!htb]
        \centering
        \subfigure [Aluminum (Al): Boumang Expt.]
        {
         \includegraphics[width=0.46\textwidth]{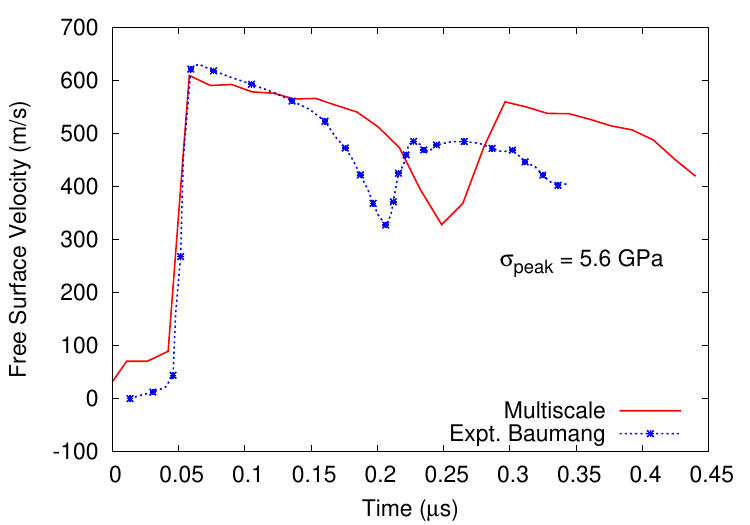}
%% source directory/path for thesisChap4Fig4pt4dSubst.eps
%         \includegraphics[width=0.48\textwidth]{Figures/Spall_SC_MScale/HydroRunsAl/KanelJAPVol9Jul2001PG136-143Run/Run/thesisChap4Fig4pt4dSubst} 	%% House PC
%         \includegraphics[width=0.48\textwidth]{
%/home/madhavpc2008/03082015_General/Lammps/BiCrystal/RsltFrom20TfMachines_Al_NAG_Spall/CopyOfCode_Runs_FromPendriveUsedForGnuplot/MD-PSO-HydroResult/Code/KanelJAPVol9Jul2001PG136-143Run/Run/thesisChap4Fig4pt4dSubst}     %% #Office Data mini pc
        \label{chap4Fig5aPofTAl1}
        }
        \subfigure [Aluminum (Al): Asay Expt.28]
        {
        \includegraphics[width=0.46\textwidth]{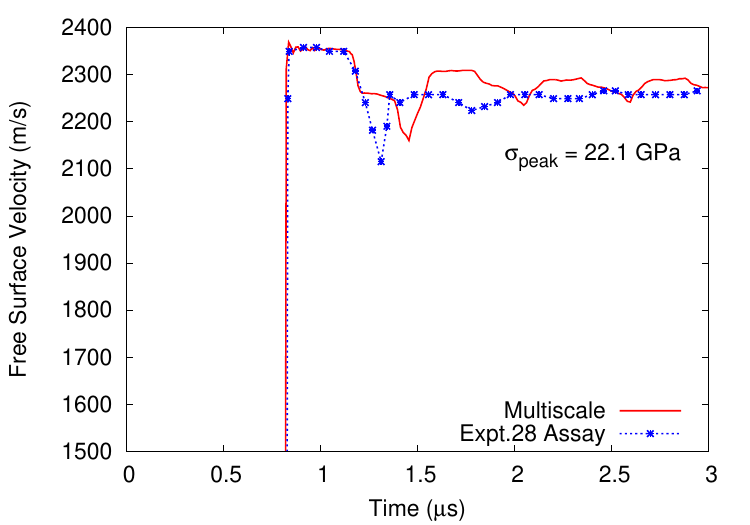}
        \label{chap4Fig5bPofTAl2}
        }
        \subfigure [Aluminum (Al): Asay Expt.8]
        {
        \includegraphics[width=0.46\textwidth]{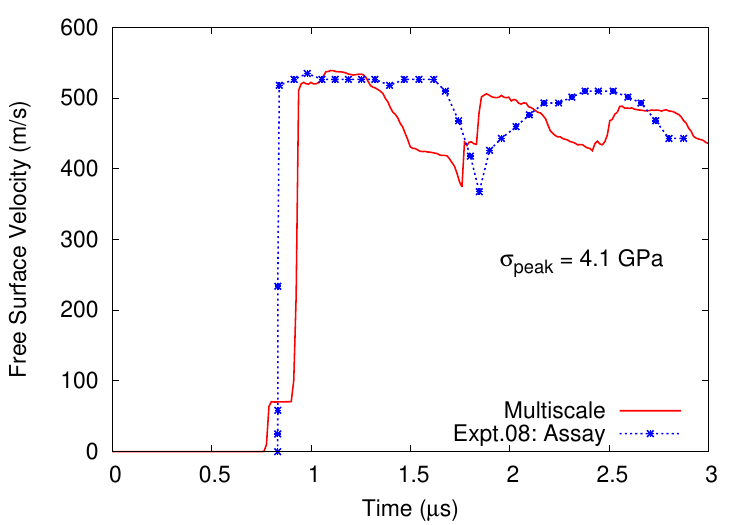}
        \label{chap4Fig5cPofTAl3}
        }
\vspace{-0.2cm}
 \caption
        [FSV profile for SC (Cu, Nb, Mo \& Al): Compared with Expt.]
        {Comparison of Temporal evolution of Free Surace Velocity (FSV) from `Multiscale' simulations
	  for single crystal of Al in 3 different configurations:
          (a) with Baumang-expt. in ref.~\cite{Kanel2001JAPSingleCrysFoilAl};
	  (b) \& (c): with Assay expt. 28, 8 as in ref.~\cite{chen_asay} }
    \label{FSVAl3Cases}
\end{figure}

For Al, three different cases of experiments are simulated.
One is for the experimental
parameters of Kanel~\cite{Kanel2001JAPSingleCrysFoilAl}, while the other two are for the
experimental parameters of Asay~\cite{chen_asay}. Time varying free surface velocities for
these three cases of Al, are shown in figures~\ref{chap4Fig5aPofTAl1}-~\ref{chap4Fig5cPofTAl3}.

The temporal free surface velocity profile calculated from the multiscale method
for single crystal metals matched reasonably well when compared with
the experimental results. The deviation in the spall strength value estimated
from these multiscale simulations when compared with experiments is ranging
between 2\% to 15.4\% among these four materials. For copper, with two
experimental configurations (B70 \& B71) the deviations are 9.8\% and 1\%
respectively. It may be noted that the variation in the target thickness among
the experiments, B70 and B71 is 4.5\% only. Reported spall strength, corresponding
to these experiments are 4.49 GPa and 3.95 GPa, which is about 12\% difference
among them. The main difference between these two experiments is that the target
%is annealed for 2 hours at 900$^o$ C for B71: pg.B-71 of \cite{Anton_spalldata}
is annealed for 2 hours at 900$^o$ C for B71~\cite{Anton_spalldata}.
%may be referred
Annealing is not considered in the simulation and it starts
at room temperature. The only difference between these two configurations in the
simulations is the excess 2 mm target thickness in B71 in comparison to that of
B70. Though, qualitatively no significant difference is seen in the pattern of
the FSV profiles of B70 and B71 of Cu, but the pull-back velocity (${\Delta}U$)
resulted for these two cases have some differences. Using
eqn.~\ref{eqn_sp_strengthChap4}, calculated spall strength, for the cases of Cu:
B70 and B71 differ by $\sim10\%$ which is consistent with the difference seen in
their FSV profiles of the multiscale calculations. In the case of Niobium, the spall
strength obtained from the multiscale method deviates by $\sim$3.5\%,
in comparison to the experiments marked as B99 and B101~\cite{Anton_spalldata}.
For Mo, calculated spall strength  deviated by 15.4\% for B80 and 6.9\% for B82,
when compared to respective experiments~\cite{Anton_spalldata}.

Comparison of multiscale result from Rawat~\cite{rawat_thesis} for Cu, with the
experiment ID B71~\cite{Anton_spalldata}, the spall strength was found to be over
estimated by $\sim21.5\%$. Whereas, this work has improved the multiscale calculation
in a way that such a large deviation of over 20\% reported by Rawat~\cite{rawat_thesis}
is reduced to just 1\%. With the experimental configurations of Nb (B99), V.R.
Ikkurthi {\it et al.}~\cite{ramana2_hydro} used multiscale to calculate the 
spall strength of niobium that under-estimated the spall strength by 11.8\%
in comparison to the experiment. Spall strength obtained for niobium from this
research work deviates from that of the experiment only by 3.5\%.

For aluminum with three different configurations adapted from Kanel-Baumang, Chen-Asay-28,
and Chen-Asay-8, the deviations are 9\%, 8.6\% and 8\% respectively. To the best of
our knowledge this is the first time, a multiscale method is applied to a single
crystal aluminum to find its NAG fracture coefficients and spall parameters. The previous
literature works have successfully demostrated this methodology for Cu~\cite{rawat_thesis},
Nb and Mo~\cite{ramana2_hydro}.

The spall thickness resulted from the experiments are reported~\cite{Anton_spalldata} for
Cu, Nb and Mo.  For Cu B70, the spall thickness also matched well and the deviations
are well with in 3\%, while for B71 the deviation is the spall thickness is 22\%. For
Nb:B99 and B101, the deviations are found to be 13\% and 16.7\% respectively. For Mo with
two configurations (B80 and B82), the deviation is found to be 23\% and 24\% respectively.
For single crystal aluminum, spall thickness is not reported by any of the three
experiments cited here. When the FSV profile resulting from the simulations are compared
qualitatively with respective experiments, it is found that the time at which the pull
back velocity changes its direction is different. The time period of oscillations are
also different when compared with experiments. This indicates that the spall thickness
obtained from the simulations are different from that of the experiments. For aluminum
the spall thickness values are mentioned in table~\ref{tableSingleCrystalSpallVal}.
These values are calculated using the eqn.~\ref{eqn_sp_thickChap4}, where the values
of $T_f$ are obtained from the temporal FSV profile of corresponding experiments
and simulations. The reason for such a large deviation in the spall thickness is not
clear from the present study. It needs further studies to know the underlying
physics and to account for such phenemona in the multiscale model.
%physics that makes such deviations.
%% Spall strength
% Cu B70 (4.5-3.96)/4.5*100 = 12 %
% Cu B71 (3.95-3.71)/3.95*100 = 6.08 %
% Nb B99 (8.66-8.88)/8.66*100 = 2.54 %
% Mo B80 (5.4-5.95)/5.4*100 = 10.18%
% Mo B87 (4.6-4.83)/4.6 = 5 %)
%Al Baumang:  (2.2-2.01)/2.2  = 8.6 % 0.0863636363636365
%Al Asay 28:  (1.74-1.47)/1.74 = 15.5 % 0.155172413793103
%Al Asay 28:  (1.591-1.47)/1.74 = 15.5 % 0.085632183908046
%Al Asay  8:  (1.12-1.21)/1.12 = 8    % -0.0803571428571427
%% Spall thickness

\section{Conclusion}
The multi-scale model is developed for simulating flyer-plate shock-impact experiments.
This method is executed sequentially in three parts. First, molecular dynamics method
is used to obtain temporal void volume fraction and pressure data for a metallic single
crystal under tri-axial tensile force. MD simulations are performed using a open source
LAMMPS at atomic length and time scales. Results from the MD simulations are post processed
%by a PSO method.
using voxel approach and the 
NAG fracture model parameters are calculated for aluminum and molybdenum using the PSO code.
These NAG parameters are used in one dimension hydrodynamic code,
to simulate flyer impact on a target at high velocities and the spall parameters
are calculated from the FSV profiles obtained from the HD simulations.
Results from multiscale method for single crystals of four metals
are presented and discussed in this chapter. A comparison with experiments are also
presented. Following are the conclusions drawn from this work.
\begin{itemize}
\vspace{-0.2cm}
\item MD simulations of tri-axial tensile deformation are perfomed and time varying void volume
	fraction and pressure data are obtained. This method simulated the void nucleation,
	void growth and coalascence at a constant strain rate ($\dot{\epsilon}$ = 5$\times{10}^9\ s^{-1}$).
\vspace{-0.2cm}
\item The PSO method developed for this work is found to be efficient and fast in convergence. The NAG fracture
	parameters obtained for the single crystals are in the right range.
\vspace{-0.2cm}
\item For copper, niobium and molybdenum, the spall strength and spall thickness are estimated using
      multiscale model.
      The FSV profiles obtained from this work match the experiments better than other MD results
	published in the literature~\cite{rawat_thesis, ramana2_hydro}.
\vspace{-0.2cm}
\item The multiscale method is appplied to aluminum single crytal for the first time.
For aluminum, NAG parameters are obtained by PSO fitting and the temporal Free Surface Velocity (FSV) profile
      is obtained from hydrodynamic calculation.
The spall strength estimated from FSV for `Al', is found to be matching well with experiments.
      The calculated spall strength is found to deviate around 8\% in comparison to the experiments.
\vspace{-0.2cm}
%\item Multiscale method is found to meet the objective of spall strength calculation for the
%single crystals (Cu, Nb, Mo \& Al).
\item The multiscale method calculates the spall strength of single crystals with an accuracy in the range 85\% (for Mo)
 to 98\% for (Nb). However, the deviation found in the period of oscillation of the FSV profile are large,
 when compared with experiments. Further investigations are required to understand the reasons 
 for such deviations.
\vspace{-0.2cm}
\end{itemize}

% Al by obtaining the spall strength and spall thickness and verifying with experiments. The new PSO algorithm and the 1-D hydrodynamic code too have been validated by comparing the results from these with the earlier published results from our multi-scale modeling of Mo, Nb and Cu. Atomistic simulations using the LAMMPS code was used to generate temporal pressure and void volume changes of single crystals subjected to isotropic tensile pressures. Particle Swarm Optimization was used to fit the NAG model parameters to the void volume changes with pressure observed in the MD simulations. The NAG parameters so obtained for single crystals of  Mo, Nb, Cu and Al, are used in locally developed 1D hydro-dynamic FORTRAN program to obtain the spall strength and spall thickness, under impact dynamics. The results are found to match well with the available experimental results.

\clearchapter

% BiCrystal_Setup : Chapter 5. Introduction to BiC, creation using MD, GBE validation (figures, table for STGB and STwGB)
%  \include{BiCrystal_Setup}
% MD_Spall_BiC : Chapter 5. Spallation by impact shock using MD only for Al. (paper under preparation: MS&E:A)
  %\chapter{Spallation by impact shock using MD in Al Bi-Crystals~\label{chapter_spallMDonly}}
\chapter{MD Simulations of Spallation at Aluminum Bi-crystal Boundaries\label{chapter_spallMDonly}}
%\nopagebreak
%\makeatletter
%{\renewcommand{\@makefntext}[1]{\noindent #1}%
%\footnotetext{%
%  \fontsize{10pt}{12pt}\selectfont%
%  The content of this chapter is partly adapted from the author's published work~\cite{ImplastMatPr12023}.}}
%\makeatother

%% Section 1
\section{Background of the problem~\label{sec1-Chap5}}
As already mentioned in chapter~\ref{chapter_Intro},  several groups have
studied the behavior of materials under static and dynamic loading conditions that
includes experiments~\cite{JC_fracture, fortov1991, spall_moshe1998, spall_moshe2000,
Kanel2001JAPSingleCrysFoilAl} and computational methods~\cite{seamanCompmodelDuctBrit_JAP,
curran_nag1, johnsonPulseDurationSpallFrac_JAP, ikkurthi1_nag1, RINGDALENVATNE2013306,
IJF_AlBucklingSM}. Molecular Dynamics (MD) methods are widely used to study the dynamic
response of materials under impact shock and the fracture parameters are
calculated~\cite{srinivasan_MdSpallSingleNiXl, srinivasan2007, rawat1, JIANG2022114474,
Influence_Fensin, ZHU2022110923}. Specifically, bi-crystals with two grains separated
by a single grain boundary are easy to set up in the MD domain to study the effect of
shock-propagation across grain boundaries~\cite{ADLAKHA2014345, LIN201294, Luo_bicrys_asymm_sigma3_110TiltGB,
Han_spall_shockCuBiC, Fensin_GBonplasticdefor, ZHOU2014116, Barrles_MSEA}.

Srinivasan {\it et al.} have observed in a single crystal Ni, that a compressive shock
wave changes the microstructure followed by the creation of multiple grain boundaries
due to a tensile relaxation wave~\cite{srinivasan_MdSpallSingleNiXl}. Ductile failure
occurs when tensile waves interact near a grain boundary because the grain boundary serves
as void nucleation sites~\cite{srinivasan_MdSpallSingleNiXl}. Grain boundary structure
has its role in the formation of dislocation. By providing a varying degree of mobility,
the interface atoms assist plastic deformation~\cite{ADLAKHA2014345}. For a grain boundary
interface, Yanguang {\it et al.}~\cite{ZHOU2014116} have reported different modes of crack
propagation. The mode of crack propagation at critical temperature changes from
brittle to ductile for a certain grain boundary orientation. Many aspects of the
interaction of dislocation with grain boundary and its mechanical response are still
to be explained despite extensive research held over the years~\cite{Barrles_MSEA}.
Theoretically, a critical stress $G/2\pi$ is required to generate dislocations inside
a perfect crystal, where G is the shear modulus of the material.
% Whereas it is observed that
The critical stress to generate dislocations at the proximity of a grain boundary is
observed to be smaller than that for a perfect crystal~\cite{Barrles_MSEA}. Shock
propagation across individual bicrystal grain boundaries is studied using MD simulations to
observe shock-induced phase transition in Fe~\cite{xueyang_GBonShockInduPhTransinFeBiCryst},
spallation in Cu twist GB~\cite{LONG2020109411} and effect of various STGB and STwGB on the
shock-induced spallation in Al polycrystal~\cite{madhavan_spall_multiscale}. Details of the
multiscale spall-simulation of Al polycrystal~\cite{madhavan_spall_multiscale} with various
symmetric tilt and twists (STGB \& STwGB) are given in chapter~\ref{chapter_BiCSpall-Multiscale}.

Spall parameters are calculated by experiments using the temporal Free Surface Velocity (FSV)
profile~\cite{zhernokletovBook}. %% Pg 246
But in many of the published MD spall simulations, calculated peak tensile stress (${\sigma}_{T}$) 
is mentioned as the spall strength, for example in (i) a single crystal Al~\cite{JIANG2022114474},
(ii) bicrystal Cu in tilt~\cite{Influence_Fensin} and twist~\cite{LONG2020109411} GB systems.
In Cu tilt GB system the influence of grain boundary on the spall strength (${\sigma}_{sp}$)
is studied using MD simulations by calculating only the peak ${\sigma}_{T}$~\cite{Influence_Fensin}.
However, some researchers have reported the temporal FSV profile in MD simulations
%In these works, FSV profiles are used to estimate the spall strength
and used the FSV profiles to estimate the spall strength ${\sigma}_{sp}$~\cite{srinivasan2007,
ZHU2022110923, LIN201294}. In MD simulation of single crystal Ni, the effect of strain rates on the
spall strength is also reported~\cite{srinivasan2007}. These studies~\cite{srinivasan2007,
ZHU2022110923,LIN201294} compare the values of ${\sigma}_{sp}$ obtained from both FSV and
peak tensile stress (${\sigma}_{T}$). In the case of nano Cu/Ni bimetallic multilayer
%\ac{PTS} (${\sigma}_{T}$). In the case of nano Cu/Ni bimetallic multilayer
case~\cite{ZHU2022110923}, the spall strength (${\sigma}_{sp}$) obtained from FSV was almost equal
to the values of ${\sigma}_{T}$ in a piston based shock creation system. For single crystal
Ni~\cite{srinivasan2007} and Cu bicrystal with $\Sigma$3 asymmetric tilt GB~\cite{LIN201294} at
various impact loading, the spall strength (${\sigma}_{sp}$) obtained from FSV are higher than
${\sigma}_{T}$ values, for instance ${\sigma}_{sp} \simeq 1.5 \times$ peak~$\sigma_{T}$~\cite{LIN201294}.

MD spall calculations are available in the literature for Cu bicrystal and single
crystal Al, Ni \& Cu. In this work, we focus on the spall phenomena in the Al bicrystal
system using MD simulations. The primary motivation is to understand the effect of individual
STGB~\cite{ImplastMatPr12023} and STwGB on the spall fracture parameters of aluminum (Al) 
and thereby examine the merits of using the peak tensile stress (${\sigma}_{T}$) or the FSV profile
in predicting spall strength.
Our earlier work to estimate the spall strength of Al single crystal
%using various
%% MADHAVAN2022111543 in paper = madhavan_spall_multiscale in the Thesis bib
%bicrystal GB
%was based on the multiscale modelling~\cite{madhavan_spall_multiscale}.
was based on the multiscale modelling~\cite{icons2018_madhavan}
described
%Multiscale modelling is carried out in multiple steps as mentioned
in chapter~\ref{chapter_SCSpall-Multiscale}.
Whereas in this work, the spall strength is calculated directly from MD simulations
by recording both the peak tensile strength and the temporal variation of the free surface.
%that is accelerated by the impact shock mechanism.
For this, Al bicrystals with different grain boundary (GB) orientations are
created. The sizes of the target and the flyer are chosen such that
the reflected rarefaction waves meet at the bicrystal grain boundary.
Only symmetric tilt (STGB) and symmetric twist (STwGB) cases are considered.
Apart from estimating spall strength (${\sigma}_{sp}$), peak tensile stress (${\sigma}_{T}$) is also calculated.
The main objective is to compare the ${\sigma}_{sp}$ calculated from FSV
with the values of ${\sigma}_{T}$, since some literature report calculated
${\sigma}_{T}$ as the spall strength (${\sigma}_{sp}$).
Calculated spall strength (${\sigma}_{sp}$) from direct MD simulation is compared 
with experimentally available spall data of Al polycrystalline.
Details of the MD simulations and their results are discussed in this chapter.
Concluding remarks from this study are presented in section~\ref{chap5conclusion}.

%\section{Computational details\label{chap5comp-details}}
\section{Computational aspects\label{chap5comp-details}}
\nopagebreak
%% Section 2
Three-dimensional atomistic MD impact simulations are performed using Large-scale Atomic Molecular
Massively Parallel Simulator~\cite{lammps} for ten different symmetric
tilt and twist grain boundaries of aluminum (Al) bi-crystals (bi-C) setup.

\subsection{Creation of Bi-crystal set up and validation}
\label{chap5sec2.1}
Five symmetric tilt (STGB) and five symmetric twist (STwGB) bi-crystals of Al are created following
the methods of Tschopp~\cite{tschopp1}. The range of misorientation angles for STGB is from
$0^\circ$ to $90^\circ$, while it is $0^\circ$ to $45^\circ$, for the STwGB cases. This range
covers reasonably the calculated grain boundary energy (GBE) of Al bi-crystal system for validation.
Bi-crystal MD simulations, require a certain minimum length in each of the directions to satisfy
the periodicity of atomic arrangement~\cite{tschopp1}. Accordingly, the number of unit cells
in the three directions is chosen. The Embedded Atom Method (EAM) potential by Mishin
{\it et al.}~\cite{Michin_Al99.eam.alloy} is used here. The potential is validated by comparing the
values obtained for the elastic constants, vacancy formation energy and cohesive energy etc.,
with experiments which are listed in table~\ref{elasticTable}, chapter~\ref{chapter_compMethods}.
Using iterative minimization methodology, the grain boundary energies
(GBE) are calculated and found to match the values available in the literature
Tschopp~\cite{TiGBEtschopp2} for STGB and QingYin {\it et al}~\cite{al_twist_GBE_QingYin} for STwGB.
GBE values obtained from the MD simulations are tabulated for comparison in
table~\ref{table_GBE_TiltAndTwist}. Other details of setting up a bicrystal system and calculating
grain boundary energy using the MD method can be obtained from the
literature~\cite{madhavan_spall_multiscale,tschopp1}.
\begin{table}[!ht]
\centering
\caption
[Calculated GBE (MD-BARC) for various symmetric tilt \& twist angles of Al bicrystal,
are compared with the values available in the literature]
{Calculated GBE (MD-BARC) for various symmetric tilt \& twist angles of Al bicrystal,
are compared with the values available in the MD-literature~\cite{TiGBEtschopp2, al_twist_GBE_QingYin}
(STGB MD data:Ref.~\cite{ImplastMatPr12023})}
\begin{footnotesize}
\begin{tabular}{cccccc}
\toprule
\multirow{2}{*}{\makecell {\bf GB \\  \bf Type}} &
\multirow{2}{*}{\makecell {\bf Sl. \\  \bf No.}} &
\multicolumn{2}{c}{\bf Grain Boundary} &
\multicolumn{2}{c}{\bf Grain boundary Energy (mJ/m$^2$)} \\
\cmidrule(lr){3-6}
&
&
{\bf Plane} &
{\bf Angle} &
{\bf MD-BARC} &
{\bf MD-Literature} \\
\midrule
\multirow{1}{*}{\makecell {STGB~\cite{TiGBEtschopp2}} }
& 1 & $($10 1 0$)$ & ${11.4}^\circ$ & 419 & 418.6 \\
& 2 & $($5 1 0$)$ & ${22.6}^\circ$ & 489 & 488.9 \\
& 3 & $($3 1 0$)$ & ${36.9}^\circ$ & 465 & 464.8 \\
& 4 & $($2 1 0$)$ & ${53.2}^\circ$ & 494 & 494.0 \\
& 5 & $($5 4 0$)$ & ${77.3}^\circ$ & 351 & 351.4 \\
\midrule
 \multirow{1}{*}{\makecell {STwGB~\cite{al_twist_GBE_QingYin}}} 
& 1 & $($8 1 0$)$ & ${14.2}^\circ$ & 318 & 318.7 \\
& 2 & $($5 1 0$)$ & ${22.6}^\circ$ & 343 & 343.5 \\
& 3 & $($4 1 0$)$ & ${28.0}^\circ$ & 353 & 352.7\\
& 4 & $($3 1 0$)$ & ${36.9}^\circ$ & 326 & 326.7 \\
& 5 & $($5 2 0$)$ & ${43.6}^\circ$ & 349 & 348.5 \\
\bottomrule
\end{tabular}
\end{footnotesize}
\label{table_GBE_TiltAndTwist}
\end{table}

\subsection{Replication of bi-crystal set-up for MD impact simulation}
\label{chap5sec2.2}
The bicrystal set-up made from the above methods has only a minimum number of unit cells in
three orthogonal directions. To perform shock propagation simulation and to
ensure single planar shock propagation without any reflection from edges,
these bi-crystal sets are to be replicated appropriately in all three
directions. This replication increases the simulation domain.

From our earlier work~\cite{UsUp_Madhavan2022}, we have seen that the system with unit cells
(i) 99$\times$40$\times$40, (ii) 99$\times$30$\times$30 and (iii) 99$\times$20$\times$20 did not
produce variation in the (a) shock propagation, (b) $U_s$-$U_p$ and  (c) $P$-$V$ for
aluminum single crystal system.
%Here the shock propagation direction is made to have
%the largest number of unit cells. This kind of system size ensures the elimination of the finite size
%effects~\cite{rawat1}. Thus
The computational domain for this work
is made out of 20$\times$45$\times$20 units cells in X, Y and Z directions respectively.
The plane wave shock is made to propagate along the Y direction, hence the number of unit cells
in the Y direction is higher.
\begin{figure}[!htb]
     \centering
         \includegraphics[width=0.79\textwidth]
	 {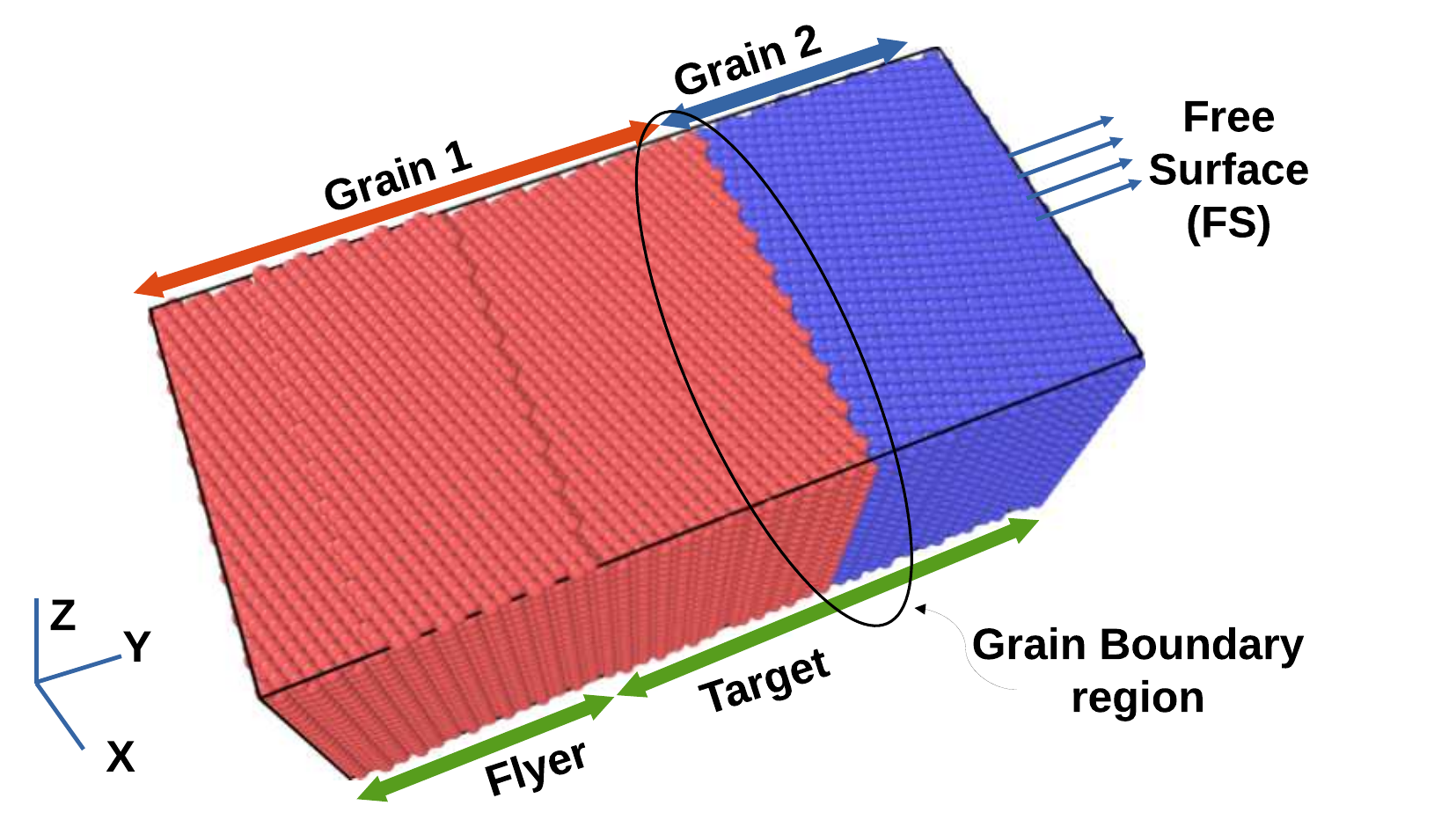}
         \caption
	[Arrangement of the atoms showing the flyer-target system and a grain boundary]
	{Initial arrangement of the atoms: The flyer, the target with a grain
        boundary. The planar shock generated by the impact is in the XZ plane and it propagates
        along the Y-direction. (Fig:Ref.~\cite{ImplastMatPr12023})}
	\label{MDgrain_initial}
\end{figure}

For the shock impact simulations, the atoms belonging to one-third of the Y-axis
length are designated as flyer atoms and initialized with a specific velocity
(valued between 1 and 3.5 km/s) and directed towards the other atoms that form
the target. The grain boundary
is made to be at the center of the target along the shock propagation direction (Y).
A three-dimensional view of the position of the atoms is shown in figure~\ref{MDgrain_initial},
at the time of impact. It shows the flyer, target and an encircled grain boundary
region that demarcates the two grains. In the target, one grain to the left side of the
grain boundary is represented by red-colored atoms while the other grain which is on the
 right side of the grain boundary is represented by blue colored atoms. The flyer consists of
only the red colored atoms whereas the target consists of both red and blue colored atoms,
with the red-blue interface lying at its center.

\subsection{MD impact simulation of a bi-crystal system}
\label{Chap5Sec2.3}
MD-NPT simulations are performed on the flyer-target system at 0 bar and 300 K with a
time-step of 1 femtosecond.  Periodic boundary conditions (PBC) are applied along X,
Y and Z directions to equilibrate the system. The positions of the atoms after the
equilibrium are used as the initial positions in the impact simulations. The impact
simulations are performed using constant \ac{NVE} ensemble.  For the impact simulations, the
boundaries in the Y-direction are set to move freely by the force resulting from the
impact, while the boundaries in the X and Z directions are set to be periodic (PBC).
The time step used for impact simulations is 0.1 femtosecond. In this work the
flyer and the target are created with Al atoms to simulate Al-Al impact.

When a planar shockwave propagates, it causes a one-dimensional strain along the
direction of propagation. The strain rate due to shock propagation can be calculated
from the spatial particle-velocity gradient. The strain rate at any instant of
time $t$ is calculated from the particle velocity gradient between any two
distinct locations that enclose the shock front. For the shock propagation along
Y-direction, the time-varying strain rate $\dot{\epsilon}(t)$ is calculated as,
\begin{equation}
\dot{\epsilon}(t) = \left(\frac{V_{y1}(t)-V_{y_2}(t)}{Y_1(t)-Y_2(t)}\right)
\label{velocity_strain_eqn}
\end{equation}
where $Y_1$ and $Y_2$ are two different locations along the Y direction, while
$V_{y1}$ and $V_{y2}$ correspond to the Y-component velocity at the locations
$Y1$ and $Y2$ respectively. In this work, the two locations are chosen such
that $Y_1$ and $Y_2$ are near respective ends of the target material along the Y-direction.

When the shock wave reaches the free surface of the target, the atoms near the
free surface start moving. To record the temporal evolution of Free Surface
Velocity (FSV), a cubical volume from the computational domain is designated as
a `free-surface region' for the target. The choice of the length in the Y-direction for
the `free-surface-region' must be a reasonable value. For instance, in
terms of the lattice parameter `a', it can be `a or 2a or 3a'. It is observed
that the temporal FSV calculated from the `free-surface-region' formed by
`a', `2a' and `3a' do not vary. For this work, `3a' is chosen. The width and
height for this `free-surface-region' are the computational domain size specified
in the X and Z directions respectively. Space and time-averaged Y directional
velocities of the atoms enclosed by this volume are calculated and stored.

\section{Results and Discussion}
%% Section 3
\label{chap5Resultssec}
Different shock loading conditions are simulated using MD for a set of
five STGB and five STwGB cases. These ten cases are individually subjected to impact
velocities in the range of 1 to 3.5 km/s in steps of 0.5 km/s. The flyer plate and the target
materials used in this work are made of Al atoms. Due to the same shock-impedance
`$\rho_0 U_s$', the magnitude of the particle velocity gained by the flyer and target is
equal to half of the impact velocity, $U_p$=$U_{flyer}$/2. The results from the MD
shock impact simulation and discussion are presented here in terms of the particle
velocity ($U_p$). Wherever the notation `$U_p$' is used in this work, it has a relationship
with flyer velocity ($U_{flyer}$) as mentioned above. When the shock wave travels and reaches the
free surface, the free surface starts moving with a velocity $V_{fs}$ which is
theoretically equal to double the value of $U_p$~\cite{meyersBook}. This can be
understood from the fact that the total stress at any free surface is zero. At the free
surface, a negative-reflected/rarefaction wave originates during the same time the
positive-incident shock wave reaches there. Here, positive and negative indicate the
direction of propagation for the incident shock wave and the reflected rarefaction wave
respectively. On the surface where these two oppositely directed waves meet, corresponding
stresses cancel each other to keep the net-stress zero~\cite{JamesASmith2016}.
Initially, the particles at the free boundary acquire a velocity $U_p$ upon the arrival of a shock wave.
At the same time, a rarefaction wave is formed due to the shock wave reflection from the free
surface. The rarefaction wave pushes the free surface away from it further. Thus with the formation
of a rarefaction wave, the particles at the free surface acquire additional velocity.
%It can be explained as given below.
In the case of a free surface, the incident wave is transmitted from a denser
(metal) to a rarer (air) medium. It enhances the particle velocity ($U_p$) at the
free surface~\cite{meyersBook}. Based on this principle, the velocity of the particles at the free surface
becomes double the magnitude of the incident particle velocity~\cite{meyersBook, JamesASmith2016}.
Thus, $V_{fs}$ = 2$U_p$ = $U_{flyer}$. The peak FSV is nearly equal to double the value of respective particle
velocity as shown in figures~\ref{1-22.6tilt_fsvoft}-\ref{2-22.6twist_fsvoft}, ~\ref{1-tilt_fsvoft}
-\ref{6-tilt_fsvoft} and~\ref{1-twist_fsvoft}-\ref{6-twist_fsvoft}.
\begin{figure}[!htb]
     \centering
	\subfigure
	[FSV for STGB $22.6^{\circ}$ (Fig:Ref.~\cite{ImplastMatPr12023})]
	{
         \includegraphics[width=0.475\textwidth]
           {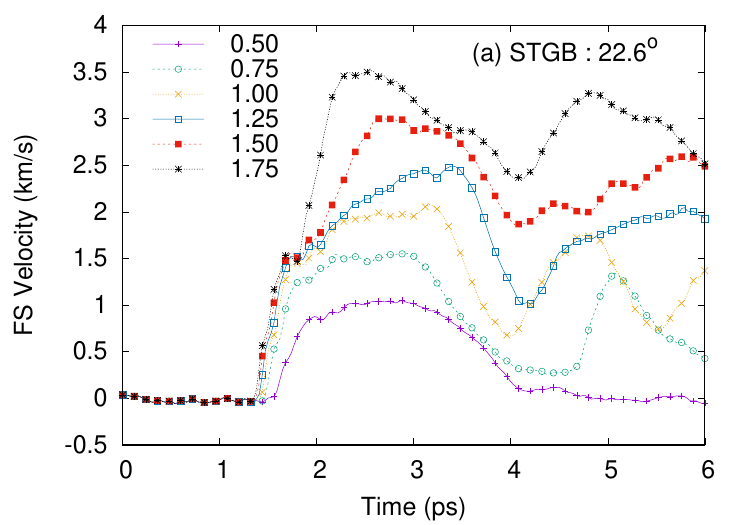}
	        \label{1-22.6tilt_fsvoft}
	}
	\subfigure
	[FSV for STwGB $22.6^{\circ}$]
	{
         \includegraphics[width=0.475\textwidth]
           {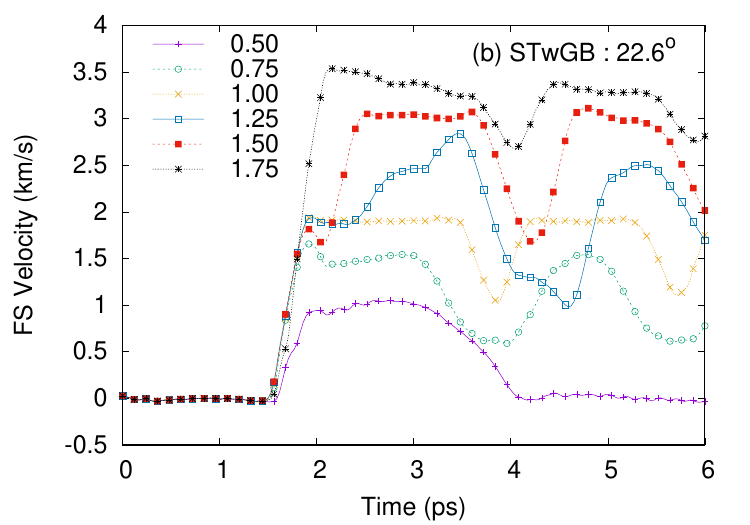}
	       \label{2-22.6twist_fsvoft}
	}
        \caption
	[MD impact shock simulations :
        Temporal free surface velocities for STGB \& STwGB]
	{MD impact shock simulations :
        Temporal variation of free surface velocities (FSV) is shown
	for GB angle, $22.6^{\circ}$ in the case of (a) STGB, (b) STwGB with
	various particle velocities ($U_p$ in km/s).
	}
	 \label{fig:spall_TiltAndTwist}
\end{figure}

\subsection{Temporal FSV for Tilt and Twist GB, $22.6^o$ with different levels of impact shock loading}
\label{chap5sec3.1}
Typical temporal variation of Free Surface Velocities (FSV) obtained from MD simulations
are shown for misorientation angle $22.6^o$ in figures~\ref{1-22.6tilt_fsvoft}
and~\ref{2-22.6twist_fsvoft} for STGB and STwGB respectively. These two figures show
various forms of FSV that is resulted for the various particle velocities when a
shock waves travel across respective grain boundaries. The particle
velocities ($U_p$) range between 0.5 to 1.75 km/s in steps of 0.25 km/s.
Using these curves, the spall strength (${\sigma}_{sp}$) is
obtained~\cite{madhavan_spall_multiscale} from the relation,
\begin{equation}
{\sigma}_{sp} = \left(\frac{1}{2}\right) {\rho}_0 C_0 {\Delta}U \label{eqn_sp_strength}
\end{equation}
where ${\rho}_0$ is the normal density and $C_0$ is the bulk sound speed of the
target material. ${\Delta}U$ is the pull-back velocity. It is the difference in
the velocity values at the 1st peak and the subsequent minima, recorded from the
temporal FSV profile.

\subsection{Comparison of Temporal FSV for 5 STGB and 5 STwGB for various particle velocities:
\label{chapsec3.2}}
Figures~\ref{1-tilt_fsvoft} to~\ref{6-tilt_fsvoft} show the temporal FSV profile for
various STGB cases corresponding to the particle velocity, $U_p$ = 0.5 to 1.75 km/s,
in steps of 0.25 km/s respectively. Similarly figures~\ref{1-twist_fsvoft} to~\ref{6-twist_fsvoft}
show the same for the various STwGB cases.
}
\begin{figure}[!htb]
     \centering
     	 \subfigure
	[FSV for $U_p$ 0.5 km/s]
	 {
         \includegraphics[width=0.45\textwidth]
           {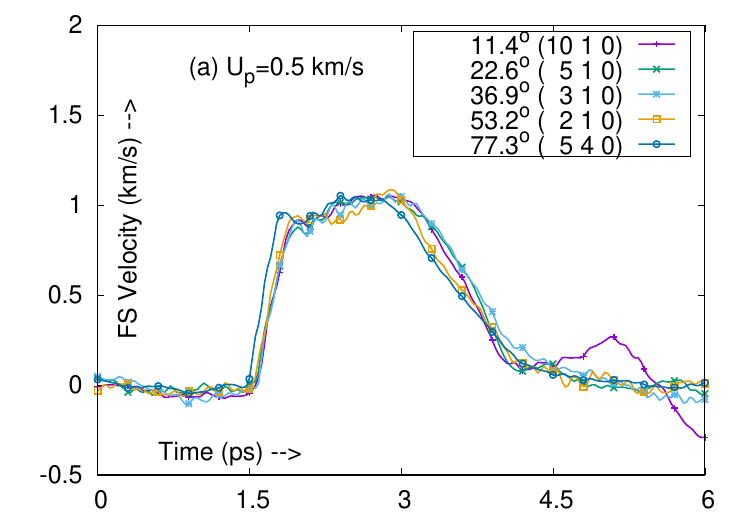}
	       \label{1-tilt_fsvoft}
	 }
         \subfigure
	[FSV for $U_p$ 0.75 km/s]
	 {
         \includegraphics[width=0.45\textwidth]
           {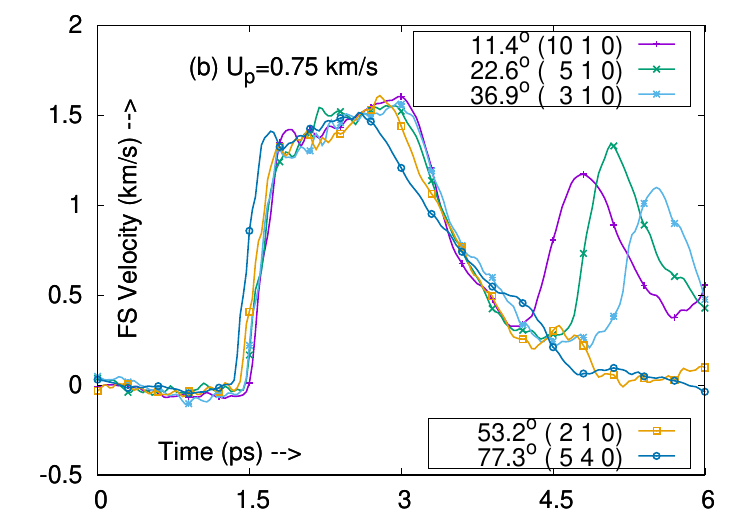}
	       \label{2-tilt_fsvoft}
	 }
         \subfigure
	[FSV for $U_p$ 1.0 km/s]
	 {
         \includegraphics[width=0.45\textwidth]
           {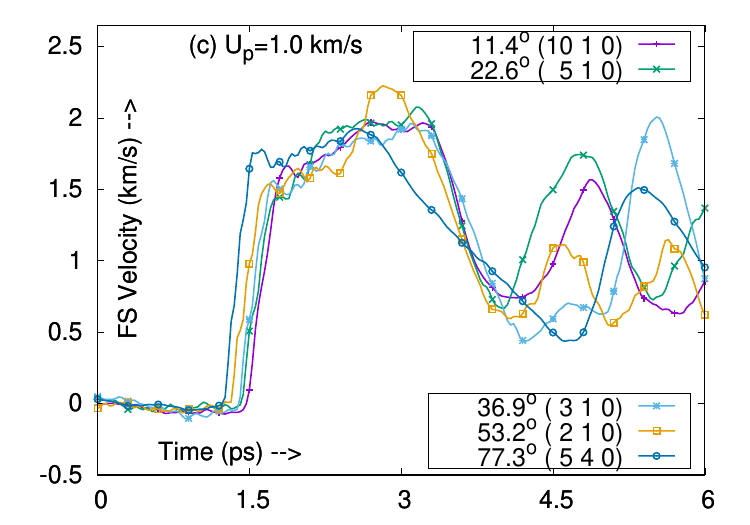}
	       \label{3-tilt_fsvoft}
	 }
         \subfigure
	[FSV for $U_p$ 1.25 km/s]
	 {
         \includegraphics[width=0.45\textwidth]
           {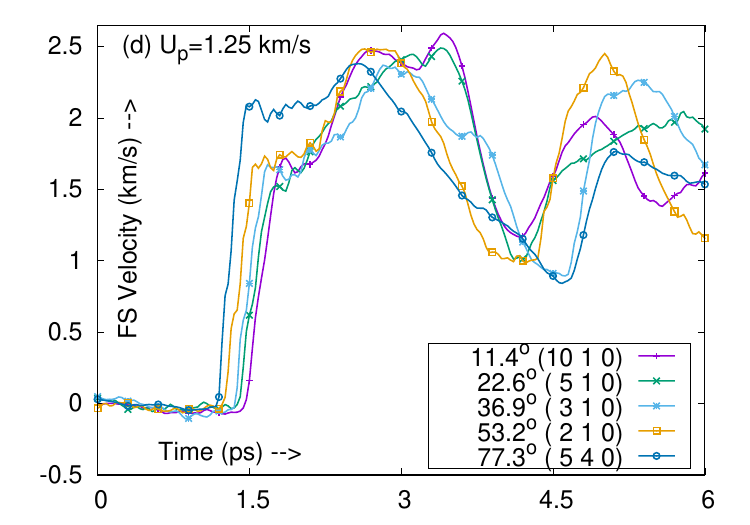}
	       \label{4-tilt_fsvoft}
	 }
         \subfigure
	[FSV for $U_p$ 1.5 km/s]
	 {
         \includegraphics[width=0.45\textwidth]
           {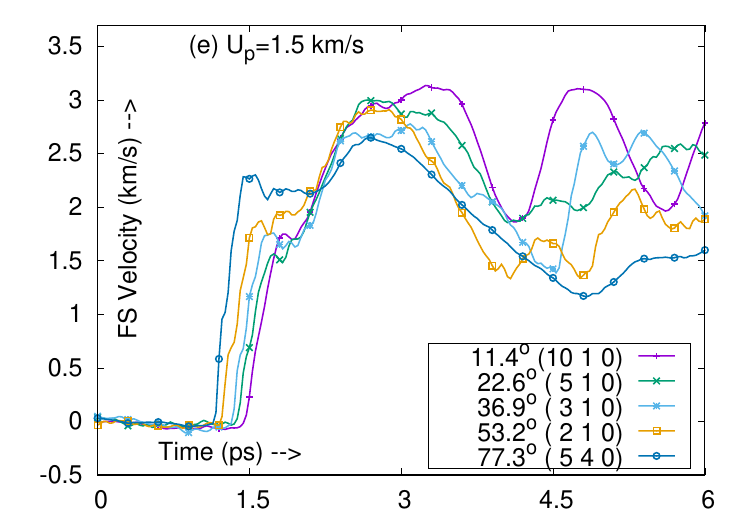}
	       \label{5-tilt_fsvoft}
	 }
         \subfigure
	[FSV for $U_p$ 1.75 km/s]
	 {
         \includegraphics[width=0.45\textwidth]
           {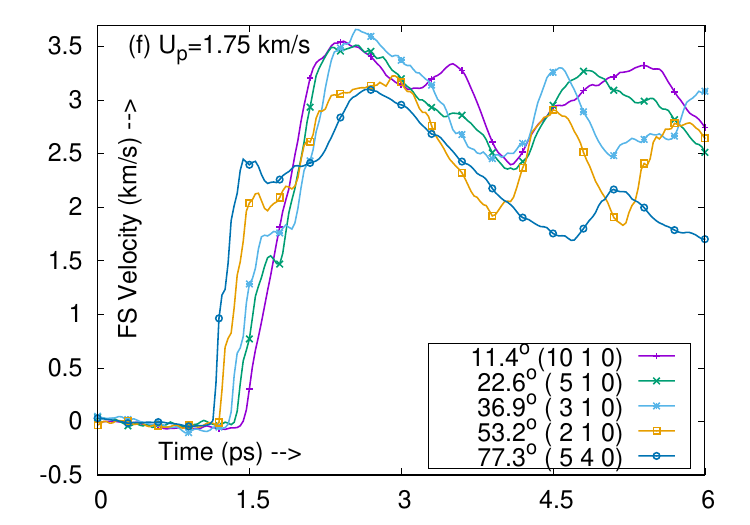}
	       \label{6-tilt_fsvoft}
	 }
        \caption
[FSV as a function of time for different STGB with various $U_p$]
{FSV as a function of time for different STGB with various particle velocities ($U_p$) (Fig:Ref.~\cite{ImplastMatPr12023})}
	\label{fsv_allTilts}
\end{figure}

\begin{figure}[!htb]
     \centering
	 \subfigure
	 [FSV for $U_p$ 0.5 km/s]
	 {
         \includegraphics[width=0.45\textwidth]
           {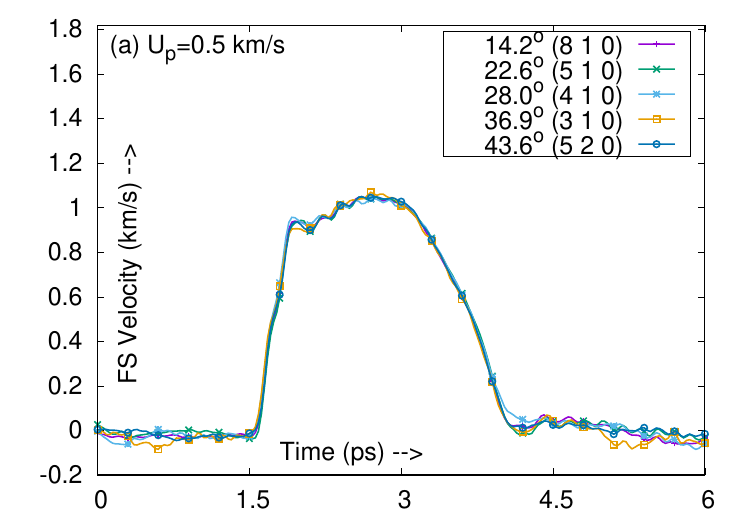}
	       \label{1-twist_fsvoft}
	 }
	 \subfigure
	 [FSV for $U_p$ 0.75 km/s]
	 {
         \includegraphics[width=0.45\textwidth]
           {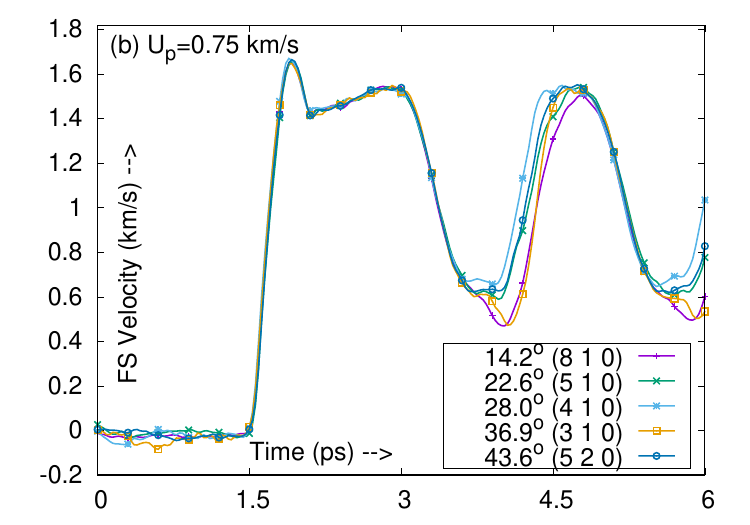}
	       \label{2-twist_fsvoft}
	 }
	 \subfigure
	 [FSV for $U_p$ 1.0 km/s]
	 {
         \includegraphics[width=0.45\textwidth]
           {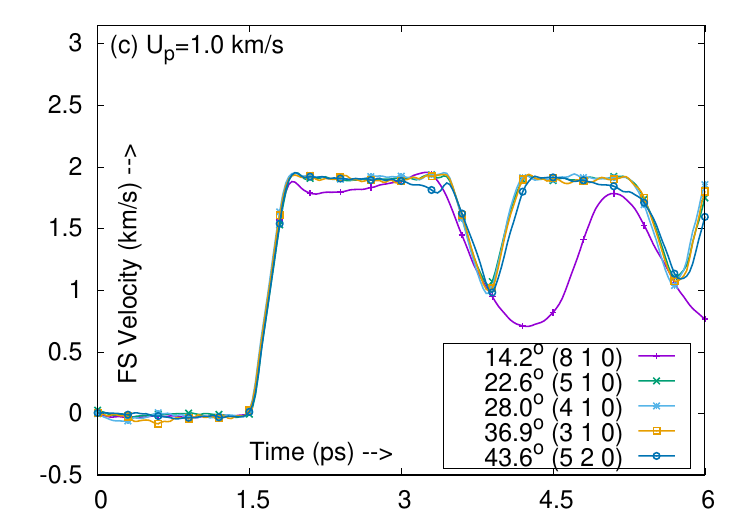}
	       \label{3-twist_fsvoft}
	 }
	 \subfigure
	 [FSV for $U_p$ 1.25 km/s]
	 {
         \includegraphics[width=0.45\textwidth]
           {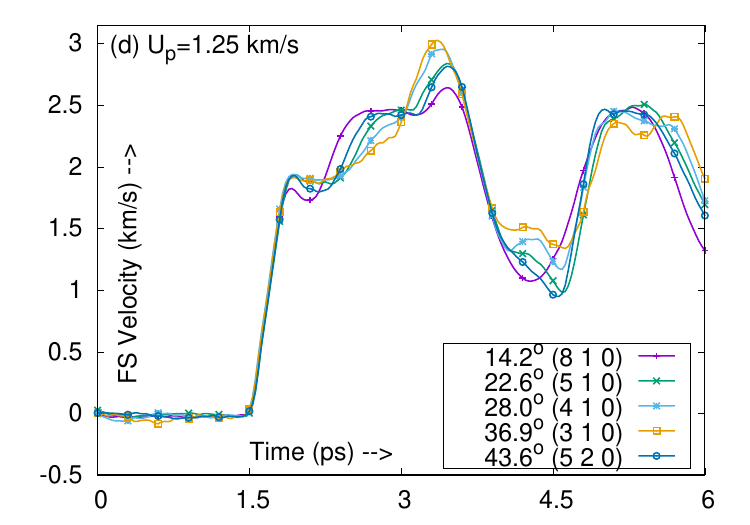}
	       \label{4-twist_fsvoft}
	 }
	 \subfigure
	 [FSV for $U_p$ 1.5 km/s]
	 {
         \includegraphics[width=0.45\textwidth]
           {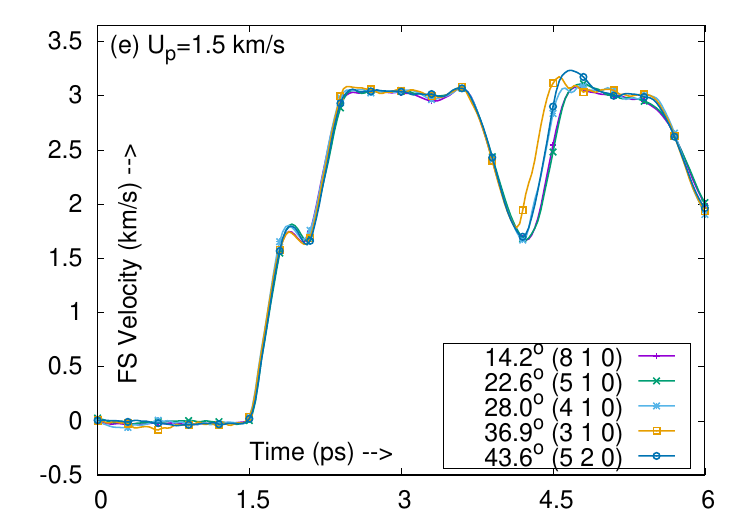}
	       \label{5-twist_fsvoft}
	 }
	 \subfigure
	 [FSV for $U_p$ 1.75 km/s]
	 {
         \includegraphics[width=0.45\textwidth]
           {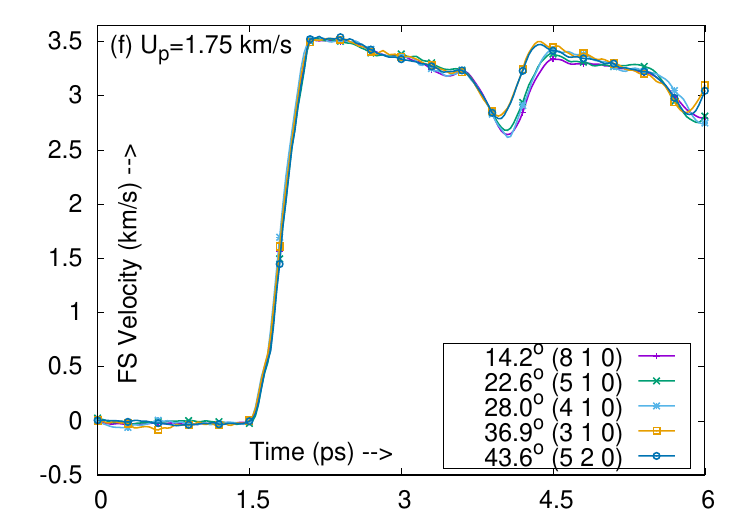}
	       \label{6-twist_fsvoft}
	 }
        \caption
[FSV as a function of time for different STwGB with various $U_p$]
{FSV as a function of time for different STwGB with various particle velocities ($U_p$)}
	\label{fsv_allTwist}
\end{figure}

\subsubsection{Trend of FSV profile obtained from MD simulations for the STGB and
STwGB cases.} % at impact velocities ($U_p$ =0.5 to 1.75 km/s) }
From the temporal FSV profile in figure~\ref{fsv_allTilts} it is inferred that,
\vspace{-3mm}
\begin{itemize}
\item No spall occurred for $U_p$ = 0.5 km/s for all the five STGBs
\vspace{-3mm}
\item For $U_p$ = 0.75 km/s, no spall observed for higher misorientation angle GBs
        ($53.2^o$ and $77.3^o$) and
\vspace{-3mm}
\item For all other cases spall occurred.
\end{itemize}
\vspace{-3mm}
However the time of occurrence of spall and the pull back velocities ($\Delta{U}$) are
different among themselves as seen in figures~\ref{2-tilt_fsvoft}-\ref{6-tilt_fsvoft}.
This indicates that the spallation parameters are sensitive to grain boundary misorientation
at the atomistic scales of length ($\AA$) and time ($10^{-12}$-$10^{-15}$ ps).
As desired, MD simulations indicate that the spall originates near the grain boundary in most cases.
For the cases of STwGBs, the frames in figure~\ref{fsv_allTwist} indicate that,
\vspace{-3mm}
\begin{itemize}
\item No spall occurred for $U_p$ = 0.5 km/s.
\vspace{-3mm}
\item Unlike STGB cases, for $U_p \geq 0.75$ km/s, spall occurred for all STwGBs.
\vspace{-3mm}
\item FSV profiles that are obtained for different STwGBs with the same impact conditions
are not very different.
\end{itemize}
%\vspace{-3mm}

\noindent More importantly, the free surface starts moving almost at the same time for a given
impact velocity. This is because the grain orientation differs only in the `XZ' plane
and the periodic arrangement of atoms along the `Y' direction is disrupted lesser in
STwGB when compared to STGB. For STGB, the grain orientation is at an angle along
the `Y' direction and it influences the shock propagation speed. This angle
varies with the STGB cases, hence the shock reaches the free surface at
different times, for a given impact velocity. This effect is visible in
figures~\ref{1-tilt_fsvoft}-\ref{6-tilt_fsvoft}, between the times 1.2 to 1.7 ps.

\subsection{Averaged ${\sigma}_{sp}$ and the peak $\sigma_{T}$
for various STGB and STwGB~\label{chap5sec.3.3}}
\begin{figure}[!htb]
     \centering
	\subfigure
	[$\sigma_{fsv}$ Vs $U_p$ (Fig:Ref.~\cite{ImplastMatPr12023})]
	{
         \includegraphics[width=0.46\textwidth]
	 {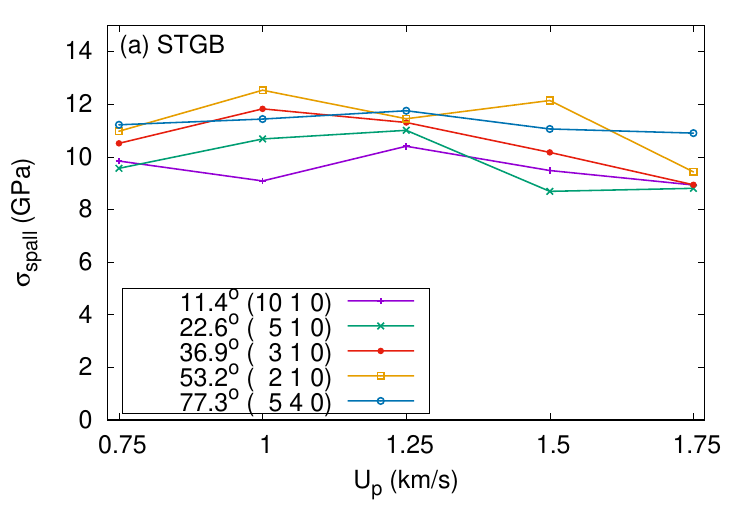}
	\vspace{-1.8\baselineskip}
	\label{spallFSVvsUpStgb}
	}
    \vspace{2mm}
	\subfigure
	[$\sigma_{fsv}$ Vs $U_p$]
	{
         \includegraphics[width=0.46\textwidth]
	 {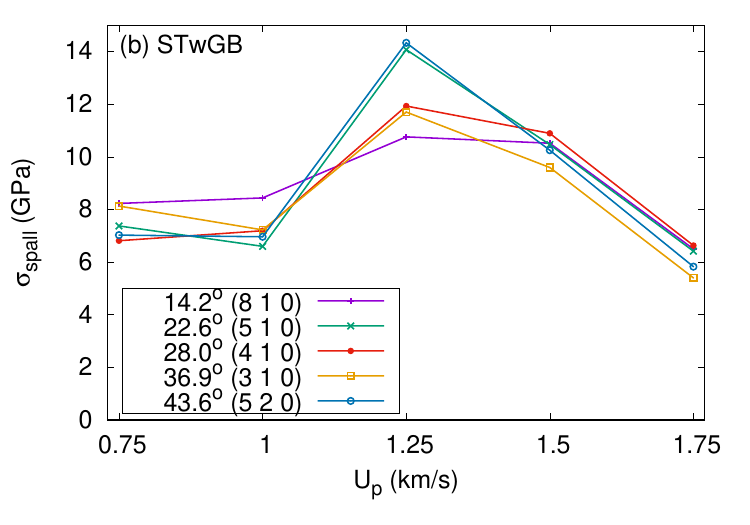}
	\vspace{-1.8\baselineskip}
	\label{spallFSVvsUpStwgb}
	}
	\subfigure
	[$\sigma_{fsv}$ Vs GB Angle]
	{
         \includegraphics[width=0.46\textwidth]
	 {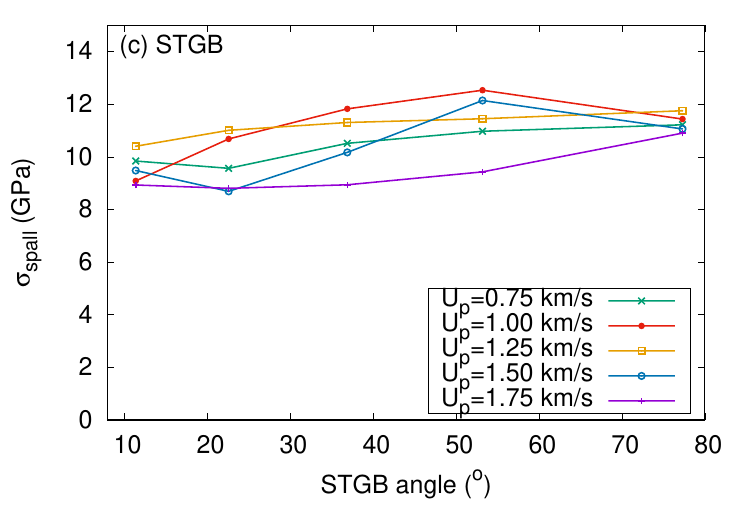}
	\vspace{-1.8\baselineskip}
         \label{SpallFSVvsStgbAng}
	}
    \vspace{2mm}
	\subfigure
	[$\sigma_{fsv}$ Vs GB Angle]
	{
         \includegraphics[width=0.46\textwidth]
	 {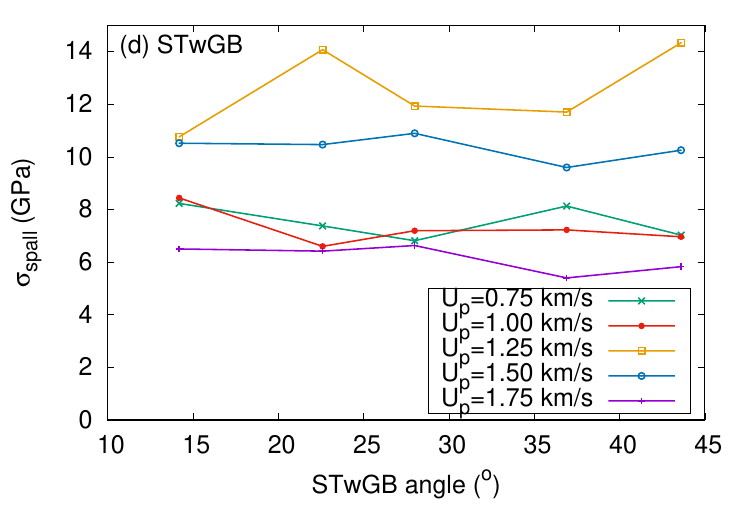}
	\vspace{-1.8\baselineskip}
	 \label{SpallFSVvsStwgbAng}
	}
    \vspace{2mm}
	\subfigure
	[$\sigma_{fsv}$ Vs GBE]
	{
         \includegraphics[width=0.46\textwidth]
	 {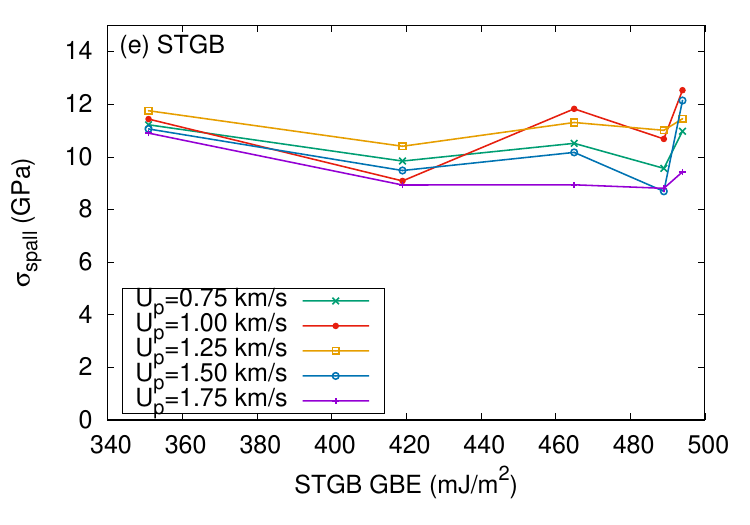}
	\vspace{-1.8\baselineskip}
	\label{spallFSVvsUpStgbE}
	}
	\subfigure
	[$\sigma_{fsv}$ Vs GBE]
	{
         \includegraphics[width=0.46\textwidth]
	 {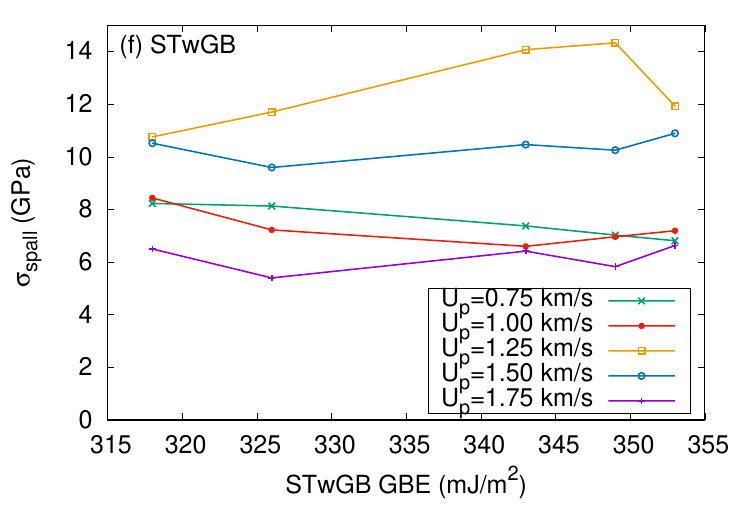}
	\vspace{-1.8\baselineskip}
	\label{spallFSVvsUpStwgbE}
	}
     \caption
	[Comparison of  the spall strength (${\sigma}_{sp}$)]
	{Comparison of  the spall strength (${\sigma}_{sp}$)
      obtained from MD simulations for various STGBs in figures a,c,e
	\& STwGBs in figures b,d,f.}
	\label{multi_fig5}
\end{figure}

\begin{figure}[!htb]
	\subfigure
	[$\sigma_{fsv},\ \sigma_{T}$  Vs $U_p$]
	{
         \includegraphics[width=0.48\textwidth]
	 {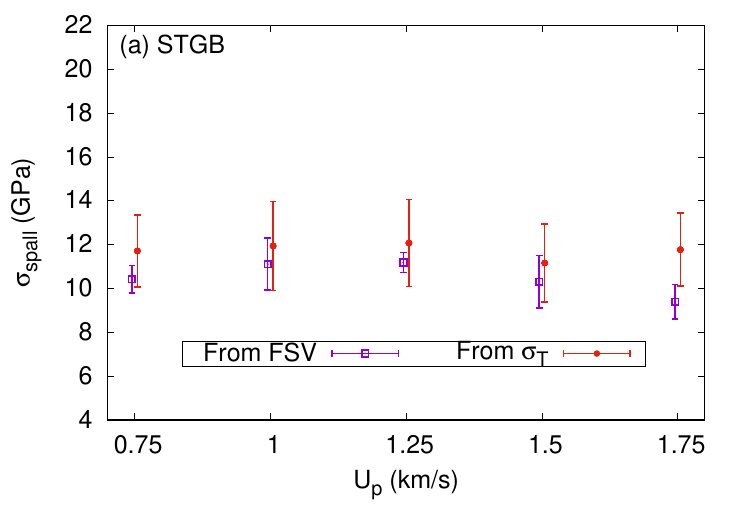}
	\vspace{-1.8\baselineskip}
	\label{spall12_FSVvsUpStgb}
	}
    \vspace{2mm}
	\subfigure
	[$\sigma_{fsv},\ \sigma_{T}$  Vs $U_p$]
	{
         \includegraphics[width=0.48\textwidth]
	 {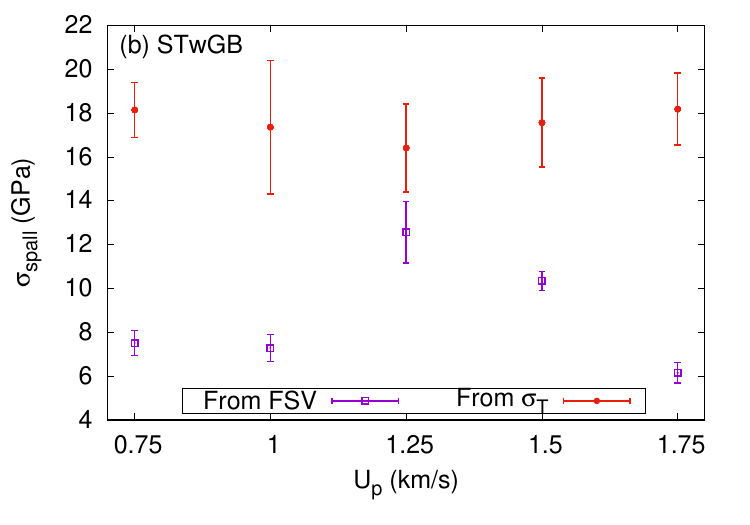}
	\vspace{-1.8\baselineskip}
	\label{spall12_FSVvsUpStwgb}
	}
	\subfigure
	[$\sigma_{fsv},\ \sigma_{T}$  Vs GB Angle]
	{
         \includegraphics[width=0.48\textwidth]
	 {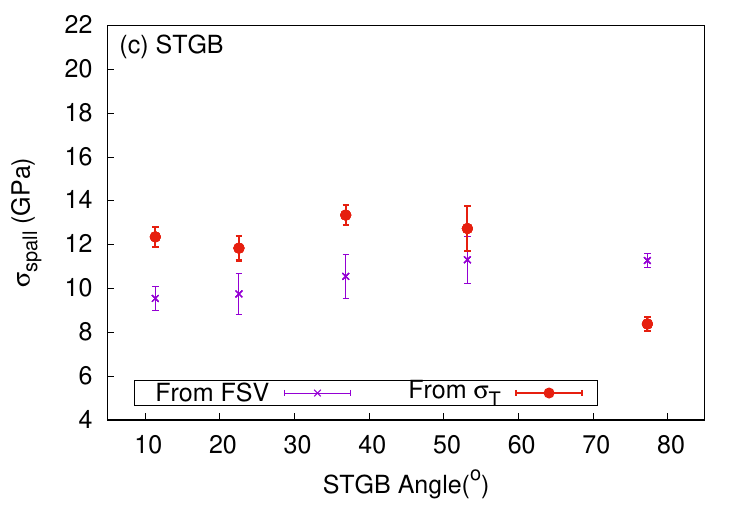}
	\vspace{-1.8\baselineskip}
         \label{spall12_FSVvsStgbAng}
	}
    \vspace{2mm}
	\subfigure
	[$\sigma_{fsv},\ \sigma_{T}$  Vs GB Angle]
	{
         \includegraphics[width=0.48\textwidth]
	 {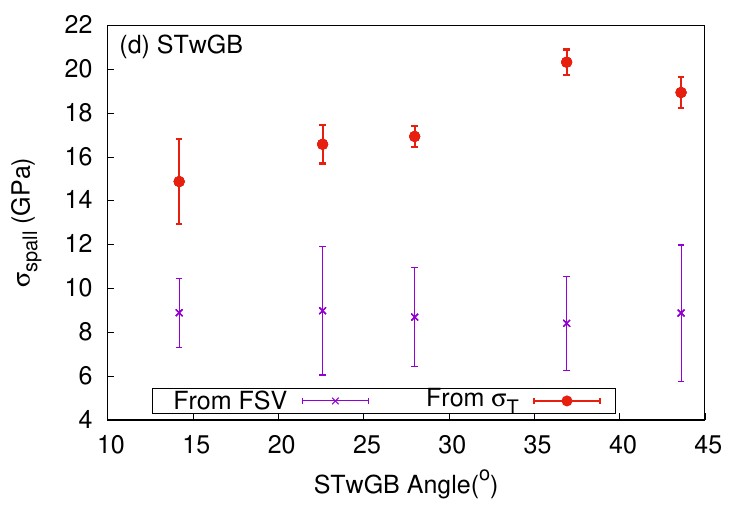}
	\vspace{-1.8\baselineskip}
	 \label{spall12_FSVvsSTwgbAng}
	}
    \vspace{2mm}
	\subfigure
	[$\sigma_{fsv},\ \sigma_{T}$  Vs GBE]
	{
         \includegraphics[width=0.48\textwidth]
	 {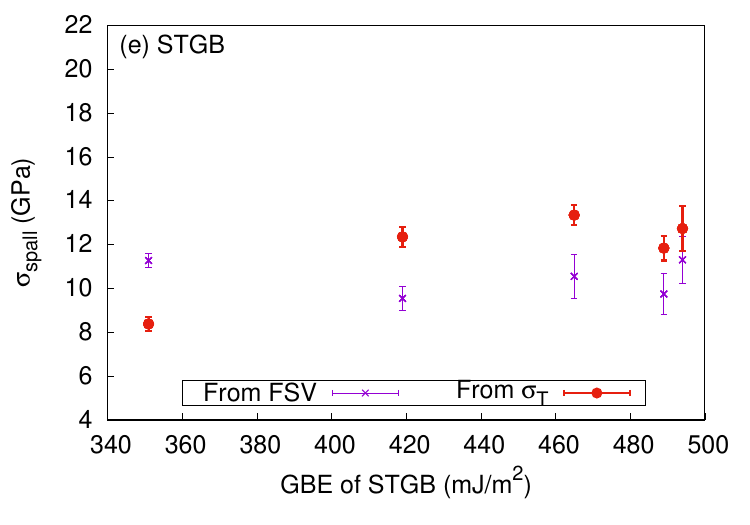}
	\vspace{-1.8\baselineskip}
	\label{spall12_FSVvsUp_StgbE}
	}
	\subfigure
	[$\sigma_{fsv},\ \sigma_{T}$  Vs GBE]
	{
         \includegraphics[width=0.48\textwidth]
	 {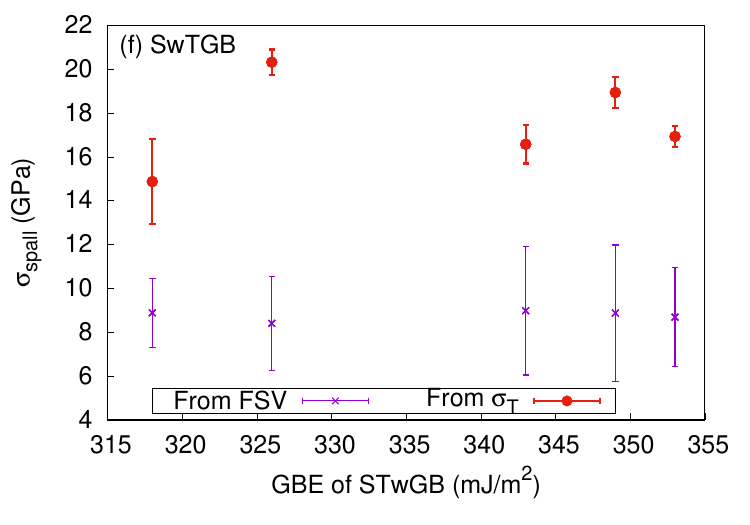}
	\vspace{-1.8\baselineskip}
	\label{spall12_FSVvsUp_stwgbE}
	}
     \caption
	[Arithmetic mean values of $\sigma_{sp}$ obtained]
	{Arithmetic mean values of $\sigma_{sp}$ obtained
	 from MD simulations with their respective RMS deviations are
	 shown in figures a,c,e for STGB and b,d,f for STwGB.
	 Figures a to f also show respective values of $\sigma_{T}$
	 mean and their RMS values.}
	\label{multi_fig6}
\end{figure}

Spall strength (${\sigma}_{sp}$) calculated using equation~\ref{eqn_sp_strength}
are shown in figure~\ref{multi_fig5} for STGB and STwGB cases.
Calculated spall strength values (${\sigma}_{sp}$) from
MD simulations are shown as a function
of (i) $U_p$ for various GB angles, (ii) GB angles \&
(iii) Grain boundary energy (GBE) for every $U_p$.

Corresponding figure numbers are:~\ref{spallFSVvsUpStgb},~\ref{SpallFSVvsStgbAng}
\&~\ref{spallFSVvsUpStgbE} for STGB and~\ref{spallFSVvsUpStwgb}
,~\ref{SpallFSVvsStwgbAng} and~\ref{spallFSVvsUpStwgbE} for STwGB cases, respectively.

Notable variations are observed in the spall strength (${\sigma}_{sp}$)
values calculated from the FSV profile, as shown in figure~\ref{multi_fig5}.
For each of these curves, arithmetic mean values and their
%root mean square deviation (RMSD)
\ac{RMSD} are calculated. These are shown in figure~\ref{multi_fig6}.
Figures~\ref{spall12_FSVvsUpStgb} to~\ref{spall12_FSVvsUp_stwgbE} correspond,
respectively to the same parameters as used in figures~\ref{spallFSVvsUpStgb}
to~\ref{spallFSVvsUpStwgbE}. From MD simulations for all these cases, the peak
value of the tensile strength ($\sigma_{T}$) is also calculated. A wide range
is observed in the calculated values of $\sigma_{T}$. All these values are the results
of the combination of  various $U_p$, grain boundary angles (GB Angle)
and grain boundary energies (GBE). For each combination, arithmetic mean and
RMSD is calculated for $\sigma_{T}$ also. These are shown in
figures~\ref{spall12_FSVvsUpStgb} to~\ref{spall12_FSVvsUp_stwgbE} along with
RMSD values of ${\sigma}_{sp}$ values for comparison.  Here ${\sigma}_{sp}$
values are calculated from the free surface velocity (FSV).

\subsubsection{Variation of ${\sigma}_{sp}$ in terms of (i) varying $U_p$, (ii) GB angles and
(iii) GB Energy \label{chap5sec.3.3.1}}
(i) Figure~\ref{spallFSVvsUpStgb} indicates that for various impact conditions
($U_p$), the value of ${\sigma}_{sp}$ varies over a range of 8.7 to 12.5 GPa for STGB.
For a given STGB angle, calculated ${\sigma}_{sp}$ do not vary much with different
impact conditions as seen in figure~\ref{spallFSVvsUpStgb}. For STwGB,
figure~\ref{spallFSVvsUpStwgb} shows the calculated ${\sigma}_{sp}$ has a wide range,
5.4 to 14.6 GPa but no trend is observed. However the range becomes narrow, 5.4 to 8.2
GPa between the three $U_p$ values viz. 0.75, 1 and 1.75 km/s. For the mid-range of
$U_p$(=1.25, 1.5 km/s), higher values between 10 to 14.6 GPa are obtained for ${\sigma}_{sp}$.

(ii) Spall strength (${\sigma}_{sp}$) for each of the $U_p$ values are shown in
figures~\ref{SpallFSVvsStgbAng} and~\ref{SpallFSVvsStwgbAng}, as a function of
STGB and STwGB angles respectively. In the case of STGB angles, for a given
$U_p$ calculated ${\sigma}_{sp}$ do not vary much. The maximum variation
is in the range of 8.6 to 12 GPa for $U_p$ = 1.5 km/s, followed by
1 km/s. While in the case of STwGB the largest variation is only 10.5 to 14.6 GPa
for $U_p$ = 1.25 km/s. The nature of the curves does not follow any trend.
The random distribution of spall strength ${\sigma}_{sp}$ in terms of
$U_p$ and GB angles or GB energies for Al bicrystal is similar to the
case of silicon carbide~\cite{LI201951}. In the case of single crystal
silicon carbide for $U_p$ up to 2 km/s, no trend was seen for the crystal
orientations [001] and [110] at elevated temperatures~\cite{LI201951}.

(iii) Figures~\ref{spallFSVvsUpStgbE} and~\ref{spallFSVvsUpStwgbE} show
the variation of spall strength (${\sigma}_{sp}$) as a function of
GBE values, for both STGB and STwGB respectively. Among the
figures~\ref{spallFSVvsUpStgb} to~\ref{spallFSVvsUpStwgbE}, only
figure~\ref{spallFSVvsUpStgbE} shows some trend. For different values of $U_p$
(0.75 to 1.75 km/s) the scatter in ${\sigma}_{sp}$ is found to increase with
increasing GBE, in the case of STGB.

\subsection{Statistical scatter of calculated ${\sigma}_{sp}$ and ${\sigma}_{T}$}
Correspond to the figures~\ref{spallFSVvsUpStgb}-\ref{spallFSVvsUpStwgbE}, the scatter
in ${\sigma}_{sp}$ are shown in figures~\ref{spall12_FSVvsUpStgb}-\ref{spall12_FSVvsUp_stwgbE}.
Scatter in the observed peak ${\sigma}_{T}$ are also shown respectively, in these figures.
The values do not follow any pattern. However, the following points are inferred.

(i) Except one, in all cases the mean values of ${\sigma}_{T}$ are higher in comparison to the
mean values of ${\sigma}_{sp}$. For STGB angle $77.3^o$ that has 351 $mJ/m^2$ GBE,
mean values of peak ${\sigma}_{T}$ is lower than the mean of ${\sigma}_{sp}$ (fig.~\ref{spall12_FSVvsStgbAng}).
This reverse trend can be understood from the fact that ${\sigma}_{T}$ influences
only the nucleation of voids but not its growth, whereas the free surface
velocity includes the effect of both the nucleation of voids as well as their growth.
It may be noted that from the MD simulations of
nano Cu/Ni bimetallic multilayer~\cite{ZHU2022110923},
the spall strength (${\sigma}_{sp}$) obtained from FSV was numerically lower than
the values of ${\sigma}_{T}$ for various duration of piston driven shock system.
While in the case of Cu bicrystal~\cite{LIN201294} asymmetric tilt GBs,
the spall strength (${\sigma}_{sp}$) obtained from FSV were much higher
then peak ${\sigma}_{T}$ (${\sigma}_{sp} \simeq 1.5 \times$ peak $\sigma_{T}$).
MD simulations of shock-induced fracture in single crystal silicon carbide~\cite{LI201951}
have shown that for some values of $U_p$, the spall strength
calculated from FSV (${\sigma}_{sp}$) are higher than the peak tensile strength
${\sigma}_{T}$; For some other $U_p$, peak tensile strength values
were higher than ${\sigma}_{sp}$.

(ii) Only in the case of fig~\ref{spall12_FSVvsUp_StgbE} the scatter grows from
smaller to bigger, as the GBE is increased. This is observed for both ${\sigma}_{sp}$
and ${\sigma}_{T}$. In all other figures, the nature of
scatter does not follow any pattern in respect of the parameters on the x-axis
(figs.~\ref{spall12_FSVvsUpStgb}-\ref{spall12_FSVvsSTwgbAng} \&~\ref{spall12_FSVvsUp_stwgbE}).

(iii) In the case of STGB, figures~\ref{spall12_FSVvsUpStgb},~\ref{spall12_FSVvsStgbAng}
\&~\ref{spall12_FSVvsUp_StgbE} show that the scatter for both ${\sigma}_{sp}$ and peak ${\sigma}_{T}$
are smaller. Also in these three cases, the mean values of ${\sigma}_{sp}$
and ${\sigma}_{T}$ differ by less than 3 GPa.

(iv) In the case of STwGB (figs.~\ref{spall12_FSVvsUpStwgb},~\ref{spall12_FSVvsUp_StgbE}
\&~\ref{spall12_FSVvsUp_stwgbE}) the scatter goes up 6 GPa for both
${\sigma}_{sp}$ and peak ${\sigma}_{T}$. Also the difference between the mean
values of ${\sigma}_{sp}$ and peak ${\sigma}_{T}$ are larger than the case of corresponding
STGB. The mean values of ${\sigma}_{sp}$ as a function of STwGB angles
or GBE are almost the same (${\sigma}_{sp}$=9 GPa), whereas the mean of peak ${\sigma}_{T}$ varies
from 14 to 16 GPa as a function of $U_p$ (fig.~\ref{spall12_FSVvsUpStwgb}). But,
peak ${\sigma}_{T}$ as a function of STwGB angles or GBE, vary from 15-20 GPa
(fig.~\ref{spall12_FSVvsSTwgbAng} \&~\ref{spall12_FSVvsUp_stwgbE}).

(v) The value of  ${\sigma}_{sp}$ obtained from FSV is different from the
value of peak ${\sigma}_{T}$. The peak ${\sigma}_{T}$ is found to be large for Al bicrystal with
different tilt and twist GB. It indicates that unlike the case of single
crystal Al~\cite{JIANG2022114474} and bicrystal Cu~\cite{LONG2020109411,
Influence_Fensin}, the value of peak ${\sigma}_{T}$ cannot be claimed
as the value of spall strength (${\sigma}_{sp}$) for an Al bicrystal.
%%% added on 22 Jan 2023
As it is already mentioned above (in i),
the spallation depends on void nucleation and growth. ${\sigma}_{T}$ influences
only the nucleation of voids but does not contribute to void growth.
While the FSV profile contains both the effects of void nucleation and growth.
The spall strength derived from FSV has all the physical phenomena
in comparison to ${\sigma}_{T}$. Hence for `Al', it is desirable to
calculate the spall strength from the temporal FSV profile.

\subsection{On the Mean \& RMSD values for $\sigma_{sp}$ and peak $\sigma_{T}$ for STwGB}
For aluminum, Qing Yin {\it et.al}~\cite{al_twist_GBE_QingYin} reported that
for lower twist angles $<$ $15^o$, the peak tensile stress is lower than other twist
angles. It is also mentioned that for twist angle $=$ $36.9^o$, the tensile strength
is maximum. The peak tensile strength obtained from this work (MD), is shown in
figure~\ref{spall12_FSVvsSTwgbAng}. Similar to the observation of Qing Yin,
figure~\ref{spall12_FSVvsSTwgbAng} shows the lowest and highest values of the peak
tensile stress are at the STwGB angles  $14.2^o$ and $36.9^o$ respectively.

\subsection{Effect of Strain rate on the spall strength for various STGB and STwGB
subjected to various impact loading}
\label{chap5sec3.3}
Spall strength (${\sigma}_{sp}$) calculated using equation~\ref{eqn_sp_strengthChap4}
for various $U_p$ are shown in figure~\ref{formstrainRates2Fns_TiltTwists}.
It is shown in terms of various strain rates ($\dot{\epsilon}$) corresponding to
the particle velocities
 $U_p$ = 0.5 to 1.75 km/s, in steps of 0.25 km/s. The value of spall
strength is indicated separately for STGB and STwGB cases. In
figures~\ref{strainRatesFn1_TiltTwists} and ~\ref{strainRatesFn2_TiltTwists}
the spall strength (${\sigma}_{sp}$)
reported from four different experiments at various strain rates
($\dot{\epsilon}$) are also included. Experiment results are taken
from the literature: (i) Fortov-91~\cite{fortov1991}, (ii) Moshe-98~\cite{spall_moshe1998},
(iii) Moshe-00~\cite{spall_moshe2000} for polycrystalline Al, (iv)
Kanel-PF and (v) Kanel-SC are respectively for pure foil and single crystal
Al~\cite{Kanel2001JAPSingleCrysFoilAl}.
\begin{figure}[!htb]
     \centering
         \subfigure
         [Fn1 (eqn.~\ref{srinivasanfunformstrateFn1}): Spall strength Vs $\dot{\epsilon}$]
	{
         \includegraphics[width=0.475\textwidth]
	{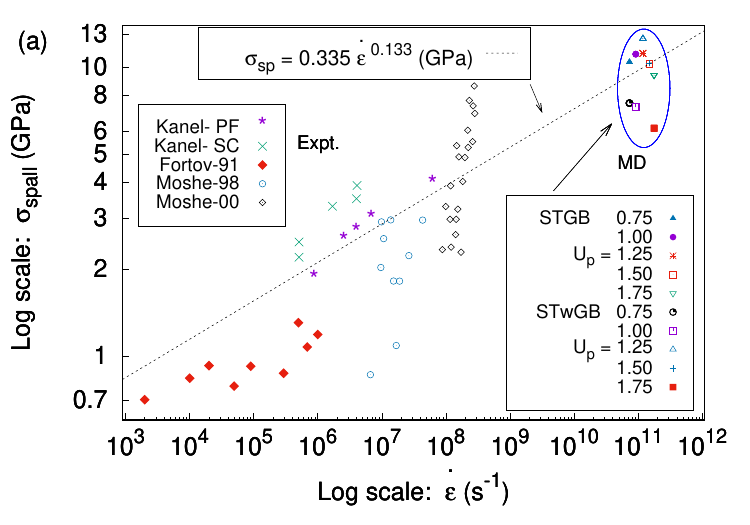}
	\label{strainRatesFn1_TiltTwists}
	}
    \hfill 
         \subfigure
         [Fn2 (eqn.~\ref{Moshe98funformstrateFn2}): Spall strength Vs $\dot{\epsilon}$]
	{
         \includegraphics[width=0.475\textwidth]
	 {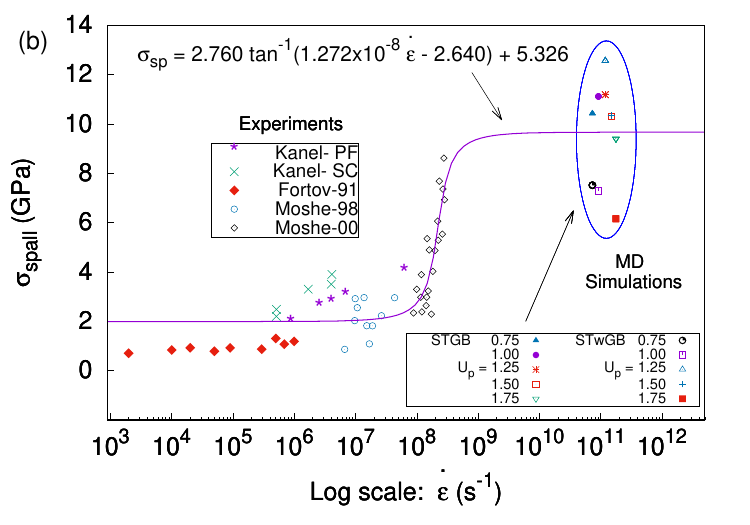}
	\label{strainRatesFn2_TiltTwists}
	}
     \caption
        [Spall strength as a function of Strain rates: MD Results compared with Expt.]
	{
	(a): Log-Log scale with the functional form 1-(eqn.~\ref{srinivasanfunformstrateFn1}).
	(b): Log-Linear scale with the functional form 2-(eqn.~\ref{Moshe98funformstrateFn2}).
     Calculated spall strengths (${\sigma}_{sp}$) for 
     various strain rates ($\dot{\epsilon}$) from MD simulations are
     compared with experiments. The various strains rates in MD simulations
     correspond to different particle velocities ($U_p$) and are indicated as
     0.5, 0.75, 1.0, 1.25, 1.5 \& 1.75 in km/s. The experiment results at
     various $\dot{\epsilon}$ are taken from (i) Kanel-PF for pure Al
     foil~\cite{Kanel2001JAPSingleCrysFoilAl}; (ii) Kanel-SC for single
     crystal Al~\cite{Kanel2001JAPSingleCrysFoilAl};
     (iii) Fortov-91~\cite{fortov1991}, (iv) Moshe-98~\cite{spall_moshe1998}
	and (v) Moshe-00~\cite{spall_moshe2000} for polycrystalline Al.
	\label{formstrainRates2Fns_TiltTwists}
}
\end{figure}
\subsubsection{Spall strength as a function of strain rate}
Strain rates $({\dot\epsilon})$ obtained from MD simulations are in the range
$5\times 10^{10}$ to $17\times 10^{10}\ s^{-1}$. Similar strain rates are
common in MD simulations.
MD simulations are suitable for extreme experimental conditions,
which can calculate the system evolution even at high strain rates due to the
sub-nanometer spatial and picosecond (ps) temporal scales. For
instance in MD simulation of single crystal silicon carbide, ${\dot\epsilon}$ up to
$18\times 10^{10}\ s^{-1}$ was observed for $U_p$=1.75 km/s~\cite{LI201951}.

To accommodate the strain rates of various orders of magnitude,
fig.~\ref{formstrainRates2Fns_TiltTwists} is shown in the log scale for the ${\dot\epsilon}$.
Averaged values from various STGB are calculated and shown in
terms of respective strain rates corresponding to the different flyer velocities
(or $U_p$). Similarly, the spall strength values obtained from various STwGB are averaged
and shown in the same figure. Spall strength values calculated from MD simulations are
in the same order of magnitude as obtained from various impact
experiments~\cite{Kanel2001JAPSingleCrysFoilAl, fortov1991, spall_moshe1998,
spall_moshe2000} at strain rates exceeding $10^{8}\ s^{-1}$. Two functional forms
are attempted here, to relate the spall strength `${\sigma_{sp}}$' obtained from the
MD simulations with experiments. These two functions are chosen from the published
works of Srinivasan~\cite{srinivasan2007} and Moshe-98~\cite{spall_moshe1998}.

Srinivasan {\it et al.} have shown for single crystal `Ni', calculated MD spall strength
(${\sigma_{sp}}$) values are higher and associated strain rates (${\dot\epsilon}$) are
also higher when compared to experimental spall strength in polycrystalline Ni~\cite{srinivasan2007}.
They presented the variation of ${\sigma_{sp}}$ as a function of strain rate in a log-log
scale~\cite{srinivasan2007}, because the strain rate varies over several orders of magnitude.
Spall strength variation with strain rate is assumed to be a functional form,
\begin{equation}
{\sigma_{sp}}(\dot\epsilon) = a\times({\dot\epsilon})^b
\label{srinivasanfunformstrateFn1}
\end{equation}
Here, `$a$' and `b' are material-dependent coefficients. Using the spall data available from
the experimental and MD simulations, the coefficients $a$ and $b$ can be obtained by functional
fitting. Srinivasan et al. have compared the `$\dot\epsilon-\sigma_{sp}$' data obtained from
polycrystal experiments with the MD results for `Ni'. The values of the coefficients $a$ and $b$
are found to be 0.11 $GPa$-$sec^{b}$ and 0.25 respectively~\cite{srinivasan2007}.

Suppose the form of a function f(x) can be expressed in terms of one or more unknown coefficients,
then the value of the function can be calculated with in the applicable range of the independant
variable `x', by the method of `fitting'. For `fitting', some of the values of function at discreate
values of `x' must be known. GNUplot is a software that has the feature of fitting a curve to a
function~\cite{GNUplot}. The fit is bassed on the sum of the squared differences between the input
data points and the function values. The residuals are `weighted' by the input data errors before
being squared. Hence it is called \ac{WSSR}.
%weighted sum of the squared residuals (WSSR).

The functional form represented by the equation~\ref{srinivasanfunformstrateFn1} is applied to `Al' data.
By fitting `$\dot\epsilon-\sigma_{sp}$' data obtained from this work (MD simulations)
and experiments~\cite{Kanel2001JAPSingleCrysFoilAl, fortov1991, spall_moshe1998, spall_moshe2000},
the values of the coefficients `a' and `b' are calculated to be 0.335 $GPa$-$sec^{b}$ and
0.133 respectively. The function obtained by the method of fitting is plotted in
figure~\ref{strainRatesFn1_TiltTwists}. The Weighted Sum of Squared Residuals
(WSSR) found to be 1.36542 with ${\chi}^2$ = 1.86438, using `GNUploT' fitting~\cite{GNUplot}.

The second functional form was adapted from the literature work by Moshe in 1998~\cite{spall_moshe1998}.
The function is represented by equation~\ref{Moshe98funformstrateFn2}.
\begin{equation}
{\sigma_{sp}}(\dot\epsilon) = P_0\ tan^{-1}({\tau}_0 {\dot\epsilon} - {\epsilon}_0) + P_1
\label{Moshe98funformstrateFn2}
\end{equation}
Here, $P_0$ and $P_1$ have the dimension of stress. ${\tau}_0$ has the dimension of time while
${\epsilon}_0$ has the dimension of strain. The values of these parameters are calculated by
fitting the functional form to `$\sigma_{sp}-\dot\epsilon$' data.
The values of the coefficients resulting from `GNUplot' fitting~\cite{GNUplot} are
$P_0$ = 2.76 GPa, $P_1$ = 5.326 GPa, ${\epsilon}_0$ = 2.64 (dimensionless)
and ${\tau}_0$ = 1.272 $\times\ 10^{-8}$ sec.
The function obtained by the method of fitting is shown in
figure~\ref{strainRatesFn2_TiltTwists}. For this fitting, the Weighted Sum of Squared Residuals
(WSSR) is found to be 1.03733 with ${\chi}^2$ = 1.07606.

Overall, calculated spall strength is found to be higher when compared to the experiments.
It is reported by Moshe et al~\cite{spall_moshe1998} that their experimental results
suggest a critical value of strain rates ($10^7\ s^{-1}$) above which the spall
strength approaches the theoretical value (17.1 GPa or 171 kbar) calculated from
the cohesive energy ($U_{coh}$), bulk modulus ($B_0$) and density ($\rho_0$) of the polycrystal
($\sigma_{sp}=\sqrt{\rho_0U_{coh}B_0/8}$)~\cite{spall_moshe1998}. The spall strength
values calculated from the MD simulations lie between the experimental and theoretical limits.
It may be noted that the strain rates involved in MD simulations are also consistently higher.
Experimental high-velocity impact systems produce strain rates in the order
$10^5$-$10^7\ s^{-1}$~\cite{meyersBook}. But much higher strain rates
($5$-$17$ $\times$ $10^{10}$ $s^{-1}$) are observed from the MD simulations in this work. This is
not clearly understood now and needs further investigation.
	    
\clearpage
\subsection{Effect of stuctural phase transition in the evolution of free surface velocity}
The Phase transition from Face Centered Cubic (FCC) to \ac{BCC} is observed in the MD simulations.
Such phase transition has been reported for single crystal
Al~\cite{JIANG2022114474} under shock loading conditions. It has been observed
by DD Jiang {\it et.al}~\cite{JIANG2022114474} that due to phase transition
crystal defects occur and it leads to a sudden change in spall strength.
In the study, it is mentioned for Al single crystal, spall strength
($\sigma_{sp}$) increases to $\simeq$10 GPa for $U_p$ = 1.5 km/s after
plastic deformation. Subsequently at higher $U_p$ up to 2.1 km/s
value of $\sigma_{sp}$ shown to be decreasing~\cite{JIANG2022114474}.
MD results shown in figures~\ref{spall12_FSVvsUpStgb} and~\ref{spall12_FSVvsUpStwgb}
indicate for $U_p$=1.25 km/s, calculated spall
strength from FSV ($\sigma_{sp}$) is the maximum value at
11.1 and 12.5 GPa for STGB and STwGB respectively. For higher
$U_p$, calculated values of $\sigma_{sp}$ show decreasing trends
which are 10.3, 9.4 GPa for STGB while they are 10.3 and 6.2 GPa for STwGB.

\begin{figure}[!htb]
         \centering
	\subfigure
	 [$U_p$ 0.5: FCC 95\%]
	 {
         \includegraphics[width=0.42\textwidth]
	{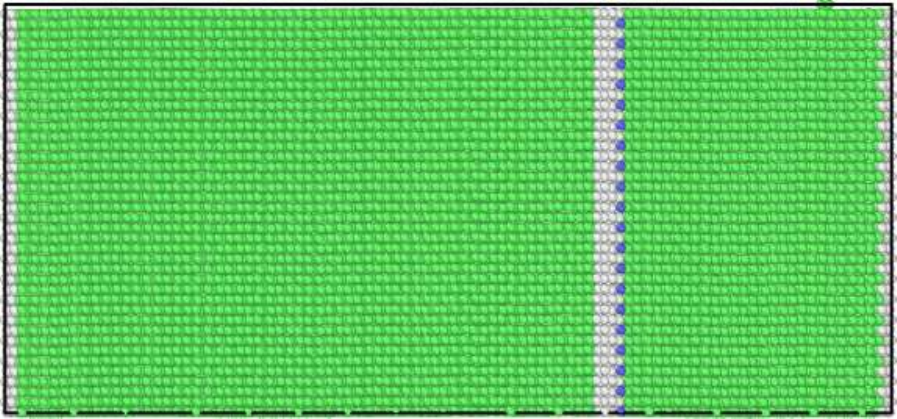}
	\label{phaseChange22.6STGB-01}
	}
	\subfigure
	 [$U_p$ 0.75: FCC 93.6\%]
	 {
         \includegraphics[width=0.42\textwidth]
	{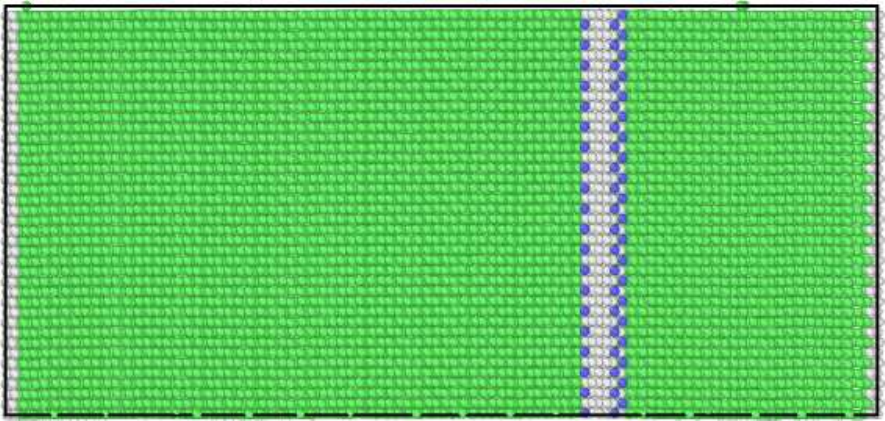}
	\label{phaseChange22.6STGB-02}
	}
	\subfigure
	 [$U_p$ 1.0: FCC 84.9\%]
	 {
         \includegraphics[width=0.42\textwidth]
	{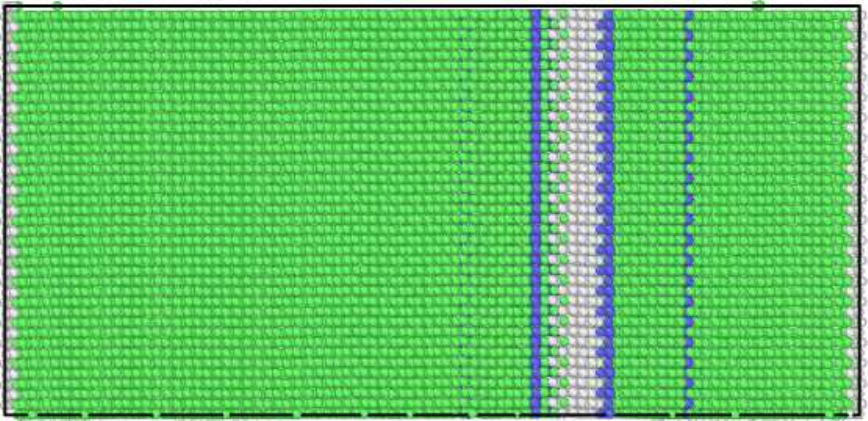}
	\label{phaseChange22.6STGB-03}
	}
	\subfigure
	 [$U_p$ 1.25: FCC 58.8\%]
	 {
         \includegraphics[width=0.42\textwidth]
	{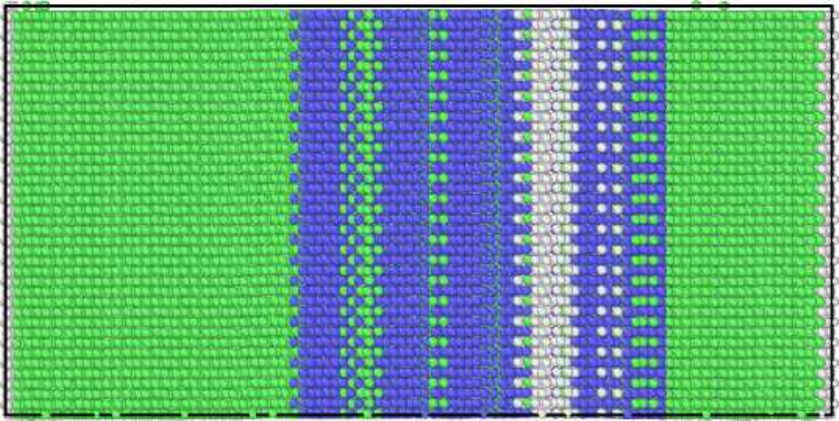}
	\label{phaseChange22.6STGB-04}
	}
	\subfigure
	 [$U_p$ 1.5: FCC 45.4\%]
	 {
         \includegraphics[width=0.42\textwidth]
	{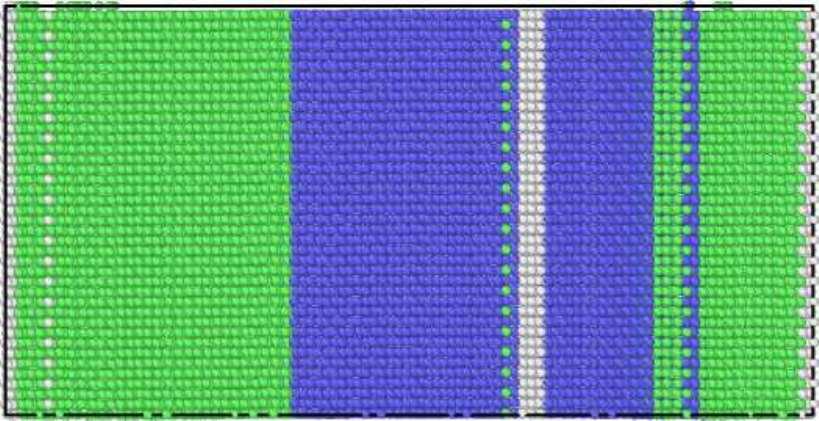}
	\label{phaseChange22.6STGB-05}
	}
	\subfigure
	 [$U_p$ 1.75: FCC 38.6\%]
	 {
         \includegraphics[width=0.42\textwidth]
	{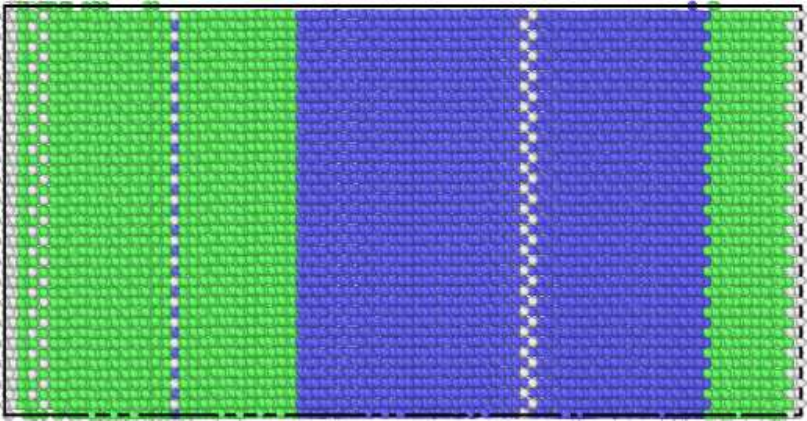}
	\label{phaseChange22.6STGB-06}
	}
     \caption
[OVITO-images: Phase transition in STGB (FCC$\longrightarrow$BCC)]
{OVITO-based visualization images in YZ plane, showing different phases:
FCC (green), BCC (blue) and Unknown (grey) for STGB ($22.6^o$) at t=1.65 picosec.
Figures {\it a} to {\it f} correspond to $U_p$=0.5 to 1.75 km/s
in steps of 0.25 km/s respectively. The shock propagates along the horizontal Y-axis.}
\end{figure}

\begin{figure}[!htb]
         \centering
	\subfigure
	 [$U_p$ 0.5: FCC 94.8\%]
	 {
         \includegraphics[width=0.42\textwidth]
	{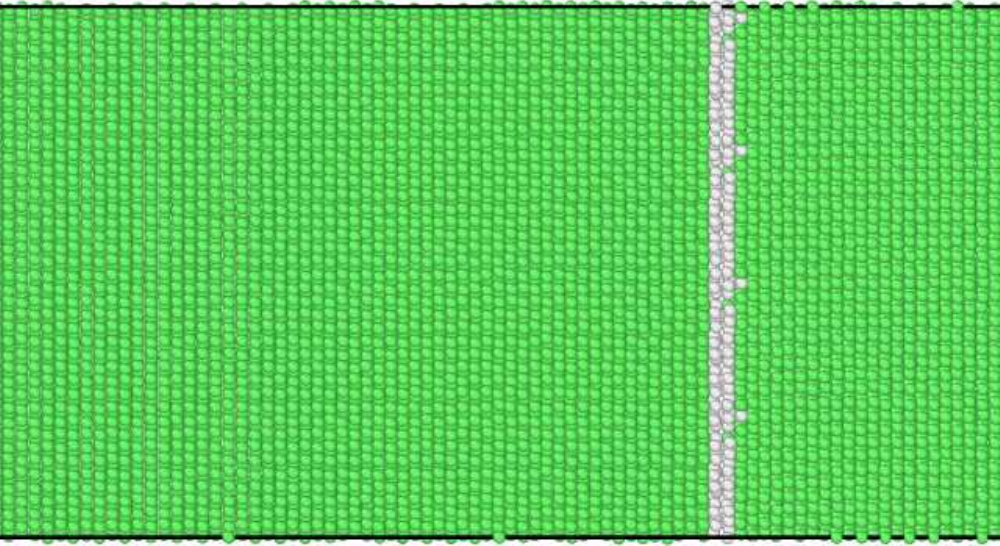}
	\label{phaseChange22.6STwGB-01}
	}
	\subfigure
	 [$U_p$ 0.75: FCC 92.5\%]
	 {
         \includegraphics[width=0.42\textwidth]
	{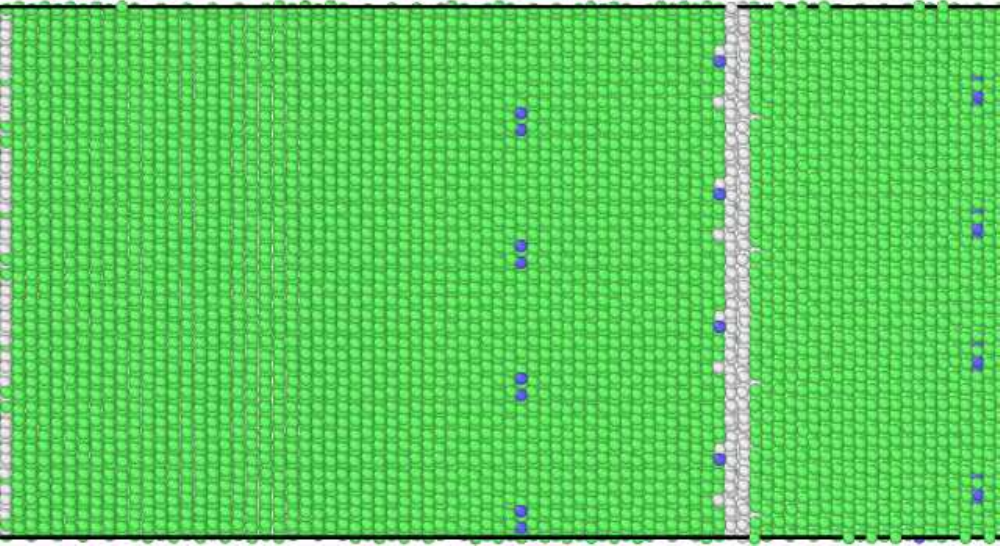}
	\label{phaseChange22.6STwGB-02}
	}
	\subfigure
	 [$U_p$ 1.0: FCC 80.7\%]
	 {
         \includegraphics[width=0.42\textwidth]
	{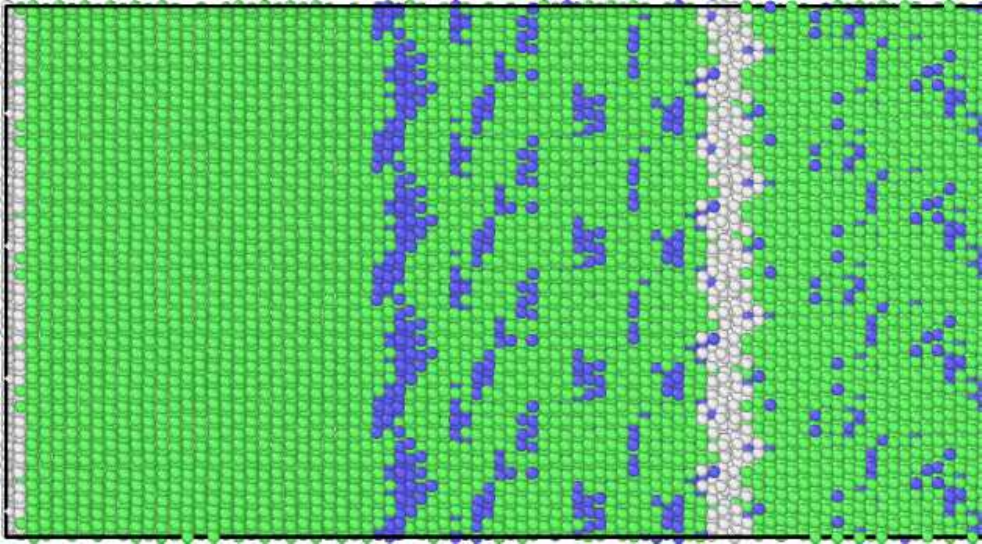}
	\label{phaseChange22.6STwGB-03}
	}
	\subfigure
	 [$U_p$ 1.25: FCC 55.9\%]
	 {
         \includegraphics[width=0.42\textwidth]
	{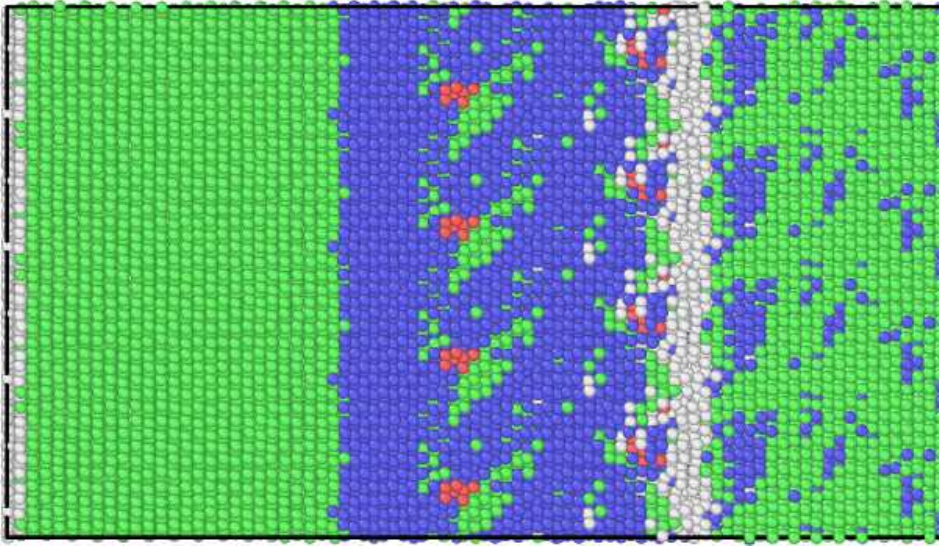}
	\label{phaseChange22.6STwGB-04}
	}
	\subfigure
	 [$U_p$ 1.5: FCC 40.3\%]
	 {
         \includegraphics[width=0.42\textwidth]
	{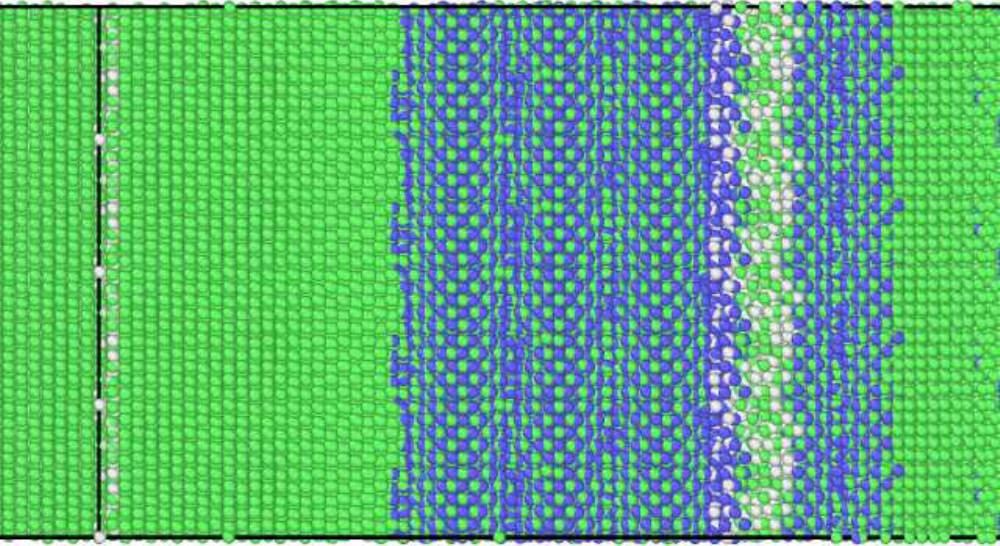}
	\label{phaseChange22.6STwGB-05}
	}
	\subfigure
	 [$U_p$ 1.75: FCC 31.0\%]
	 {
         \includegraphics[width=0.42\textwidth]
	{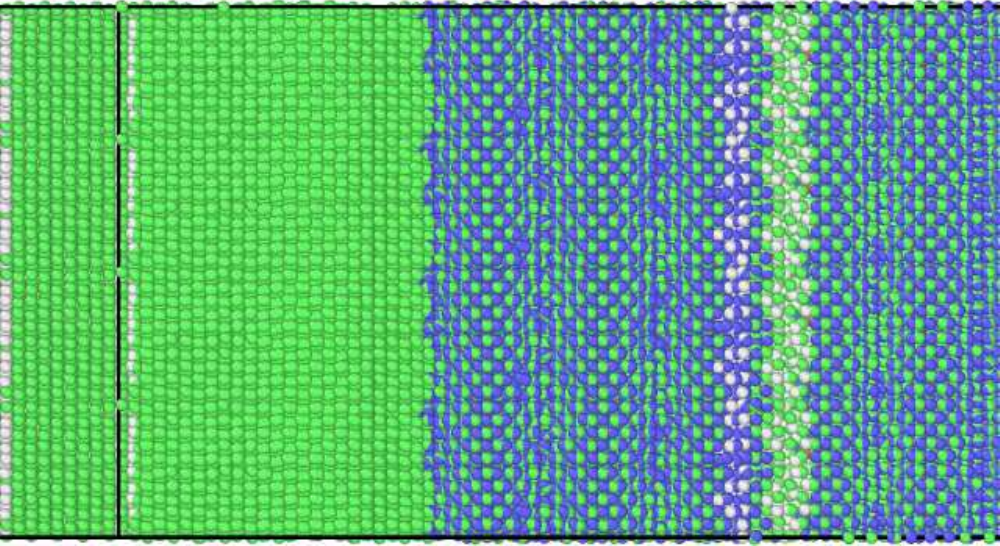}
	\label{phaseChange22.6STwGB-06}
	}
     \caption
[OVITO images: Phase transition in STwGB (FCC$\longrightarrow$BCC)]
{OVITO-based visualization images in YZ plane showing different phases:
FCC (green), BCC (blue),  HCP (red) and unknown (grey) for STwGB ($22.6^o$) at t=1.65 picosec.
Figures {\it a} to {\it f} correspond to $U_p$=0.5 to 1.75 km/s
in steps of 0.25 km/s respectively. The shock propagates along the horizontal Y-axis.}
\end{figure}

As mentioned in sec.~\ref{chap5Resultssec}, the free surface velocity should
reach a maximum velocity that is equal to the impact velocity
($V_{fs}$ = 2$U_p$ = $U_{flyer}$ ref.~\cite{meyersBook}). But when the phase
change occurs the free surface velocity does not reach its maximum
value in a single step but in a few steps. First, it reaches an intermediate
value that corresponds to the \ac{HEL}. Subsequently
as described by Mayers~\cite{meyersBook} it gradually increases
to a value that corresponds to phase transition status. In our MD simulations, such
a step increase is observed due to phase transition.
Figures~\ref{1-tilt_fsvoft}-\ref{6-tilt_fsvoft} and
figures~\ref{1-twist_fsvoft}-\ref{6-twist_fsvoft} clearly show such effect
in STGB for $U_p$ $\geq$ 0.75 km/s and STwGB for $U_p$ $\geq$ 1.25 km/s
respectively.

Table~\ref{table_PT_FCC_BCC_HCP_Others} list the proportions of
FCC, BCC, \ac{HCP} and others for STGB and STwGB for GB angle $22.6^o$.
Corresponding to the values in table~\ref{table_PT_FCC_BCC_HCP_Others},
figs.~\ref{phaseChange22.6STGB-01} to~\ref{phaseChange22.6STGB-06}
show the various proportion of different structural phases (FCC, BCC etc.)
in the case of STGB. Similarly, these are shown for STwGB cases in
figs.~\ref{phaseChange22.6STwGB-01} to~\ref{phaseChange22.6STwGB-06}.
Atoms position from MD simulations are viewed using an open source package,
\ac{OVITO}~\cite{Stukowski_2009}. The structural phases are identified using the 
\ac{CNA} algorithm that is available in OVITO~\cite{Stukowski_2009}.

\begin{table}[!ht]
\centering
\caption
[Observed proportion of various phases of Al STGB and STwGB ($22.6^o$) 
using CNA]
{Observed proportion of various phases of Al STGB and STwGB ($22.6^o$) 
using CNA, after 1.65 ps from the impact time.}
\begin{tabular}{lllllll}
\toprule
 {\bf GB}   & {\bf Sl.} & ${\bf U_p}$ & {\bf FCC} & {\bf BCC} & {\bf HCP} & {\bf Others} \\
 {\bf Type} & {\bf No.} & {\bf (km/s)}  & {\bf  \%} & {\bf \%}  & {\bf \%}  & {\bf \%}     \\
\midrule
 \multirow{1}{*}{\makecell {STGB}} & 1 & 0.5 & 95.0 & 0.3 & 0.0 & 4.7 \\
 & 2 & 0.75 & 93.6 & 0.7  & 0.0 & 5.7 \\
 & 3 & 1.0  & 84.9 & 7.8  & 0.0 & 7.3 \\
 & 4 & 1.25 & 58.8 & 33.0 & 0.0 & 8.2 \\
 & 5 & 1.5  & 45.4 & 47.8 & 0.0 & 6.8 \\
 & 6 & 1.75 & 38.6 & 53.6 & 0.0 & 7.8 \\
\midrule
 \multirow{1}{*}{\makecell {STwGB}} & 7 & 0.5\  & 94.8 & 0.0 & 0.0 & 5.2 \\
 & 8  & 0.75 & 92.5 & 1.7  & 0.0 & 5.8 \\
 & 9  & 1.0  & 80.7 & 12.7 & 0.0 & 6.6 \\
 & 10 & 1.25 & 55.9 & 35.7 & 0.6 & 7.8 \\
 & 11 & 1.5  & 40.3 & 52.2 & 0.0 & 7.5 \\
 & 12 & 1.75 & 31.0 & 62.1 & 0.0 & 6.9 \\
\bottomrule
\end{tabular}
\label{table_PT_FCC_BCC_HCP_Others}
\end{table}

%% Below sections are verified with the manuscript content - 10 Feb 2023
\subsection{Observed dislocation in twist GB}
Figure~\ref{3-twist_fsvoft} for STwGB angle $14.2^o$ shows different behavior
for $U_p$ = 1 km/s than other GB angles at the same velocity. The pullback
event occurs for a longer duration in this case. This is attributed to dislocation formation
at this particle velocity. Thus the calculated $\sigma_{sp}$ for STwGB angle
$14.2^o$ is found to be higher in comparison to other GB angles at $U_p$ = 1 km/s.
\begin{figure}[!htb]
   \centering
%   \begin{tabular}{cc}
	\subfigure
	 [1.2 ps]
	{
         \includegraphics[width=0.3\textwidth]
	{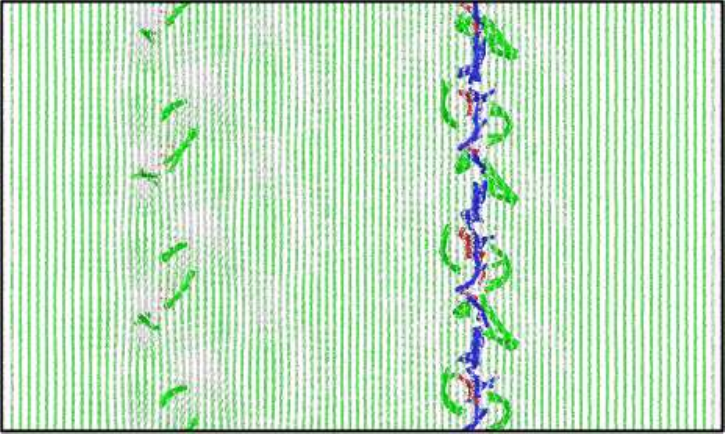}
	\label{DislocLines14.2STwGB-01}
	}
	\subfigure
	 [1.5 ps]
	 {
         \includegraphics[width=0.3\textwidth]
	{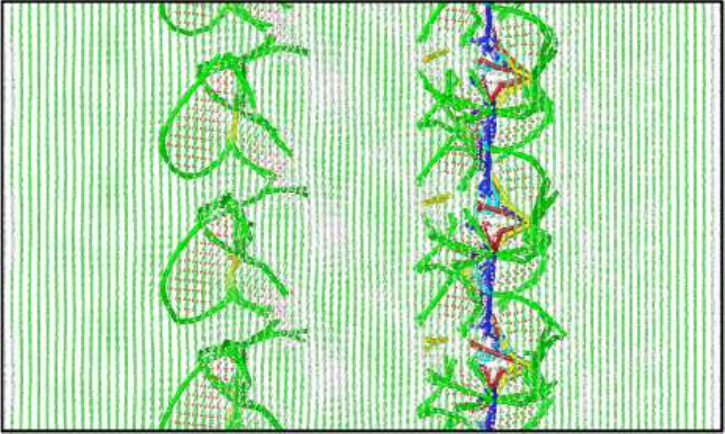}
	\label{DislocLines14.2STwGB-02}
	}
	\subfigure
	 [1.8 ps]
	 {
         \includegraphics[width=0.3\textwidth]
	{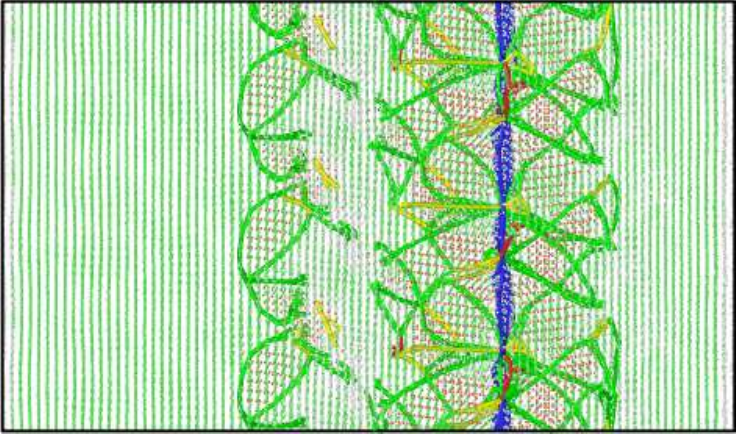}
	\label{DislocLines14.2STwGB-03}
	}
	\subfigure
	 [2.1 ps]
	 {
         \includegraphics[width=0.3\textwidth]
	{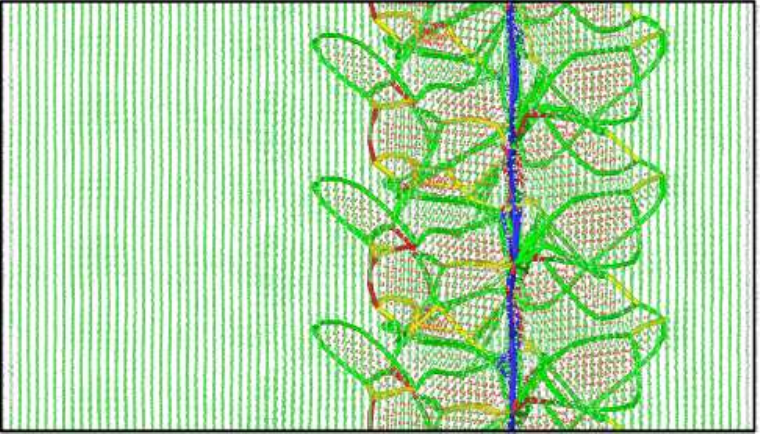}
	\label{DislocLines14.2STwGB-04}
	}
	\subfigure
	 [2.4 ps]
	 {
         \includegraphics[width=0.3\textwidth]
	{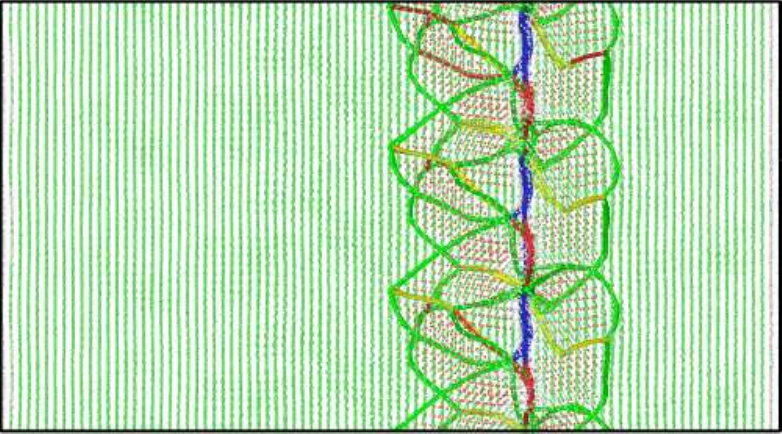}
	\label{DislocLines14.2STwGB-05}
	}
	\subfigure
	 [2.7 ps]
	 {
         \includegraphics[width=0.3\textwidth]
	{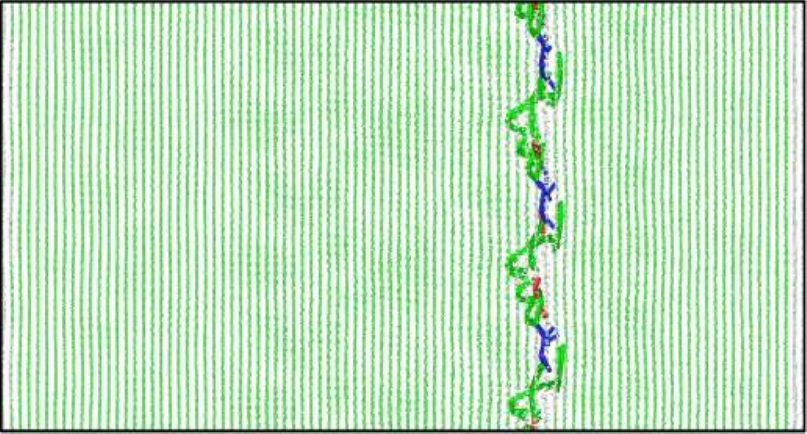}
	\label{DislocLines14.2STwGB-06}
	}
	\subfigure
	 [3.0 ps]
	 {
         \includegraphics[width=0.3\textwidth]
	{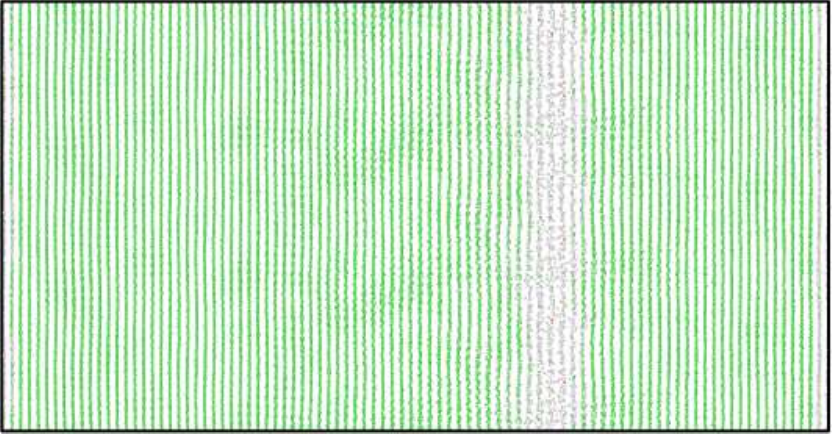}
	\label{DislocLines14.2STwGB-07}
	}
	\subfigure
	 [3.3 ps]
	 {
         \includegraphics[width=0.3\textwidth]
	{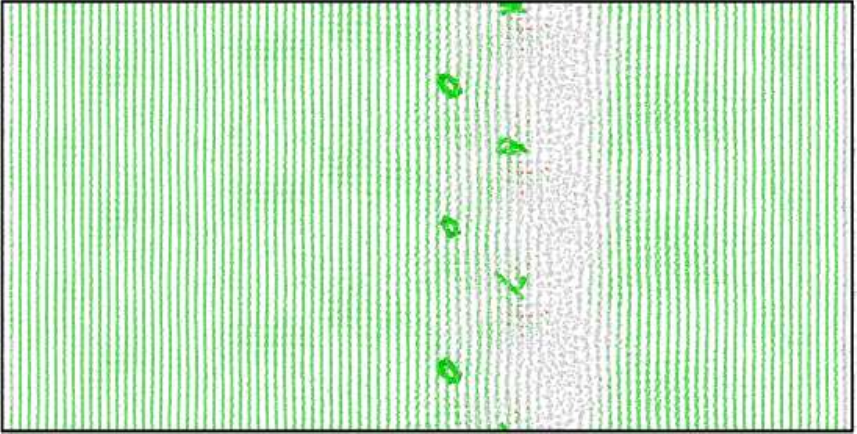}
	\label{DislocLines14.2STwGB-08}
	}
%\end{figure}
%\begin{figure}[htb]\ContinuedFloat
%\centering
	\subfigure
	 [3.6 ps]
	 {
         \includegraphics[width=0.3\textwidth]
	{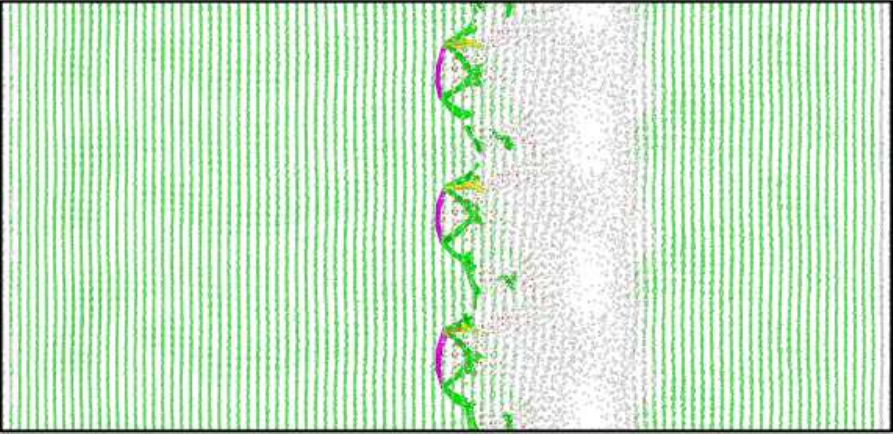}
	\label{DislocLines14.2STwGB-09}
	}
	\subfigure
	 [3.9 ps]
	 {
         \includegraphics[width=0.3\textwidth]
	{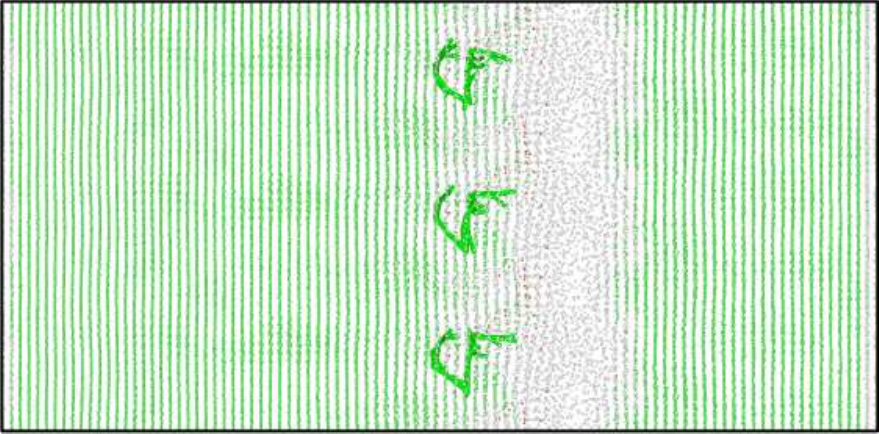}
	\label{DislocLines14.2STwGB-10}
	}
	\subfigure
	 [4.2 ps]
	 {
         \includegraphics[width=0.3\textwidth]
	{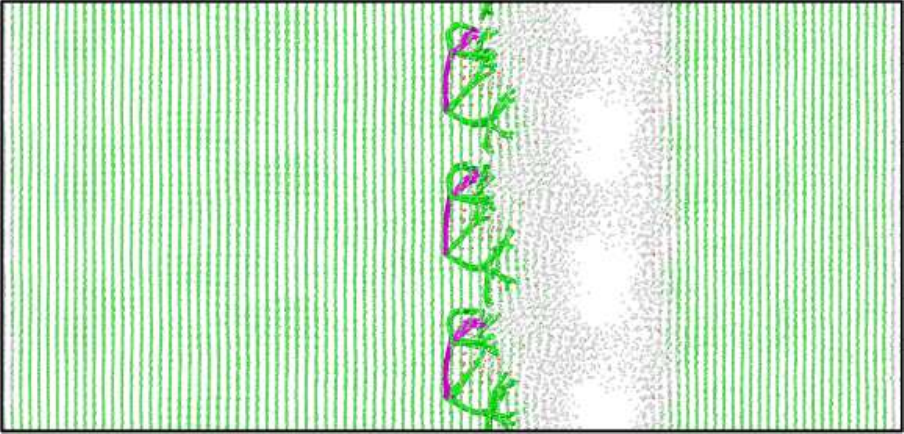}
	\label{DislocLines14.2STwGB-11}
	}
	\subfigure
	 [4.5 ps]
	 {
         \includegraphics[width=0.3\textwidth]
	{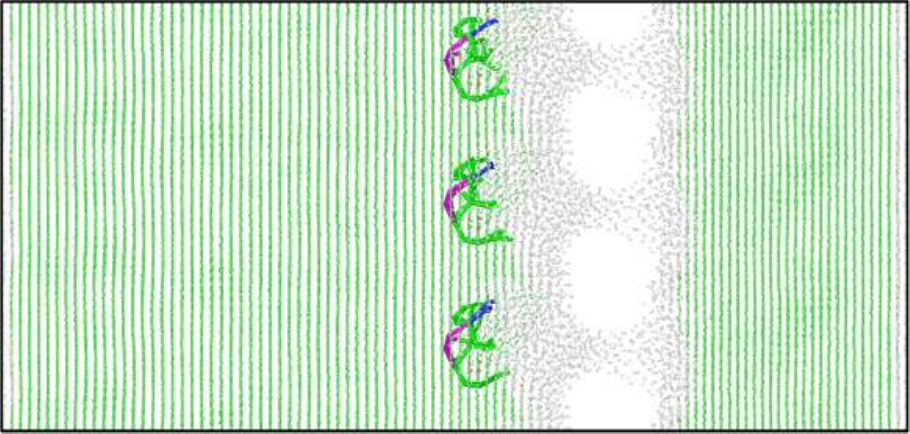}
	\label{DislocLines14.2STwGB-12}
	}
%   \end{tabular}
     \caption
[OVITO images in YZ plane, to show the dislocation evolution]
{OVITO-based visualization images in YZ plane, to show the dislocation
formation for STwGB ($14.2^o$) at various times: corresponding $U_p$ is 1 km/s
(flyer velocity = 2 km/s). Shock propagates along the horizontal Y-axis.}
\end{figure}

\begin{figure}[!htb]
     \centering
	\includegraphics[width=0.65\linewidth]
	{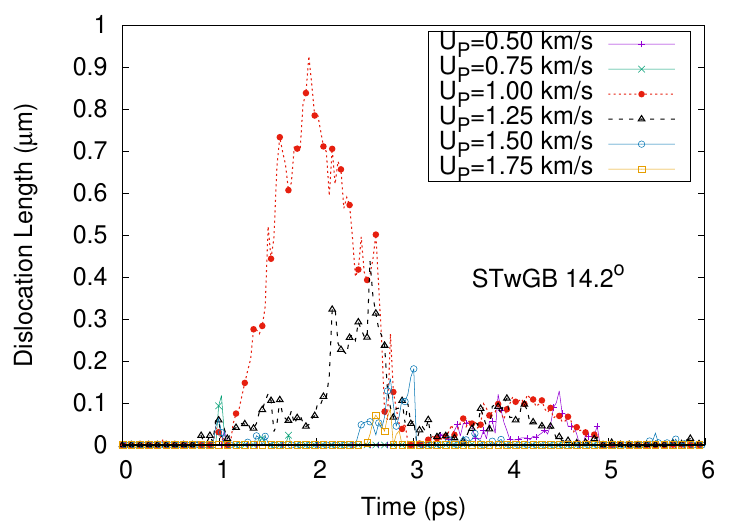}
     \caption
	[Total line length of dislocations for
          STwGB ($14.2^o$) at various $U_p$]
	{Total line length of dislocations for
	\label{DLlength_sumOfAllTypes14.2STwGB}
STwGB ($14.2^o$) at various $U_p$, calculated by OVITO-based DXA tool.
}
\end{figure}

\ac{DXA}~\cite{Stukowski_2009, 
ovito_dislocation} is applied to find if a dislocation occurs.
It is also used to calculate the dislocation line length. 
Figures~\ref{DislocLines14.2STwGB-01}-\ref{DislocLines14.2STwGB-12} show
the temporal formation of dislocation for STwGB angle $14.2^o$. Corresponding
time values for each of the snapshots (fig~\ref{DislocLines14.2STwGB-01}-\ref{DislocLines14.2STwGB-12})
are 1.2 to 4.5 ps in steps of 0.3 ps respectively. Dislocation dominates between 1.8 ps to 2.4 ps
and is located in and around the grain boundary. Among all the 10 GB angles
used in this MD simulation only for STwGB $14.2^o$, dislocation is observed
for $U_p$ = 1 and 1.25 km/s. The calculated dislocation line length is higher for
$U_p$ = 1 km/s among these two $U_p$, which is shown in
figure~\ref{DLlength_sumOfAllTypes14.2STwGB}. It is also observed
from this figure that the dislocation reaches zero around 3 ps from its first
peak values, but started growing again to reach its second peak around 4 ps.
However, the second peak is much smaller than the first peak for both
the particle velocities ($U_p$ =  1 and 1.25 km/s). In both cases,
spallation occurs around 4 to 4.2 ps. The first peak occurs before
the spallation and hence the spallation is prolonged.
Dislocations lead to grain
boundary plasticity. The value of $\sigma_{sp}$ depends on whether
GB plasticity is produced or not during the compression~\cite{LONG2020109411}.
If GB plasticity occurs the value of $\sigma_{sp}$ gets increased.
This effect is seen in figure~\ref{3-twist_fsvoft}. For the case 
of GB angle $14.2^o$, $\sigma_{sp}$ calculated from FSV is higher than others
due to the formation of dislocation.

\subsubsection{Possible reason for dislocation observed in $14.2^o$ STwGB case}
When the bicrystal interface is subjected to shear stress, atomic shuffling
occurs~\cite{SANSOZ20051931}. It occurs before the emission of the first
partial dislocation~\cite{DESpearot2005}. The atoms are locally rearranged
due to atomic shuffling at the bicrystal interface. This assists the nucleation
process. In all cases of STGB and STwGB due to shear stress, atomic shuffling
likely has occurred and led to the local rearrangement of atoms.
This eventually assists the nucleation process. Thus dislocation is not mostly
seen. This effect is more to have occurred for all STGB cases because the
shock passes through the grain boundary surface at a substantial angle to
the grain boundary plane respectively for the various GB angles. In the case of
STwGB the orientation of grain boundary varies between the two planes of the
interface and hence the atomic shuffling should have occurred lesser than STGB.
However, among the various STGB angle, $14.2^o$ is relatively smaller and
the effect of rearrangement of atoms due to shear stress should be lesser.
Thus, in this case, the nucleation of voids is not greatly assisted by atomic
rearrangement. But this could have assisted the dislocation to grow.

\section{Conclusion}
\label{chap5conclusion}
Ten Al bicrystals with different tilt and twist angles have been created and equilibrated with MD NPT simulations.
Finite size effects are taken care of by using the appropriate MD domain size. This avoids any
spurious reflection from the boundaries. Both the boundaries along the Y direction are allowed
to move freely along which the plane-wave shock propagates. Periodic boundary conditions are
applied along X and Z directions. The target and the flyer sizes were chosen so that the
tensile spall occurs at/near the grain boundary. Flyer velocities in the range of 1 to 3.5
km/s  are used in this work. From the temporal FSV profile obtained from the MD
simulations, the following are the observation:
\vspace{-2mm}
\begin{itemize}
\item No spall occurs at $U_p$ = 0.5 km/s for all STGB and STwGB cases.
	For $U_p$ = 0.75 km/s, no spall is observed for higher misorientation angle STGBs
	($53.2^o$ and $77.3^o$). For all the cases STwGB at this $U_p$, spall is observed.
\vspace{-3mm}
\item Overall, the values of the spall strength obtained for the individual STGB and STwGB
	cases are higher than that of the polycrystalline Al.
	The calculated strain rates from the MD simulations are in the range $10^{10}$-$10^{11}$ $s^{-1}$.
	The pattern seen in the spall strength calculated from MD simulations at these strain
	rates is consistent with experimental results of polycrystalline Al at lower
	strain rates.
\vspace{-3mm}
\item Coefficients of two different functional forms that relate spall strength with strain rate are estimated
	by fitting the ${\sigma_{sp}}(\dot\epsilon)$ data of polycrystal Al from experiments
	and MD simulation of Al bi-crystal.
\vspace{-3mm}
\item Phase transition (PT) occurs near GB upon shock propagation and it influences the free
	surface velocity profile. At higher particle velocities ($U_p$) due to phase transition,
	the spall strength gets reduced.
\vspace{-3mm}
\item If dislocations occur due to shock-induced grain boundary plasticity, pull-back 
	is prolonged and the time to spall is delayed.
	This leads to higher spall strength value (eg. STwGB, $14.2^o$ for $U_p$ at 1 km/s).
\vspace{-3mm}
\item Spall strength obtained from FSV was found to be lower than that obtained from the peak tensile
	strength. Values of spall strength and the peak tensile strength are not the same for Al bicrystal
	systems.
\end{itemize}
\vspace{-2mm}
The time of occurrence of spall and the pullback velocities ($\Delta{U}$) are different among
the various STGB cases for the same particle velocity. This is seen in figures~\ref{2-tilt_fsvoft}
-\ref{6-tilt_fsvoft}. This indicates that the spall parameters are sensitive to grain
boundary misorientation at the atomistic scales of length ($\AA$) and time ($10^{-12}$-$10^{-15}$ ps).
As desired by the choice of the flyer and the target sizes, the spall originates near the grain
boundary in most cases.

\clearchapter

% MD_MS_Spall_BiC : Chapter 6. Multiscale spall calculation for BiC Al : Com.Mater.Sci-2022 paper
  \chapter
%[Spall parameters calculation using Multiscale Method]
%{Multiscale method to calculate spall parameters for polycrystalline Al~\label{chapter_BiCSpall-Multiscale}}
{Multiscale Modeling to obtain the Dynamic Spall Strength of Polycrystal Al~\label{chapter_BiCSpall-Multiscale}}
\nopagebreak
\makeatletter
{\renewcommand{\@makefntext}[1]{\noindent #1}%
\footnotetext{%
  \fontsize{10pt}{12pt}\selectfont%
  The content of this chapter is fully adapted from the author's published work~\cite{madhavan_spall_multiscale}.
  A note on \copyright\ is available in page number~\pageref{copyrightNotePage}. }}
\makeatother

\section{Background of the problem~\label{sec1-Chap6}}
In the chapters~\ref{chapter_Intro}-~\ref{chapter_compMethods}, the multi-scale method is
briefly discussed. The multiscale method is applied and validated  for single crystals of 
Cu, Nb \& Mo in chapter~\ref{chapter_SCSpall-Multiscale}. Subsequently, 
for Al single crystal, the spall parameters are estimated. However, the metal used in
real life is polycrystalline. A polycrystal metal contains various grains of different sizes with
different grain boundary orientations. The importance of shock propagation
through individual grain boundaries is studied and presented for Al bicrystals
(5 STGB \& 5 STwGB, chapter~\ref{chapter_spallMDonly}) in which, the free surface
velocity under impact-shock propagation is directly calculated from MD simulations.
The spall parameters obtained from the MD simulations of individual grain boundaries
(chpater~\ref{chapter_spallMDonly}) were found to be higher than the poly-crystallites 
reported from the experimental works. Thus a better method is desirable to calculate
spall parameters from the MD for the polycrystalline material.

Bi-crystal MD simulations are available in the literature. A few are mentioned
here in the present context of shock propagation across a bi-crystal grain boundary.
In the case of a bi-crystal, a grain boundary plays a main role in two aspects,
(i) dislocation propagation in the material and (ii) void nucleation and growth.
MD simulations of the tensile deformation for a nano-sized `Al' crystal are reported
by Swygenhoven {\it et al.} ~\cite{HVSwygenhoven2005}. The Voronoi method was used by
them to create nano-crystals. Effects of GB on the nucleation of dislocation and 
its propagation in nano-crystallites are reported in it. MD simulations for three
STGBs of Al bicrystals are reported by Spearot {\it et al.}~\cite{DESpearot2005}
in which, uni-axial tension stress is applied to Al. They showed that the dislocation
loops are nucleated at the interfaces of two grains.
Dremov {\it et al.} ~\cite{VDremov_2006} used a flyer-target impact system for poly and
nano-crystals using MD methods.  The loading direction of the shock plays the main role
in the microstructures of the spall planes in monocrytals~\cite{VDremov_2006}.
Along the grain boundaries, void nucleates in a polycrytal~\cite{VDremov_2006}.
Bi-crystal of Cu has been studied using MD-shock simulations by Long
{\it et al}~\cite{LONG2020109411} in a flyer-target impact system.
A detailed step-by-step process to set up a grain boundary with different orientations
for MD simulations, are available from the works of Tschopp {\it et al.}~\cite{tschopp1}.
%The size of the flyer and targets were such that the rarefaction waves
%are positioned to meet near the grain boundary.

The MD simulation studies involving grains and grain boundaries
help to understand certain aspects of setting up grain boundaries and initiate
shock propagation across a grain boundary.
Along with this knowledge and also with the experience of successfully executing
multiscale modeling to obtain spall parameters for a single crystal, 
the multiscale method is found to be a promising option for studying the shock
and spall phenomena in an `Al' bi-crystal.
For this work, eleven STGB and twelve STwGB bi-crystals of Al are created based
on the methodology of Tschopp~\cite{tschopp1}. Spall parameters are obtained by a
novel `averaging scheme'.

The methodology involved in creating various grain boundaries
and the isotropic deformation using MD is described in sec.~\ref{comp_methods}.
The results are presented with experimental comparison in sec.~\ref{ResultsAndDiscussion}.
The concluding remarks are listed in sec.~\ref{conclusion-6.4}.

\section{Computational aspects\label{comp_methods}}

%\begin{figure}[!htb]
%\centering
%%\includegraphics[width=1.0\linewidth]{Figures/Spall_MScale/01_fig_multi_scale_flow}
%\includegraphics[width=0.7\linewidth]{Figures/Spall_MScale/01_fig_multi_scale_flow}
%%\caption
%[Multi-scale modelling work flow]
%{Workflow of Multi-scale modelling consisting of MD simulations of 
%triaxial deformation of bi-crystals to obtain the void nucleation and growth as 
%a function of tensile pressure. This information is used by a particle swarm 
%optimization (PSO) code to fit the parameters of the macroscopic void 
%nucleation and growth (NAG) model. The NAG model is used in hydrodynamic 
%simulations of flyer plate impact on targets to obtain the free surface 
%velocities and therefore the spall strength}
%\label{flowchart_multi-scale}
%\end{figure}

A flowchart of the multi-scale method is shown in figure~\ref{subfig:StepsInMSModel} in
chapter~\ref{chapter_compMethods}. The flow starts with assigning the values needed for
setting up bi-crystals and executing MD simulations. Required inputs are 
(i) the grain boundary angle in terms of the orientation planes of the two grains
for eg. (3 1 0) (3, -1 0), (ii) lattice parameter, (iii) time step,
(iv) total runtime, etc. Initially, 11 symmetric tilt (STGB) and 12 symmetric twist (STwGB)
bi-crystal structures are created and relaxed using NPT MD simulations.
%Grain boundary
%energy (GBE) is calculated from the MD simulation. These are compared with the results
%from the literature~\cite{tschopp1,al_twist_GBE_QingYin} to validate the grain
%boundary set up.

\subsection{MD simulations of void nucleation and growth in bi-crystals}
\label{subsec.lab1}

\subsubsection{MD : Setting up the Grain Boundary \& Calculating GBE}
\label{MD-step1_GBEAndValidation}
The GB angles covered in this work range between 0$^{\circ}$ and 90$^{\circ}$ for the 
STGB and 0$^{\circ}$ and 45$^{\circ}$ for the STwGB. The range for the STwGB is
restricted to 45$^{\circ}$ since GBE as a function of GB angle has a mirror symmetry
at 45$^{\circ}$ as seen in figure \ref{twist_GBE}. The required number of unit cells depends upon the
system size in each direction. The length of the system in the three orthogonal
directions X, Y \& Z are denoted as $L_x$, $L_y$, $L_z$. Their values used for the GBE
calculation are listed in table~\ref{MD_geometry_table1}. The table also contains
the system size used for the MD simulation of tri-axial tensile deformation.
Periodic boundary condition (PBC) is applied for all the boundaries. For Al,
EAM potential by Y. Mishin, {\it et al.}~\cite{Michin_Al99.eam.alloy} is used.
The EAM potential is validated in-house, by comparing the values calculated
for the cohesive energy, the vacancy formation energy
and the elastic parameters ($Y_0$, $B_0$, $G_0$, $\mu$) of aluminum with experiments as listed
in table~\ref{elasticTable} of chapter~\ref{chapter_compMethods}.
Figure~\ref{tilt_GBE} shows the GBE calculated for STGB angles, while figure~\ref{twist_GBE}
shows the same for STwGB angles. For comparison, the GBE values calculated by
Tschopp ~\cite{TiGBEtschopp2} and QinYin~\cite{al_twist_GBE_QingYin} are included
in figures~\ref{tilt_GBE} and~\ref{twist_GBE}, respectively.
%For comparison, figure~\ref{tilt_GBE} includes the GBE values
%calculated by Tschopp ~\cite{TiGBEtschopp2} and figure~\ref{twist_GBE} includes
%the GBE values calculated by QinYin ~\cite{al_twist_GBE_QingYin}.

After validating the bi-crystal setup, they are used for the MD simulations of
tri-axial tensile deformation, but with a larger volume. Volume is increased by
replicating the length in all three directions. The replication is done such
that the computational domain is nearly cubical. The
replication is also carefully done to eliminate finite size effects upon subsequent 
tri-axial deformation, as recommended by Rawat {\it et al.}~\cite{rawat1}.

\begin{table}[!ht]
\centering
\caption [Geometry details of MD simulations used for GBE calculation]
{Geometry details of MD simulations used for Grain Boundary
Energy (GBE) minimization and Triaxial Deformation respectively. (Table:Ref.~\cite{madhavan_spall_multiscale})}
\begin{footnotesize}
\begin{tabular}{cccccccccc}
\toprule
\multicolumn{3}{c}{\multirow{2}{*}{\makecell {\bf Grain \\ \bf Boundary}}} &

\multicolumn{6}{c} {\bf System Size ($\AA$) } &

\multirow{2}{*}{\makecell {\bf Replication \\ \bf Factor}}
\\
\cmidrule(lr){4-9}
 
& & &

\multicolumn{3}{c} {\bf GBE Minimization } &
\multicolumn{3}{c} {\bf Triaxial Deformation } 
& %{\bf Factor } 
\\
\cmidrule(lr){1-10}
{\bf Type} &
{\bf Plane} &
{\bf Angle}
& {${\bf L_x}$}
& {${\bf L_y}$}
& {${\bf L_z}$}
& {${\bf L_x}$}
& {${\bf L_y}$}
& {${\bf L_z}$}
& {$\bf R_x$ x $\bf R_y$ x $\bf R_z$ }
\\
\midrule

\parbox[t]{1mm}{\multirow{11}{*}{\rotatebox[origin=c]{90}{Symmetric Tilt}}} &
  (20 1 0) &\ 5.7$^{\circ}\ $  &
 81.10 & 324.34 &\ 4.05 & 
 325.56 & 325.50 & 325.15 &\ 4 x 1 x 80 \\
& (10 1 0) & 11.4$^{\circ}\ $  &
  40.70 & 244.10 &\ 4.05 &
 245.20 & 245.09 & 248.05 &\ 6 x 1 x 61 \\
& ( 5 1 0) & 22.6$^{\circ}\ $ &
  20.65 & 248.24 &\ 4.05 &
 248.65 & 249.08 & 247.88 & 12 x 1 x 61  \\
& ( 4 1 0) & 28.1$^{\circ}\ $  &
  16.69 & 265.71 &\ 4.05 &
 268.20 & 266.74 & 268.33 & 16 x 1 x 66 \\
& ( 3 1 0) & 36.9$^{\circ}\ $  &
  12.80 & 256.76 &\ 4.05 &
 256.96 & 257.58 & 255.97 & 20 x 1 x 63 \\
& ( 9 4 0) & 47.9$^{\circ} $  &
  39.88 & 317.79 & 16.20 &
 320.20 & 318.89 & 325.12 &\ 8 x 1 x 20\\
& ( 2 1 0) & 53.1$^{\circ} $  &
 \ 9.06 & 254.60 &\ 4.05 &
 254.22 & 255.26 & 255.80 & 28 x 1 x 63\\
& ( 7 4 0) & 59.5$^{\circ}\ $  &
  32.65 & 260.06 & 16.20 &
 262.09 & 260.94 & 260.07 &\ 8 x 1 x 16\\
& ( 3 2 0) & 67.4$^{\circ}\ $  &
  14.60 & 261.87 &\ 4.05 &
 263.75 & 262.78 & 264.16 & 18 x 1 x 65\\
& ( 5 4 0) & 77.3$^{\circ}\ $  &
  25.93 & 257.87 & 16.20 &
 259.20 & 259.20 & 259.20 & 10 x 1 x 16\\
& ( 9 8 0) & 83.3$^{\circ}\ $  &
  48.76 & 290.51 & 32.40 &
 293.71 & 291.61 & 292.70 & 6 x 1 x 9\\
\midrule
\parbox[t]{1mm}{\multirow{12}{*}{\rotatebox[origin=c]{90}{Symmetric Twist}}} &
%############ Twist data
  (11 1 0) & 10.3$^{\circ}\ $ &
  44.73 & 243.36 & 44.73 &
 269.44 & 244.31 & 269.44 & 6 x 1 x 6 \\
& ( 9 1 0) & 12.6$^{\circ}\ $ &
  36.67 & 243.37 & 36.67 &
 257.62 & 244.22 & 257.62 & 7 x 1 x 7 \\
& ( 8 1 0) & 14.2$^{\circ}\ $ &
  32.65 & 243.39 & 32.65 &
 262.22 & 244.33 & 262.22 & 8 x 1 x 8 \\
& ( 7 1 0) & 16.2$^{\circ}\ $ &
  28.63 & 243.41 & 28.63 &
 258.70 & 244.32 & 258.70 & 9 x 1 x 9 \\
& ( 6 1 0) & 18.9$^{\circ}\ $ &
  24.63 & 243.43 & 24.63 &
 247.23 & 244.31 & 247.23 & 10 x 1 x 10 \\
& ( 5 1 0) & 22.6$^{\circ}\ $ &
  20.65 & 243.46 & 20.65 &
 248.72 & 244.35 & 248.72 & 12 x 1 x 12 \\
& ( 4 1 0) & 28.0$^{\circ}\ $ &
  16.69 & 243.49 & 16.69 &
 234.66 & 244.41 & 234.66 & 14 x 1 x 14 \\
& ( 7 2 0) & 31.8$^{\circ}\ $ &
  29.48 & 243.52 & 29.48 &
 236.68 & 244.36 & 236.68 & 8 x 1 x 8 \\
& ( 3 1 0) & 36.9$^{\circ}\ $ &
  12.80 & 243.56 & 12.80 &
 244.12 & 244.35 & 244.12 & 19 x 1 x 19 \\
& ( 8 3 0) & 41.1$^{\circ}\ $ &
  34.60 & 243.55 & 34.60 &
 243.08 & 244.42 & 243.08 & 7 x 1 x 7 \\
& (13 5 0) & 42.0$^{\circ}\ $ &
  56.41 & 243.55 & 56.41 &
 283.09 & 244.45 & 283.09 & 5 x 1 x 5 \\
& ( 5 2 0) & 43.6$^{\circ}\ $ &
  21.80 & 243.55 & 21.80 &
 240.81 & 244.47 & 240.81 & 11 x 1 x 11 \\
\bottomrule
\end{tabular}
\end{footnotesize}
\label{MD_geometry_table1}
\end{table}

% Multiplot 1
\begin{figure}[!htb]
     \centering
\subfigure[GBE for various STGB angles]
            {
         \includegraphics[width=0.47\textwidth]
           {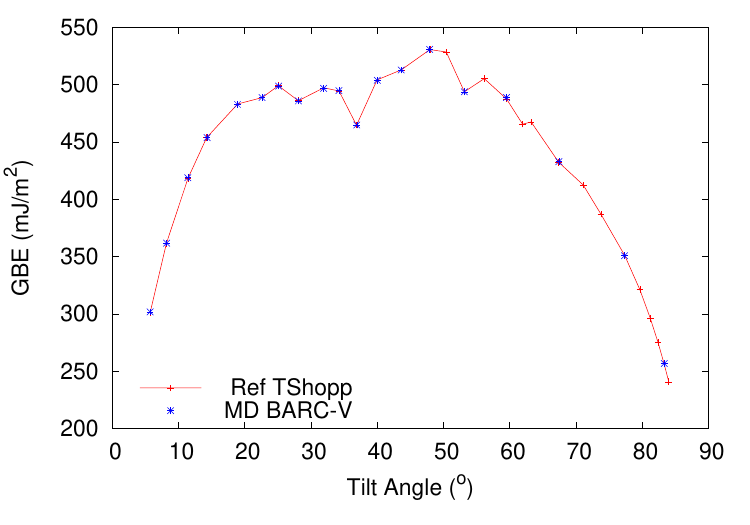}
	        \label{tilt_GBE}
            }
\subfigure[GBE for various STwGB angles]
            {
         \includegraphics[width=0.47\textwidth]
           {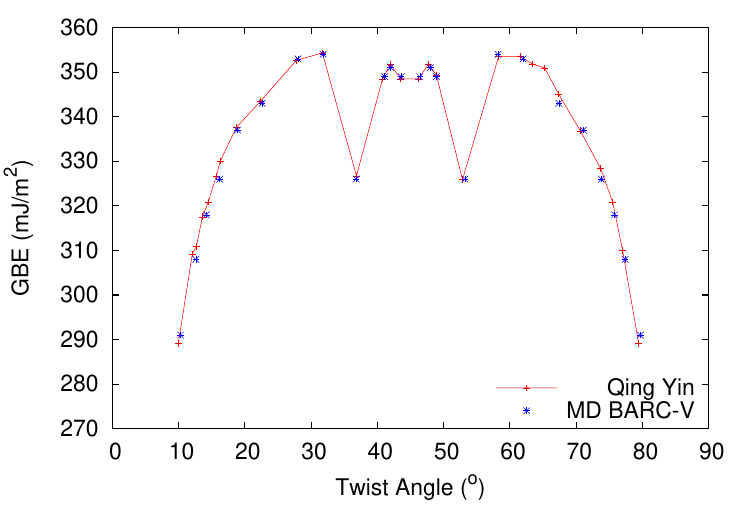}
	       \label{twist_GBE}
            }
%%     \begin{subfigure}[b]{0.49\textwidth}
%         \centering
%         \includegraphics[width=0.495\textwidth]
%           {Figures/Spall_MScale/02afig_tschopp_Al_tilt_GBE}
%%         \caption{GBE for various STGB angles}
%	        \label{tilt_GBE}
%%     \end{subfigure}
%    \hfill 
%%     \begin{subfigure}[b]{0.49\textwidth}
%         \centering
%         \includegraphics[width=0.495\textwidth]
%           {Figures/Spall_MScale/02bfig_Yin_Al_twist_GBE_0to90}
%         \caption{GBE for various STwGB angles}
%	       \label{twist_GBE}
%%    \end{subfigure}
	 \caption
[Comparison of grain boundary energy calculated for Al: STGB \& STwGB]
{Comparison of grain boundary energy calculated from our 
	 simulation, with published results for Al, for various symmetric tilt 
	 ~\cite{TiGBEtschopp2} and symmetric twist 
	 ~\cite{al_twist_GBE_QingYin} angles.\ \  (Fig:Ref.~\cite{madhavan_spall_multiscale})}.
	 \label{fig:GBE_TiltAndTwist}
\end{figure}

\subsubsection{Multiscale methods: MD, PSO \& 1D-HD}
\label{subsubsec.lab1Chapter5}
As described earlier (chapters \ref{chapter_Intro}, \ref{chapter_compMethods} \&
\ref{chapter_SCSpall-Multiscale}), the multiscale simulation is carried out in sequence.
The order of sequence is the execution of MD, PSO and 1D-HD. Results from these
simulations are discussed in the following section.
%\begin{itemize}
%\item {MD high-strain rate simulations are based on the computational
%aspects that are discussed in chapter~\ref{chapter_compMethods}, section~\ref{lammpsMDsec2.1} and
%the aspects discussed in chapter~\ref{chapter_SCSpall-Multiscale}, section~\ref{chap4:sec4.2.2}.}
%
%\item {PSO details used in this work are already explained in 
%chapter~\ref{chapter_compMethods}, section~\ref{PSOsec.2.2}.  }
%
%\item {NAG parameter description and calculation of its coefficients for Al are already
%described in chapter~\ref{chapter_compMethods}, section~\ref{NAG_intro} and
%in chapter~\ref{chapter_SCSpall-Multiscale}, section~\ref{chap4:sec4.2.2}.}
%
%\item {1D-HD as explained in 
%chapter~\ref{chapter_compMethods}, sections~\ref{1Dhydroec.2.3} and
%the aspects discussed in chapter~\ref{chapter_SCSpall-Multiscale}, section~\ref{1dHD-Chap4.2.3}
%are used for is work.}
%
%\end{itemize}

\section{Results and Discussion}
\label{ResultsAndDiscussion}

\subsection{MD simulations of tri-axial expansion of bi-crystals}

Figure~\ref{prsTime_dislocation_Tilt} shows the results from the MD simulations 
of high strain rate triaxial expansion of Al bi-crystal representing an STGB 
59.5$^{\circ}$ as several sub-figures which help in understanding void 
nucleation and growth in a bi-crystal ~\cite{rawat2_hydro,ramana2_hydro}. The 
strain rate applied is ~$5\times10^9$ s$^{-1}$. Figure~\ref{prsTime_740tilt} 
shows the variation of the pressure with time due to void nucleation, growth 
and coalescence, while Figure~\ref{Voidnum_MDTilt740} shows the number of voids 
in the system as a function of time which can be seen in conjunction with 
Figure~\ref{prsTime_740tilt} to mark the system pressure at which nucleation, 
growth and coalescence occur. In this figure, the corresponding pressure is 
also plotted to aid the reader in interpreting the pressure changes due to
nucleation and growth of voids. The magnitude of the tensile pressure 
increases continuously up to 8 GPa after which it turns around and decreases 
with further expansion. The turn-around of pressure is preceded by the
nucleation of voids and is the result of the stress relaxation due to
nucleation and growth of voids.

Figure~\ref{dxa740_tiltTime1} shows the atomic positions in the XY plane around 
the time the voids nucleate (~10.7 ps). Note also that the voids nucleate at 
the grain boundary, which is expected since the bi-crystal boundary is where 
the bonds are weakest. Since periodic boundary conditions were applied in all 
three directions, partial voids at both ends of the y-axis are also seen in 
these figures. The rest of the sub-figures show the growth and coalescence of 
the voids at different times of the simulation using the open visualization 
toolkit ~\cite{Stukowski_2009,ovito_dislocation}. At 11.9 ps the voids have 
grown and initial dislocation lines are visible. Subsequently, the coalescence 
of voids and an increase in dislocation density occurs. From 
figs~\ref{dxa740_tiltTime1}-\ref{dxa740_tiltTime6} it is clear that triaxial 
expansion initially nucleates voids at the grain boundaries and then
dislocations nucleate due to the growth of voids.

\begin{figure}[!htb]
        \centering
	\subfigure [Pressure(t)]
	{
	\includegraphics[width=0.47\textwidth] {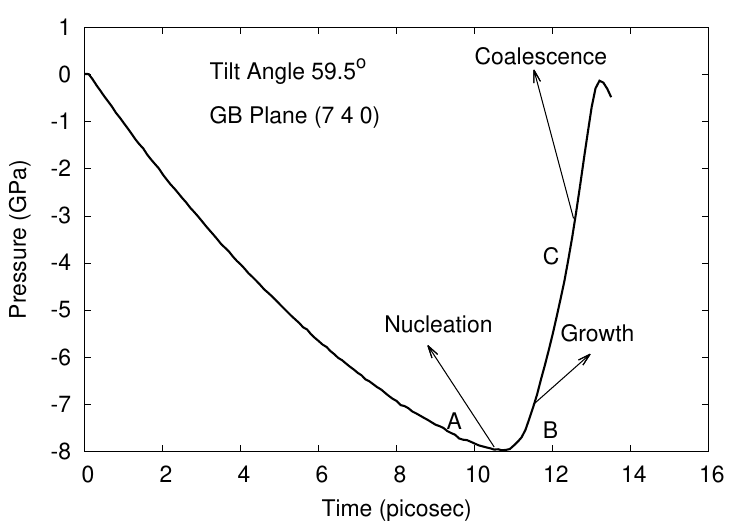}
		 \label{prsTime_740tilt}
	}
	\subfigure [Number of voids as a function of time(t)]
	{
	\includegraphics[width=0.47\textwidth]{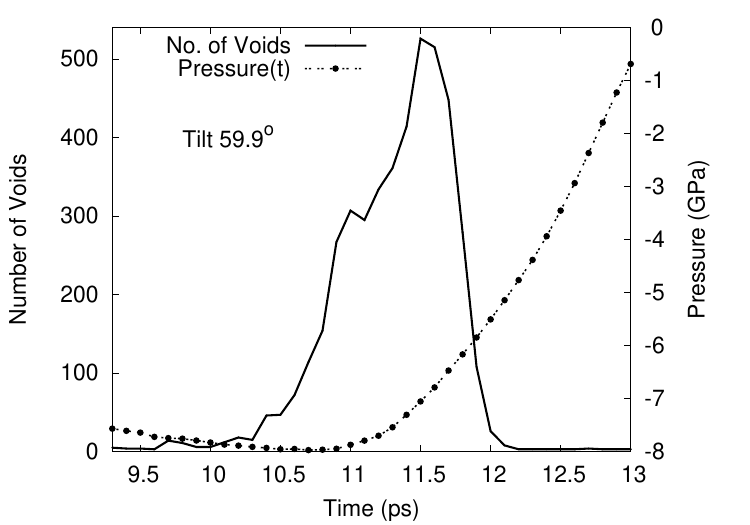}
	\label{Voidnum_MDTilt740}
	}
	\subfigure [t=10.7 ps]
	{
        \includegraphics[width=0.3\textwidth] {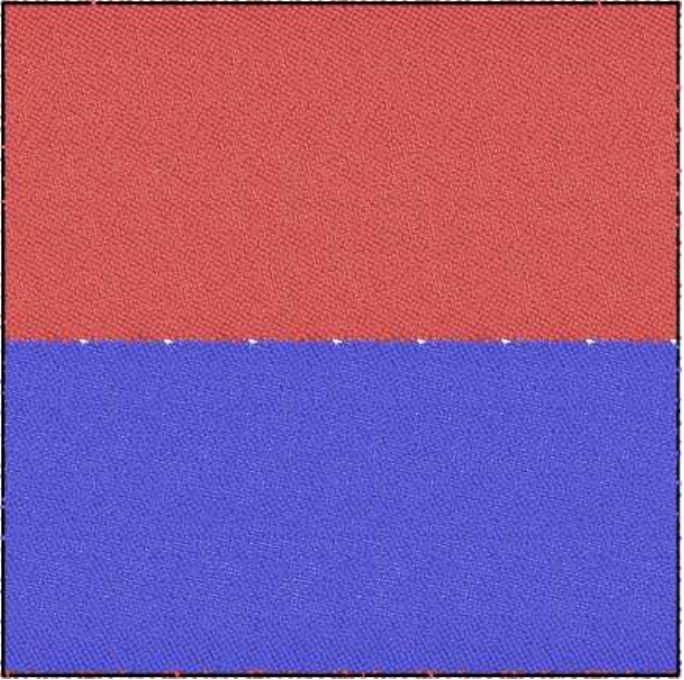}
	    \label{dxa740_tiltTime1}
	}
	\subfigure [t=11.9 ps]
	{
        \includegraphics[width=0.3\textwidth] {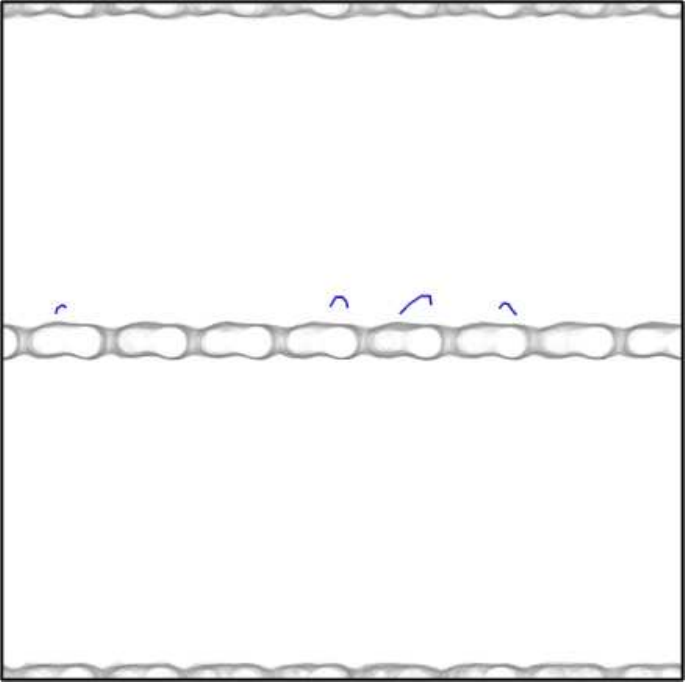}
	    \label{dxa740_tiltTime2}
	}
	\subfigure [t=12.3 ps]
	{
        \includegraphics[width=0.3\textwidth] {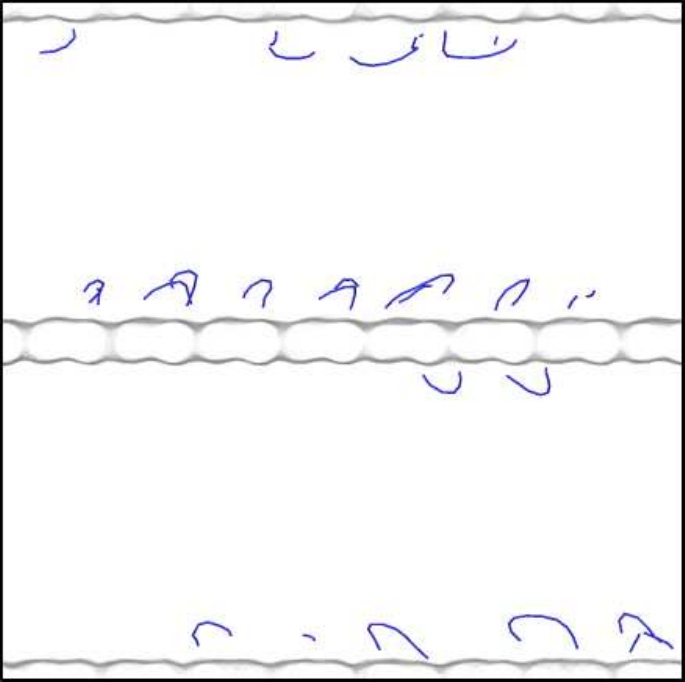}
	    \label{dxa740_tiltTime3}
	}
	\subfigure [t=12.5 ps]
	{
        \includegraphics[width=0.3\textwidth] {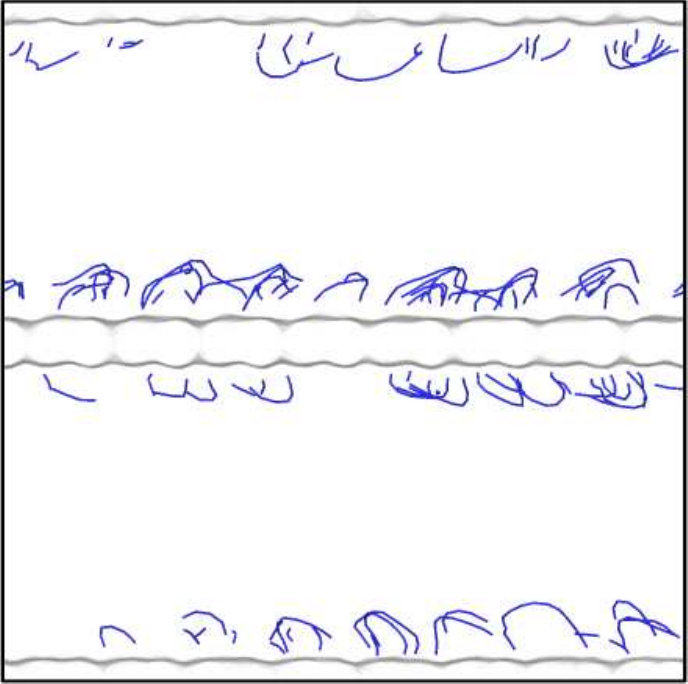}
	    \label{dxa740_tiltTime4}
	}
	\subfigure [t=12.7 ps]
	{
        \includegraphics[width=0.3\textwidth] {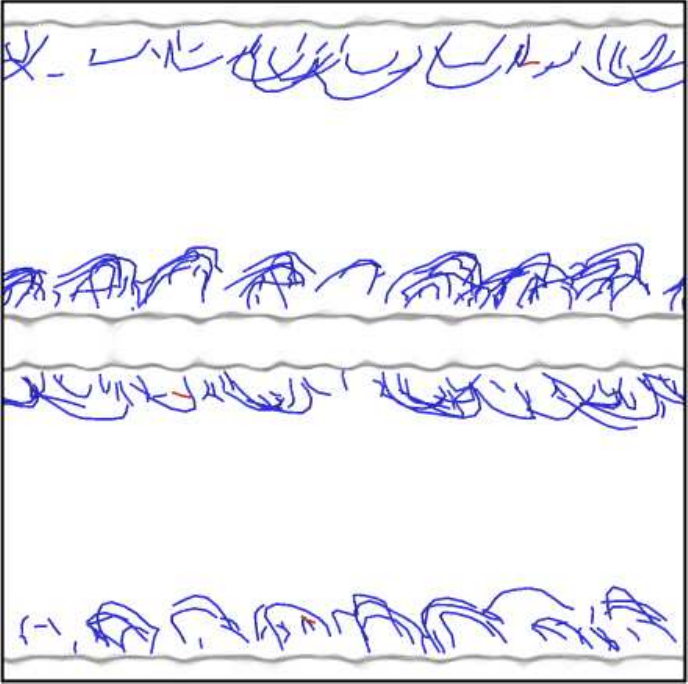}
	    \label{dxa740_tiltTime5}
	}
	\subfigure [t=12.9 ps]
	{
        \includegraphics[width=0.3\textwidth] {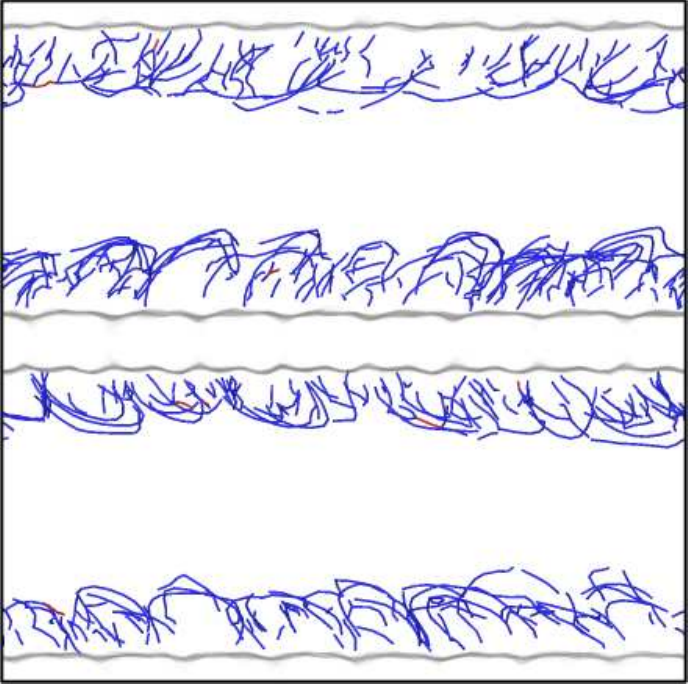}
	    \label{dxa740_tiltTime6}
	}
	\caption
	[MD simulation of high strain rate triaxial expansion of Al (STGB)]
	{MD simulation of high strain rate triaxial expansion of Al 
	bi-crystal representing an STGB 59.5$^{\circ}$: (fig. a) Temporal pressure 
	profile, (fig. b) Temporal evolution of the number of voids and (figs. c-h)
	the MD computational domain in XY Plane showing the nucleation,
	growth and coalescence of voids and dislocation lines. The Al atoms are
	not shown in figs. (d-h) to display the voids and dislocations clearly.\ \ (Fig:Ref.~\cite{madhavan_spall_multiscale}).}
    \label{prsTime_dislocation_Tilt}
\end{figure}

Similar to figure~\ref{prsTime_dislocation_Tilt}, the results for MD simulation 
of high strain rate triaxial expansion of Al bi-crystal representing an STwGB 
31.8$^{\circ}$ is shown in figure~\ref{prsTime_dislocation_Twist}. The
qualitative behavior of the tensile pressure, the number of voids, void nucleation
and growth, etc. are the same as for the tilt case. However the turnaround of the
tensile pressure occurs at a higher tensile pressure ($\simeq -9$ GPa) compared
to the tilt case ($\simeq -8$ GPa), i.e. the turnaround occurs later than in
the twist case. This trend is observed for all the twist cases when compared to
the tilt cases. The higher turnaround value of the tensile pressure is due to
the lower GBEs in the twist cases as seen in fig,~\ref{fig:GBE_TiltAndTwist}.

\begin{figure}[!htb]
        \centering
	\subfigure [Pressure(t)]
	{
	\includegraphics[width=0.47\textwidth] {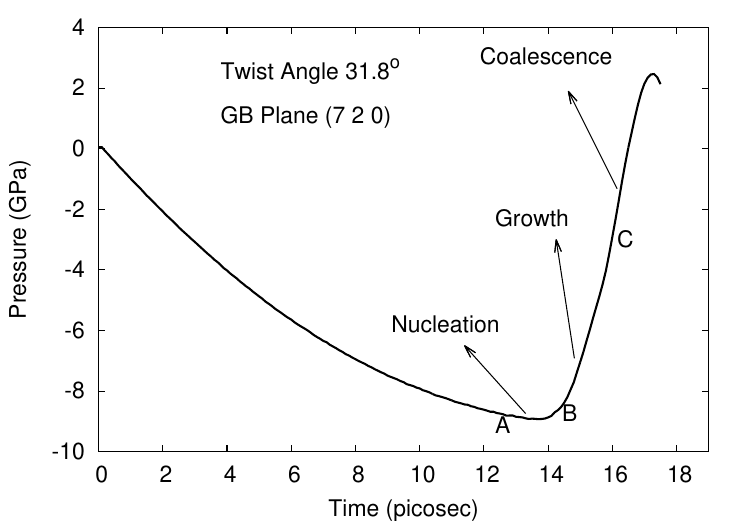}
                \label{prsTime_720twist}
	}
	\subfigure [Number of voids as a function of time(t)]
	{
	\includegraphics[width=0.47\textwidth] {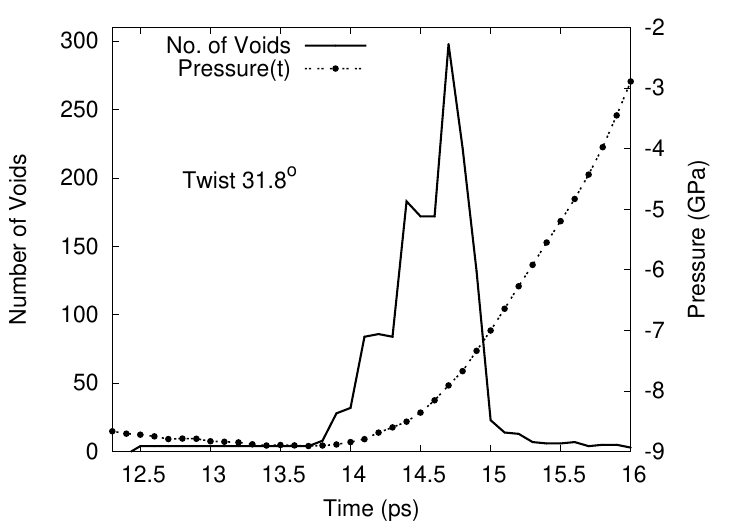}
	\label{Voidnum_MDTwist720}
	}
	\subfigure [t=14.5 ps]
	{
	 \includegraphics[width=0.3\textwidth]{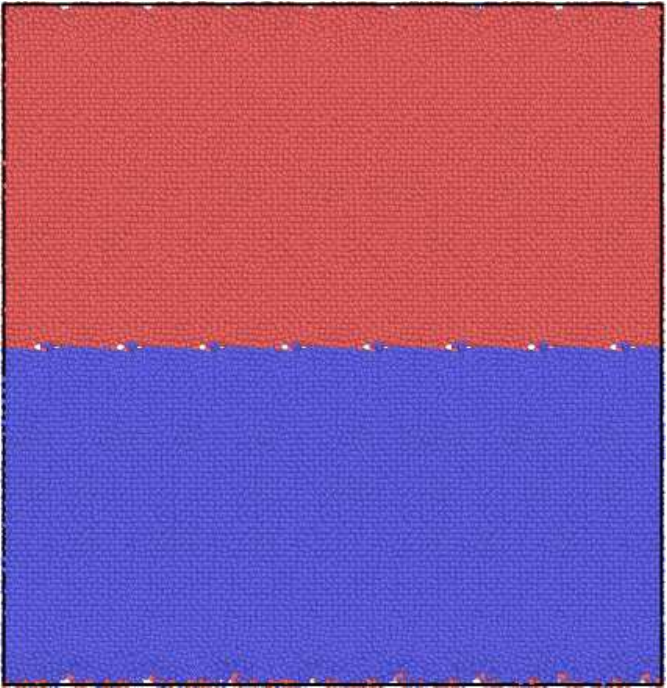}
         \label{dxa720_twistTime1}
	}
	\subfigure [t=15.1 ps]
	{
        \includegraphics[width=0.3\textwidth] {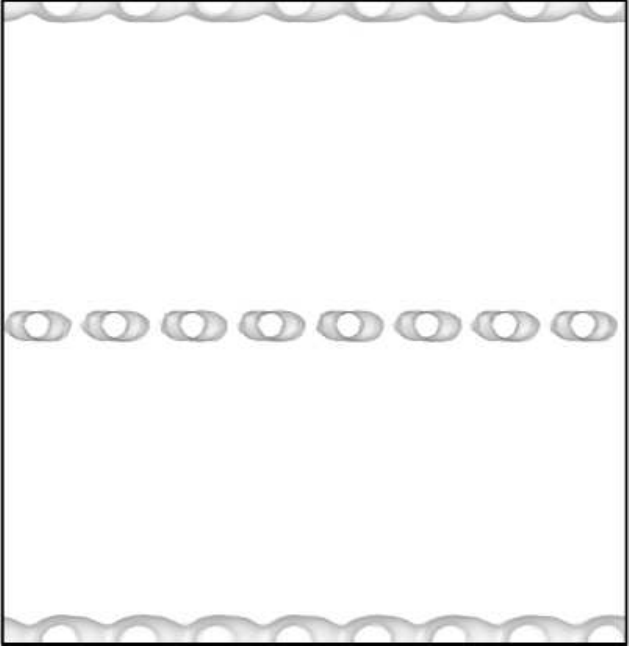}
	\label{dxa720_twistTime2}
	}
	\subfigure [t=15.7 ps]
	{
        \includegraphics[width=0.3\textwidth] {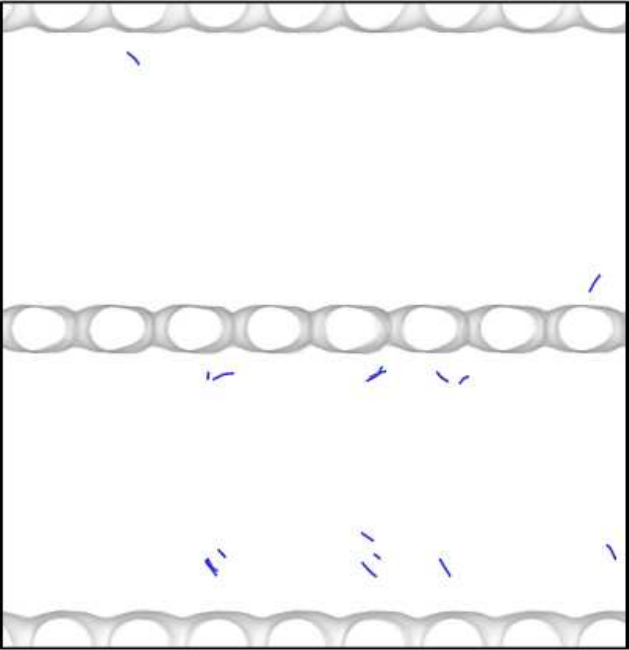}
	\label{dxa720_twistTime3}
	}
	\subfigure [t=16.1 ps]
	{
        \includegraphics[width=0.3\textwidth] {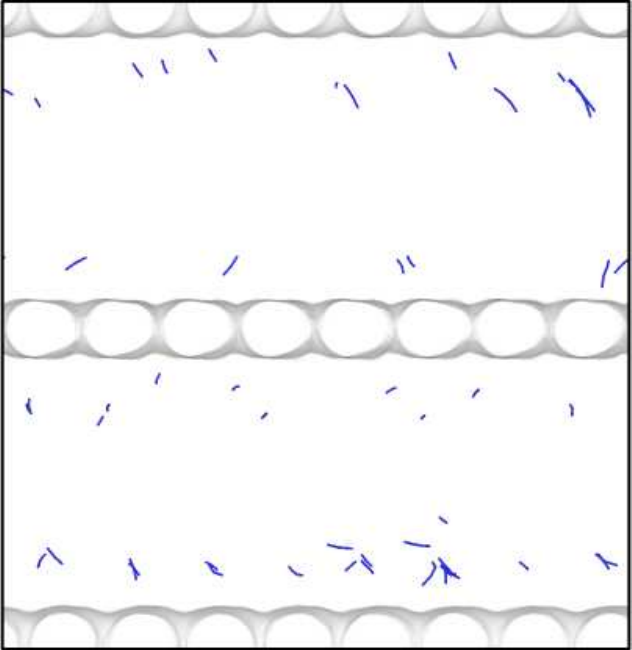}
	\label{dxa720_twistTime4}
	}
	\subfigure [t=16.5 ps]
	{
        \includegraphics[width=0.3\textwidth] {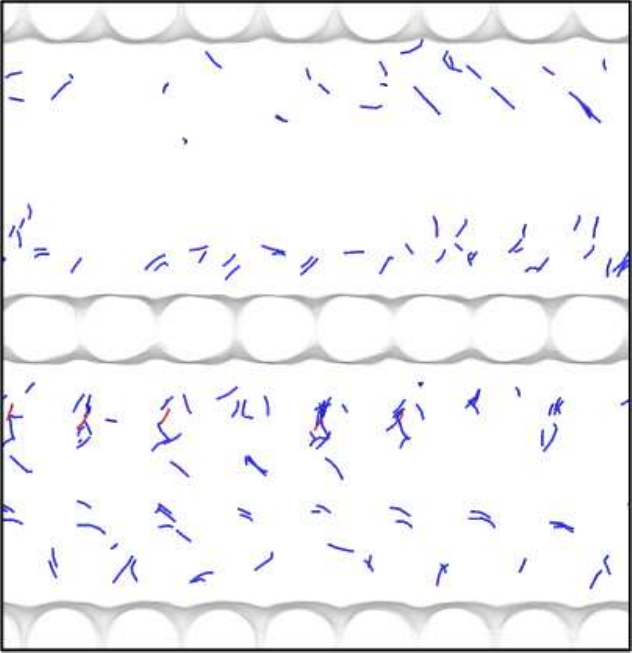}
	\label{dxa720_twistTime5}
	}
	\subfigure [t=16.9 ps]
	{
        \includegraphics[width=0.3\textwidth] {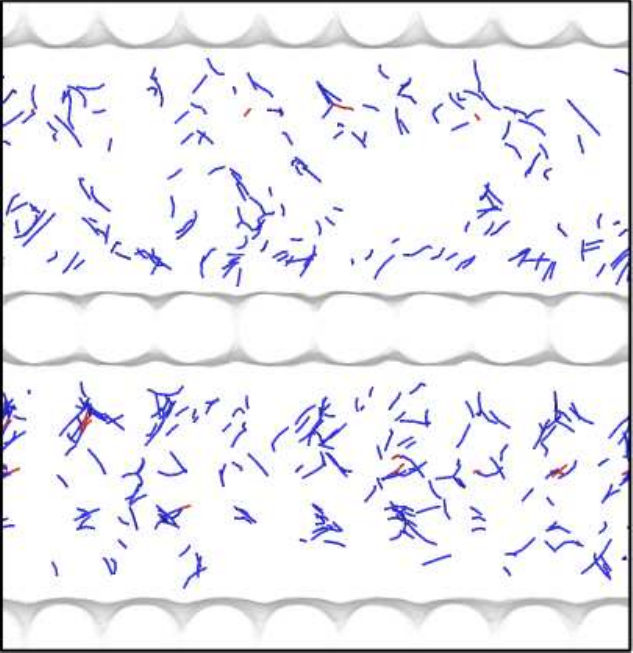}
	\label{dxa720_twistTime6}
	}
	\caption
	[MD simulation of high strain rate triaxial expansion of Al (STwGB)]
	{MD simulation of high strain rate triaxial expansion of Al
        bi-crystal representing an STwGB 31.8$^{\circ}$: (fig. a) Temporal
        pressure profile, (fig. b) Temporal evolution of the number of voids
        and (figs. c-h) the MD computational domain in XY Plane showing the
        nucleation, growth and coalescence of voids and dislocation lines.
        The Al atoms are not shown in figs. (d-h) to display the voids and
	dislocations clearly.\ \ (Fig:Ref.~\cite{madhavan_spall_multiscale}).}
    \label{prsTime_dislocation_Twist}
\end{figure}

The DXA (dislocation extraction algorithm) ~\cite{Stukowski_2009, 
ovito_dislocation} is applied to calculate the dislocation density for
misorientation angles 59.5$^{\circ}$ STGB (7 4 0) and 31.8$^{\circ}$ STwGB (7 2 0).
The time evolution of these is shown in figure~\ref{rhoD_MDTilt740Twist720}.
Dislocation line length and number of dislocation segments at different times are
shown in figures~\ref{DXA_numbers_tilt} and~\ref{DXA_numbers_twist} for
tilt and twist respectively.
Most of the dislocations segments have the
Burger's vectors value $1/2<$011$>$. `Common Neighbor Analysis' (CNA) also shows
that for the applied strain rate, the atoms do not undergo any structural phase
change.

\begin{figure}[!htb]
         \centering
	\subfigure [Dislocation density of Al]
	{
         \includegraphics[width=0.47\textwidth]
         {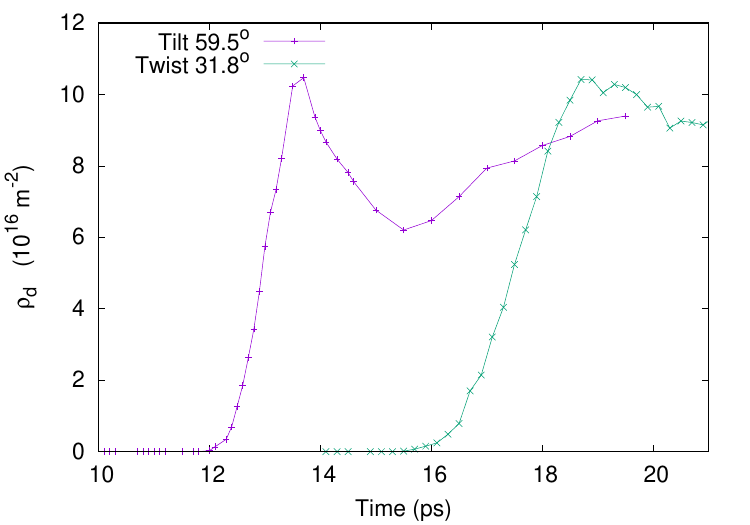}
	     \label{rhoD_MDTilt740Twist720}
	}
	\subfigure[Dislocation (STGB 59.5$^{\circ}$)]
	{
         \includegraphics[width=0.47\textwidth]
         {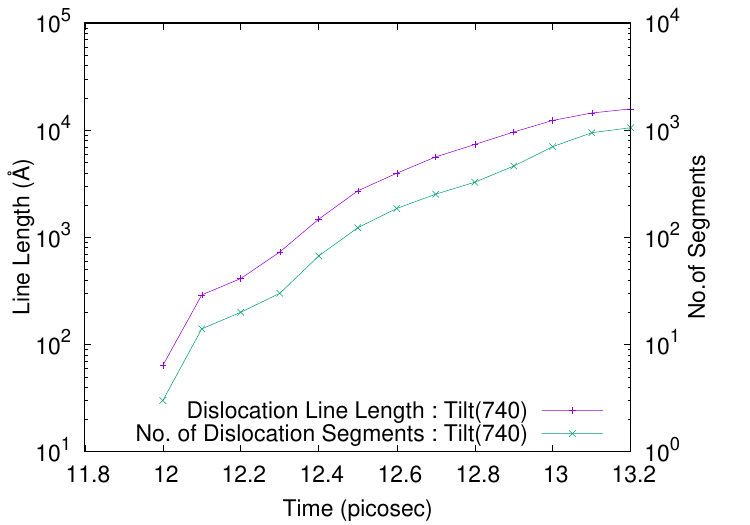}
	     \label{DXA_numbers_tilt}
	}
	\subfigure[Dislocation (STwGB 31.8$^{\circ}$)]
	{
         \includegraphics[width=0.49\textwidth]
         {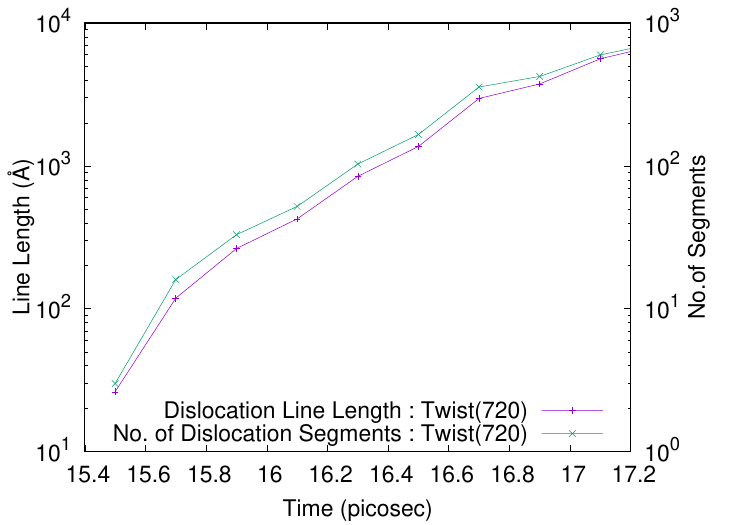}
	     \label{DXA_numbers_twist}
	}
%     \end{subfigure}
	 \caption
	 [Dislocation line length and number of dislocations segments from
	 MD simulations Al: STGB \& STwGB]
	 {(a) Temporal evolution of dislocation density - tilt \& twist; 
	 Dislocation line length and number of dislocations segments obtained from 
	 MD triaxial deformation simulations are shown in fig.(b) for
	 misorientation angle 59.5$^{\circ}$ STGB (7 4 0) and in fig.(c) for
	 31.8$^{\circ}$ STwGB (7 2 0).\ \ (Fig:Ref.~\cite{madhavan_spall_multiscale})}.
	 \label{fig:VoidNumRhoDinTime}
\end{figure}

\subsection{Fitting NAG model using PSO}
The total volume of voids as a function of the pressure of the system obtained 
from the MD simulations of tensile deformation on the 11-tilt and 12-twist 
grain boundaries were fitted to obtain the NAG parameters for void growth using 
PSO. These parameters for (i) Symmetric tilt grain boundaries (STGB) and (ii) 
Symmetric twist grain boundaries (STwGB) are listed in 
table.~\ref{nagvalues_table1}. Comparison of the NAG fit with the MD results 
for void volume fraction as a function of pressure is shown for various tilt 
angles in fig.~\ref{fig:NAGfiterr_tilt} and for various twist angles in 
fig.~\ref{fig:NAGfiterr_twist}.

\begin{table}

\centering
\caption [Coefficients of NAG fracture model parameters obtained by PSO]
{Coefficients of NAG fracture model parameters obtained by PSO method on MD 
data, for various tilt and twist GBs.\ \  (Table:Ref.~\cite{madhavan_spall_multiscale})}

\begin{footnotesize}
\begin{tabular}{ccccccccc}
\toprule

\multicolumn{3}{c}{\multirow{2}{*}{\makecell {\bf Grain \\ \bf Boundary}}} &
\multicolumn{6}{c} {\bf NAG parameters obtained by PSO Fitting on the MD output} \\
\cmidrule(lr){4-9}

& & &

${\bf \dot{N}_0\ (10^{36})}$
& ${\bf P_{1}}$ 
& ${\bf P_{n0}}$ 
& ${\bf P_{g0}}$ 
& ${\bf \eta\ (10^{-3})}$ 
&
\multirow{2}{*}{\makecell {\bf Relative \\ \bf Error (\%)}}
\\
\cmidrule(lr){1-3}
{\bf Type} &
{\bf Plane} &
{\bf Angle}

& ${\bf (m^{-3}s^{-1})}$
& ${\bf (GPa)}$
& ${\bf (GPa)}$
& ${\bf (GPa)}$
& ${\bf (Pa-s)}$
&
\\
\midrule
\parbox[t]{1mm}{\multirow{11}{*}{\rotatebox[origin=c]{90}{Symmetric Tilt}}} &

  (20 1 0) &\ 5.7$^{\circ}\ $  &
 8.36816 & -0.11381 & -6.17830 & -1.63150 & 2.250700 &\ 9.82 \\
& (10 1 0) & 11.4$^{\circ}\ $  &
 8.00696 & -0.01403 & -6.21211 & -2.37083 & 0.720881 &\ 5.28 \\
& ( 5 1 0) & 22.6$^{\circ}\ $ &
 3.60722 & -0.11753 & -6.52964 & -2.14020 & 0.612641 & 26.52 \\
& ( 4 1 0) & 28.1$^{\circ}\ $  &
 4.05373 & -0.05601 & -6.70000 & -3.45412 & 0.642362 & 21.57 \\
& ( 3 1 0) & 36.9$^{\circ}\ $  &
 9.96193 & -0.17645 & -6.80000 & -4.29694 & 0.397970 & 22.32 \\
& ( 9 4 0) & 47.9$^{\circ} $  &
 7.87692 & -0.38946 & -6.54308 & -4.16546 & 0.611128 & 13.87 \\
& ( 2 1 0) & 53.1$^{\circ} $  &
 9.05817 & -0.14422 & -7.23866 & -1.70001 & 0.704807 & 20.25 \\
& ( 7 4 0) & 59.5$^{\circ}\ $  &
 2.63232 & -0.15671 & -7.28917 & -2.01479 & 1.111740 & 18.30 \\
& ( 3 2 0) & 67.4$^{\circ}\ $  &
 8.38800 & -0.10753 & -7.00960 & -3.37014 & 0.868266 & 12.68 \\
& ( 5 4 0) & 77.3$^{\circ}\ $  &
 7.44183 & -0.18027 & -6.59521 & -3.49204 & 0.590714 &\ 8.95 \\
& ( 9 8 0) & 83.3$^{\circ}\ $  &
 9.88908 & -0.13119 & -7.26090 & -4.20800 & 0.891031 & 17.69 \\

\midrule
\parbox[t]{1mm}{\multirow{12}{*}{\rotatebox[origin=c]{90}{Symmetric Twist}}} &
  (11 1 0) & 10.3$^{\circ}\ $ &
 4.95057 & -0.37323 & -7.69390 & -5.98259 & 0.342235 & 19.31 \\
& ( 9 1 0) & 12.6$^{\circ}\ $ &
 8.54382 & -1.98829 & -7.13818 & -1.80495 & 1.288590 & 23.97 \\
& ( 8 1 0) & 14.2$^{\circ}\ $ &
 1.16989 & -0.22827 & -7.51852 & -4.50005 & 0.616554 & 16.28 \\
& ( 7 1 0) & 16.2$^{\circ}\ $ &
  7.90004 & -1.01043 & -7.23602 & -3.08728 & 1.085590 & 21.81 \\
& ( 6 1 0) & 18.9$^{\circ}\ $ &
  2.77167 & -0.43557 & -7.18365 & -4.13188 & 0.609862 & 21.85 \\
& ( 5 1 0) & 22.6$^{\circ}\ $ &
  5.58651 & -0.28320 & -7.10936 & -4.89349 & 0.522616 & 31.89 \\
& ( 4 1 0) & 28.0$^{\circ}\ $ &
  5.52753 & -0.21105 & -7.65841 & -2.56325 & 0.813525 & 25.95 \\
& ( 7 2 0) & 31.8$^{\circ}\ $ &
  4.63374 & -0.37402 & -7.60683 & -5.76872 & 0.355182 & 24.61 \\
& ( 3 1 0) & 36.9$^{\circ}\ $ &
  9.30121 & -0.35873 & -7.99872 & -5.31194 & 0.519195 & 17.38 \\
& ( 8 3 0) & 41.1$^{\circ}\ $ &
  2.32459 & -0.34623 & -7.36754 & -3.82911 & 0.794952 & 31.44 \\
& (13 5 0) & 42.0$^{\circ}\ $ &
  3.00334 & -0.20979 & -7.53975 & -4.06506 & 0.658519 & 15.37 \\
& ( 5 2 0) & 43.6$^{\circ}\ $ &
  6.01735 & -0.35906 & -8.19276 & -3.87510 & 0.646097 & 28.11 \\
\bottomrule
\end{tabular}
\end{footnotesize}
\label{nagvalues_table1}
\end{table}

In the case of tilt GBs, the errors in the fit are mostly in the range 5 to 20 \%. 
For five tilt angles, the fitting errors are less than 15\% and for the other five 
tilt angles, the fitting errors are around 20\%. However, the highest deviation 
in the fitting is 26.52 \% for tilt angle $22.6^o$. In the case of twist GBs, the
errors in the fit are in the range 15-20\% for 4 cases and 21-25\% for 5 cases.
The highest deviation in fitting is around $\sim$31\% for the twist angles
$22.6^o$ and $41.1^o$. These relatively large errors are due to (i) the void
coalescence is also present in the MD results, which is not handled by the NAG
model, (ii) we are fitting a model developed for the macroscopic to
atomistic scale data.

\begin{figure}[!htb]
     \centering
	 \subfigure
	[Tilt angle $11.4^o$, Fitting Error 5.3\%]
	{
         \includegraphics[width=0.47\textwidth]{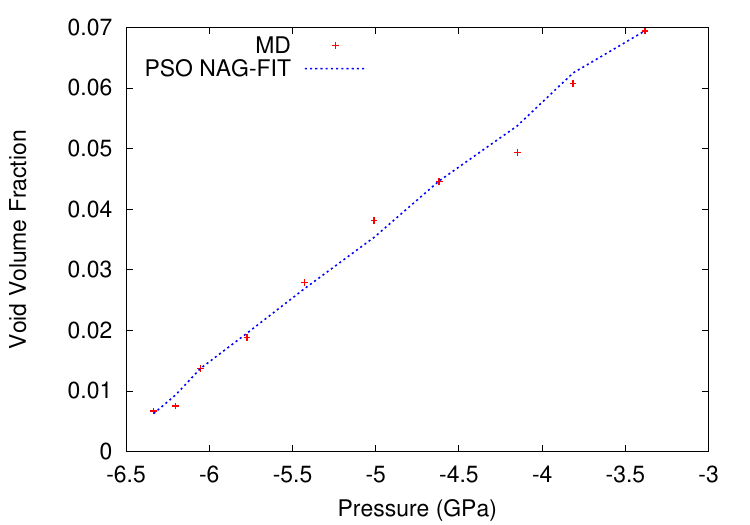}
	        \label{NAGfiterr540_tilt}
	}
         \subfigure
	[Tilt angle $77.3^o$, Fitting Error 8.9\%]
	 {
         \includegraphics[width=0.47\textwidth]{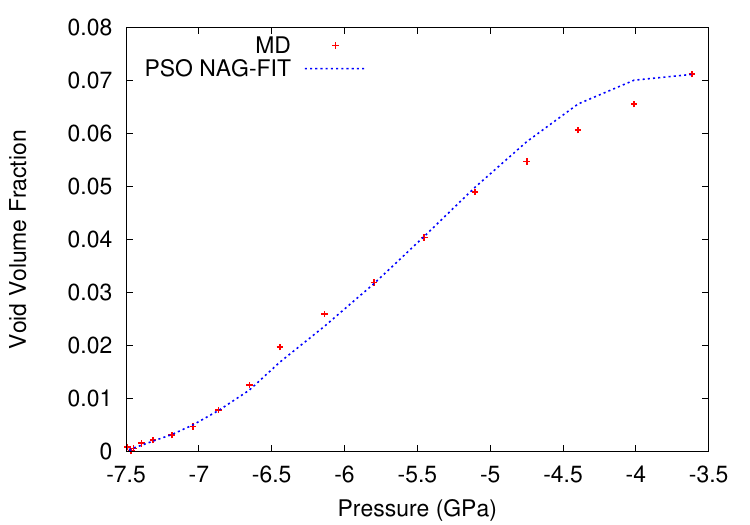}
	        \label{NAGfiterr410_tilt}
	}
         \subfigure
	[Tilt angle $47.9^o$, Fitting Error 13.9\%]
	 {
         \includegraphics[width=0.47\textwidth]{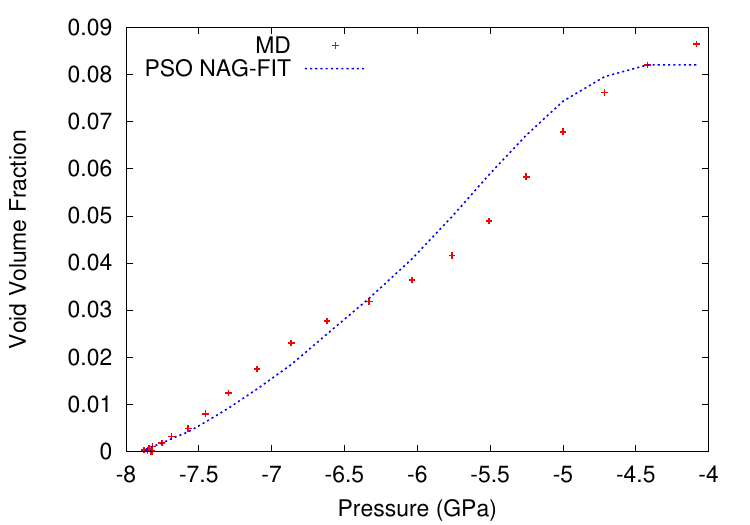}
	        \label{NAGfiterr740_tilt}
	}
         \subfigure
	[Tilt angle $83.3^o$, Fitting Error 17.7\%]
	 {
         \includegraphics[width=0.47\textwidth]{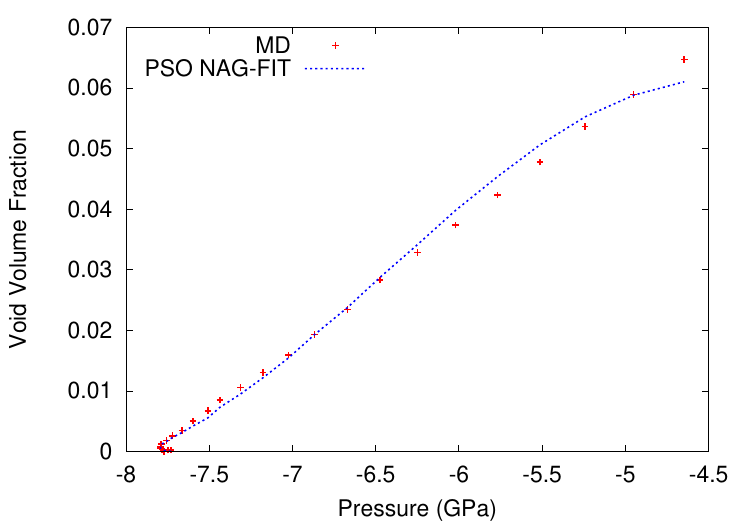}
	        \label{NAGfiterr510_tilt}
	}
     \caption[Void Volume Fraction obtained by MD data fitting for STGB]
{{\it Void Volume Fraction} obtained using NAG parameter fitting are compared with MD simulations results
		for various tilt angles as indicated. \ \ (Fig:Ref.~\cite{madhavan_spall_multiscale})}.
     \label{fig:NAGfiterr_tilt}
\end{figure}

\begin{figure}[!htb]
     \centering
         \subfigure
	[Twist angle $14.2^o$, Fitting Error 16.3\%]
	 {
         \includegraphics[width=0.47\textwidth]{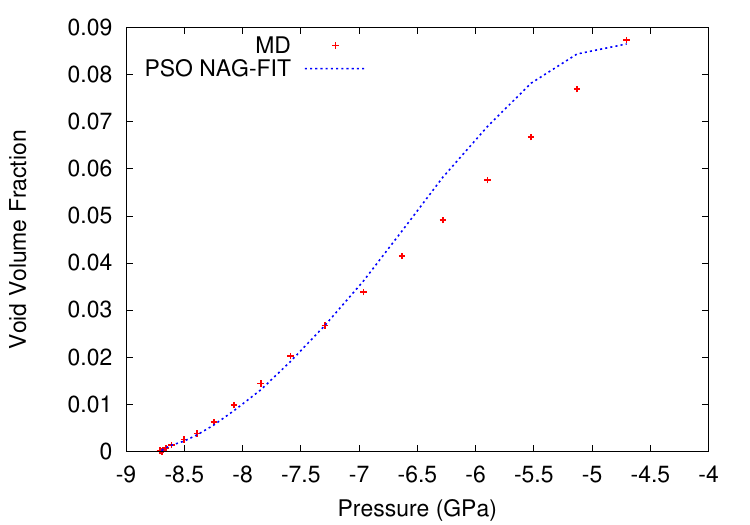}
	        \label{NAGfiterr1350_twist}
	}
         \subfigure
	[Twist angle $10.3^o$, Fitting Error 19.3\%]
	 {
         \includegraphics[width=0.47\textwidth]{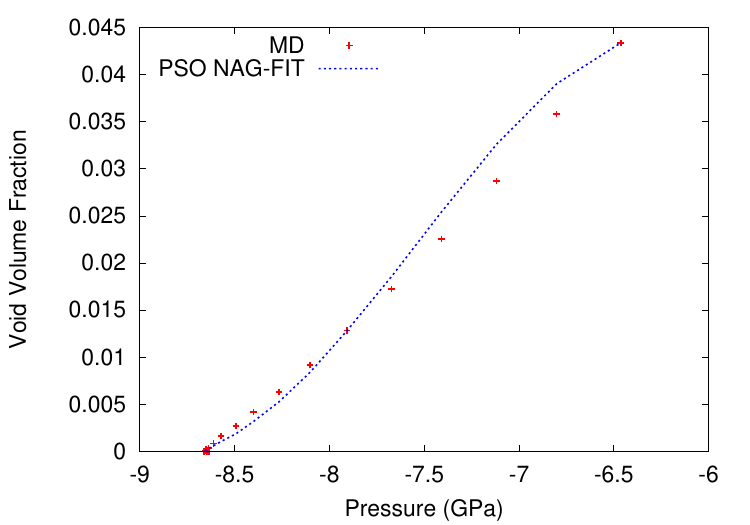}
	        \label{NAGfiterr810_twist}
	}
         \subfigure
	[Twist angle $16.2^o$, Fitting Error 21.9\%]
	 {
         \includegraphics[width=0.47\textwidth]{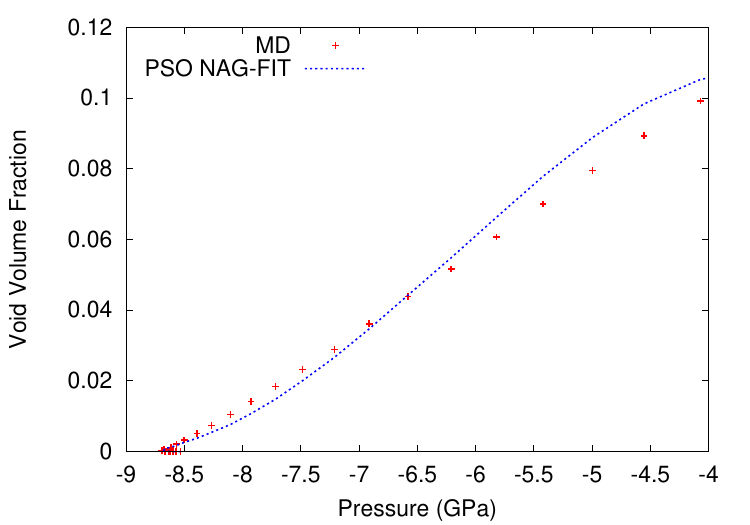}
	        \label{NAGfiterr610_twist}
	}
         \subfigure
	[Twist angle $43.6^o$, Fitting Error 28.1\%]
	 {
         \includegraphics[width=0.47\textwidth]{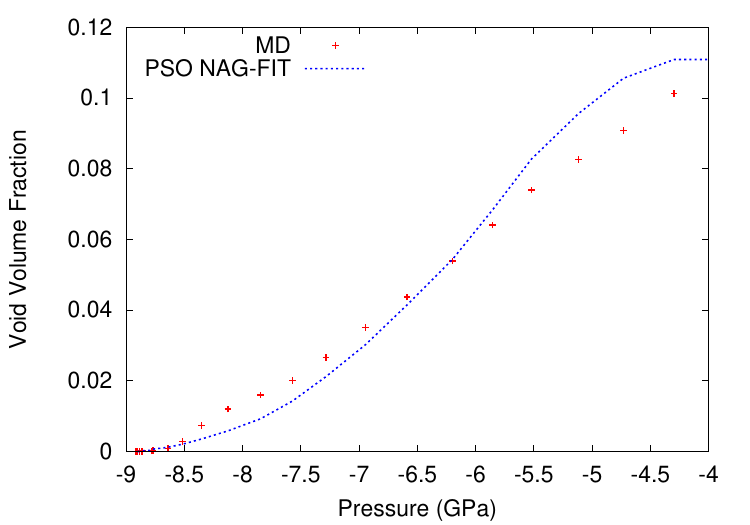}
	        \label{NAGfiterr0720_twist}
	}
     \caption[Void Volume Fraction obtained by MD data fitting for STwGB]
	{{\it Void Volume Fraction} obtained using NAG parameter fitting 
	 are compared with MD simulations results for various twist angles as 
	 as indicated. \ \ (Fig:Ref.~\cite{madhavan_spall_multiscale})}.
     \label{fig:NAGfiterr_twist}
\end{figure}

\subsection{Results from Hydrodynamic simulations}
\label{1dhydroresults}
One-dimensional hydrodynamic (1D-hydrodynamic) plate impact simulations are 
carried out with the NAG parameters obtained above for simulating 
Al-Al impact experiments published separately by Owen {\it et al}
in 2017~\cite{owen_expt} and Chen {\it et al} in 2006~\cite{chen_asay}.
The free surface velocity obtained from hydrodynamics simulations is compared
with the results, specifically experiments 1 from Owen {\it et al}~\cite{owen_expt},
expt. no. 19 and 26 from  Chen {\it et al}~\cite{chen_asay}. These correspond to
the impact velocities of 518 m/s, 1588 m/s and 2275 m/s respectively.
In the case of the experiment by Owen~\cite{owen_expt}, the flyer
thickness was 2.92 mm and the target thickness was 3.99 mm. This is represented
as set-1, in this work. For the experiments of Chen~\cite{chen_asay} the target
thickness was $\simeq$5.8 mm. The flyer thicknesses were 3 mm and 2 mm for the
 expt. number 19 and 26 respectively. These are represented as set-2 and set-3, in this
work.

The NAG parameters obtained from MD and PSO are corresponding to particular twist and
twin symmetric grain boundaries. The spall experimental results are not available in
the literature for materials with such a specific arrangement. In any case, the bulk material
is formed with various combinations of STGB, STwGB, the \ac{ASTGB}, and the \ac{ASTwGB}.
Nucleation and growth of voids happen in the presence of many such
combinations. So, it is desirable, in any case, to obtain NAG parameters applicable
for bulk material consisting of many such GBs. For this, we propose a novel approach
which can be called as ``Average Void Growth in Fluid Element (AVGFE)'' method.  In
this method for each hydrodynamic time step, the increment in void volume fraction ($V_v$)
that appears in the equation~\ref{V_vEqn0} is calculated using each set of NAG
parameters resulting from various STGB and STwGB independently. The arithmetic mean of
the calculated values of void volume fraction ($\Delta{V_{v}}$) is used as the incremental
void volume due to nucleation ($\Delta{V_{vn}}$) and growth ($\Delta{V_{vg}}$) for the fluid element
in that time step. The method is simple to implement in the hydrodynamic domain which is spatially
discretized into several solid/fluid elements. Such discretized domain supports flow properties
flow properties when the material is subjected to very high strain rate shock impact. A hydrodynamic
fluid element has its volume of the order O($\text{mm}^3$). In this volume, a million grains with
different tilt and twist GB can be accommodated. Thus calculating the average void volume fraction
from the NAG parameters of a set of various grain boundaries is a good approximation.
%A hydrodynamic fluid element has its volume of the order O($\text{nm}^3$). In this volume,
%a million grains with different tilt and twist GB be accommodated. Thus calculating the average
%void volume fraction from the NAG parameters of a set of various grain boundaries is a good approximation.

%\vspace{-0.5mm}
\begin{figure}
     \centering{}
     \includegraphics[width=0.6\textwidth]{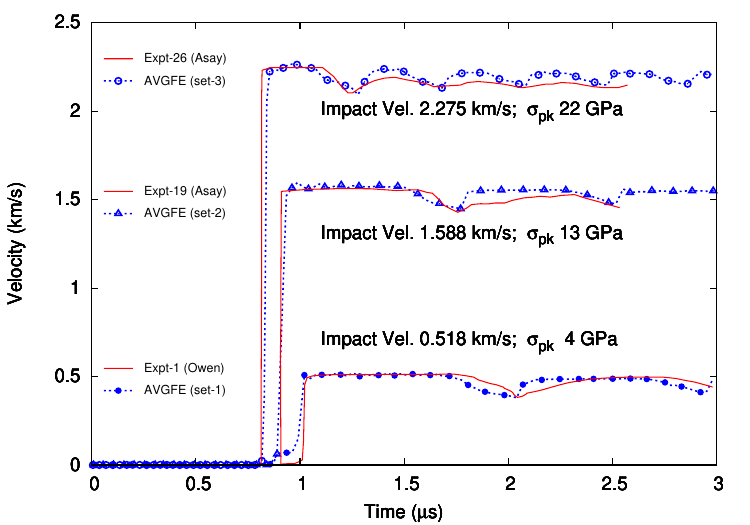}
	\vspace{-0.5mm}
     \caption
	[Multiscale model : Free surface velocity calculated from MD; Comparison with experiments (sets 1-3)]
	{Multiscale model : Calculated time-varying free surface velocity
	obtained for set-1, set-2 \& set-3 described in sec. \ref{1dhydroresults}.
	For comparison respective results from Expt. No.1 of Owen,
	ref.~\cite{owen_expt} and Expt. No. 19 \& 26 of Asay,
	ref.~\cite{chen_asay} are also shown. \ \ (Fig:Ref.~\cite{madhavan_spall_multiscale})}.
	\label{fsv_case5and6}
\end{figure}

\begin{table}
\centering
\caption [Comparison of spall parameters obtained from Multiscale (AVGFE) method with experiments]
{Comparison of spall parameters obtained from the multiscale model simulation with experiments (Table:Ref.~\cite{madhavan_spall_multiscale})}
\begin{tabular}{ccccccccc}
\toprule
\multirow{3}{*}{\parbox{0.6cm}{\centering{Set No.}}}  &
\multirow{3}{*}{\parbox{0.6cm}{\centering{Flyer Vel. (m/s)}}} 
 & \multicolumn{2}{c} {Peak Stress}
 & \multicolumn{2}{c} {${\Delta}{U}_{fs}$}
 & \multicolumn{2}{c} {${\sigma}_{sp}$} 
 & \multirow{3}{*}{\parbox{0.8cm}{\centering{Rel. error (\%)}}} \\
 &
 & \multicolumn{2}{c} {(GPa)}
 & \multicolumn{2}{c} {(m/s)}
 & \multicolumn{2}{c} {(GPa)}
 & \\
\cmidrule(lr){3-8}
 & & Expt. & Simu. & Expt. & Simu. & Expt. & Simu. & \\
\midrule
1 &\ 518 &\ 3.9 &  4.0 & 128 & 131 & 0.93 & 0.95 & 2.15 \\
2 & 1588 & 13.7 & 14.6 & 138 & 134 & 1.07 & 1.04 & 2.89 \\
3 & 2275 & 21.2 & 22.3 & 149 & 158 & 1.16 & 1.23 & 6.06 \\
\bottomrule
\end{tabular}
\label{hydroResultFreeSurfExpt19-26}
\end{table}
%\vspace{-0.2mm}

Using the AVGFE method for the nucleation and growth of voids, 1D hydrodynamic
simulations are performed for the set-1, set-2 and set-3 cases. The strain rates
calculated from the 1D-hydrodynamic simulation 
are 8.2${\times}10^6$ s$^{-1}$, 1.3${\times}10^7$ s$^{-1}$ \&
2.6${\times}10^7$ s$^{-1}$ respectively.
The time histories of the free surface velocity are shown in the figure 
~\ref{fsv_case5and6} from multiscale simulations for set-1 (Expt.1~\cite{owen_expt}), set-2 and
set-3 (Expt.19 \& 26~\cite{chen_asay}). Peak stress, pull-back velocity, spall strength and spall
thickness obtained from the simulations are listed in
table ~\ref{hydroResultFreeSurfExpt19-26}, along with experimental results. 
Computed peak pressures match very well with experimental values. The spall
strengths calculated are 0.95 GPa, 1.04 GPa and  1.23 GPa for the three cases
respectively, whereas experimentally observed values are 0.93 GPa, 1.07 GPa and
1.16 GPa respectively. The computed and experimental spall strength
values match very well within the relative error (deviation) of 2.15\% for set-1,
2.8\% for set-2 and 6\% for set-3.

In the case of set-1 (expt. 1) the period of oscillation of the free surface
velocity, from figure~\ref{fsv_case5and6} is 940 ms, while the same from the
multiscale model is 903 ms. These correspond to the spall thickness
2.49 mm and 2.39 mm respectively for the experiment and simulation
(calculated using eqn.~\ref{eqn_sp_thickChap4}).
In the case of set-2 (expt. 19) the period of oscillation of the free surface 
velocity, from figure~\ref{fsv_case5and6} is 779 ms while the same from
the simulation is 749 ms. The corresponding spall thicknesses are 2.25 \& 2.16 mm, 
respectively for the experiment and simulation. In the case of set-3 (expt. 26),
the periods of oscillation are 412 \& 402 ms for the experiment and simulation
respectively and the corresponding spall thicknesses are 1.19 \& 1.16 mm.
The relative deviation in spall thickness obtained from multiscale
modelling are 4.11\%, 4\% and 2.4\% for set-1, set-2 \& 3 respectively,
when compared with the experiments.
The match between HD simulations results and experimental results proves the validity
of NAG parameters obtained from our multiscale approach involving MD, PSO and AVGFE
methods.

Note that the AVGFE method organically gives more weightage to bicrystals
which contribute more to void nucleation and growth. To prove its
efficacy we compare it with two alternate models, (i) where a single value for each
of the NAG coefficients can be obtained by calculating the arithmetic mean (AMEAN)
of the corresponding NAG parameters listed in table~\ref{nagvalues_table1}, (ii)
NAG values for single crystal (SC) aluminum~\cite{icons2018_madhavan}. 
Fig~\ref{fsv_avgfeMEANSingleXl} shows this comparison. The values of the spall
parameter obtained from these three methods using the free surface velocity profile
are listed in table~\ref{FreeSurf3methods}. It may be noted that the results from
the AVGFE show a much better match than the two alternate models. `No spall' is
predicted by both the alternate methods for the impact velocity, 518 m/s. They
also show a higher period of oscillation in the free surface velocity which indicates
the spall thicknesses predicted by these methods are higher than that predicted by
the AVGFE method which matches the experimental results well.
\vspace{-2.5mm}
\begin{figure}
     \centering{}
     \includegraphics[width=0.6\textwidth]{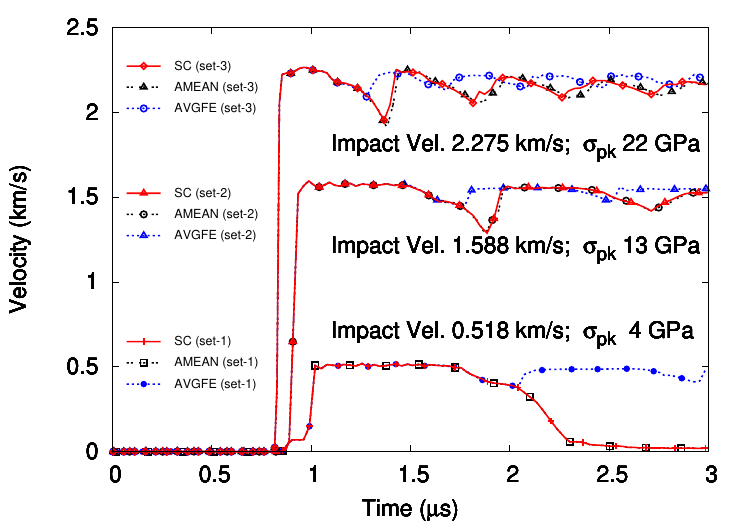}
\vspace{-0.5mm}
     \caption
	[Comparison of Free surface velocity obtained from AVGFE, AMEAN and Single Crystal 
	   for sets 1-3]
	{Comparison of Free surface velocity obtained from AVGFE, AMEAN and Single Crystal for
	set-1, set-2 \& set-3.\ \ (Fig:Ref.~\cite{madhavan_spall_multiscale})}.
	\label{fsv_avgfeMEANSingleXl}
\end{figure}
\vspace{-0.5mm}

\begin{table}
\centering
\caption [Comparison of the spall data obtained from AVGFE \& AMEAN methods]
{Table of spall strength obtained from Multiscale model using
AVGFE, MEAN methods \& Single Crystal (SC) in comparison with experiments.
(Table:Ref.~\cite{madhavan_spall_multiscale})}
\begin{tabular}{cccccc}
\hline
\multirow{2}{*}{\parbox{1.0cm}{\centering{Set No.}}}  &
\multirow{2}{*}{\parbox{2.0cm}{\centering{Flyer Vel. (m/s)}}}
 & \multicolumn{4}{c} {${\sigma}_{sp}$ (GPa)} \\
%\hline{3-6}
\cmidrule(lr){3-6}
 & & Expt & AVGFE & AMEAN & (SC) \\
\hline
1 &\ 518 & 0.93 & 0.95 &  No Spall & No Spall  \\
2 & 1588 & 1.07 & 1.04 & 2.07 & 2.15 \\
3 & 2275 &1.16 & 1.23 & 2.66 & 2.35 \\
\hline
\end{tabular}
\label{FreeSurf3methods}
\end{table}
\vspace{-0.5mm}
%%!!!

%\vspace{-2.5mm}
\section{Conclusions}
\vspace{-0.5mm}
\label{conclusion-6.4}
A multi-scale model to obtain spall parameters at high strain rates for 
polycrystalline materials using MD simulations at the atomistic scales and 
hydrodynamic simulations at the macro-scales has been developed. We had earlier 
successfully applied this model to single crystals in Cu, Al, Mo and Nb to 
obtain the spall strength and spall thickness in flyer impact experiments 
\cite{icons2018_madhavan, rawat2_hydro, ramana2_hydro}. We extend this model to 
poly-crystallites by assuming that the grain boundaries are the weakest 
location and voids can nucleate there when subjected to tensile deformation. 
The grain boundaries in the MD simulations are mimicked using eleven symmetric 
tilt (STGB) and twelve symmetric twist (STwGB) boundaries. The nucleation, 
growth and coalescence of voids due to isotropic expansion were monitored as a 
function of pressure in the MD simulation. In all the cases the voids nucleated 
at the grain boundaries as expected since the bonds are weakest at the grain 
boundaries.

Using particle swarm optimization to fit a nucleation and growth model of voids 
with the MD results the NAG parameters were obtained for each of the 23 
bi-crystals. An average void growth in the fluid element model (AVGFE) is proposed
to simulate the effect of the various grain boundaries in a simulation cell of the 
hydrodynamic simulations. Using these NAG parameters, one-dimensional hydrodynamic
simulations are performed for flyer impact simulations at three different flyer
velocities, 518, 1588 and 2275 m/s (set-1, set-2 \& 3) reported in experiments for
pure Al. The free surface velocity curve vs time calculated from the multiscale model
is found to match the experiments. The spall parameters calculated from the  multiscale
simulations showed excellent match with the experiments for three different impact
velocities. The deviations for the spall strength are within 2.15\%, 2.9\% and 6.0\%,
while for the spall thicknesses the deviations are within 4\%, 4\% \& 2.39 respectively\%.

%%%%%%%%%%%%%%%%%%%%%%%%%%%%%%%%
%The present study does not indlude the effect of various grain size. Grain size effects
%can be included in the hydrodynamic simulations by choosing an appropriate material model like
%Zerilli-Armstrong~\cite{zerilliArm}}. Calculated strain rates for the impact velocities
%from the HD simulations are in the range 8.2${\times}10^6$ s$^{-1}$ to 2.6${\times}10^7$ s$^{-1}$.
%In the present study (MD), 5${\times}10^9$ s$^{-1}$ was used due to computational
%constraints. Future study is planned to include lower strain rates.
%%%%%%%%%%%%%%%%%%%%%%%%%%%%%%%%
%Future studies to include the effect of various grain sizes, by choosing an appropriate material model
%like Zerilli-Armstrong~\cite{zerilliArm}}. The strain rates corresponding to the impact velocities
%518-2275 m/s, are in the range 8.2${\times}10^6$ s$^{-1}$ to 2.6${\times}10^7$ s$^{-1}$, but
%5${\times}10^9$ s$^{-1}$ was used in MD methods due to the computational constraints. Future studies
%also will include lower strain rates.
%%%%%%%%%%%%%%%%%%%%%%%%%%%%%%%%
Future studies to include the effect of various grain sizes with an appropriate material model like
Zerilli-Armstrong~\cite{zerilliArm}. The strain rates for the impact velocities 518-2275 m/s are between
8.2${\times}10^6$ s$^{-1}$ to 2.6${\times}10^7$ s$^{-1}$, but 5${\times}10^9$ s$^{-1}$ is used here.
Future study will include lower strain rates.
%The range of the strain rates corresponding to the impact
%velocities 518-2275 m/s is between 8.2${\times}10^6$ s$^{-1}$ to 2.6${\times}10^7$ s$^{-1}$, but
%5${\times}10^9$ s$^{-1}$ is used in this work (MD) due to the computational constraints.
%Future studies will include lower strain rates.

\clearchapter

% Summary : Chapter 7. Concluding results from this research work. Work done and work for future.
  \chapter{Concluding summary and Future scope\label{chapterSummary}}
A computational study with the title ``Multi-scale modeling of high strain rate deformation
and spall fracture in poly-crystalline metals'' has been carried out.
%The outcome of this
%research work is summarized here. This chapter also outlines the possible further study
%that can follow the current studies.

A physical system can be represented by mathematical models. The underlying physical
phenomena are studied either in the macroscale or the microscale range. The macroscale models
are not accurate enough to account for every minute detail while the microscale models are not
efficient enough and may offer too much information. By combining various scales, a multiscale model
gives reasonable accuracy and is also efficient. A multiscale model simultaneously uses different
models/methods at different lengths and different time scales to describe physical phenomena that involve
different spatial and temporal scales of resolution. Different spatial scales involved in a multiscale
model are ranging from macroscale (m, cm, mm), microscale ($\mu$m), nanoscale ($10^{-9}$ m) and
atomic scales ($10^{-10}$ m). The various time scales extend from macroscopic (s, $10^{-3}$ s) to
microscale ($10^{-6}$ s) and further down to much lower time scales ($10^{-12}$-$10^{-15}$ s).

In this research work, high-strain rate responses of metallic single and by-crystals are simulated using LAMMPS
code at atomic length (angstrom units) and time scales (pico-second). High strain rates are induced by
subjecting the metallic crystals to a tri-axial tensile deformation. Temporal pressure and void volume
data are obtained from this atomic simulation. These data are used in an optimizer (PSO method) to fit the
unknown coefficients of a fracture model (Nucleation and Growth). These coefficients represent the collective
mechanical/fracture properties of the materials in the macroscopic scales. Such properties are responsible
for material failure under tensile forces. This failure is called `spallation'. To predict the spall 
parameters (spall strength and spall thickness), hydrodynamic (HD) simulations are performed in the 
macroscopic scales (mm \& $\mu$s). These steps are executed in sequence following the multiscale formalism.

%\section{MD \& HD: Promising tools to generate shock-spall database}
%\section{Multi-scale model: A promising tool to generate shock \& spall database and for predictive simulations}
\section{Multi-scale model: A promising tool to generate shock \& spall database}
A flyer-target impact system produces shock waves in both the flyer and the target.
The shock propagates in the flyer-target materials but in opposite directions.
Upon reflection from both the free boundaries, the shock wave travels back into the
respective materials as a rarefaction wave. In this duration, the material relaxes
and sets into tension. When two rarefaction waves meet, the tensile stress is increased
further due to constructive interference. A threshold value exists for the `tensile stress'
beyond which a material fails by spall-fracture. % This failure process is called `spallation'.
Though experimental spall data is available in the literature for many metals and
composites, it is necessary to understand the physical phenomena involved in
the spallation process at the atomic spatial and time scales.
% The above para is checked in 'Grammarly'.

The process of materials spallation can be studied using the hydrodynamic method.
This method requires the equation of state (EOS), a dynamic strength model and a
suitable dynamic fracture model. These models contain various coefficients
that are material-dependant.
These coefficients can be obtained from the experimental data. The experiment data
includes the actual physical phenomena involved.  However, there are limitations
in the accessibility of such data.  Acquisition of experimental data involves
challenging diagnostic methods and measurement techniques.  These experiments are
expensive and time-consuming because each shock data point on the shock Hugoniot
equations requires a separate experiment at a particular impact velocity and an
associated strain rate. Also, such data may not be available for specific materials
or composites of users' interest. Ab-initio/first principles (FP) calculation based
on fine details of atoms is an alternate choice. FP calculations involve high data
storage and computing power. The computational cost involved in a system with 100
atoms, is very high. For a metal of nano-sized materials, FP simulations are
difficult to accomplish even with the availability of present-day's computing power.
Thus there are limitations in obtaining shock data from both experimental and FP
methods. The reliability of the prediction from hydrodynamic methods depends on the
model coefficients (EOS, dynamic strength \& fracture). The data should include all
the physical phenomena to make the prediction by a hydrodynamic code, useful. With fast-growing
computational hardware and software, calculations based on molecular dynamics (MD) are
helpful to generate the material database for various materials.

In the present research work, MD simulations are performed in a flyer-target arrangement
to obtain the dynamic shock response of single crystals (Al, Cu, Ni Mo and Nb) and bi-crystals
(Al). The multiscale method is employed to obtain free surface velocity (FSV) and the
spall parameters for both single and bi-crystal setups. FSV and spall strength are
calculated at the atomic scales by direct MD simulations. The results obtained from the
direct MD method are compared with experimental and multi-scale results.

\section{Outcome of this research study}
Four problems listed below, are considered in this research study.% Results obtained from each of them are described in this section:
\begin{enumerate}
\vspace{-0.2cm}
\item Dynamic shock response of single crystals (Cu, Al and Ni): MD calculations for the shock-Hugoniot at a pressure exceeding 100 GPa.
This work has been published in a refereed journal~\cite{UsUp_Madhavan2022}.
\vspace{-0.2cm}
\item Determination of NAG fracture model coefficients and spall parameters of single crystals (Cu, Nb, Mo and Al) by multiscale approach.
This work was presented at ICONS-2018, an Intl. conference. It is published in its proceedings~\cite{icons2018_madhavan}
\vspace{-0.2cm}
\item MD-based study on the dynamic response of a bi-crystal grain boundary under impact shock propagation.
	This work was presented at IMPLAST-2022, an Intl. Conf.; Subsequently a part of it was peer-reviewed
	and published in a refereed journal~\cite{ImplastMatPr12023}.
\vspace{-0.2cm}
\item Determination of NAG fracture model coefficients and spall parameters for a poly-crystalline by including
the effects of various symmetric tilt and twist grain boundaries: A multiscale approach.
This work has been published in a refereed journal~\cite{madhavan_spall_multiscale}
\vspace{-0.2cm}
\end{enumerate}

\subsection{MD calculation of shock-Hugoniot for single crystals of Cu, Al and Ni}
The flyer-target impact system is created in the MD domain for the three FCC crystals
(independent simulation for Cu, Al and Ni). With various flyer velocities, impact
simulations are carried out on the metallic target.  Each simulation is performed
independently for Cu, Al and Ni following the same procedure. Shock velocity and
thermodynamic pressure are estimated.  The following are the salient points
resulting from this study.

\begin{itemize}
\vspace{-0.2cm}
\item{Isothermal compressibility and shock-Hugoniot are obtained for FCC single
crystals (independently for Cu, Al \& Ni).}
\vspace{-0.2cm}
\item{The curve relating $P$ and $V$ is generated at (i) isothermal equilibrium (static state)
and (ii) the dynamic shocked-state conditions, using the MD method.
These are compared with published experimental and FP results.}
\vspace{-0.2cm}
\item{$U_s$ and $U_p$ shock-Hugoniot data is generated and a straight-line fit is
constructed.
%\item{
%For all cases of FP calculations (Al, Cu \& Ni), the relative error
%is \textless 6\%, when compared with experiments.
For both Cu and Al, the
relative errors obtained from MD simulations are \textless 6\%, when
compared with experiments~\cite{meyersBook, lasl_shockdata, alcuta_usup_Mitchell}.
%\item{The relative error for the shock hugoniot parameter $S$ from MD simulation of Ni,
The relative error for the shock Hugoniot parameter $S$ from MD simulation of Ni,
16\%. Stiffening the pair interaction component did not reduce the relative error for
Foiles' EAM potential.}
\vspace{-0.2cm}
\item{MD simulations using Foiles {\it et al.}~\cite{CuNi_eam} EAM potential for Ni,
found to give the shock data that has the closest match with experiments, in comparison
to other EAM potentials used by Choi {\it et al.}\cite{Ni_md_Choi},
by Jarmakani {\it et al.}~\cite{Ni_md_Jarmakani}
and Liu Hai {\it et al.}~\cite{Ni_md_Liu_Hai}.}
\vspace{-0.2cm}
\item {It is concluded  that the EAM potentials used for Cu~\cite{CuNi_eam}
and Al~\cite{Al_eam}, can be readily used for MD simulations at high
strain-rates, $10^7$-$10^8\ s^{-1}$.
For Ni, a trend of deviation is observed at higher impact velocity,
in comparison to experiments and FP calculation for the $P$-$V$ relation.
However, the shock-Hugoniot parameters obtained using Foiles EAM
potentials for Ni from this study show the closest match with the experiments
at higher compression region (fig.~\ref{fig:Ni_PVHugo_DiffEAMS}).}
\vspace{-0.2cm}
\item {It has specific implication: At higher-compression shock region, the
pressure reaches Mbar in magnitude. Compared to six other potentials, Foiles
EAM potential seems to give the least error relative to the experimental results}
\vspace{-0.2cm}
\end{itemize}

\subsection{Multiscale method to find NAG coefficients and spall parameters for
single crystals of Cu, Nb, Mo \& Al}
To the perfect single crystal set-up, a half plane of atoms is removed to introduce crystal
defect (line defect in this case). High strain rate 
($\dot{\epsilon}$ = 5$\times{10}^9\ s^{-1}$) tensile deformation
is imposed upon the atoms in the MD computational domain. Presure-Void volume
fraction as a function of time is recorded. Using the PSO method NAG fracture parameters are
obtained by optimization. These NAG fracture parameters when used in one-dimensional
hydrodynamic (1D-HD) simulations, the spall parameter of single crystals of Cu, Nb and Mo
are calculated and compared with earlier works available in the
literature~\cite{rawat2_hydro,ramana2_hydro}. The same
method is followed to obtain the spall parameter of an aluminum (Al) single crystal and the results
are compared with experiments. Following conclusions are drawn from this study.

\begin{itemize}
\vspace{-0.2cm}
\item The PSO method developed for this work is found to be efficient and fast in convergence. The NAG fracture
        parameters obtained for the single crystals are in the right range.
\vspace{-0.2cm}
\item For copper, niobium and molybdenum, the spall strength and spall thickness estimated from this work are
	found to be in the right range. The FSV profiles obtained from this work match the experiments better than other MD results
        published in the literature~\cite{rawat2_hydro, ramana2_hydro}
%Calculated FSV profile from this work found to match better with experiment,
%in comparison to the previsouly published data~\cite{rawat_thesis, ramana2_hydro}.
\vspace{-0.2cm}
\item For Al, the NAG parameters are obtained by PSO fitting to the MD data.
      Time-varying free surface velocity (FSV) profile is obtained for the single
	crystal `Al', from the hydrodynamic calculation.
\vspace{-0.2cm}
\item Spall strength estimated from FSV for `Al', is found to be matching well with experiments.
      Calculated spall strength was found to deviate around 8\% in comparison to the experiments.
\vspace{-0.2cm}
\item The deviation found in the period of oscillation of the FSV profile is large
      when compared with experiments. Further investigations are required to understand the reasons
        for such deviations.
\vspace{-0.2cm}
\end{itemize}

\subsection{Shock propagation through bi-crystal grain boundary of Al: FSV, Spall parameters
calculation from MD simulations~\label{sec7.2.3}}
%calculated directly from MD simulations}
The multiscale model involves various steps from MD to HD. It takes hours of computational
time. Also, the data produced by the MD method for optimization is high in quantity.
Alternately, MD simulation of shock propagation can be performed directly using a plat-impact
set up and associated temporal FSV data can be recorded. This method does not involve
optimization and hydrodynamic simulations. With this idea, five symmetric-tilt and five
symmetric-twist grain boundaries of `Al' are created, and their GBE are validated. To this
setup, a shock wave is launched by an impact mechanism. Shock wave propagation is studied 
and FSV is recorded to calculate the spall strength. Following are the conclusion derived
from the MD simulations. 
\begin{itemize}
\vspace{-0.2cm}
\item {Spallation did not occur for $U_p$ = 0.5 $km/s$ for all STGB and STwGB cases.}
\vspace{-0.2cm}
\item {For $U_p$ = 0.75 $km/s$, no spall is observed for higher misorientation angle STGBs
        ($53.2^o$ and $77.3^o$), whereas for all cases STwGB at this $U_p$, spall is observed.}
\vspace{-0.2cm}
\item {Overall, the values of the spall strength obtained for the individual STGB and STwGB
        cases are higher than that of the polycrystalline Al.}
\vspace{-0.2cm}
\item {The calculated strain rates from the MD simulations are in the range $10^{10}$-$10^{11}$ $s^{-1}$.
        The pattern seen in the spall strength calculated from MD simulations at these strain
        rates is consistent with the experimental results of polycrystalline Al at lower
        strain rates.}
\vspace{-0.2cm}
\item {Coefficients of a functional form that relates spall strength with strain rate are estimated
        by fitting the ${\sigma_{sp}}(\dot\epsilon)$ data of polycrystal Al from experiments
        and MD simulation of Al bi-crystal.}
\vspace{-0.6cm}
\item {Phase transition (PT) occurs near GB upon shock propagation and it influences the free
        surface velocity profile.}
\vspace{-0.2cm}
\item {At higher particle velocities ($U_p$), due to the phase transition, spall strength is found to get reduced.}
\vspace{-0.2cm}
\item {If the dislocation occurs due to shock-induced grain boundary plasticity time to spall is delayed.
        This leads to higher spall strength value (eg. STwGB, $14.2^o$ at 1 km/s $U_p$).}
\vspace{-0.2cm}
\item {Values of spall strength and the peak tensile strength are not the same for Al bicrystal
	systems. Spall strength obtained from FSV was found to be lower than the peak tensile strength.}
\vspace{-0.2cm}
\item {The spall strength obtained by this method is higher than in experiments. Thus, it is 
	concluded that the multiscale
	method, described in chapter~\ref{chapter_BiCSpall-Multiscale} for bi-crystals is better
	for the spall calculation of polycrystal materials.}
\vspace{-0.2cm}
\end{itemize}

\subsection{Multiscale method to find NAG coefficients and spall parameters for polycrystals: Al}
In addition to the various grain boundaries generated, as mentioned in
the previous section-\ref{sec7.2.3}, more bi-crystal grain
boundaries are used in this work. A total of twenty-three (23) grain boundaries are
generated, eleven for STGB and twelve for STwGB. GBE is calculated and validated for
each of these 23 GBs. The procedure described for single-crystal multi-scale
simulation is repeated for each of the 23 GBs independently. We generated a 23 set
of NAG parameters. A novel modal, namely ``Average void growth in fluid element model''
(AVGFE), is implemented to account for the combined effects of various grain
boundaries on the spallation process. Accordingly, we obtain FSV-time history
from 1D-HD simulation for many impact velocities. Concluding remarks from the
multi-scale modeling of shock propagation across various grain boundaries are
listed below.

\begin{itemize}
\vspace{-0.2cm}
\item {Due to isotropic expansion, the nucleation, growth and coalescence of voids occurred near the grain boundaries.
 Because this is the weakest region}
\vspace{-0.2cm}
\item {1D-HD simulations are performed with flyer velocities: 518, 1588 and 2275 m/s}
\vspace{-0.2cm}
\item {The spall strengths are calculated for these 3 cases. Results obtained from the multiscale
simulations showed an excellent match with experiments. The deviations are within 2.15\%, 2.9\% and 6.0\%}
\vspace{-0.2cm}
\item {The frequency of oscillations obtained from the FSV profile matched with experiments.
 Thus the spall thickness calculated from multiscale is reliable. Compared with experiments, the
spall thicknesses were found to match very well. The relative errors involved in the comparison
are 4\%, 4\% \& 2.39\%, for three different impact velocities}
\vspace{-0.2cm}
\end{itemize}

\section{Future scope in continuation of the present study}
\begin{enumerate}
\vspace{-0.2cm}
\item {For the case of single crystal Ni at higher impact velocities, the shock data ($S$ in $U_p$-$U_s$ relationship) obtained from the MD
simulations are deviated by $\sim$16\% in comparison to the experiments
(Plz. ref. chapter-\ref{chapter_SCImpactShock}). To improve the MD results the pair part of the EAM potential
was stiffened. However this did not reduce the deviation. Hence, the embedded part of the EAM potential can
be explored to see the dynamic response of Ni under impact-shock effects at higher impact velocities.}
\vspace{-0.2cm}
\item {A constant strain rate ($\dot\epsilon$ = $5\times{10}^9\ s^{-1}$), was used in the
multiscale simulations reported here for single as well as bi-crystals. The dynamic
response at a lower strain rate range ($\dot\epsilon: {10}^6-{10}^8\ s^{-1}$) may give
more details of the nucleation process of voids. With lower strain rates, it is expected
that the nucleation of voids could reduce the tensile pressure by equilibration in a
time that is lesser than the time of the strain rate and may delay or prevent other
local nucleation of voids. Nevertheless, such lower strain rate deformation could give more
physical insight into the tensile-force-induced spallation process.}
\vspace{-0.2cm}
\item {MD alone bicrystal shock-propagation simulations, be repeated for other
 materials such as Ni, Mo, Nb etc. Such a method may be suitable for composites,
hence MD simulations may be extended for studying shock propagation effects across
the boundaries of materials in composites~\label{7-iii}}
\vspace{-0.2cm}
\item{Grain size effect - multiscale simulations of shock propagation across
the bi-crystal boundary: This can be incorporated into the hydrodynamic code by choosing an
appropriate material model like Zerilli-Armstrong~\cite{zerilliArm}}
\vspace{-0.2cm}
\item{Asymmetic grain boundary: In this research work only selected symmetric tilt
and twist grain boundaries are considered. Most of the symmetric grain boundaries are
considered in this work.  Hence, it may be sufficient to obtain the NAG and spall
parameters for `Al' polycrystalline materials. However, in the interest of research,
a study on the effect of shock propagation across ASymmetric Tilt Grain Boundaries
(ASTGB) and ASymmetric Twist Grain Boundaries (ASTwGB) may give
a better picture of the nucleation and growth and associated spallation process in
a polycrystalline. Hence for the work listed above (points ii-iv), asymmetric grain
boundaries may also be included in addition to the present study.}
\vspace{-0.2cm}
\end{enumerate}

\clearchapter

%%%%%%%%%%%%%%%%%%%%%%%%%%%
%\phantomsection
%\bibliographystyle{elsarticle-num}
%\addcontentsline{toc}{chapter}{Bibliography}
%\bibliography{allIn1}
%\begin{appendices}
%  \input{EAMParametersAppendix}
%  \input{reprints}
%\end{appendices}
%%%%%%%%%%%%%%%%%%%%%%%%%%%
\clearpage
\phantomsection
\setcounter{page}{76}	% with twoside pages
\addcontentsline{toc}{chapter}{Bibliography}
\bibliographystyle{elsarticle-num}
\bibliography{allIn1}

\begin{thebibliography}{100}
\setlength{\itemsep}{1pt}
\expandafter\ifx\csname url\endcsname\relax
  \def\url#1{\texttt{#1}}\fi
\expandafter\ifx\csname urlprefix\endcsname\relax\def\urlprefix{URL }\fi
\expandafter\ifx\csname href\endcsname\relax
  \def\href#1#2{#2} \def\path#1{#1}\fi

\bibitem{grahamBook}
R.~Graham, Solids Under High-Pressure Shock Compression, Springer, 1993.

\bibitem{meyersBook}
M.~A. Meyers, Dynamic Behavior of Materials, John Wiley \& Sons, Inc., New
  York, 1994.

\bibitem{JC_fracture}
G.~Johnson, W.~Cook, Fracture characteristics of three metals subjected to
  various strains, strain rates, temperatures and pressures, Eng. Fract. Mech.
  21 (1985) 31--48.

\bibitem{fortov1991}
V.~E. Fortov, V.~V. Kostin, S.~Eliezer, Spallation of metals under laser
  irradiation, Journal of Applied Physics 70 (1991) 4524.
\newblock \href {http://dx.doi.org/10.1063/1.349087}
  {\path{doi:10.1063/1.349087}}.

\bibitem{spall_moshe1998}
E.~Moshe, S.~Eliezer, E.~Dekel, A.~Ludmirsky, Z.~Henis, M.~Werdiger, I.~B.
  Goldberg, An increase of the spall strength in aluminum, copper, and metglas
  at strain rates larger than $10^7$ $s^{-1}$, Jnl. of Applied Physics 83
  (1998) 4004.

\bibitem{spall_moshe2000}
E.~Moshe, S.~Eliezer, Z.~Henis, M.~Werdiger, E.~Dekel, Y.~Horovitz, S.~Maman,
  I.~B. Goldberg, D.~Eliezer, Experimental measurements of the strength of
  metals approaching the theoretical limit predicted by the equation of state,
  Applied Physics Letters 76 (2000) 1555--1557.

\bibitem{Kanel2001JAPSingleCrysFoilAl}
G.~I. Kanel, S.~V. Razorenov, K.~Baumung, J.~Singer, Dynamic yield and tensile
  strength of aluminum single crystals at temperatures up to the melting point,
  Jnl. of App. Phy. 90 (2001) 136--143.
\newblock \href {http://dx.doi.org/doi.org/10.1063/1.1374478}
  {\path{doi:doi.org/10.1063/1.1374478}}.

\bibitem{seamanCompmodelDuctBrit_JAP}
L.~Seaman, D.~R. Curran, D.~A. Shockey, Computational models for ductile and
  brittle fracture, Jnl. Appl. Phys. 47 (1976) 4814.

\bibitem{curran_nag1}
D.~R. Curran, L.~Seaman, D.~A. Shockey, Dynamic failure of solids, Phys. Rep.
  147~(5-6) (1987) 253--388.

\bibitem{johnsonPulseDurationSpallFrac_JAP}
J.~N. Johnson, G.~Gray, N.~Bourne, Effect of pulse duration and strain rate on
  incipient spall fracture in copper, Jnl. Appl. Phys. 86 (1999) 4892--4901.

\bibitem{ikkurthi1_nag1}
V.~R. Ikkurthi, S.~Chaturvedi, Use of different damage models for simulating
  impact-driven spallation in metal plates, Int. J. Impact Engg. 30 (2004)
  275--301.

\bibitem{RINGDALENVATNE2013306}
I.~R. Vatne, A.~Stukowski, C.~Thaulow, E.~{\O}stby, J.~Marian,
  Three-dimensional crack initiation mechanisms in {BCC-F}e under loading modes
  {I, II and III}, Materials Science and Engineering: A 560 (2013) 306--314.
\newblock \href {http://dx.doi.org/10.1016/j.msea.2012.09.071}
  {\path{doi:10.1016/j.msea.2012.09.071}}.

\bibitem{IJF_AlBucklingSM}
S.~Madhavan, V.~Mehra, S.~Pahari, S.~Ghosh, C.~D. Sijoy, S.~Chaturvedi,
  Buckling and longitudinal cracks in electromagnetically accelerated hollow
  cylinders, International Journal of Fracture 193 (2015) 1--16.
\newblock \href {http://dx.doi.org/10.1007/s10704-015-0010-9}
  {\path{doi:10.1007/s10704-015-0010-9}}.

\bibitem{Anton_spalldata}
T.~H. Antoun, L.~Seaman, D.~R. Curran,
  \href{{https://apps.dtic.mil/sti/pdfs/ADA362759.pdf}}{Dynamic failure of
  materials: Volume 2-compilation of russian spall data}, Tech. Rep.
  DSWA-TR-96-77-V2, { URL} active as on 30th Mar 2023 (1998).
\newline\urlprefix\url{{https://apps.dtic.mil/sti/pdfs/ADA362759.pdf}}

\bibitem{shockExptChallanges}
H.~E. Lorenzana, J.~F. Belak, K.~S. Bradley, {Shocked materials at the
  intersection of experiment and simulation}, Scientific Modelling and
  Simulations. 15 (2008) 159--186.
\newblock \href {http://dx.doi.org/10.1007/s10820-008-9107-z}
  {\path{doi:10.1007/s10820-008-9107-z}}.

\bibitem{VISAR1974}
J.~R. Asay, L.~M. Barker, Interferometric mmeasurement of shock-induced
  particle velocity and spatial variations of particle velocity, Journal of
  Applied Physics 45 (1974) 2540--2546.

\bibitem{ORVIS1983}
D.~D. Bloomquist, S.~A. Sheffield, Optically recording interferometer for
  velocity measurements with subnanosecond resolution, Journal of Applied
  Physics 54 (1983) 1717.
\newblock \href {http://dx.doi.org/10.1063/1.332222}
  {\path{doi:10.1063/1.332222}}.

\bibitem{duff1957}
R.~E. Duff, F.~S. Minshall, {Investigation of a Shock-Induced Transition in
  Bismuth}, Physical Review 108~(5) (1957) 1207--1212.

\bibitem{srinivasan_MdSpallSingleNiXl}
S.~G. Srinivasan, M.~I. Baskes, G.~J. Wagner, Spallation of single crystal
  nickel by void nucleation at shock induced grain junctions, Journal of
  Materials Science 41 (2006) 7838--7842.

\bibitem{srinivasan2007}
S.~G. Srinivasan, M.~I. Baskes, G.~J. Wagner, Atomistic simulations of shock
  induced microstructural evolution and spallation in single crystal nickel,
  Journal of Applied Physics 101 (2007) 043504.
\newblock \href {http://dx.doi.org/10.1063/1.2423084}
  {\path{doi:10.1063/1.2423084}}.

\bibitem{rawat1}
S.~Rawat, M.~Warrier, S.~Chaturvedi, V.~Chavan, Effect of material damage on
  the spallation threshold of single crystal copper: a molecular dynamics
  study, Modelling and Simulation in Material Science and Engineering 20 (2012)
  015012.

\bibitem{JIANG2022114474}
D.~D. Jiang, J.~L. Shao, B.~Wu, P.~Wang, A.~M. He, Sudden change of spall
  strength induced by shock defects based on atomistic simulation of single
  crystal aluminum, Scripta Materialia 210 (2022) 114474.
\newblock \href
  {http://dx.doi.org/https://doi.org/10.1016/j.scriptamat.2021.114474}
  {\path{doi:https://doi.org/10.1016/j.scriptamat.2021.114474}}.

\bibitem{Influence_Fensin}
S.~J. Fensin, S.~M. Valone, E.~K. Cerreta, G.~T. Gray, Influence of grain
  boundary properties on spall strength: Grain boundary energy and excess
  volume, Journal of Applied Physics 112 (2012) 083529.

\bibitem{ZHU2022110923}
Y.~Zhu, J.~Hu, S.~Huang, J.~Wang, G.~Luo, Q.~Shen, Molecular dynamics
  simulation on spallation of [111] {C}u/{N}i nano-multilayers: Voids evolution
  under different shock pulse duration, Computational Materials Science 202
  (2022) 110923.
\newblock \href {http://dx.doi.org/10.1016/j.commatsci.2021.110923}
  {\path{doi:10.1016/j.commatsci.2021.110923}}.

\bibitem{hemp_wilkins}
M.~L. Wilkins, Computer Simulation of Dynamic Phenomena, Springer, 1999.

\bibitem{Murphy_2010}
W.~J. Murphy, A.~Higginbotham, G.~K. et~al., The strength of single crystal
  copper under uniaxial shock compression at 100 {GP}a, Journal of Physics:
  Condensed Matter 22~(6) (2010) 065404.
\newblock \href {http://dx.doi.org/10.1088/0953-8984/22/6/065404}
  {\path{doi:10.1088/0953-8984/22/6/065404}}.

\bibitem{rawat2_hydro}
S.~Rawat, V.~R. Ikkurthi, M.~Warrier, S.~Chaturvedi, Multiscale simulations of
  damage of perfect crystal {C}u at high strain rates, {PRAMANA} Journal of
  Physics 83~(2) (2014) 265--272.

\bibitem{ramana2_hydro}
V.R.Ikkurthi, H.Hemani, R.Sugandhi, S.Rawat, P.Pahari, M.Warrier,
  S.Chaturvedia, Multi-scale computational approach for modelling spallation at
  high strain rates in single-crystal materials, Procedia Engineering 173
  (2017) 1177--1184.

\bibitem{icons2018_madhavan}
S.~Madhavan, V.~R. Ikkuthi, P.~V. Laxminarayana, M.~Warrier, Multiscale
  modelling to estimate spall parameters in metallic single crystals, in: Proc.
  ICONS2018, International Conference on Structural Integrity, IITM, Chennai,
  India, 2018.
\newblock \href {http://dx.doi.org/https://doi.org/10.48550/arXiv.2112.00368}
  {\path{doi:https://doi.org/10.48550/arXiv.2112.00368}}.

\bibitem{madhavan_spall_multiscale}
S.~Madhavan, H.~Hemani, P.~V. Lakshminarayana, V.~R. Ikkurthi, M.~Warrier,
  Effect of symmetric tilt and twist grain boundaries on the void nucleation,
  growth and spall in polycrystalline {A}l : Multiscale modelling,
  Computational Materials Science 211 (2022) 111543.
\newblock \href {http://dx.doi.org/10.1016/j.commatsci.2022.111543}
  {\path{doi:10.1016/j.commatsci.2022.111543}}.

\bibitem{UsUp_Madhavan2022}
S.~Madhavan, V.~Mishra, P.~V.~L. Narayana, M.~Warrier, {Dynamic Response of
  Single Crystal Al, Cu \& Ni Upon Impact : MD and Ab-Initio Calculations},
  Journal of Dynamic Behavior of Materials 9(1) (2023) 24--35.
\newblock \href {http://dx.doi.org/10.1007/s40870-022-00356-5}
  {\path{doi:10.1007/s40870-022-00356-5}}.

\bibitem{brad}
B.~L. Holian, Modeling shock-wave deformation via molecular dynamics, Physical
  Review A 37~(7) (1988) 2562--2568.
\newblock \href {http://dx.doi.org/10.1103/PhysRevA.37.2562}
  {\path{doi:10.1103/PhysRevA.37.2562}}.

\bibitem{bringa}
E.~M. Bringa, J.~U. Cazamias, P.~Erhart, et. al, Atomistic shock {H}ugoniot
  simulation of single-crystal copper, Journal of Applied Physics 97~(7) (2004)
  3793--3799.
\newblock \href {http://dx.doi.org/10.1063/1.1789266}
  {\path{doi:10.1063/1.1789266}}.

\bibitem{Ni_md_Jarmakani}
H.~N. Jarmakani, E.~M. Bringa, P.~Erhart, B.~A. Remington, Y.~M. Wang, N.~Q.
  Vo, M.~A. Meyers, Molecular dynamics simulations of shock compression of
  nickel: From monocrystals to nanocrystals, Acta Materialia 56 (2008)
  5584--5604.

\bibitem{Ni_md_Liu_Hai}
L.~Hai, H.~Jie, Z.~Zhi-xuan, M.~Zhao-xia, Atomistic simulations of
  elastic-plastic deformation in nickel single crystal under shock loading,
  Procedia Engineering 204 (2017) 397--404.

\bibitem{Ni_md_Choi}
J.~Choi, S.~Yoo, S.~Song, J.~S. Park, K.~Kang, Molecular dynamics study of
  {H}ugoniot relation in shocked nickel single crystal, Journal of Mechanical
  Science and Technology 32~(7) (2018) 7983.

\bibitem{SUGANDHI2015113}
R.~Sugandhi, M.~Warrier, S.~Chaturvedi, Identification of best parameters of
  void nucleation and growth model using particle swarm technique., Applied
  Soft Computing 35 (2015) 113--122.

\bibitem{ADLAKHA2014345}
I.~Adlakha, M.~A. Tschopp, K.~N. Solanki, The role of grain boundary structure
  and crystal orientation on crack growth asymmetry in aluminum, Materials
  Science and Engineering: A 618 (2014) 345--354.
\newblock \href {http://dx.doi.org/10.1016/j.msea.2014.08.083}
  {\path{doi:10.1016/j.msea.2014.08.083}}.

\bibitem{xueyang_GBonShockInduPhTransinFeBiCryst}
X.~Zhang, K.~Wang, W.~Zhu, J.~Chen, M.~Cai, S.~Xiao, H.~Deng, W.~Hu, Effect of
  grain boundaries on shock-induced phase transformation in iron bicrystals,
  Jnl. of Applied Physics 123 (2018) 045105.
\newblock \href {http://dx.doi.org/10.1063/1.5003891}
  {\path{doi:10.1063/1.5003891}}.

\bibitem{LONG2020109411}
X.~Long, X.~Liu, W.~Zhang, Y.~Peng, G.~Wang, Shock deformation and spallation
  of {C}u bicrystals with (111) twist grain boundaries, Computational Materials
  Science 173 (2020) 109411.
\newblock \href {http://dx.doi.org/10.1016/j.commatsci.2019.109411}
  {\path{doi:10.1016/j.commatsci.2019.109411}}.

\bibitem{LIN201294}
E.~Q. Lin, H.~J. Shi, L.~S. Niu, E.~Z. Jin, Shock response of copper bicrystals
  with a ${\Sigma}$3 asymmetric tilt grain boundary, Computational Materials
  Science 59 (2012) 94--100.
\newblock \href
  {http://dx.doi.org/https://doi.org/10.1016/j.commatsci.2012.02.025}
  {\path{doi:https://doi.org/10.1016/j.commatsci.2012.02.025}}.

\bibitem{Luo_bicrys_asymm_sigma3_110TiltGB}
S.~N. Luo, T.~C. Germann, D.~L. T.~Q. An, Shock wave loading and spallation of
  copper bicrystals with asymmetric ${\Sigma}$3 $<$110$>$ tilt grain
  boundaries, Jnl. Appl Phys. 108 (2010) 093526.
\newblock \href {http://dx.doi.org/10.1063/1.3506707}
  {\path{doi:10.1063/1.3506707}}.

\bibitem{Han_spall_shockCuBiC}
W.~Z. Han, Q.~An, S.~N. Luo, T.~C. Germann, D.~L. Tonks, W.~A. Goddard,
  Deformation and spallation of shocked {C}u bicrystals with ${\Sigma}$ 3
  coherent and symmetric incoherent twin boundaries, Phys Rev B 85 (2012)
  024107.
\newblock \href {http://dx.doi.org/10.1103/Physrevb.85.024107}
  {\path{doi:10.1103/Physrevb.85.024107}}.

\bibitem{Fensin_GBonplasticdefor}
S.~J. Fensin, S.~M. Valone, E.~K. Cerreta, et.al., Effect of grain boundary
  structure on plastic deformation during shock compression using molecular
  dynamics, Model Simul Mater Sci. 21 (2013) 015011.
\newblock \href {http://dx.doi.org/10.1088/0965-0393/21/1/015011}
  {\path{doi:10.1088/0965-0393/21/1/015011}}.

\bibitem{ZHOU2014116}
Y.~Zhou, Z.~Yang, Z.~Lu, Dynamic crack propagation in copper bicrystals grain
  boundary by atomistic simulation, Materials Science and Engineering: A 599
  (2014) 116--124.

\bibitem{Barrles_MSEA}
L.~A. Barrales-Mora, Y.~Tokuda, D.~A. Molodov, S.~Tsurekawa, On incipient
  plasticity in the vicinity of grain boundaries in aluminum bicrystals:
  Experimental and simulation nanoindentation study, Materials Science and
  Engineering: A 828 (2021) 142100.
\newblock \href {http://dx.doi.org/10.1016/j.msea.2021.142100}
  {\path{doi:10.1016/j.msea.2021.142100}}.

\bibitem{wwpang2014}
W.~W. Pang, P.~Zhang, G.~Zhang, A.~Xu, X.~Zhao, Dislocation creation and void
  nucleation in {FCC} ductile metals under tensile loading: A general
  microsopic picture, Scientific Reports 4 (2014) 6981.

\bibitem{HVSwygenhoven2005}
H.~V. Swygenhoven, P.~M. Derlet, A.~G. Froseth, Nucleation and propagation of
  dislocatiosn in nanocrystalline fcc metals, Acta Materialia 54 (2005)
  1975--1983.

\bibitem{DESpearot2005}
D.~E. Spearot, K.~L. Jacob, D.~L. McDowell, Nucleation of dislocations from
  [001] bicrystal interfaces in aluminum, Acta Materialia 53 (2005) 3579--3589.

\bibitem{VDremov_2006}
V.~Dremov, A.~Petrovtsev, P.~Sapozhnikov, M.~Smirnova, Molecular dynamics
  simulations of the initial stages of spall in nanocrystalline copper, Phy.
  Rev. B 74 (2006) 144110.

\bibitem{lammps}
S.~Plimpton, Fast parallel algorithms for short range molecular dynamics,
  Journal of Computational Physics 117 (1995) 1--19.

\bibitem{rawat_thesis}
S.~Rawat,
  \href{http://www.hbni.ac.in/phdthesis/phys/PHYS01200704031.pdf}{Behavior of
  solids under high strain rate deformation}, Ph.D. thesis, (2012) pp 32-38
  chapter 3.3; {URL} active as on 30th Mar 2023.
\newline\urlprefix\url{http://www.hbni.ac.in/phdthesis/phys/PHYS01200704031.pdf}

\bibitem{lammpsBenchmarkpg}
Lammps benchmarks, \url{https://www.lammps.org/bench.html}, {URL} active as on
  30th Mar 2023.

\bibitem{CuNi_eam}
S.~M. Foiles, M.~I. Baskes, M.~S. Daw, Embedded-atom-method functions for the
  fcc metals {Cu, Ag, Au, Ni, Pd, Pt} and their alloys, Physical Review B 33
  (1986) 7983.

\bibitem{potential_website}
Interatomic potentials overview, \url{https://www.ctcms.nist.gov/potentials},
  {URL} active as on 30th Mar 2023.
\newblock \href {http://dx.doi.org/10.18434/m37} {\path{doi:10.18434/m37}}.

\bibitem{Michin_Al99.eam.alloy}
Y.~Mishin, D.~Farkas, M.~J. Mehl, D.~A. Papaconstantopoulos, Interatomic
  potentials for monoatomic metals from experimental data and ab initio
  calculations, Physical Review B 59~(5) (1999) 3393--3407.

\bibitem{Zhou_MoEAM}
X.~W. Zhou, R.~A. Johnson, H.~N.~G. Wadley, {Misfit-energy-increasing
  dislocations in vapor-deposited CoFe/NiFe multilayers}, {Physical Review B}
  69~(14) (2004) 144113.
\newblock \href {http://dx.doi.org/{10.1103/physrevb.69.144113}}
  {\path{doi:{10.1103/physrevb.69.144113}}}.

\bibitem{sspKittel}
C.~Kittel, Introduction to Solid State Physics, 8th Edition, John Wiley \&
  Sons, 2005.

\bibitem{mechEnggHBDanBManghitu}
D.~B. Marghitu, {Mechanical Engineer's handbook}, Academic Press, 2001.

\bibitem{vacancyFormEnrAl}
{P. Tzanetakis and J. Hillairet and G. Revel}, {The Formation Energy of
  Vacancies in Aluminium and Magnesium}, physica status solidi B 75~(2) (1976)
  433--439.

\bibitem{MCGERVEY197353}
J.~McGervey, W.~Triftshuser, Vacancy-formation energies in copper and silver
  from positron annihilation, Physics Letters A 44~(1) (1973) 53--54.
\newblock \href
  {http://dx.doi.org/https://doi.org/10.1016/0375-9601(73)90957-2}
  {\path{doi:https://doi.org/10.1016/0375-9601(73)90957-2}}.

\bibitem{PhysRevB.80.224104}
T.~R. Mattsson, N.~Sandberg, R.~Armiento, A.~E. Mattsson, Quantifying the
  anomalous self-diffusion in molybdenum with first-principles simulations,
  Phys. Rev. B 80~(22) (2009) 224104.
\newblock \href {http://dx.doi.org/10.1103/PhysRevB.80.224104}
  {\path{doi:10.1103/PhysRevB.80.224104}}.

\bibitem{WOLFF19974759}
J.~Wolff, M.~Franz, J.-E. Kluin, D.~Schmid, {Vacancy formation in nickel and
  $\alpha$-nickel-carbon alloy}, Acta Materialia 45~(11) (1997) 4759--4764.
\newblock \href
  {http://dx.doi.org/https://doi.org/10.1016/S1359-6454(97)00112-2}
  {\path{doi:https://doi.org/10.1016/S1359-6454(97)00112-2}}.

\bibitem{Al_eam}
K.~W. Jacobsen, J.~K. Norskov, M.~J. Puska, Interatomic interactions in the
  effective-medium theory, Physical Review B 35 (1987) 7423.

\bibitem{KaiNotesPkg2timestepMD}
K.~Nordlund, Basics of md, initialization, time step choice, speedup methods,
  \url{http://beam.helsinki.fi/~knordlun/atomistiset/lecture2.ps.gz}, {URL}
  active as on 30th Mar 2023 (2000).

\bibitem{Yshi_PSO}
Y.~Shi, R.~Eberhart, A modified particle swarm optimizer, in: Proc. of IEEE
  international conference on evolutionary computation. IEEE world congress on
  computational intelligence (Cat. No. 98TH8360), IEEE, 1998, pp. 69--73.

\bibitem{book_optimisation}
X.~S. Yang, Test problems in optimization, John Wiley \& Sons, 2010.

\bibitem{wilkinsMethodsInCompPhys}
M.~L. Wilkins, Calculation of elastic-plastic flow, in: B.~Alder (Ed.), Methods
  in Computational Physics, Vol.~3, 1964, pp. 211--263.

\bibitem{SG_strength}
D.~J. Steinberg, S.~G. Cochran, M.~W. Guinan, A constitutive model for metals
  applicable at high-strain rate, Journal of Applied Physics 51~(3) (1980)
  1498--1504.

\bibitem{STEINBERG1987603}
D.~J. Steinberg, Constitutive model used in computer simulation of
  time-resolved, shock-wave data, International Journal of Impact Engineering
  5~(1) (1987) 603--611, hypervelocity Impact Proceedings of the 1986
  Symposium.
\newblock \href
  {http://dx.doi.org/https://doi.org/10.1016/0734-743X(87)90075-3}
  {\path{doi:https://doi.org/10.1016/0734-743X(87)90075-3}}.

\bibitem{mcqueen1970equation}
R.~G. McQueen, S.~P. Marsh, J.~W. Taylor, J.~N. Fritz, W.~J. Carter, The
  equation of state of solids from shock wave studies, High velocity impact
  phenomena 293 (1970) 294--417.

\bibitem{lasl_shockdata}
S.~P. Marsh, {LASL} Shock {H}ugoniot Data, University of california press,
  Berkeley, 1980.

\bibitem{zel2002physics}
Y.~Zel'dovich, Y.~Raizer,
  \href{https://books.google.co.in/books?id=zVf27TMNdToC}{{Physics of Shock
  Waves and High-Temperature Hydrodynamic Phenomena}}, Dover Books on Physics,
  Dover Publications, New York, 2002.
\newline\urlprefix\url{https://books.google.co.in/books?id=zVf27TMNdToC}

\bibitem{alcuta_usup_Mitchell}
A.~C. Mitchell, W.~J. Nellis, Shock compression of aluminum, copper, and
  tantalum, Journal of Applied Physics 52~(5) (1981) 3363--3374.

\bibitem{alcu_high_prs_eos}
A.~D. Chijioke, W.~J. Nellis, I.~F. Silvera, High-pressure equations of state
  of {Al, Cu, Ta and W}, Journal of Applied Physics 98 (2005) 073526.

\bibitem{Martin}
M.~O. Steinhauser, S.~Hiermaier, A review of computational methods in materials
  science: Examples from shock-wave and polymer physics, Int. J. Mol. Sci.
  10~(12) (2009) 5135--5216.

\bibitem{al_md_shock_melting}
Y.~Y. Ju, Q.~M. Zhang, Z.~Z. Gong, G.~F. Ji, L.~Zhou, Molecular dynamics
  simulation of shock melting of aluminum single crystal, Journal of Applied
  Physics 114 (2013) 093507.

\bibitem{md_shockcompress_metals_confProc}
A.~A. Selezenev, V.~K. Golubev, A.~Y. Aleinikov, O.~I. Butney, R.~A. Barabanov,
  B.~L. Voronin, Molecular dynamics simulation of shock wave compression of
  metals, AIP Conference Proceedings 620 (2002) 374--377.

\bibitem{eos_alcu_etc}
A.~C. Mitchell, W.~J. Nellis, J.~A. Moriarty, R.~A. Heinle, N.~C. Holmes, R.~E.
  Tipton, G.~W. Repp, Equation of state of {Al, Cu, Mo, and Pb} at shock
  pressures up to 2.4 {TP}a (24 {M}bar), Journal of Applied Physics 69 (1991)
  2981--2986.

\bibitem{eos1stprins_Joshi}
K.~D. Joshi, S.~C. Gupta, S.~Banerjee, Shock {H}ugoniot of osmium up to 800
  {GP}a from first principles calculations, J. Phys. Condens. Matter 21 (2009)
  415402(1--6).

\bibitem{vinayak_eos_papers}
J.~Gyanchandani, V.~Mishra, G.~K. D. S.~K. Sikka, Super heavy element
  copernicium: Cohesive and electronic properties revisited, Solid State
  Communications 269 (2018) 16.

\bibitem{velocity_error_estimation}
J.~R. Taylor, An introduction to error analysis, 2nd Edition, University
  Science Books, Sausalito, California, 1997, Ch.~3, pp. 51--53.

\bibitem{Cu_vol_prs_phyRewB70}
A.~Dewaele, P.~Loubeyre, M.~Mezouar, Equations of state of six metals above 94
  {GP}a, Physical Review B 70 (2004) 094112.

\bibitem{Ni_vol_prs_phyRewB78}
A.~Dewaele, M.~Torrent, P.~Loubeyre, M.~Mezouar, Compression curves of
  transition metals in the mbar range: Experiments and projector augmented-wave
  calculations, Physical Review B 78 (2008) 104102.

\bibitem{Cu_vol_prs_phyRewlet}
Y.~Akahama, M.~Nishimura, K.~Kinoshita, H.~Kawamura, Y.~Ohishi, Evidence of a
  fcc-hcp transition in aluminum at multimegabar pressure, Physical Review
  Letters 96 (2006) 045505.

\bibitem{ziegler1985stopping}
J.~F. Ziegler, J.~P. Biersack, The stopping and range of ions in matter, in
  {T}reatise on {H}eavy-{I}on {S}cience, Springer (1985) 93--129.

\bibitem{Belashchenko_EAM}
D.~K. Belashchenko, Computer simulation of nickel and the account for electron
  contributions in the molecular dynamics method, High Temp. 58~(1) (2020)
  64--77.

\bibitem{dolgoborodov_nNi_porous}
A.~Y. Dolgoborodov, S.~Y. Ananev, V.~V. Yakushev, Shock {H}ugoniot of porous
  nanosized nickel, J. Appl. Phys. 131 (2022) 125902.

\bibitem{SAK_high_porosityNi}
S.~A. Kinelovskii, K.~K. Maevskii, Estimation of the thermodynamic parameters
  of a shock-wave action on high-porosity heterogeneous materials, Technical
  Phys 61~(8) (2016) 1244--1249.

\bibitem{shock_repository_rusbank_Online}
The international shock-wave database ({ISW}db), { URL} active as on 30th Mar
  2023.

\bibitem{intro_hydro_zukas}
J.~A. Zukas, Introduction to Hydrocodes, Elsevier, Amsterdam, 2004.

\bibitem{eliezer}
S.~Eliezer, An Introduction to Equations of State: Theory and Applications,
  Cambridge University Press, Cambridge, 1986.

\bibitem{condmat4030071}
S.~V.~G. Menon, B.~Nayak, An equation of state for metals at high temperature
  and pressure in compressed and expanded volume regions, Condensed Matter
  4~(3) (2019) article no. 71.
\newblock \href {http://dx.doi.org/10.3390/condmat4030071}
  {\path{doi:10.3390/condmat4030071}}.

\bibitem{chen_asay}
X.~Chen, J.~R. Asay, S.~K. Dwivedi, D.~P. Field, Spall behavior of aluminum
  with varying microstructures, Journal of Applied Physics 99 (2006) 023528.
\newblock \href {http://dx.doi.org/10.1063/1.2165409}
  {\path{doi:10.1063/1.2165409}}.

\bibitem{Fellinger_NbEAM}
M.~R. Fellinger, H.~Park, J.~W. Wilkins, Force-matched embedded-atom method
  potential for niobium, Physical Review B 81~(14) (2010) 144119.
\newblock \href {http://dx.doi.org/10.1103/physrevb.81.144119}
  {\path{doi:10.1103/physrevb.81.144119}}.

\bibitem{void_vol_harsh}
H.~Hemani, M.~Warrier, N.~Sakthivel, S.~Chaturvedi, Voxel based parallel post
  processor for void nucleation and growth analysis of atomistic simulations of
  material fracture, Journal of Molecular Graphics and Modeling 50 (2014)
  134--141.

\bibitem{zhernokletovBook}
M.~V. Zhernokletov, B.~L. Glushak, Material Properties under Intensive Dynamic
  Loading, Springer, 2006.

\bibitem{ImplastMatPr12023}
S.~Madhavan, P.~V. Laxminarayana, M.~Warrier, Spall fracture in aluminum
  bicrystals: Molecular dynamics study, Materials Today: Proc. 87~(1) (2023)
  164--169.
\newblock \href {http://dx.doi.org/10.1016/j.matpr.2023.03.281}
  {\path{doi:10.1016/j.matpr.2023.03.281}}.

\bibitem{tschopp1}
M.~A. Tschopp, S.~P. Coleman, D.~L. McDowell, Symmetric and asymmetric tilt
  grain boundary structure and energy in {C}u and {A}l (and transferability to
  other fcc metals), Integrating Materials and Manufacturing Innovation 4
  (2015) 176--189.
\newblock \href {http://dx.doi.org/10.1186/s40192-015-0040-1}
  {\path{doi:10.1186/s40192-015-0040-1}}.

\bibitem{TiGBEtschopp2}
M.~A. Tschopp, D.~L. Mcdowell, Asymmetric tilt grain boundary structure and
  energy in copper and aluminium, Philosophical Magazine 87(25) (2007)
  3871--3892.

\bibitem{al_twist_GBE_QingYin}
Q.~Yin, Z.~Wang, R.~Mishra, Z.~Xi, Atomic simulations of twist grain boundary
  structures and deformation behaviors in aluminum, AIP Advances 7 (2017)
  015040.

\bibitem{JamesASmith2016}
J.~Smith, J.~Lacy, D.~Levesque, J.~Monchalin, M.~Lord, {Use of the Hugoniot
  elastic limit in laser shockwave experiments to relate velocity
  measurements}, Vol. 1706, AIP Conf. Proc., 2016, p. 080005.
\newblock \href {http://dx.doi.org/10.1063/1.4940537}
  {\path{doi:10.1063/1.4940537}}.

\bibitem{LI201951}
W.~Li, E.~N. Hahn, X.~Yao, T.~Germann, X.~Zhang, Shock induced damage and
  fracture in {SiC} at elevated temperature and high strain rate, Acta
  Materialia 167 (2019) 51--70.
\newblock \href
  {http://dx.doi.org/https://doi.org/10.1016/j.actamat.2018.12.035}
  {\path{doi:https://doi.org/10.1016/j.actamat.2018.12.035}}.

\bibitem{GNUplot}
T.~Williams, C.~Kelley, et.al., Gnuplot 4.6: An interactive plotting program
  {(April 2013)}, \url{http://gnuplot.sourceforge.net}, {URL} active as on 30th
  Mar 2023.

\bibitem{Stukowski_2009}
A.~Stukowski, Visualization and analysis of atomistic simulation data with
  {OVITO}{\textendash}the open visualization tool, Modelling and Simulation in
  Materials Science and Engineering 18~(1) (2009) 015012.

\bibitem{ovito_dislocation}
A.~Stukowski, V.~V. Bulatov, A.~Arsenlis, Automated identification and indexing
  of dislocations in crystal interfaces, Modelling and Simulation in Materials
  Science and Engineering (2012) 085007.

\bibitem{SANSOZ20051931}
F.~Sansoz, J.~F. Molinari, Mechanical behavior of ${\Sigma}$ tilt grain
  boundaries in nanoscale {C}u and {A}l: A quasicontinuum study, Acta
  Materialia 53~(7) (2005) 1931--1944.

\bibitem{owen_expt}
C.~D. Owen, D.~J. Chapman, G.~Whiteman, S.~M. Millett, S.~Johnson, Spall
  behavior of single crytal aluminum at three principal orientations, Journal
  of Applied Physics 122 (2017) 155102.

\bibitem{zerilliArm}
F.~J. Zerilli, R.~W. Armstrong, Dislocation mechanics based constitutive
  relations for material dynamics calculations, Journal of Applied Physics 61
  (1987) 1816--1825.
\newblock \href {http://dx.doi.org/doi.org/10.1063/1.338024}
  {\path{doi:doi.org/10.1063/1.338024}}.

\bibitem{WangNiAlPotParam2018}
G.~Wang, Y.~Xu, {Embedded-atom potential for Ni-Al alloy}, Vol. 452, IOP Conf.
  Series: Materials Science and Engineering, 2018, p. 022025.
\newblock \href {http://dx.doi.org/10.1088/1757-899X/452/2/022025}
  {\path{doi:10.1088/1757-899X/452/2/022025}}.

\end{thebibliography}

\clearpage
\setcounter{page}{86}	% with twoside pages
\appendices

\chapter{Parameters of EAM potentials used in this work\label{appendixEAMparameters}}
\singlespacing
%\chapter{Parameters of EAM potentials used in this work: \label{appendixEAMparameters}}
%of Al, Cu, Mo \& Ni \label{appendixEAMparameters}}
EAM potential can be expressed in parametric functional forms. This part lists the parametric
functions and the values of parameters involved in them. The EAM potential
function used in this work have different functional forms for,
\begin{enumerate}
\item{Single crystal \ac{Cu}, \ac{Al} and \ac{Ni} in shock Hugoniot calculation.}
\item{Single crystal \ac{Cu}, \ac{Al} and \ac{Mo} in multiscale spall calculation.}
\item{Bi-crystals of \ac{Al} in both shock Hugoniot and multiscale spall calculation.}
\end{enumerate}
For aluminum \ac{Al}, the EAM potential parameters of Jacobsen's is used for single
crystal (shock and spall study) and Mishin's EAM potential is used for bi-crystal
aluminum (shock and spall study). Jacobsen's potential data is available 
in LAMMPS~\cite{lammps} distribution (Aug. 2018), whose coefficients
are not explcitely given therein. This appendix lists the parameters of,
\begin{itemize}
\item Foiles' EAM for copper and nickel
in section~\ref{parametersCuNiEAM} page~\pageref{parametersCuNiEAM} and
\item Zhou's EAM for molybdenum
in section~\ref{parametersMoEAM} page~\pageref{parametersMoEAM}.
\item Mishin's EAM for aluminum
in section~\ref{parametersAlMishinEAM} page~\pageref{parametersAlMishinEAM}.
\end{itemize}

\section
[EAM Parameters for Copper and Nickel]
{EAM Parameters for Copper and Nickel~\cite{CuNi_eam} \label{parametersCuNiEAM}}
%\vspace{-1.0cm}
The first-principles calculations do give the following important information about the
general behavior of these functions. The embedding energy (defined relative to the free-atom
energy) must go to zero for zero electron density and should have a negative slope and
positive curvature for the background electron densities found in metals. The pair-interaction
term $\phi(r)$ is purely repulsive. This shows that the pair interaction between
two different species can be approximated by the geometric mean of the pair interaction for
the individual species. This observation, along with the term, suggest Coulombic origin of
the pair-interaction writing the pair interaction between atoms of types $i$ and $i$ in terms of
effective charges as,
\begin{equation}
	\phi_{ij}(r)=Z_i(r) Z_j(r)/r \label{appendixCuNifoilesEqnPhi}
\end{equation}

According to Foiles' EAM potential model, the RHS of the equation~\ref{eampotEqn}
(Chapter.~\ref{chapter_compMethods}) can be expressed in terms equation~\ref{appendixCuNifoilesEqnPhi}.
A parametrized form is used for $Z(r)$ given by,
\begin{equation}
	Z(r) = Z_0 ( 1 + \beta R^\nu) e^{-\alpha R}  \label{appendixCuNifoilesEqnZr}
\end{equation}

The electron density term $\rho_j$ in eqn.~\ref{eampotEqn} can be expressed as,
\begin{equation}
	\rho_i(r) = n_s \rho_s(r) + n_d \rho_d(r) \label{appendixCuNifoilesEqnElectronRho}
\end{equation}
where $n_s\ \text{and}\ n_d$ are the number of outer $s$ and $d$ orbit electrons. ${\rho}_s$ and ${\rho}_d$
are the densities associated with $s$ and $d$ wave functions~\cite{CuNi_eam}. There are wave functions
available for various atomic configurations, i.e., different occupations of the $s$ and $d$ orbitals.
The configurations used in the calculation~\cite{CuNi_eam} are indicated with a denotation
A-Conf$^{*}$
in table~\ref{tableRefEAMCuNi}, where the values of the parameters that appear in
equations~\ref{appendixCuNifoilesEqnPhi} and~\ref{appendixCuNifoilesEqnElectronRho} are listed.
\vspace{-0.3cm}
\begin{table}[H] \centering
\caption
[Value of EAM parameters of Cu and Ni]
{Value of the parameters of the EAM potential~\cite{potential_website} for copper and nickel~\cite{CuNi_eam}:
The potential file names are `Cu\_u3.eam' and `Ni\_u3.eam' for copper and nickel respectively that
are available in the LAMMPS~\cite{lammps} distribution dated 22$^{nd}$ August, 2018.}
\vspace{0.3cm}
\begin{tabular}{|c|c|c|c|}
\toprule
Parameters & Cu & Ni & Unit \\
\midrule
$Z_0$  & 11.0   & 10.0 & Coulomb\\
$\alpha$ & 1.7227 & 1.8633 & ${\AA}^{-1}$\\
$\beta$  & 0.1609 & 0.8957 & ${\AA}^{-1}$\\
$\nu$    & 2 & 1 & -- \\
$n_s$  & 1.0 & 1.5166 & -- \\
A-Conf$^{*}$ & $3d^{10}4s^1$ & $3d^{8}4s^2$ & --\\
\bottomrule
\end{tabular}
\label{tableRefEAMCuNi}
\end{table}

\section
[EAM Parameters for Molybdenum]
{EAM Parameters for Molybdenum~\cite{Zhou_MoEAM} \label{parametersMoEAM}}
For a pure element $a$, the EAM potential is composed of
three functions: the pair energy $\phi$, the electron density $\rho$ and
the embedding energy $f$. For an alloy, the EAM potential
contains not only the three functions  $\phi$, $\rho$ and $f$ for each of
the constituent elements, but also the pair energy $\phi_{ij}$ be tween different elements
$i$ and $j$ ($i\ne{j}$).

The RHS of the equation~\ref{eampotEqn} (Chapter.~\ref{chapter_compMethods}) can be expressed in
terms of twenty unknowns parameters given by,
{\small
\begin{equation}
\phi{(r)} =  \frac{A\ exp[-\alpha(r/r_e-1)]}{1+(r/r_e-\kappa)^{20}} -\frac{B\ exp[-\beta(r/r_e-1)]}{1+(r/r_e-\lambda)^{20}} \label{appeqn.Mo1}
\end{equation}
}
\normalfont
where $r_e$ is the equilibrium spacing between nearest neighbors,
$A$, $B$, $\alpha$ and $\beta$ are four adjustable parameters. $\kappa$
and $\lambda$ are two additional parameters for the cutoff. The electron
density function is taken with the same form as the
attractive term in the pair potential with the same values of $\beta$
and $\lambda$, i.e.,
{\small
\begin{equation}
f(r) = \frac{f_e\ exp[-\beta(r/r_e-1)]}{1+(r/r_e-\lambda)^{20}}\label{appeqn.Mo2} \\
\end{equation}
The pair potential between different species $i$ and $j$ is then constructed as,
\begin{equation}
	\phi_{ij}{(r)}= \frac{1}{2}\left[\frac{f_j(r)}{f_i(r)}\phi_{ij}(r) + \frac{f_i(r)}{f_j(r)}\phi_{ij}(r)\right] \label{appeqn.Mo3}
\end{equation}
}
\normalfont

Embedding energy functions that work well over a wide range of electron density
require that three equations be used to separately fit three different electron
density ranges. For a smooth variation of the embedding energy, these equations
are required to match values and slopes at their junctions. These equations are,
{\small
\begin{eqnarray}
f(\rho) & = & \sum_{i=0}^3 f_{ni} \left(\frac{\rho}{\rho_n}-1 \right)^i,\ \ \rho<\rho_n,\ \ \rho_n=0.85 \rho_e \\ \label{appeqn.Mo4}
f(\rho) & = & \sum_{i=0}^3 f_{i}  \left(\frac{\rho}{\rho_n}-1 \right)^i,\ \ \rho_n\leq\rho<\rho_0,\ \ \rho_0=1.15 \rho_e \\ \label{appeqn.Mo5}
f(\rho) & = & f_{e} \left[ 1 - \text{ln} \left(\frac{\rho}{\rho_s}\right)^{\eta} \right] \left(\frac{\rho}{\rho_s}\right)^{\eta},\ \ \rho_0\leq\rho \label{appeqn.Mo6}
\end{eqnarray}
}
\normalfont
%% Rererence:
%10.1103/physrevb.69.144113
%Misfit-energy-increasing dislocations in vapor-deposited CoFeÕNiFe multilayers
%X. W. Zhou, R. A. Johnson, and H. N. G. Wadley
%Physical Review B Vol 69 (2004) 144113

The parameters used for Molydbenum are listed in table~\ref{MoparamTable}.
\vspace{-0.3cm}
\begin{table}[H] \centering 
\caption
[Value of EAM parameters of Mo]
{Value of the 20 parameters of EAM potential~\cite{potential_website} for Molybdenum: Ref.~\cite{Zhou_MoEAM}}
\vspace{0.3cm}
\begin{tabular}{|l|l|l|l|} 
\toprule
$r_e$     = 2.728100    & $f_e$   = 2.723710  & ${\rho}_e$  = 29.354065 & ${\rho}_s$  = 29.354065 \\
$\alpha$  = 8.393531 & $\beta$    = 4.476550  & $A$         = 0.708787  & $B$         = 1.120373 \\
$\kappa$  = 0.137640 & $\lambda$  = 0.275280  & $ F_{n0}$   = -3.692913 & $F_{n1}$    = -0.178812 \\
$F_{n2}$  = 0.380450 & $F_{n3}$   = -3.133650 & $F_{0}$     = 3.71      & $ F_{1}$    = 0 \\
$F_{2}$   = 0.875874 & $F_{3}$    = 0.776222  & $\eta$      = 0.790879  & $F_{e}$     = -3.712093\\
\bottomrule
\end{tabular} 
\label{MoparamTable}
\end{table} 

\section
[EAM Parameters for Aluminum]
%{EAM Parameters for Aluminum~\cite{Michin_Al99.eam.alloy,WangNiAlPotParam2018} \label{parametersAlMishinEAM}}
{EAM Parameters for Aluminum~\cite{Michin_Al99.eam.alloy} \label{parametersAlMishinEAM}}
For this potential, similar to the EAM functional form written for molybdenum in page~\pageref{appeqn.Mo6},
the embedding energy function is expressed as in the equation~\ref{appeqn.Mo6}.
The function $\phi(r)$ is expressed in polynomial, while the atomic electron density functions for the
embedding part is expressed in exponential form.  These forms are given below.
{\small
\begin{eqnarray}
 \phi{(r)} = ( k_1 r^{-12} + k_2 r^{-6} + k_3 r^{-3} + k_4 r^{-2} + k_5 r^{-1} + k_6 + k_7 r + k_8 r^{6}) \Psi\left(\frac{r-r_c}{h_1}\right) \label{appeqn.AlMishin1} \\ 
 f(r) = exp(p_1 + p_2 r + p_3 r^{2} + p_4 r^{3} + p_5 r^{4} + p_6 r^{5}) \Psi\left(\frac{r-r_c}{h_2}\right)  \label{appeqn.AlMishin3}
\end{eqnarray}
}
where $\Psi$ is a cut off function defined as,
\begin{equation}
\Psi(x) = \left\{ \begin{array}{ll} \frac{x^4}{1+x^4}, \quad x < 0. \\ 0 \quad x \geq 0. \end{array} \right.  \label{appeqn.AlMishin2}
\end{equation}
The unknown parameters that appear in equation~\ref{appeqn.AlMishin1}, are $k_i$(i = 1,2,..8), $r_c$ \& $h_1$.
For the equation~\ref{appeqn.AlMishin3}, the unknown parametes are $p_i$(i = 1,2,..6), \& $h_2$.
All these unknown parameters are obtained by fitting. The parameters used in this work for aluminum (Al) are listed in
table~\ref{AlMishinparamTable}.
\vspace{-0.3cm}
\begin{table}[H] \centering 
\caption
[Value of EAM parameters of Al]
{Value of the 19 parameters of EAM potential~\cite{potential_website} for Aluminum: Ref.~\cite{WangNiAlPotParam2018}.
The parameters $F_0,\ k_i\ \& \ p_i$ are in eV, while $h_1,\ h_2\ r\ \&\ r_c$ are in ${\AA}$.
The parameter $\eta$ that appear in equation~\ref{appeqn.Mo6} is dimensionless.}
\vspace{0.3cm}
\begin{tabular}{|l|l|l|l|} 
\toprule
%\multicolumn{4}{c}{The parameters $F_0,\ k_i\ \& \ p_i$ are in eV, while $h_1,\ h_2\ r\ \&\ r_c$ are in ${\AA}$}\\
%\multicolumn{4}{c} {The parameter $\eta$ that appear in equation~\ref{appeqn.Mo6} is dimensionless} \\
%\midrule
$k_1$ = -5.0034 x   $10^3$         & $k_2$ = 9.8949  x $10^2$   & $k_3$ = -6.0747 x $10^2$ & $k_4$ = 4.53630 x  $10^2$ \\
$k_5$ = -1.3564 x   $10^2$         & $k_6$ = 1.8998  x $10^1$   & $k_7$ = -1.05203         & $k_8$ = 2.3724 x  $10^{-6}$ \\
$h_1$ = 1.60313 x   $10^{-1}$      & $r_c$ = 5.97079            & $p_1$ = 1.1100 x $10^2$  & $p_2$ = -1.9558 x $10^{2}$ \\
$p_3$ = 1.28529 x   $10^{2}$       & $p_4$ = -4.1079 x $10^1$   & $p_5$ = 6.30792          & $p_6$ = -3.7433 x $10^{-1}$ \\
$h_2$ = 0.15        & $F_0$ = 2.89 & $\eta$ = 0.39             & \\
%$r_e$     = 2.728100 & $f_e$   = 2.723710  & ${\rho}_e$  = 29.354065 & ${\rho}_s$  = 29.354065 \\
%$\alpha$  = 8.393531 & $\beta$    = 4.476550  & $A$         = 0.708787  & $B$         = 1.120373 \\
%$\kappa$  = 0.137640 & $\lambda$  = 0.275280  & $ F_{n0}$   = -3.692913 & $F_{n1}$    = -0.178812 \\
%$F_{n2}$  = 0.380450 & $F_{n3}$   = -3.133650 & $F_{0}$     = 3.71      & $ F_{1}$    = 0 \\
%$F_{2}$   = 0.875874 & $F_{3}$    = 0.776222  & $\eta$      = 0.790879  & $F_{e}$     = -3.712093 \\
\bottomrule
\end{tabular} 
\label{AlMishinparamTable}
\end{table} 

%{\noindent}The parameters $k_i,\ p_i\ \&\ F_0$ are in eV, while $h_1,\ h_2\ r \&\ r_c$ are in ${\AA}$
%The parameters {\noindent}$k_6,\ p_1\ \&\ F_0$ are in eV. Other $k_i$'s (than $k_6$) have the unit, eV.${\AA}^{-w}$,
%where $w$ is the power of $r$ in respective terms. Other $p_i$'s (than $p_1$) are in ${\AA}^{-q}$, where $q$ is the power of $r$ in respective terms.
%The parameters, $h_1,\ h_2\ \&\  r_c$ are in ${\AA}$, while $\eta$ that appear in equation~\ref{appeqn.Mo6} is dimensionless.

%\clearchapter
%\doublespacing

\chapter{Visual index of published articles \label{Reprints}:}
\vspace{-0.5mm}
\noindent The thumbnails shown below are provided solely as a visual index to the author's publications referenced in this thesis. They are included for identification purposes only; the publisher's official versions should be consulted for the complete articles.\\
\vspace{1.0mm}
%\chapter{First page of Published Research Papers \label{Reprints}}
\vspace{-0.15in}

%{\normalsize
%\noindent \textbf{I. Refereed Journals (Reverse chronological order: latest - oldest):} \label{ReprintsRJ}
%\begin{enumerate}
%\item {On the relationship between shock and particle velocities in single and bicrystal systems of Aluminum: A molecular dynamics study, May 2023} pg.~\pageref{MTP23May-1}
%\item {Spall Fracture in Aluminum Bicrystals : Molecular Dynamics Study, Mar.2023} pg.~\pageref{MTP23-1}
%\item {Dynamic Response of Single Crystal Al, Cu \& Ni Upon Impact: MD
%       and Ab-Initio Calculations, Nov.2022} pg.~\pageref{JDBM22-1}
%\item {Effect of Symmetric Tilt and Twist Grain Boundaries on the Void Nucleation,
%       Growth and Spall in polycrystalline Al : Multiscale modelling, Jun.2022} pg.~\pageref{CMS22-1}
%\end{enumerate}
%
%\vspace{0.2cm}
%\noindent \textbf{II. Proceedings of Conferences \& Symposia: (Reverse chronological order)} \label{ReprintsCS}
%\begin{enumerate}
%\item {Modelling of Shock-Hugoniot for single crystal Ni using Molecular
%       Dynamics: Comparison with ab-initio and Experimental results, Dec.2021} pg.~\pageref{DAESSPS21-1}
%\item {Multiscale Modelling to Estimate Spall Parameters
%       in Metallic Single Crystals, Dec.2018} pg.~\pageref{ICONS18-1}
%\end{enumerate}
%}
%
%%\begin{landscape}
%\Chapter{Visual Index of Published Research Papers}
\label{app:visual_gallery}

\begingroup
\centering

% Define hardcoded label tracking targets to protect your existing page references
\refstepcounter{figure}\label{MTP23May-1}
\refstepcounter{figure}\label{MTP23-1}
\refstepcounter{figure}\label{JDBM22-1}
\refstepcounter{figure}\label{CMS22-1}
\refstepcounter{figure}\label{DAESSPS21-1}
\refstepcounter{figure}\label{ICONS18-1}

%\vspace{1.0cm}

% Row 1: Papers 1 - 3
\begin{minipage}[b]{0.3\textwidth}
  \centering
  \fbox{\includegraphics[bb=0 0 612 792,width=\textwidth,height=6.0cm,keepaspectratio=false]{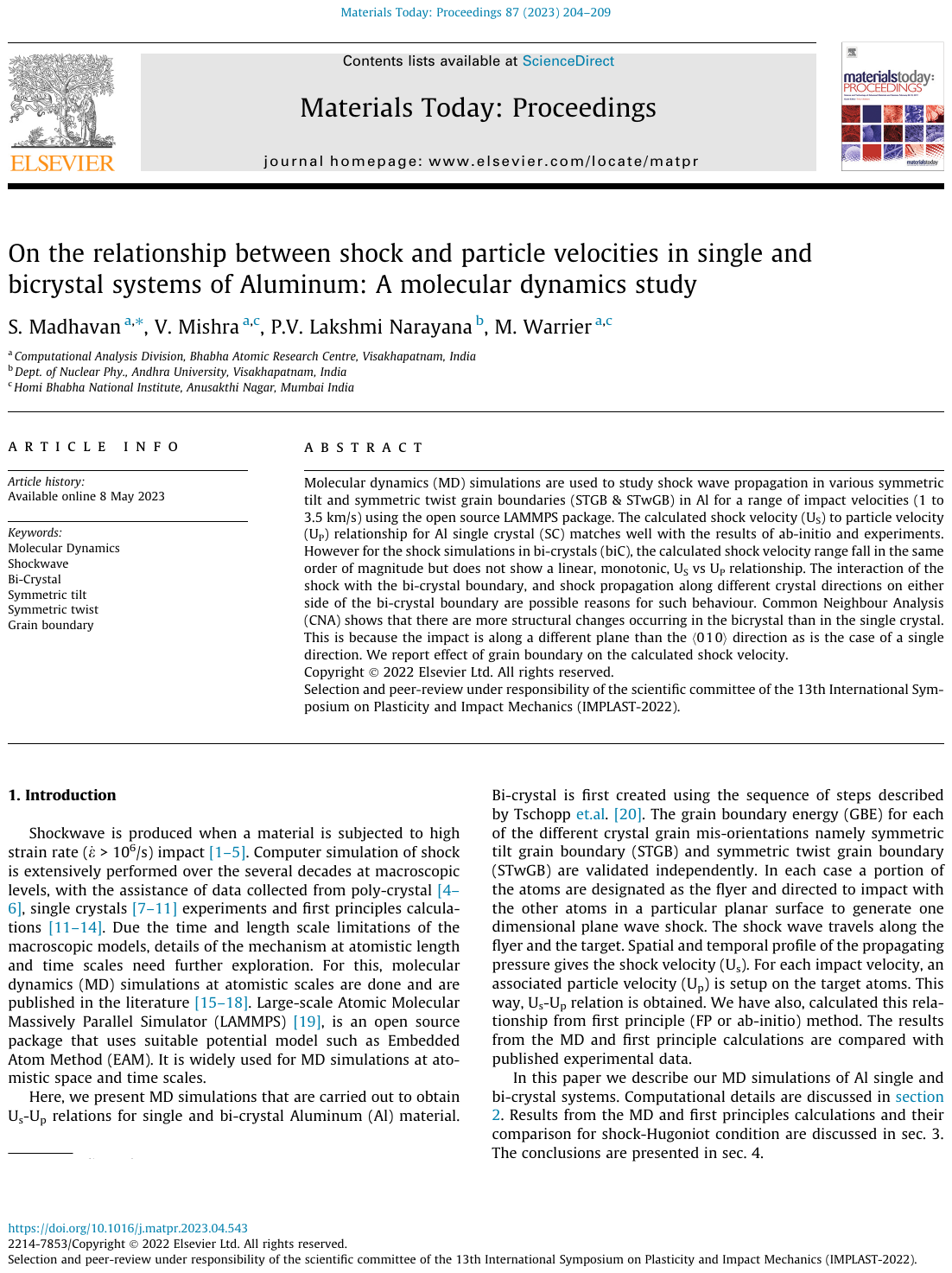}}
  \\ {\footnotesize (a) Mater. Today Proc. (May 2023)}
\end{minipage}
\hfill
\begin{minipage}[b]{0.3\textwidth}
  \centering
  \fbox{\includegraphics[bb=0 0 612 792,width=\textwidth,height=6.0cm,keepaspectratio=false]{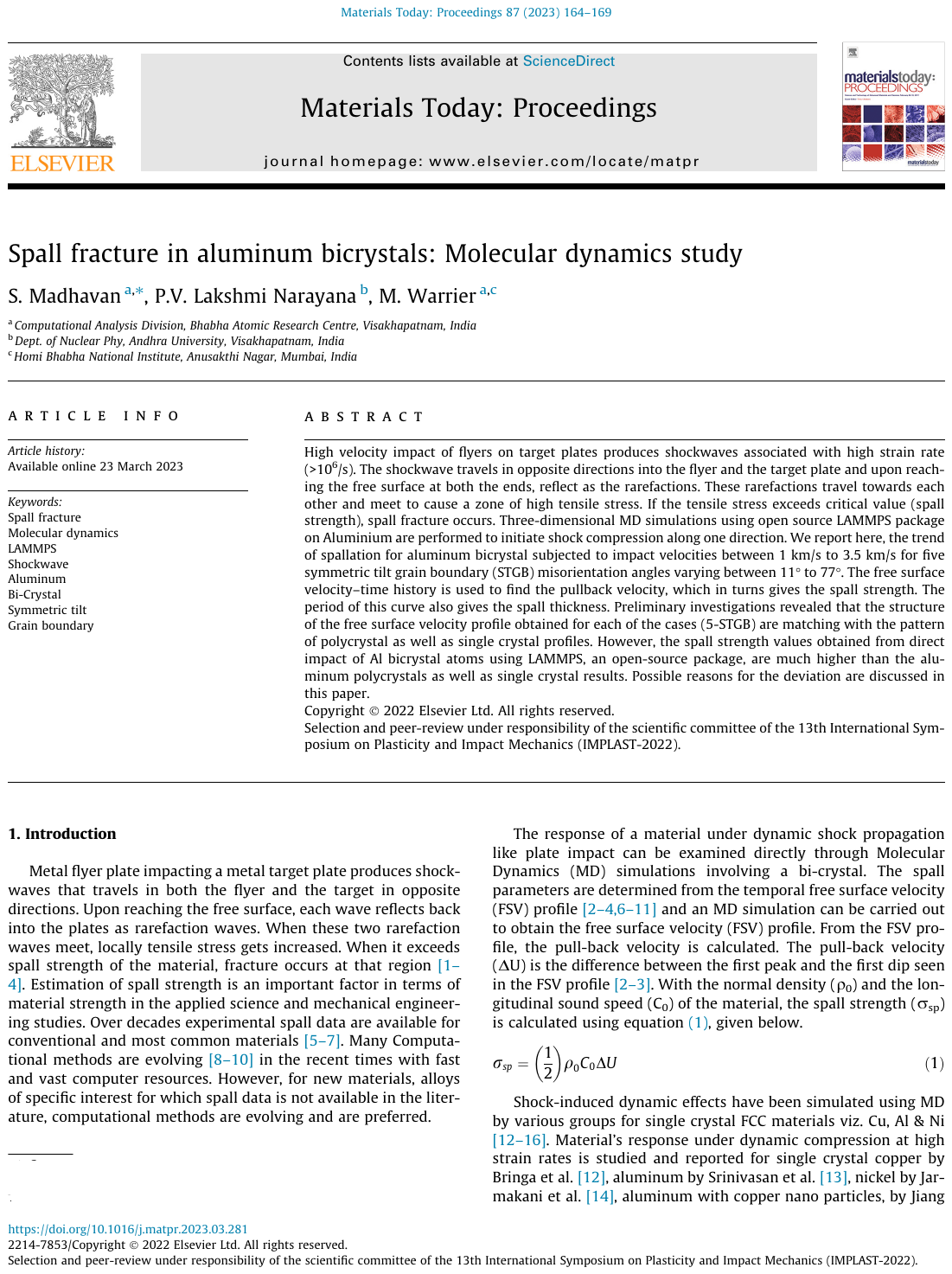}}
  \\ {\footnotesize (b) Mater. Today Proc. (Mar. 2023)}
\end{minipage}
\hfill
\begin{minipage}[b]{0.3\textwidth}
  \centering
  \fbox{\includegraphics[bb=0 0 612 792,width=\textwidth,height=6.0cm,keepaspectratio=false]{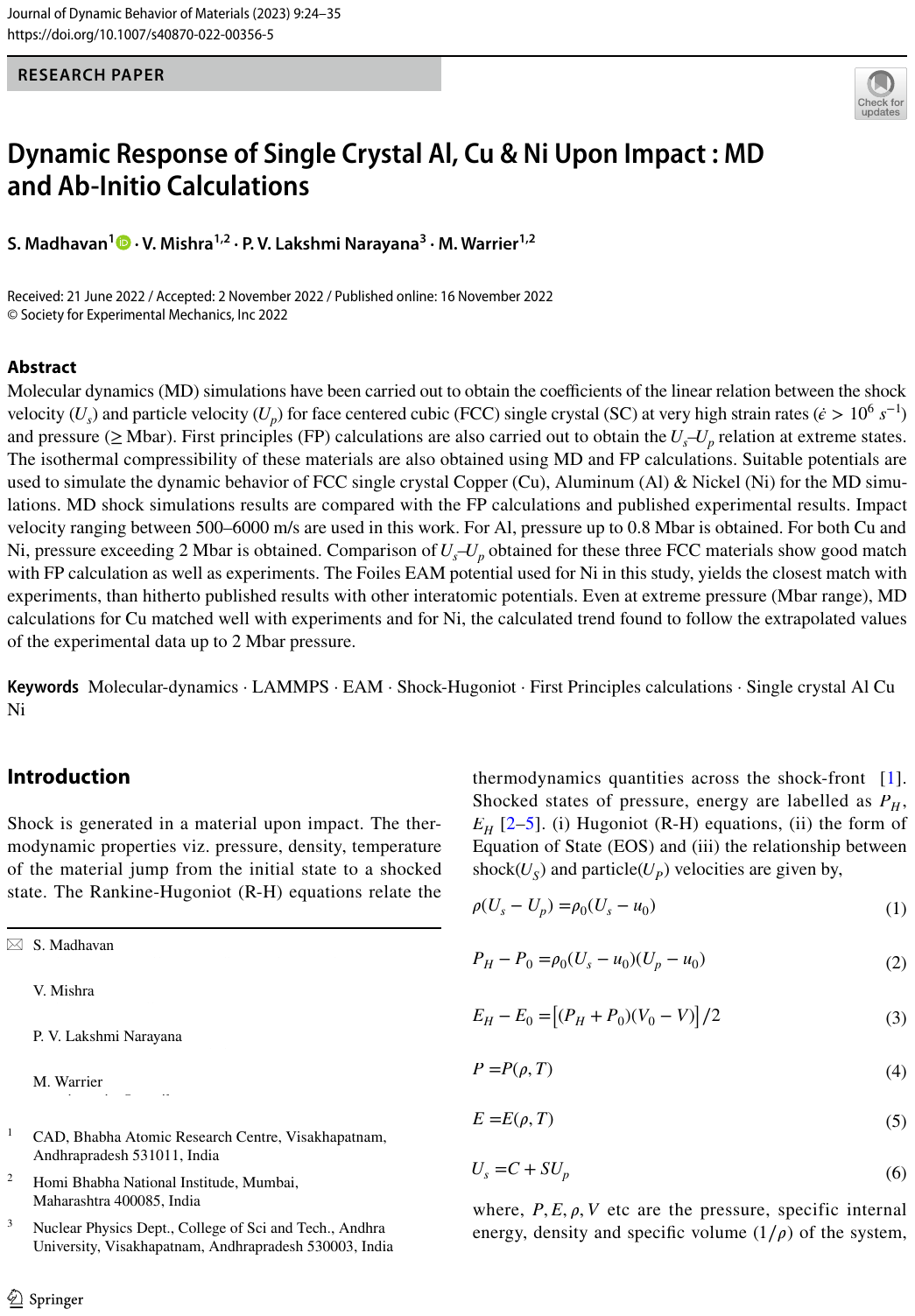}}
  \\ {\footnotesize (c) J. Dyn. Behav. Mater. (2022)}
\end{minipage}

\vspace{1.0cm}
% Row 3: Papers 4 - 6
\begin{minipage}[b]{0.3\textwidth}
  \centering
  \fbox{\includegraphics[bb=0 0 612 792,width=\textwidth,height=6.0cm,keepaspectratio=false]{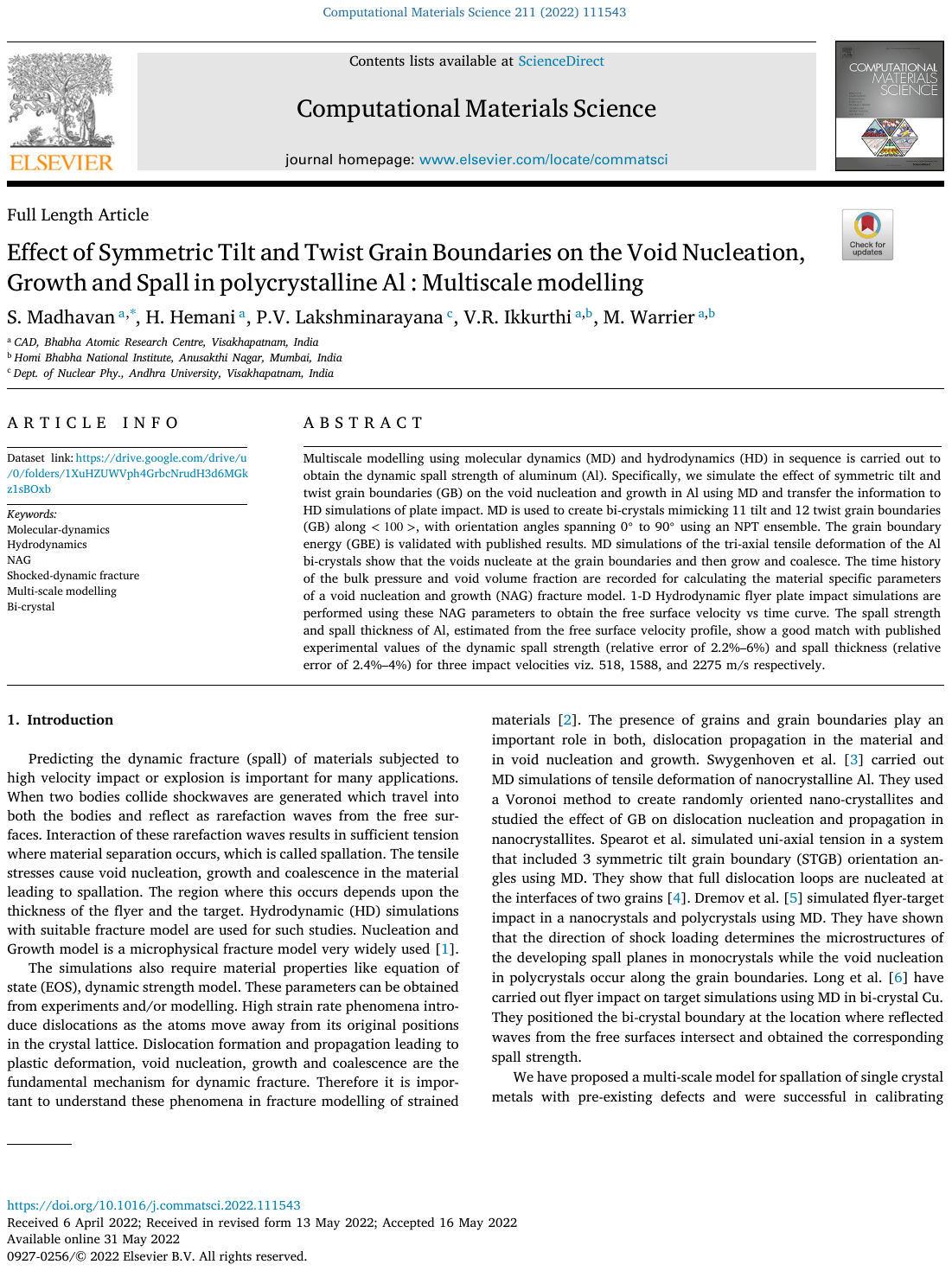}}
  \\ {\footnotesize (d) Comput. Mater. Sci. (Jun. 2022)}
\end{minipage}
\hfill
\begin{minipage}[b]{0.3\textwidth}
  \centering
  \fbox{\includegraphics[bb=0 0 612 792,width=\textwidth,height=6.0cm,keepaspectratio=false]{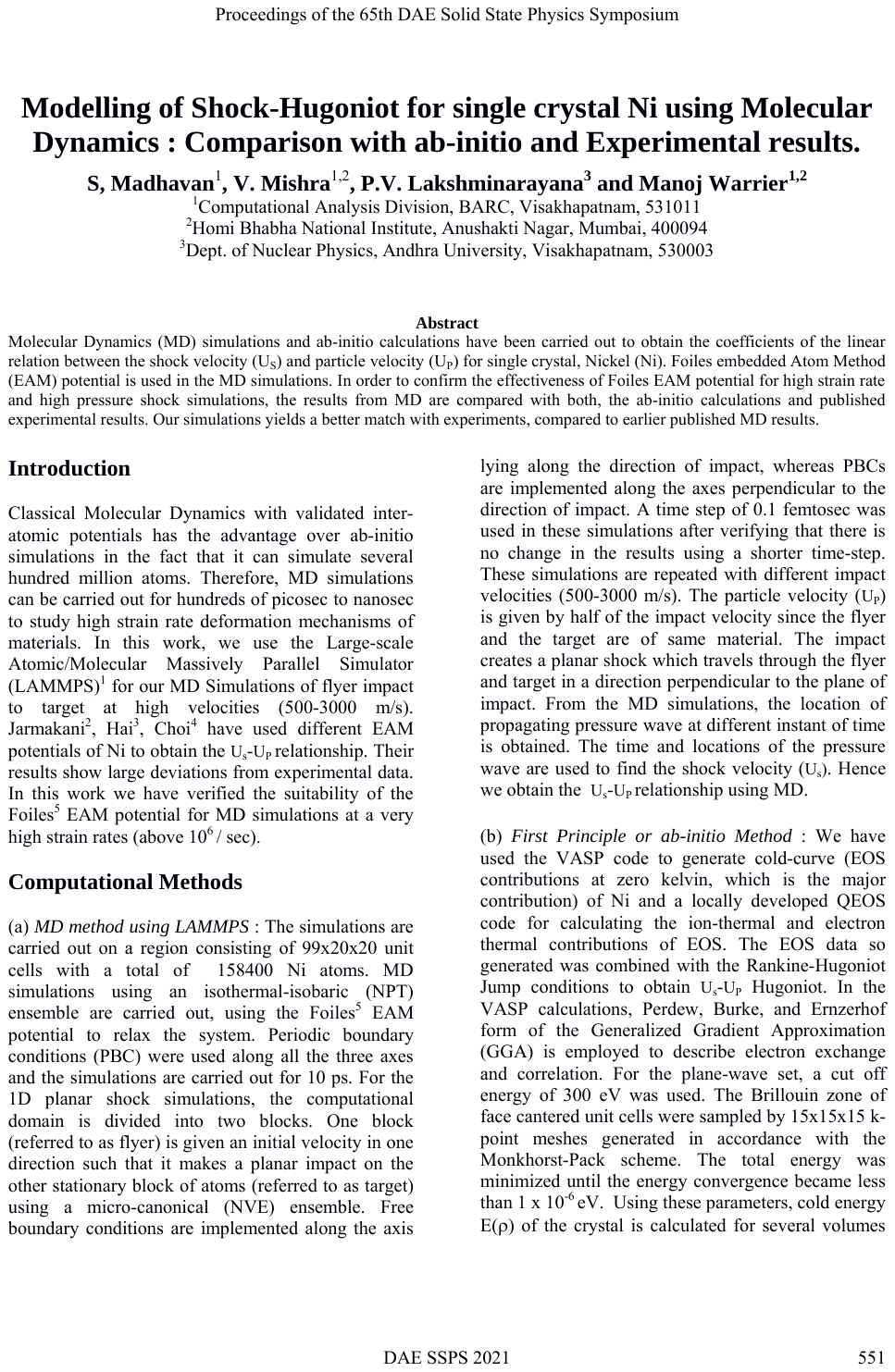}}
  \\ {\footnotesize (e) DAE SSPS (Dec. 2021)}
\end{minipage}
\hfill
\begin{minipage}[b]{0.3\textwidth}
  \centering
  \fbox{\includegraphics[bb=0 0 612 792,width=\textwidth,height=6.0cm,keepaspectratio=false]{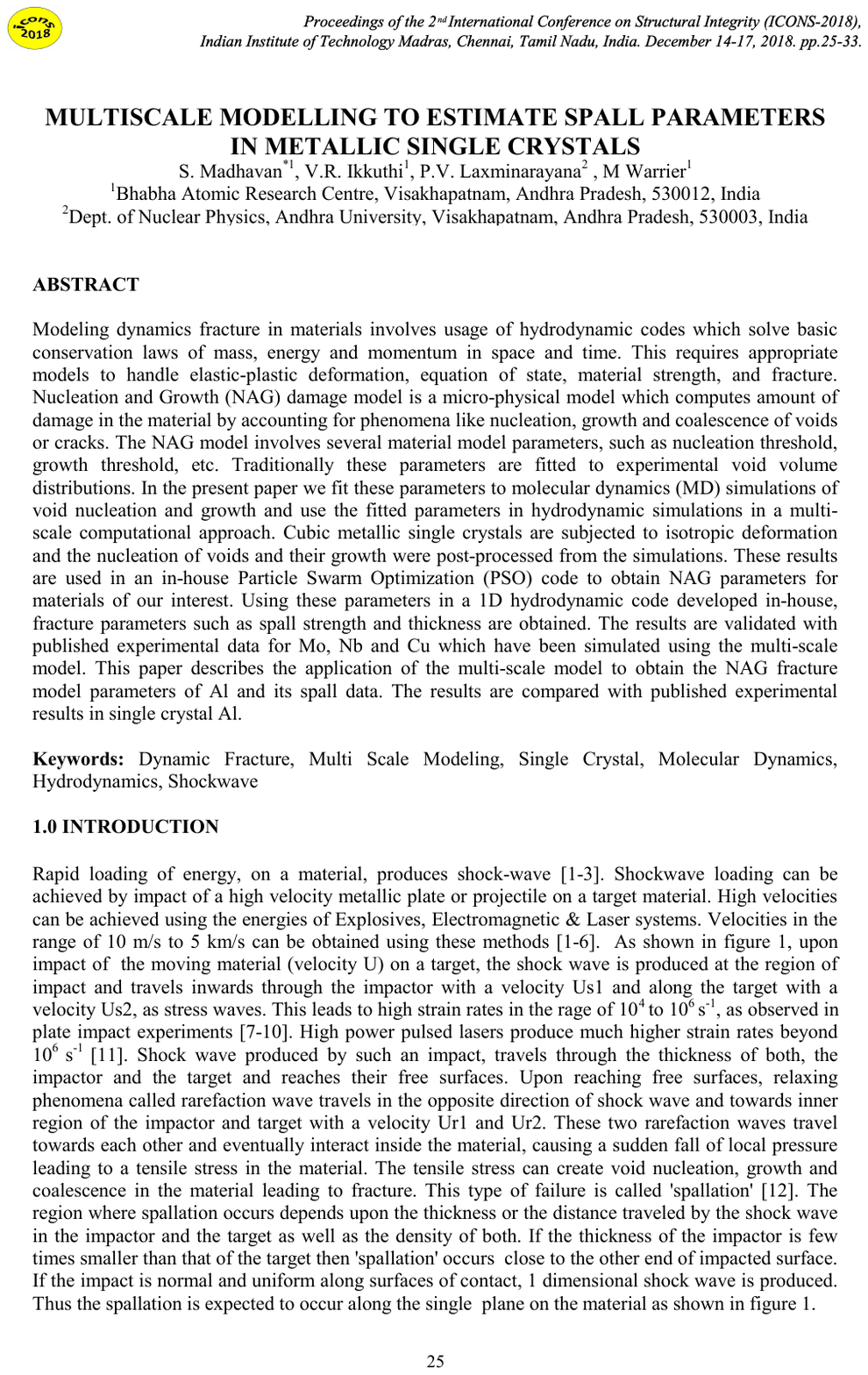}}
  \\ {\footnotesize (f) ICONS (Dec. 2018)}
\end{minipage}
%\end{landscape}

%******************%*

\endgroup

\AddToHookNext{shipout/background}{%
  \begin{tikzpicture}[remember picture,overlay]
  \draw[gray!60,line width=0.4pt]
      ([xshift=25mm,yshift=18mm]current page.south west)
      --
      ([xshift=75mm,yshift=18mm]current page.south west);

  \node[
      anchor=south west,
      text=gray!70,
      font=\footnotesize,
      align=left
  ] at ([xshift=25mm,yshift=8mm]current page.south west)
          {A note on \copyright\ is available in page number~\pageref{copyrightNotePage}.};
  \end{tikzpicture}%
}
%%%%%%%%%%%%%%

\end{document}